\documentclass[phd, titlesmallcaps, 
copyrightpage, foronline ]{mqthesis}

\usepackage{rotating}  
\usepackage{theorem}   
\usepackage{bm}  
\usepackage{amsmath,amsfonts,amssymb}
\usepackage{booktabs}  
\usepackage{pdfsync}   
 \usepackage{color} 

\makeatletter
\@ifundefined{textcolor}{}
{%
 \definecolor{BLACK}{gray}{0}
 \definecolor{WHITE}{gray}{1}
 \definecolor{RED}{rgb}{1,0,0}
 \definecolor{GREEN}{rgb}{0,.4,0}
 \definecolor{BLUE}{rgb}{0,0,1}
 \definecolor{CYAN}{cmyk}{1,0,0,0}
 \definecolor{MAGENTA}{cmyk}{0,1,0,0}
 \definecolor{YELLOW}{cmyk}{0,0,1,0}
 \definecolor{ORANGE}{cmyk}{0,0.45,0.55,0}
 }
 
 \newcommand{\grn}{\protect\color{GREEN}}
\newcommand{\red}{\protect\color{RED}}
\newcommand{\blu}{\protect\color{BLUE}}
\newcommand{\blk}{\protect\color{BLACK}}

\newcommand{\ora}{\protect\color{ORANGE}}

\newcommand{\beq}{\begin{equation}}
\newcommand{\eeq}{\end{equation}}
\newcommand{\ket}[1] {\vert #1 \rangle}
\newcommand{\bra}[1] {\langle #1 \vert}

\newcommand{\9}{\rangle}
\newcommand{\6}{\langle}

\newcommand{\tr}{\mathop{\mathrm{tr}}}

\newcommand{\ba}{\begin{align}}
\newcommand{\eal}{\end{align}}
\newcommand{\ham}{\mathcal{H}}

\newcommand{\hH}{\hat{H}}
\newcommand{\hW}{\hat{W}}
\newcommand{\ha}{\hat{a}}
\newcommand{\hc}{\hat{a}^\dagger}
\newcommand{\hV}{\hat{V}}
\newcommand{\hU}{\hat{U}}
\newcommand{\hX}{\hat{X}}
\newcommand{\hZ}{\hat{Z}}
\newcommand{\hK}{\hat{K}}
\newcommand{\hS}{\hat{S}}
\newcommand{\het}{\hat{\eta}}
\newcommand{\hx}{\hat{x}}

\newcommand{\hA}{\hat{A}}
\newcommand{\hB}{\hat{B}}
\newcommand{\hsi}{\hat{\sigma}}

\newcommand{\hq}{\hat{q}}
\newcommand{\hp}{\hat{p}}
\newcommand{\Om}{\mathbf{\Omega}} 
\newcommand{\Ga}{\mathbf{\Gamma}} 
\newcommand{\Sy}{\mathbf{Y}} 
\newcommand{\mV}{\mathbf{V}} 
\newcommand{\mU}{\mathbf{U}}
\newcommand{\mZ}{\mathbf{Z}} 
\newcommand{\mI}{\mathbf{I}} 
\newcommand{\mAd}{\mathbf{A_{\text{d}}}}

\newcommand{\s}{\text{shad}}

\newcommand{\cs}{\ket{\text{CS}}}
\newcommand{\tcs}{\text{CS}}
\newcommand{\tsc}{\text{SC}}
\newcommand{\cz}{\hat{\text{C}}_{\hat{Z} (j,k)}}

\newcommand{\poly}{\text{poly}}

\def\af{{\sf f}}
\def\ak{{\sf k}}
\def\ab{{\sf b}}
\def\ae{{\sf e}}

\def\ma{\mathcal{M}}
\def\ta{\mathcal{T}}
\def\bma{\mathcal{B}}
\def\RR{\mathbb{R}}

\def\d{\text{d}}

\def\bof{\mathbf{f}}
\def\bk{\mathbf{k}}
\def\hbb{{\hat{\mathbf{b}}}}
\def\hbf{{\hat{\bof}}}
\def\hbk{{\hat{\bk}}}
\def\b{\beta}

\ifpdf
    \pdfcatalog { /PageMode (/UseOutlines) /OpenAction (fitbh)  }
\fi

\begin{document}

\frontmatter

\title{Geometrical and Topological Aspects of Quantum Information Systems}

\ifthenelse{\boolean{foronline}}{
  \author{\href{mailto:tommaso.demarie@gmail.com}{Tommaso Federico Demarie}}
  \department{Physics and Astronomy}
}{
  \author{Tommaso Federico Demarie}
  \department{Physics and Astronomy}
}

\submitdate{December 2013}


\titlepage

\begin{savequote}[8.5cm] 
\sffamily
``Another uninnocent elegant fall into the unmagnificent lives of adults."
\qauthor{The National - Mistaken for Strangers}
\end{savequote}

\chapter{Acknowledgements}

Since the censoring red pen of my supervisor Danny does not have access to the Acknowledgments, here I am allowed to be as poetical and flowery as I like. Ergo forgive me if, more often than I should, I will fall off the cliff of scientific standards.\\

First of all I sincerely need to express my gratitude to all the people who told me that moving to Australia was to stupidest idea ever. I am glad you have been honest, and I am glad I did not listen to you. And thank you Eva for letting me go, after all.\\

The biggest thank you goes to the two people who offered (and gave) me the chance of being here, my supervisors A/Prof. Daniel Terno and A/Prof. Gavin Brennen.\\

Danny, you have been a great boss. Thanks from the bottom of my heart for the support you provided me during these years. Apart from being a passionate and amazing physicist, you pushed me to find my own ways, encouraged me to explore new ideas and gave me space to work on what I liked the most. I am sure that more than once you would have loved to put your hands around my neck and firmly close your fingers, but you never did it. In fact, you have always been ready to help me when in need, to kick me when lazy and to congratulate me those few times I did something good. From you I learnt to love the rigor in working and not to fear taking the wrong way (for a short while) when researching. And if I was the only freak in the office who got excited about foundations, that is definitely your fault. Thanks for everything!\\

Gavin, thanks for being always patient and available to explain again, and again, and again, anything I would not feel confident about. I have always been impressed by the apparently unlimited extension of your knowledge about any concept in Physics. It was easy to see how much passion you put in your work and this motivated me to improve and do my best all the time. Thanks also for sending me around the world. It has been a real privilege working and partying with you!\\

Along the duration of my studies I have met some great people, whom I have been fortunate enough to work with. Let me say few words about them.

To start, I would like to thank Prof. Kwek for inviting me to visit his group in Singapore and taking me to eat any sort of delicious food while discussing great physics. And I am equally grateful to Prof. Murao for inviting me to pay a visit to her group in Tokyo. 

It is also my pleasure to thank Dr. Sai Vinjanampathy for the great time spent together at CQT and in the various Hawker centers of Singapore. Witnessing our discussions shifting from Star Trek to warp drive in general relativity was a clear manifestation of the beauty of being nerds. Equally kind has been Michal Hajdusek, who organized my visit to Tokyo, providing me with financial support: Dude, it was great talking physics, but it was even greater lowering the already pretty low level of the Celt one beer after the other. You guys are legends! 

Nicolas "Doc" Menicucci, you have been an example both at the whiteboard and on the dance floor. Doing calculations via Skype at 4 a.m. with you is something that, one day, I will proudly tell my kids. You taught me the importance of perfection and I will always keep in mind your lesson: It makes no sense to upload it on the \emph{arxiv} if it is not going to be the best paper out there. And needless to say, thanks for sharing your best pick-up lines with me.

A warm thank you to all the admin people who constantly helped me fixing any sort of bureaucratic trouble. Especially Lisa: It is a miracle that you never threw something at me after the $n$-th time I would give you again the wrong module. Also thanks to Michele for her constant (and patient) help with tickets and flights!\\

There are some special people I wish to thank: They are Aharon the Dude, Andrea and Hossein, Cristina culo gordo, Franceschina, Eleonora la Dude, Luchino and Marion, Mauro il Dude, Paolino, His darkness Saatana and Thorn. You have been like brothers and sisters to me. I always knew that, no matter what, with you guys around I would have been safe and sound. Together we shared so much foolery, laughters, sometimes tears, and I am extremely grateful to all of you guys for that.\\
Allow me to be particularly sentimental: Aharon, thanks for your friendship (in the shape of suggestions, bad jokes, physics, coffee, beer and food). It always soothed the mess that distance and uncertainty sometimes caused in my mind. Franceschina, I know that although you expressed disgust any time I hugged you too tight, you liked every bit of it. Thanks for offering me a roof and a cot when I needed it the most! Mauro, it has been phenomenal making this journey together. We have shared too much beer, meat and mead to tell it all here in a few lines. Good luck with the rest of the adventure, and hold me a dry spot under the bridge please.\\

Everybody knows about the tragedy of finding a \emph{proper} house in Australia\footnote{If you do not, well, read \emph{``He died with a falafel in his hand'' by J. Birmingham}.}. However, in these years I have been extremely lucky to share low standards of living and poor hygienic conditions with some amazing characters, who undoubtedly made the whole experience of living in Sydney more special. Therefore, let me thank Safi, Josh, Kamin, Gabriela and especially Marika, who took care of me like a younger brother. And a special tribute goes to Lili, who helped me to overcome my dislike of cats.\\

I should ideally thank  so many more people for too many reasons. Anyone I met in the last years gave me something special and in a way made me the person I am today. However, allow me to name ``few'' of these beings who are particularly important to me: Alika, ChrisMcMahon, Christoph, Doozie, Enzo, Federica, Geraldine, Helena, Ivan, Jacopo, Johann and Elna, Kayla and Xavier, Michael and Louise, Nieke, Nora and Jack, Richard and Marija, Robert, Sukhi, Valentina A, Valentina B, Valentina D, Vikesh, and the rest of the quantum guys. Thanks folks for being part of this adventure.\\

Thanks to my friends in Italy who have never forgotten me while I have been away and have always forgiven my unjustifiable long pauses in replying to emails and phone calls: I hug you all Alberto, Castagna, Chiara, Damiano, Elena B, Elena G, Eleonora, Francesco and Lucia, Gianni, Giulia, the Goitre, Lara, and Stefania.\\

Every time I visit the Bacic family in Brisbane and the Aso family in Tokyo I feel welcomed like a son, and it is an amazing feeling indeed. I thank you for your generosity! Knowing that I have an extended family across the borders of this planet is like sweet popcorn for my soul.\\

Thank you Orly, for both the great and the disastrous time spent together. You taught me the value of honesty and integrity the hard way and I shall not forget it.

It is people like Ai Leen that make the world shivering with hope. Thanks for being so caring, idealistic and never ready to compromise. You are a wonderful small little being, please promise me you will never change, no matter what.

Thanks Sabrina for making these last months here in Sydney special. I could have easily drowned into the deep sea of despair, but you managed to hold me up with a strong grasp and an enchanting smile. Some things happen unexpectedly, and you surely are a surprising but wonderful gift.\\

Last, but certainly not least, I wish to thank my family, for all the support and unconditional love you gave me in these years. My mum, Federica, who won her fears and flew all the way down under to see with her own eyes the places I have been telling her about. My dad, Gianpiero, who will be ecstatic to have my burden back on his shoulders. Nonna Irma, who innocently asked me how long it takes to get to Sydney by train. Thanks also to my brother Alessandro, my aunties Renata and Chiara and my cousins. A thought always goes to my grandparents who are not here anymore: Nonno Beppe, Nonna Nella and Nonno Severino, I know how proud you are that your fool of a grandson made it to the end. I dedicate this work to you.

\begin{savequote}[8cm] 
\sffamily
``Considerate la vostra semenza:\\
fatti non foste a viver come bruti,\\
ma per seguir virtute e canoscenza.''
\qauthor{Dante Alighieri - Inferno, Canto XXVI}
\end{savequote}
~\\
~\\
\chapter{List of Publications}

\begin{itemize}
\item[\cite{brodutch2011}] A. Brodutch, T.F. Demarie and D.R. Terno\\
\emph{Photon polarization and geometric phase in general relativity}.\\
Phys. Rev. D \textbf{84}, 104043 (2011).
\item[\cite{rideout2012}] D. Rideout, T. Jennewein, G. Amelino-Camelia, T.F. Demarie, B.L. Higgins, A. Kempf, A. Kent, R. Laflamme, X. Ma, R.B. Mann, E. Martin-Martinez, N.C. Menicucci, J. Moffat, C. Simon, R. Sorkin, L. Smolin, and D.R. Terno\\
\emph{Fundamental quantum optics experiments conceivable with satellites -- reaching relativistic distances and velocities}.\\
Class. Quantum Grav. \textbf{29}, 224011 (2012).
\item[\cite{demarie2012}] T.F. Demarie\\
\emph{Pedagogical introduction to the entropy of entanglement for Gaussian states} (2012).  (Pre-print arXiv:1209.2748).
\item[\cite{demarie2013}] T.F. Demarie and D.R. Terno\\
\emph{Entropy and entanglement in polymer quantization}.\\
Class. Quantum Grav. {\bf 30}, 135006 (2013).
\item[\cite{demarie2013b}] T.F. Demarie, T.Linjordet, N.C. Menicucci and G.K. Brennen\\
\emph{Detecting Topological Entanglement Entropy in a Lattice of Quantum Harmonic Oscillators} (2013).  (To appear in New J. Phys., Pre-print arXiv:1305.0409).
\end{itemize}

\chapter{Abstract}

Geometry and topology play key roles in the encoding of quantum information in physical systems. Ability to detect and exploit geometrical and topological invariants is particularly useful when dealing with transmission, protection and measurement of the fragile quantum information. In this Thesis, we study quantum information carriers associated with discrete or continuous degrees of freedom that live on various geometries and topologies.

In the first part of this study, intrigued by the possibility of implementing quantum communication protocols in space, we analyze the effects of a gravitational field on the polarization of photons. In this case, we investigate discrete degrees of freedom moving on a continuous space-time: Specifically, we look at the geometrical description of the problem. We find that for closed trajectories, in both static and general space-times, the amount of rotation, or phase, caused by the action of gravity is independent of the reference frame chosen to define the polarization vector. We also prove that similarly to other instances of a geometric phase, its value is given by the integral of the (bundle) curvature over the surface that is encircled by the trajectory.

In the second part we study a new approach to topological quantum information by using Gaussian states to construct a system that exhibits topological order. We describe a (Gaussian) continuous-variable state analog to Kitaev surface codes prepared using quantum harmonic oscillators on a two-dimensional discrete lattice, which has the distinctive property of needing only two-body nearest-neighbor interactions for its creation. We show that although such a model is gapless, it satisfies an area law. Its ground state can be simply prepared by measurements on a finitely squeezed and gapped two-dimensional cluster-state, which does not exhibit topological order. A universal signature of topologically ordered phases is the topological entanglement entropy. Due to low signal to noise ratio it is extremely difficult to observe the topological entanglement entropy in qubit-based systems, and one usually resorts to measuring anyonic statistics of excitations or non-local string operators to reveal the order. We prove that for our continuous-variable model the topological entanglement entropy can be observed simply via quadrature measurements, in contrast with qubit-based systems. This provides a practical path to observe topological order in bosonic systems using current technology.

In the third and last part we study the well-definiteness of the concept of entropy in a scheme alternative to Schr\"odinger quantization: Polymer quantization. The kinematical construction of the Hilbert space in polymer quantization is based on the discretization of the real line, which is treated as a one-dimensional graph with discrete topology. On such a setup, which resembles the more complicated construction of loop quantum gravity, we analyze whether the values of the quantum entropy computed in the Schr\"odinger and polymer quantization coincide or not. We study the convergence of the entropies of physically equivalent states in unitarily inequivalent representations of the Weyl-Heisenberg algebra and derive a general bound to relate the values of entropy.

\begin{savequote}[10cm] 
\sffamily
``Scientific truth is beyond loyalty and disloyalty."
\qauthor{Isaac Asimov - Foundation}
\end{savequote}

\chapter{Contributions}

The author and his collaborators have made the respective contributions to the results presented in this work. They are listed following the structure of the Thesis as presented in the outline.

\subsubsection{Relativistic quantum information}
These results are based on the papers published in Phys. Rev. D \textbf{84}, 104043 (2011) and Class Quantum Grav. \textbf{29}, 224011 (2012). The idea of deriving a gauge invariant phase for closed paths came from A/Prof. Daniel Terno. I did the calculations for the explicit form of the phase. The discussions on quantum experiments in space were initially suggested by A/Prof. Thomas Jennewein and Prof. Raymond Laflamme. A/Prof. Terno and I contributed to section 3 of the review paper, which accounts for the relativistic effects in quantum information theory. We performed the analysis of relative magnitudes of the effects within the parameters of the QEYSSAT mission.

\subsubsection{Gaussian states}
I have written a pedagogical review on Gaussian states, arXiv:1209.2748, where I focus on the entanglement properties of such states, analyzing in particular the form of the covariance matrix and symplectic eigenvalues for a pair of coupled quantum harmonic oscillators. 

\subsubsection{Continuous-variable topological order}
The idea of detecting topological order using continuous-variable systems was the result of fruitful discussions between A/Prof. Gavin Brennen and myself. I proposed to calculate the topological entanglement entropy for a lattice of oscillators using the Gaussian states formalism and Dr. Nicolas Menicucci introduced us to the graphical calculus for Gaussian states. I derived the form of the covariance matrix for the continuous-variable physical surface code state and did all the numerical simulations and graphs presented here. A/Prof. Brennen and I calculated the gap for the surface code state, while A/Prof. Brennen derived the bound for the decay of the correlations.  I proposed to test topological order using the topological logarithmic negativity and did the related calculations. A/Prof. Brennen, Dr. Menicucci and I constructed the noise model and derived the related covariance matrix. Dr. Menicucci and I derived the analytical bound for the topological mutual information, while A/Prof. Brennen suggested how to derive the bound for the topological entanglement entropy and I did the calculations. The experimental proposals at the end of chapter \ref{CVTO} are due to the visionary ideas of Dr. Menicucci and his collaborators. These results are presented in arXiv:1305.0409v3, while the manuscript has been submitted to the New Journal of Physics. 

\subsubsection{Entropy in polymer quantization}
The idea of studying the behavior of entropy in different quantization schemes is due to A/Prof. Terno and was the original project of my PhD. I first proposed to use Gaussian states as an example and did the calculations for the entropy of two coupled oscillators quantized in the polymer scheme, which then helped us to derive the precise entropy bonds. A/Prof. Terno introduced me to the representation theory of canonical commutation relations that were the basis for the proofs of the entropy bounds, which have been derived jointly by the two of us. These results have been published in Class. Quantum Grav. \textbf{30}, 135006 (2013).

\tableofcontents
\listoffigures

\mainmatter

\begin{savequote}[10cm] 
\sffamily
``The world isn't run by weapons anymore, or energy, or money. It's run by little ones and zeroes, little bits of data. It's all just electrons.'' 
\qauthor{Cosmo - Sneakers}
\end{savequote}

\chapter{Introduction}

Quantum mechanics was introduced at the beginning of the last century as a radical new theory, defining the space where physics takes place and the rules that physical systems are required to obey at the very cold and small scales \cite{einstein1906,heisenberg1925,born1926,pauli1926,schroedinger1926,born1926b,heisenberg1927,pauli1927}. This astonishing novel concept, together with the newly born theory of relativity \cite{einstein1907,einstein1911,einstein1912a, einstein1912b,einstein1915}, turned the rational idea of \emph{classical physics} on its head, completely changing our understanding of the laws of physics. This was the first \emph{quantum revolution}. Despite the innumerable experimental confirmations of quantum theory, foundational questions such as the resolution of the measurement problem (the macroscopic transition from quantum to classical) remain \cite{zurek2003}. 

The ``second quantum revolution'' kicked off at the end of the last century when researchers started to wonder wether it was possible to obtain a better understanding of the laws of the quantum world looking at information theory through the lenses of quantum mechanics (and vice-versa). While the seminal proposals for a quantum computer are familiarly attributed to Feynman \cite{feynman1982} and Deutsch \cite{deutsch1985}, it was not until Shor's factorization algorithm \cite{shor1997}, which gave the shivers to every cryptographers on the planet, that these ideas led to a new view of the role played by information in physics, giving birth to the fascinating field of quantum information. As a result, the past 25 years have been a prosperous time for the physicists interested in understanding how to exploit the quantum mechanical properties of nature in order to accomplish computational and operational tasks that would be incredibly onerous in terms of classical physical resources, if not practically impossible \cite{bennett1984, bennett1993, gisin2002, nielsen2000}. Today, in parallel with discoveries in quantum information theory, experimentalists have reached an extraordinary level of control over single and composite quantum systems \cite{stajic2013}, opening up the doors to the world of large scale fully quantum devices and the related applications and implications \cite{kaye2007, nielsen2000}.

In this Thesis, we pursue investigations to some of the open problems in present-day quantum information and quantum computation theory, analyzing geometrical and topological properties of both discrete and continuous quantum information systems. Roughly speaking this work is divided in three parts, preceded by a general introduction to the concepts of quantum mechanics, quantum information, relativity and loop quantum gravity.

The first part, Chapter \ref{GRchapter}, deals with photons in curved space-time \cite{peres2004, terno2005rqi}. While photons can be used as physical realizations for \emph{qubits} \cite{nielsen2000}, the basic elements in quantum information, it is still not entirely understood how the presence of a gravitational field affects the information stored in the polarization of a single photon \cite{brodutch2011}. Taking into account these effects is particularly important when designing schemes of quantum communication between Earth and orbiting satellites \cite{rideout2012}. Moreover, future tests of general relativity involving quantum systems will rely on precise estimations of the relativistic effects on the quantum degrees of freedom \cite{brodutch2011b}. After a brief account of the relativistic description of the system, we will focus on polarization rotation in closed trajectories on a curved space-time, detecting interesting invariants and suggesting possible experimental applications for these effects. 

Chapters \ref{ChapGS} - \ref{CVTO} account for the second part of the Thesis. We first introduce the notions of Gaussian states \cite{demarie2012, olivares2012} and topological order \cite{kitaev2003} and then combine them together into the new concept of continuous-variable topological order \cite{demarie2013b}. In general, topology refers to the ability of relating systems whose shape look different at a first glance. By analyzing common geometrical properties, one extracts quantities that are used to characterize the global properties of otherwise very different-looking systems. Topological quantum computation is the area of quantum information that studies the topological properties of a special class of two-dimensional system, also called \emph{topologically ordered} \cite{pachos2012, hamma2005}. These systems are particularly fascinating because they can support \emph{anyons} \cite{brennen2008}, unusual quasi-particles that are neither bosons nor fermions and whose very non-local properties enable the implementation of fault-tolerant quantum computation via braiding operations \cite{stern2008, nayak2008}. Performing quantum computation using anyons does not require the physical implementation of quantum gates, whose action is now replaced by the spatial exchange of anyons. While very appealing from a theoretical point of view, the experimental realization of topological systems based on qubits (or qudits) is highly challenging, due to the level of control needed to build a lattice of interacting spins \cite{demarie2013b}. 

We present here an alternative approach that allows for a much simpler realization in the laboratory with today's technology. Starting from the toric code state, defined as the ground state of an highly-entangled lattice of qubits, we introduce its continuous-variable surface code state analog, replacing qubits with light field modes, which are described mathematically by Gaussian states. This choice is motivated both by the easy mathematical formalism that characterizes Gaussian states, which allows for a straightforward description of the state in terms of its first and second statistical moments \cite{arvind1995}, and the availability of well-developed experimental techniques \cite{weedbrook2012}. Exploiting the Gaussianity of the state, one can prove that the continuous-variable surface code exhibits topological order by detecting a non-zero topological entanglement entropy, which is the distinctive feature of topological phases \cite{kitaev2006, levin2006}. Remarkably, this can be achieved simply performing measurements of the quadrature operators of the light modes. Furthermore, this system requires only two-body nearest-neighbor interactions for its creation, and hence opens up to scalable implementations using, for instance, quantum optics \cite{ menicucci2006, gu2009} or circuit-QED technology \cite{aolita2011, gerardo2009}. 

Since the identification of topological order in topological quantum computation theory is strictly related to the entropic properties of the system \cite{hamma2005}, in the third and last part of this Thesis, Chapter \ref{polymerico}, there will be space to investigate the well-definiteness of the concept of entropy in different quantization schemes. In particular, we will be dealing with polymer quantization \cite{ashtekar2003}, a representation of the Weyl-Heisenberg group unitarily inequivalent to the standard Schr\"odinger quantization \cite{bratelli1992,halvorson2004}, which was introduced as a toy model to study the kinematic and the dynamic of 1-dimensional quantum systems in analogy with the more complicated setup of Loop Quantum Gravity \cite{ashtekar2001, rovelli2004, thiemann2007}. 

While there are ways to demonstrate that physical observables in Schr\"odinger and polymer quantizations converge to the same value in the appropriate limit \cite{fell1960,haag1964}, analysis of the quantum entropy in the two schemes was missing. Specifically, it was not clear for a quantum state whose observable predictions in the two unitarily inequivalent quantization schemes are close, whether the values of the entropy will also be close or not \cite{demarie2013}.

Here we give an answer to this question, establishing the conditions for entropy convergence. Gaussian states happen to be particularly useful also in this context. By looking at the entropy of two coupled quantum harmonic oscillators, it is relatively easy to construct an example that confirms our findings and allows us to show explicitly how the corrections introduced by the discretization of the physical space affect the value of the entropy.

\section{Outline}
\label{outlinerefine}
Our intent is to give a uniform flow to the discussion, alternating necessary background and new results. This Thesis has three main parts, with four appendices presenting detailed calculations and additional useful theoretical background. Briefly, this is the outline of the work. 
\begin{itemize}
\item[$\bullet$] General introduction to the concepts of the Thesis:
~\\
Chapter \ref{CHintro2} summarizes the essential facts of quantum mechanics, quantum information, relativity and quantum gravity needed for an effective comprehension of this Thesis. We first start with some basic structural ideas of quantum mechanics. Then we review the properties of entropy for classical and quantum systems, and introduce entanglement together with ways to quantify it in pure and mixed states. After introducing the stabilizer formalism, this Chapter is concluded with general concepts of relativistic quantum information and a short section about loop quantum gravity. 
\item[$\bullet$] Part 1 - Geometric phase of photons:
~\\
The first part of the Thesis is based on the results from \cite{brodutch2011} and \cite{rideout2012}. We discuss the gauge-invariance of the polarization phase for photons traveling along closed trajectories in closed paths.
\begin{itemize}
\item[$-$] Chapter \ref{GRchapter}: Here we start explaining the relativistic effects on the states of a photon and their dependence on the choice of a well-suited reference frame. Then we introduce the geometric meaning of polarization phase through the derivation of an alternative equation for the phase in a static space-time, written in terms of the bundle connection. One can then prove that in a closed trajectory the change in polarization depends only on the geometrical properties of the space-time and not on the choice of reference frame. The Chapter is concluded with some remarks about experimental applications of quantum information in space. 
\end{itemize}
\item[$\bullet$] Part 2 - Topological order with Gaussian states:
~\\
The second part is based on \cite{demarie2012, demarie2013b} and analyses the interplay between Gaussian states and topologically ordered phases. 
\begin{itemize}
\item[$-$] Chapter \ref{ChapGS}: In this Chapter we rigorously introduce the mathematical description of Gaussian states, starting with the Hilbert space and phase space representations of such states, and continuing with the state evolution under symplectic transformations. Then a step-by-step derivation of the entanglement entropy formula for pure Gaussian states is presented, together with an example. To conclude, we introduce the graphical calculus for Gaussian states and give some examples of graphical representations of common experimental Gaussian procedures. This Chapter is based on the review \cite{demarie2012}.
\item[$-$] Chapter \ref{ChapTO}: This Chapter serves as an introduction to the most common physical codes used in the theory of topological quantum computation. After a brief account of quantum lattice systems, we give a concise explanation of the cluster-states and surface code states, explaining how to detect topological order by means of appropriate witnesses, both in the case of pure and mixed states.
\item[$-$] Chapter \ref{CVTO}: After the two chapters of introduction and motivation, here we present the results from \cite{demarie2013b}. We start this Chapter introducing the continuous-variable states that are used in the analysis, carefully describing all the properties of the physical continuous-variable surface code state. This is the analog of the toric code state, constructed by means of interacting light modes instead of qubits and described by Gaussian states formalism. We use the graphical calculus for Gaussian states to derive an explicit form of the covariance matrix of the state and from that we calculate the value of the topological entanglement entropy, which proves that the state exists in a topological phase. We conclude with the analysis of a noise model for the surface code and describe possible experimental implementations for this model.
\end{itemize}
\item[$\bullet$] Part 3 - Entropy in polymer quantization:
~\\
The third and last part of this Thesis follows the results from \cite{demarie2013}, where we studied entropy in different quantization schemes.
\begin{itemize}
\item[$-$] Chapter \ref{polymerico}: We start explaining the concept of algebraic quantum theory: this is done introducing $\text{C}^*$-algebras and posing the conditions for the physical equivalence of two representations of the same algebra. After reviewing polymer quantization, a representation of the Weyl-Heisenberg algebra, we use the properties of entropy to analyze its convergence in two different quantization schemes, specifically polymer and Schr\"odinger quantization. We then derive a general bound that relates entropies of physically equivalent states in unitarily inequivalent representations and conclude with an example where we employ Gaussian states to calculate the bipartite entanglement entropy of two coupled oscillators in polymer quantization. 
\end{itemize}
\item[$\bullet$] Conclusions and appendices:
\begin{itemize}
\item[$-$] Chapter \ref{concluded} presents the results of this work in a concise manner, highlighting connections between different areas and open questions.
\item[$-$] Appendix \ref{appDiffGeom} is a short summary of differential geometry. It contains explanations of all the concepts used to derive the photon polarization equations in Chapter \ref{GRchapter}.
\item[$-$] In Appendix \ref{appNullifiers} we first set the rules and then derive the nullifier sets that are used in Chapter \ref{CVTO} to describe the continuous-variable code states.
\item[$-$] An upper bound for the topological entanglement entropy in the CV surface code is derived in Appendix \ref{boundsA}, together with an explicit proof of the lower bound formula for the topological mutual information. These results are based on \cite{demarie2013b}.
\item[$-$] Appendix \ref{expPolymer} presents explicit calculations of the canonical operators expectation values in polymer quantization. These are used to calculate the value of the entropy in the coupled harmonic oscillators example illustrated in Chapter \ref{polymerico}.
\end{itemize}
\end{itemize}

\subsection{Original results}
\label{originalresults}
This is a brief list of the original results presented in this Thesis. They have been divided according to the relevant sections as explained in the outline. 

\subsubsection{Relativistic quantum information}
The calculations and results about the gauge invariance of the geometric phase presented in Section \ref{gaugesec} are original and were first published in \cite{brodutch2011}. The ideas for experimental applications of relativistic quantum information are reviewed in \cite{rideout2012}.

\subsubsection{Continuous-variable topological order and Gaussian states}
While Chapter \ref{ChapGS} consists of a general review of known results on Gaussian states, the example in \ref{ExGS} is original and was first presented in \cite{demarie2012}.
~\\
The descriptions of the physical CV codes in Section \ref{pCVcodes} are corrected versions of other descriptions appeared in the literature before. All the simulations, calculations and results in Section \ref{pCVcodes} and \ref{dtopocvcs} are novel, together with the entropy bounds presented in Appendix \ref{boundsA} and they were first derived in \cite{demarie2013b}.

\subsubsection{Entropy in polymer quantization}
The calculations and results in Chapter \ref{polymerico} are mostly original, apart from the polymer quantization description in Section \ref{polyquantiz}. They were first derived in \cite{demarie2013}.

\begin{savequote}[8.5cm] 
\sffamily
``Deep in the human unconscious is a pervasive need for a logical universe that makes sense. But the real Universe is always one step beyond logic.'' 
\qauthor{Paul Muad'Dib - Dune}
\end{savequote}

\chapter{
Thesis' Prolegomena}
\graphicspath{{Q_info}} 
\label{CHintro2}

This Thesis starts with a minimal review of all the relevant mathematical definitions and physical features that will be used in the following Chapters\footnote{We acknowledge \cite{halvorson2006} for inspiring the title of this Chapter.}. We begin with a short introduction to quantum mechanics, defining the basic concepts of Hilbert space, quantum states, physical observables and measurements \cite{nielsen2000,cohen1977}. Then we move to information theory, focusing in particular on the concept of entropy and information \cite{wehrl1978}. First we offer a parallel view of different information measures both in classical and quantum physics, secondly we specifically list the required properties of the quantum entropy. We then introduce the paradigmatic notion of entanglement \cite{horodecki2009}, clarifying the meaning of bipartite entanglement and presenting a number of quantities that are normally used to quantify the quantum correlations shared between two systems \cite{amico2008}. The next notion to be described is the stabilizer formalism \cite{gottesman1997}, a method that can be used to simplify the description of quantum many-body systems on a lattice. This method is important for us because these are the systems that will be employed in the discussion of topological order for continuous-variables \cite{demarie2013b} in Chapter \ref{CVTO}. 

After that, we take our quantum systems up to space, where in addition to the quantum theory we also need the laws of relativistic quantum information \cite{peres2004, rideout2012}. There, we investigate the behavior of quantum information enclosed into the degrees of freedom of photons traveling a curved space-time \cite{brodutch2011}. Therefore, in this Introduction we succinctly describe the physics of massless point-like particles in special relativity, in order to simplify the subsequent extension to general relativity in Chapter \ref{GRchapter}.

Eventually, we conclude this introductory Chapter spending few words about quantum gravity and specifically loop quantum gravity \cite{rovelli2004, rovelli2011, giesel2013}, one of the most promising attempts to derive a theory that encloses both quantum mechanics and general relativity. Loop quantum gravity is one of the main motivations to examine entropy in different quantization schemes \cite{ashtekar2003, demarie2013}, as we do in Chapter \ref{polymerico}.

\section{Basic elements of quantum mechanics for quantum information}
The protagonist of quantum information is the quantum bit, or \emph{qubit} \cite{nielsen2000}. As the name suggests, the qubit is the conceptual extension of a classical bit to the quantum world, where physical states can exist in a unique condition known as \emph{superposition} of states. Classically, a bit is the basic unit of information, which physically corresponds to a system that can exist in two different states, traditionally called \emph{zero} and \emph{one}. Start with a (fair) coin, and flip it in the air. It will fall showing you either the head ($0$) or the tail ($1$). This is the street version of a bit. The concept of qubit is built upon similar considerations, but it is now implemented by a 2-level quantum state and exhibits some extra features that cannot be found in classical physics. Before clarifying what the last sentence means, let me explain briefly part of the terminology we will adopt. 

Traditionally (any quantum mechanics book is a good source for the details presented here, see for example \cite{nielsen2000,cohen1977,rossetti1990}), quantum mechanics tells us that a (quantum) system is described by a unit vector, or state, that belongs to an \emph{Hilbert space}. Hilbert spaces are a class of $d$-dimensional complex vector spaces with certain additional properties. First of all they are equipped with a positive definite \emph{scalar product},
\beq
(\cdot, \cdot) : \ham \times \ham \to \mathbb{C} \,,
\eeq
which is a continuous map conventionally anti-linear in the first argument and linear in the second argument. Using the inner product one can introduce the concept of \emph{norm} of an element $h \in \ham$ as
\beq
|| h || = \sqrt{( h, h )} \,,
\eeq
Furthermore, a Hilbert space is a complete inner product space, meaning that all the Cauchy sequences in $\ham$ converge to a value in $\ham$. 

A qubit state lives in a 2-dimensional Hilbert space spanned by some (vector) basis that we choose to label
\beq
\ket{0}\, \qquad \text{and} \qquad \, \ket{1} \,,
\eeq
in analogy with the classical description of the bit states $\{0,1\}$, where the symbol $\ket{a}$ is called a \emph{ket} and is the standard depiction for a (pure) state of a quantum system. As mentioned before, quantum mechanics actually tells us more: A qubit can exist in states that are linear superpositions of the basis states, such as
\beq
\label{linsp}
\ket{\varphi} = \alpha \ket{0} + \beta \ket{1}\, ,
\eeq
where the absolute value of the coefficients, namely \emph{amplitudes}, ($\alpha, \beta \in \mathbb{C}$ and $|\alpha|^2 + |\beta|^2 = 1$) correspond to the normalized probabilities of finding the state $\ket{\varphi}$ in one of the two possible outcomes $\ket{0}$, $\ket{1}$ after an appropriate measurement. For future references, we call the basis $\{\ket{0}$, $\ket{1}\}$ with the established name of \emph{computational basis}. In Eq.(\ref{linsp}) lies (part of) the real power of quantum computation. Quantum states can exist as linear combinations of other quantum states, weighted with different amplitudes. Notice also that, although the amount of information that can be extracted from a qubit is exactly the same of a classical bit, one bit, the quantum superposition principle or more generally the linearity of quantum mechanics, allows us to perform pre-measurement operations on qubits that are in no way possible for classical systems. 

In the ket formalism, the inner product is rewritten as
\beq
(\ket{v},\,\ket{w}) = \bra{v} w \9 \in \mathbb{C} \,,\,\,\,\, \forall \,\, \ket{v}, \, \ket{w} \in \mathcal{H}\,,
\eeq
where the element $\bra{v}$ is called \emph{bra} and is the vector dual to the ket $\ket{v}$. Duality is a basic concept in algebra and in differential geometry. It will be discussed with more rigor later in the introduction and in Appendix \ref{appDiffGeom}. 
To generalize the inner product, introduce an orthonormal basis $\{ \ket{u_i} \}$ defined on some Hilbert space $\mathcal{H}$ and consider any two vectors $\ket{v} = \sum_i \alpha_i \ket{u_i},\,\ket{w} = \sum_i \beta_i \ket{u_i}$, $\ket{v},\,\ket{w} \in \mathcal{H}$, then we have that
\beq
\bra{v} w \9 = \sum_{i,j} \alpha^*_i \beta_j \bra{u_i} u_j \9 = \sum_i \alpha_i^* \beta_i \in \mathbb{C} \,.
\eeq
The following concept we require to introduce is the representation of \emph{physical observables} in quantum mechanics. In the quantum regime, physical (classical) observables, as position, momentum and energy, are associated to (quantum) operators that act on quantum states. When one represents quantum states by kets, any observable operator is given by a hermitian matrix $ \hat{A}$, where $ \hat{A}^\dagger =  \hat{A}$ and $ \hat{A}^\dagger$ is the \emph{Hermitian adjoint} of $ \hat{A}$. Every Hermitian matrix has real eigenvalues, and for an operator they correspond to the values that the associated physical observable can acquire after a measurement. The expectation value of some observable $ \hat{A}$ in the state $\ket{\psi}$ is given by 
\beq
\6 \hat{A} \9 = \bra{\psi}   \hat{A} \ket{\psi} \,,
\eeq
and if $\ket{\psi} = \sum_i \gamma_i \ket{u_i}$, the quantities
\beq
 \hat{A}_{i,j} = \bra{u_i} \hat{A} \ket{u_j} 
\eeq
are the \emph{matrix elements} of the observable $ \hat{A}$.

The temporal evolution of a closed quantum system from a time $t_1$ to a time $t_2$ is given by some unitary operator $ \hat{U}(t_1,t_2)$, such that 
\beq
\label{unievo}
\ket{\varphi_{t_2}} =  \hat{U}(t_1,t_2) \ket{\varphi_{t_1}} \, ,
\eeq
and $\hat{U}^\dagger \hat{U} =  \hat{U}  \hat{U}^\dagger = {\hat I}$, where ${\hat I}$ is the unit operator. The unitarity condition immediately implies that $ \hat{U}^\dagger \ket{\varphi_{t_2}} =  \hat{U}^\dagger  \hat{U} \ket{\varphi_{t_1}} = \ket{\varphi_{t_1}}$. 
Unitary operators can be seen as \emph{gates} acting on one or many-qubit state, evolving it accordingly to their properties. Among the most important 1-qubit gates are the \emph{Pauli operators}:
\begin{align}
\label{pauliqubit}
\hat{\sigma}^X \equiv {\bf X} \equiv
\left(
\begin{array}{cc}
 0 & 1  \\
  1 & 0 
\end{array}
\right)\,,\quad
\hat{\sigma}^Y \equiv {\bf Y} \equiv
\left(
\begin{array}{cc}
 0 & -i  \\
  i & 0 
\end{array}
\right)\,,\quad
\hat{\sigma}^Z \equiv {\bf Z} \equiv
\left(
\begin{array}{cc}
 1 & 0  \\
  0 & -1 
\end{array}
\right)\,.
\end{align}
For instance, it is trivial to see that the ${\bf X}$ operator swaps the computational basis vectors
\beq
{\bf X} \ket{0} = 
\ket{1} \,, \qquad \text{and} \qquad {\bf X} \ket{1} = \ket{0} \,.
\eeq 
The Pauli matrices, together with the unit operator $\hat{I}$, taken with multiplicative factors $\{ \pm 1, \pm i \}$ form a group closed under matrix multiplication \cite{tung1985}, called the qubit \emph{Pauli group}. Its extension to $N$-qubits is at the basis of the stabilizer formalism \cite{nielsen2000} and the generalization \cite{bartlett2002} to quantum-modes \cite{weedbrook2012}, to be specified later, plays a fundamental part in our future treatment of continuous-variable topological phases in Chapter \ref{CVTO}.

It is worth pointing out that an alternative basis to the computational basis is the conjugate basis
\beq
\ket{+} = \frac{1}{\sqrt{2}} (\ket{0} + \ket{1}) \,, \quad \ket{-} = \frac{1}{\sqrt{2}} (\ket{0} - \ket{1}) \,,
\eeq
and the two basis are related by the \emph{Hadamard} operation;
\begin{align}
{\bf H} \equiv \frac{1}{\sqrt{2}}\left(
\begin{array}{cc}
 1 & 1  \\
  1 & -1 
\end{array}
\right)\, \longrightarrow {\bf H} \ket{0} = \ket{+}\,, \quad {\bf H} \ket{1} = \ket{-} \,.
\end{align}

Measurements in quantum mechanics are a tricky business. If we write 
the desired observable $ \hat{M}$ on a orthogonal basis 
\beq
 \hat{M} = \sum_m \lambda_m  \hat{M}_m \,,
\eeq
then the outcome of a measurement can solely be one of the eigenvalues $\{ \lambda_m \}$ of the linear decomposition. Specifically, the probability to obtain $\lambda_m$ for a state $\ket{\varphi}$ is
\beq
p(\lambda_m) = \bra{\varphi}  \hat{M}_m \ket{\varphi} \,.
\eeq
The measurement operators $  \hat{M}_m$ are also known as \emph{projectors}, and they satisfy the completeness equation $\sum_m   \hat{M}_m^\dagger   \hat{M}_m =  {\bf I}$ and $  \hat{M}_m^2 =   \hat{M}_m$. After a measurement $\hat{M}$ with outcome $\lambda_m$, the state of the system, which initially was $\ket{\varphi}$ collapses into the state
\beq
\ket{\varphi_m} = \frac{ \hat{M}_m \ket{\varphi}}{\sqrt{p(\lambda_m)}} \,,
\label{postmes}
\eeq
with $\ket{\varphi_m} = 0$ if $p(\lambda_m) = 0$.

States that can be labelled by a ket, as the ones treated so far, are called \emph{pure states}. All the required information for their description is encoded in $\ket{\psi}$. Whenever we are \emph{ignorant} about the description of a state, we are only allowed to treat it statistically as an ensemble of pure states with different probability amplitudes. This is the \emph{density matrix} of a system, defined by
\beq
\label{dmatrix}
\rho = \sum_i p_i \ket{\psi_i} \bra{\psi_i} \,,
\eeq
with $p_i$ probability of the system to be in $\ket{\psi_i}$.
If there exists a base where $\{ p_1 = 1, \, p_{i>1} =0 \}$, then $\rho = \ket{\psi} \bra{\psi}$ and the state is pure. For pure states, the density matrix language is analogous to the ket language. If this condition is not satisfied, then the state is \emph{mixed} and 
cannot be represented by a ket. 

In the density matrix representation the conservation of probability is given by the obvious relation
\beq
\text{tr} \rho = 1 \leftrightarrow \sum_i p_i = 1\,,
\eeq
and each density matrix is a positive operator such that, given a ket $\ket{v}$, the following inequality always holds
\beq
\bra{v} \rho \ket{v} \ge 0 \,.
\eeq
The expectation value of an observable with respect to $\rho$ is expressed by
\beq
\text{tr} \left( \rho  \hat{A} \right)\,,
\eeq
while the action of an unitary operator $\hat{U}$ on the density matrix is 
\beq
\rho' =  \hat{U} \rho  \hat{U}^\dagger\,,
\eeq
and the time evolution of $\rho$ obeys the following equation
\beq
i \hbar \frac{\text{d}}{\text{d} t} \rho(t) = \left [ \hat{H}(t), \rho(t) \right] \,,
\eeq
with $\hat{H}(t)$ time-dependent Hamiltonian of the system.

Let me recall that this description of quantum mechanics, named after Heisenberg \cite{heisenberg1925, born1926}, is completely equivalent to the formulation introduced by Schr\"odinger \cite{schroedinger1926}, where kets are replaced by \emph{wave functions} $\phi({\bf x},t)$ and physical observables are equivalent to differential operators.

\section{Entropy and its properties}
\label{entropyprop}
Statistical thermodynamics, quantum mechanics and information theory promoted entropy from an auxiliary variable of the mechanical theory of heat \cite{clausius1851, clausius1856} to one of the most important quantities in science \cite{lieb1999}. In the field of information theory, entropy plays a most fundamental role: It quantifies the amount of uncertainty, or lack of information, the observer has over a system \cite{nielsen2000,cover1991}. Classically, entropy is a concept associated to the probability distribution of a classical variable, while in the quantum regime entropy is a function of the density matrix of the state \cite{wehrl1978}. In the following, we will first introduce different classical statistical functions and then present their quantum counterparts. 

\subsection{Classical world}
In classical information theory, the basic measure of uncertainty is the Shannon entropy \cite{bengtsson2006}. Given a random variable $X$, the Shannon entropy quantifies how much we \emph{do not know} about the variable $X$ before we learn its value. Less prosaically, if the variable $X$ can take $i=1,..., n$ different values, each of them with a certain probability $p_i$, such that $(p_1, ..., p_n)$ is the probability distribution of the variable $X$, then the Shannon entropy is defined as
\beq
H(X) \equiv - \sum_i p_i \log_2 p_i \,,
\eeq
and it is maximized by a uniform distribution with all equal $p_i$.

The Shannon entropy is the lower parametrical limit of a one parameter family of entropies known as R\'enyi-$\alpha$ entropies \cite{bengtsson2006, wehrl1978}, defined by
\beq
S_{\alpha} (X) := \frac{1}{1 - \alpha} \log_2 \left( \sum_i p_i^\alpha \right) \,.
\eeq 
In the limit $\alpha \to 1$, one recovers the Shannon entropy for $X$. Fruitfully, for each value of $\alpha$ the R\'enyi entropy vanishes for $p_i =1$ and it acquires its maximum value for the uniform distribution.


Given two variables $X$ and $Y$ with probability distributions $\{ p_i \}$ and $\{ q_i \}$, there exists a measure that quantifies how different these distributions are. This is called the \emph{relative entropy} \cite{cover1991}
\beq
S(X||Y) = \sum_{i} p_i \log_2 \frac{p_i}{q_i}\,,
\eeq
and it is related to the rate at which, in the limit of very large sampling, one can safely identify if we are sampling from the $X$ or $Y$ distribution. 

The last concept we wish to introduce at this point is a quantity that tells us something about how much information two distributions have in common. This is the \emph{mutual information} \cite{nielsen2000}. Classically, the mutual information is a quantity associated to the probability distributions of two classical variables $X$ and $Y$:
\beq
I(X,Y) = H(X) + H(Y) - H(X,Y) \,.
\eeq
Intuitively, the mutual information describes how much information about $X$ we learn by measuring the value of $Y$. If the two quantities are not correlated, then it simply reduces to the Shannon entropy of $X$.  

\subsection{Quantum world}
\label{secEEMI}
The quantum analog of the Shannon entropy is the von Neumann entropy, or quantum entropy \cite{ohya2004}. It is constructed replacing the classical probability distribution with the closest quantum concept, i.e. the density matrix (\ref{dmatrix}) of the quantum state, generically $\rho = \sum_i^N p_i \ket{\psi_i} \bra{\psi_i}$. Thus, given $\rho$, its von Neumann entropy is defined as
\beq
S(\rho) \equiv - \tr (\rho \log_2 \rho) = - \sum_i \lambda_i \log_2 \lambda_i  \,,
\label{defentro1}
\eeq
where $\{ \lambda_i \}$ is the set of the eigenspectrum of $\rho$. From this last definition, it is trivial to realize that for a pure state $\rho = \ket{\Psi} \bra{\Psi}$ since $\{ \lambda_1 = 1,\, \lambda_{i>1} = 0\, \,\forall \, i \}$ the von Neumann entropy is always identically zero. Therefore, only non-pure states have non-vanishing von Neumann entropy, whose greatest value $\log_2 N$ is given by the maximally mixed state ($p_i = 1/N$).

The quantum version of the relative entropy for two states $\rho, \sigma$ is defined as \cite{bengtsson2006}
\beq
S(\rho || \sigma) = \tr (\rho \log_2 \rho) - \tr (\rho \log_2 \sigma) \,.
\eeq
In analogy with the classical case, it says something about the statistical distinguishability between the two states $\rho$ and $\sigma$. Operationally, larger the value of the relative entropy, greater is the amount of information we can extract when performing a measurement aimed at distinguish the two states.

In complete analogy with the classical case, one can construct a family of quantum R\'enyi entropies for $\rho$ with $\alpha \ge 0$ \cite{bengtsson2006},
\beq
S_{\alpha} (\rho) := \frac{1}{1 - \alpha} \log_2 \tr (\rho^\alpha) \,,
\eeq 
where the von Neumann entropy $S(\rho)$ is given by the $\alpha \to 1$ limit. R\'enyi entropies are particularly appealing because the full family contains information about the (entanglement) spectrum of the state $\rho$, which is more than the information given by the only entanglement entropy. Furthermore, R\'enyi entropies showed to be useful in the treatment of topologically ordered systems, offering an alternative way to identify topological phases \cite{flammia2009a}.

The quantum mutual information for two states $\rho$ and $\sigma$ is defined substituting the Shannon entropy in the classical definition with the von Neumann entropy \cite{nielsen2000},
\beq
I(\rho,\sigma) := S(\rho) + S(\sigma) - S(\rho, \sigma) \,.
\eeq
The quantum mutual information is the most used measure to quantify correlations in mixed states, since it captures the total amount of information between the two subsystems, both in the classical and quantum case. It will be useful when dealing with mixed states of topologically ordered quantum systems.

Before proceeding, it is necessary to say something about the spaces where a state is defined. So far we have been dealing with density matrices $\rho$ implicitly associated to an Hilbert space $\ham$. However, we will mostly consider situations where states are defined on the tensor product of two or more Hilbert space. In this sense, systems made of $n$ subsystems each associated to an individual Hilbert spaces $\ham_i$ are described by states that live in the composite product space
\beq
\ham = \bigotimes_{i=1}^n \ham_i \,.
\eeq 
For instance, if $\rho$ lives in $\ham = \ham_{1} \otimes \ham_2$, then the subsystem defined on $\ham_1$ is described by the \emph{reduced density matrix} $\rho_1$. This is derived tracing out all the degrees of freedom associated to $\ham_2$:
\beq
\label{jointrho}
\rho_ 1 = \text{tr}_2 \rho \,,
\eeq 
and, conversely, $\rho_2 = \text{tr}_1 \rho$.

\subsection{Everything you always wanted to know about the von Neumann entropy but were afraid to ask}
At this point you should be convinced that quantum entropy is a rather significant and interesting quantity. It is not by any means an observable of the system, i.e. there is no quantum mechanical operator whose expectation value is the entropy. Quantum entropy is instead a functional of the state, just like in classical theory. From Eq.(\ref{defentro1}) it is evident that for a generic quantum state $\rho$ one has
\beq
S(\rho) \ge 0\, , \qquad \text{and} \qquad S(\rho) : [0, \infty ) \,.
\eeq
Moreover, $S(\rho) < \infty$ only if the rank of $\rho$ is finite and vice-versa. 
 
We will now list some of the basic properties of the von Neumann entropy \cite{wehrl1978} that will be useful hereinafter. A good starting point is to notice that the entropy of a quantum state is invariant under unitary evolution of the state. For a unitary transformation $ \hat{U}$, this property translates into 
\beq
\rho \to \rho' =  \hat{U} \rho  \hat{U}^\dagger \longrightarrow S(\rho') = S (\rho) \,.
\eeq
If one considers a direct product state $\rho$ given by
\beq
\rho = \otimes_{i=1}^n \rho_i\,,
\eeq
then the total entropy is equivalent to the sum of the entropies of the single constituents:
\beq
\label{Saddit}
S(\rho = \rho_1 \otimes \rho_2 \otimes ... \otimes \rho_n) = \sum_i S(\rho_i) \,.
\eeq
This property is called \emph{additivity} of the entropy. If we now limit ourselves to two Hilbert spaces such that $\rho \in \mathcal{H} = \mathcal{H}_1 \otimes \mathcal{H}_2$ (but the following is easily generalizable to more dimensions) then we can introduce another property of the entropy called \emph{subadditivity}. 

To define subadditivity, consider the joint state state $\rho$ of two lower dimensional density matrices as in Eq.(\ref{jointrho}), with
\beq
\rho_1 = \text{tr}_{\mathcal{H}_2} \rho\,, \qquad \text{and} \qquad \rho_2 = \text{tr}_{\mathcal{H}_1} \rho \,.
\eeq
Then the following inequalities hold
\begin{align}
S(\rho) \le &S(\rho_1) + S(\rho_2) = S(\rho_1 \otimes \rho_2) \,,\\
&S(\rho) \ge |S(\rho_1) - S(\rho_2)| \, ,
\end{align}
where the latter is known as the \emph{triangle} inequality.

An interesting immediate consequence of this is that for a global pure state $\rho \to S(\rho) = 0$, the quantum entropy of a subsystem $\rho_1$ of $\rho$ can still be non-zero and quantifies the degree of \emph{entanglement}, a special kind of correlations unique to quantum systems. From subadditivity, another property of the entropy follows. This is the \emph{concavity} \cite{lieb1975}, which implies that the total entropy of a mixed state is always bigger or equal to the weighted sum of the single entropies of the composing elements, i.e. for $\rho = \sum_i p_i \rho_i$, then
\beq
 S(\rho) \ge \sum_i p_i S(\rho_i) \,.
\eeq
Physically, this last inequality is telling us that by sampling from a mixture of state we also lose information by not knowing from which state we sample. Using concavity, it is possible to derive an upper-bound for $S(\rho)$ \cite{lanford1968}. This will be used later when discussing the convergence of entropy in different quantization schemes, namely
\beq
\label{ubconcavity}
S(\rho) \le \sum_i p_i S(\rho_i) + H(p_i) = \sum_i p_i S(\rho_i) - \sum_i p_i \log_2 p_i \,,
\eeq
where $H(p_i)$ is the Shannon entropy of the probability distribution $\{ p_i \}$. In the special case where the states $\rho_i$ are all one-dimensional projections, a fancy way to say that they can be written as $\rho_i = \ket{\varphi_i} \bra{\varphi_i}$, then Eq.(\ref{ubconcavity}) reduces to
\beq
S(\rho) \le - \sum_i p_i \log_2 p_i \,.
\eeq
The equality in (\ref{ubconcavity}) is satisfied when the $\rho_i$ have support on orthogonal subspaces. 

For infinitely dimensional Hilbert spaces, entropy as a function of $\rho$ is generally a discontinuous quantity and small variations of the density matrix can induce a discontinuous jump in the value of the entropy. For finite-dimensional systems this change is bounded by the Fannes' inequality \cite{fannes1973}: Taken two density matrices $\rho, \rho'$ 
\beq
|S(\rho) - S(\rho')| \le 2 T \log_2 (d) - 2T \log_2(2 T) \,,
\eeq
where $T = T(\rho, \rho') = \frac{1}{2} \tr \sqrt{(\rho-\rho')^2}$ is the trace distance between the two density matrices. A continuity property for the entropy is the \emph{lower semicontinuity}. Given a sequence of density matrices $\rho_k$ that weakly converge to the density matrix $\rho$
\beq
\rho_k \stackrel{weakly}\longrightarrow \rho \,,
\eeq
i.e. all matrix elements satisfy $\6 l | \rho_k | l \9 \to \6 l| \rho| l \9$, then the entropy $S(\rho)$ is upper bounded by
\beq
S(\rho) \le \lim \text{inf}\,S(\rho_k) \,.
\eeq
Furthermore, the weakly convergence for density matrices also implies that 
\beq
\label{stwc}
T(\rho_k, \rho) \to 0\,.
\eeq
When the trace distance relation above (\ref{stwc}) holds, together with the convergence of the energy expectation values,
\beq
\tr \rho_k \hat{H} \to \tr \rho \hat{H} \,,
\eeq
then the following continuity property is satisfied
\beq
\label{ceq}
S(\rho) = \lim S(\rho_k)\,.
\eeq
The continuity conditions can be relaxed when we consider bounded Hamiltonians,
\beq
\tr e^{- \beta \hat{H}} < \infty\, ,
\eeq
for all $\beta$ such that $ 0 < \beta < \infty\,$, in which case the entropy $S(\rho)$ is automatically continuous in the sense of Eq.(\ref{ceq}) for all density matrices $\{ \rho \}$ associated to finite energy \cite{wehrl1978}, i.e.
\beq
\tr \rho \hat{H} \le C < \infty \,.
\eeq

\section{My name is Verschr\"ankung, but you can call me Entanglement} 
In 1935, the EPR trio Einstein, Podolsky and Rosen, realized that quantum description allows for local operations performed on part of a global system to affect the state of another part of the system indefinitely separated in space \cite{einstein1935}. Whereas the ghost of the verschr\"ankung, the term coined by Schr\"odinger to describe this wonder of Nature, haunted Einstein till the end of his life, today \emph{entanglement} is widely recognized as the distinctive trademark of quantum mechanics and arguably the main ingredient in quantum information processing \cite{horodecki2009}.

Beyond its elusive nature, entanglement reveals itself as a non-local more-than-classical kind of correlations among two (or more) quantum states. Practically, we say that a state is \emph{entangled} if it is not \emph{separable} \cite{nielsen2000}. A pure state $\ket{\psi} \in \ham$ is separable if it can be written as the direct product of states of the subsystems,
\beq
\ket{\psi} = \ket{\psi_1} \otimes ... \ket{\psi_i} \otimes ... \ket{\psi_n} \,,
\eeq
where $\ham = \otimes^n_i \ham_i$ and $\ket{\psi_i} \in \ham_i$. Physically, separability has a direct connection to locality, meaning that different parts of a pure separable state can be prepared locally and the outcomes of local measurements on the subsystems are independent. The easiest example, for the two-qubit Hilbert space $\mathcal{H} = \mathcal{H}_1 \otimes \mathcal{H}_2$, consists of the \emph{Bell basis},
\beq
\ket{\tau} = \frac{1}{\sqrt{2}} (\ket{01} \pm \ket{10})\,,  \qquad \ket{\varkappa} = \frac{1}{\sqrt{2}} ( \ket{00} \pm \ket{11} ) \,.
\label{bellstate}
\eeq
These states are entangled, because they cannot be written as the product of two states
\beq
\label{bentst}
\ket{\tau}\,,\ket{\varkappa} \neq \ket{a} \otimes \ket{b}\,, \,\, \forall \ket{a} \in \mathcal{H}_1\,,\, \forall \ket{b} \in \mathcal{H}_2 \,.
\eeq
Similarly, a mixed state is separable if its global density matrix can be written as a convex sum of product states, 
\beq
\rho = \sum_i p_i \rho_1^i \otimes \rho_2^i ... \otimes \rho_n^i \, ,
\eeq
as shown in \cite{werner1989}. If this is not possible, then the state is entangled. 

Entanglement might well be considered the quintessence of quantum mechanics, or its \emph{quantessence}\footnote{I apologize for the terrible joke.}. Apart from being rather interesting as a phenomenon \emph{per se}, entanglement proved to be an additional \emph{physical resource} exclusive of the quantum regime. One of the first and most fascinating examples where entanglement was used to achieve something otherwise impossible in classical physics was quantum teleportation \cite{bennett1993}. Following that successful application, it was understood that entanglement could be used to perform tasks within quantum information, such as quantum computing, super-dense coding and quantum cryptography \cite{bennett1984, horodecki2009}. 

Interestingly enough, it is extremely challenging to quantify and identify correlations, quantum or classical, between subsystems of a larger system and a unique characterization of the degree of entanglement of a system (assuming this last sentence makes physical sense) still does not exist \cite{amico2008}. The classification of correlations and the appropriate choice of measures goes beyond the scopes of this Thesis, so they are not treated here. However, we still need to define a few measures of entanglement that will be used in the following Chapters. A criterion to determine whether a quantity is a meaningful entanglement witness or not was derived in \cite{vidal2000}. It states that any quantity that does not increase on average under local operations and classical communication (LOCC) is an \emph{entanglement monotone}, and can be meaningfully used to quantify entanglement. The physical reason behind this idea is that entanglement cannot increase when we limit ourselves to operations performed locally (a mathematical motivation follows from Nielsen's majorization theorem, see \cite{nielsen1999}). 

In the following section we start with the quantification of bipartite entanglement in the case of pure states and then extend the same concept to mixed states. 

\subsection{Bipartite entanglement of pure states}
For pure states, entanglement manifests itself as \emph{disorder} in the subsystems of the entangled system. Consider the pure state $\rho = \ket{\varphi} \bra{\varphi}$, divide it into two complementary subsystems $A$ and $B$ and hand them to the most revered quantum couple of all times, Alice and Bob (a strictly not necessary but always enjoyable step to do). The preferred measure to quantify the amount of entanglement between any $\{ A, B \}$ bipartition held by the two lovers\footnote{I believe that every quantum physicist likes to think that between Alice and Bob there is something more than just entanglement (although things get kinky when Eve gets involved).} is the \emph{entanglement entropy}, or degree of entanglement, which is equivalent to the von Neumann entropy of any of the two partitions (see \cite{nielsen2000, horodecki2009, demarie2012})
\beq
\label{entent}
E(\rho_{A,B}):=S(\rho_A = \text{tr}_B \rho) \equiv S(\rho_B = \text{tr}_A \rho)\,.
\eeq
If the subsystems $A$ and $B$ are in a product state, then no entanglement is present and hence $E(\rho_A) = 0$. Otherwise, the quantum correlations along the cut induce a positive value of the von Neumann entropy. Although the entanglement entropy is an appropriate measure of entanglement for pure states, it does not have any particular physical interpretation for mixed states.

\subsection{Negativity and logarithmic negativity}
While quantum mutual information takes in account all kind of correlations, a measure that manages to capture solely the quantum correlations for mixed states is the \emph{negativity} \cite{vidal2002}, which is based on Peres' criterion to determine if a mixed state $\rho_{A,B}$ is entangled between $A$ and $B$. The criterion says that for any (bipartite) separable state, taken the general form of the density matrix
\beq
\rho_{A,B} = \sum_{ijkl} (\rho_{A,B})_{ijkl} \ket{i_A \otimes j_B} \bra{k_A \otimes l_B} \,,
\eeq
then the partial transposition of $\rho_{A,B}$ with respect to $A$,
\beq
\rho_{A,B}^{T_A} \longrightarrow (\rho_{A,B}^{T_A})_{ijkl} \equiv (\rho_{A,B}^{T_A})_{kjil}
\eeq
is always positive definite $\rho_{A,B}^{T_A} \ge 0$ \cite{peres1996}. Consequently, the condition $\rho_{A,B}^{T_A} \ngeq 0$ is sufficient to assert that the state is entangled. Mathematically, this is equivalent to saying that the state is entangled if the partial transpose of the density matrix has at least one negative eigenvalue. This property can be used to construct a quantitative measure of the degree of entanglement for mixed states: For a generic $\rho$, the negativity of a subsystem $A$ is defined as
\beq
\mathcal{N}(\rho,A) :=\frac{ || \rho^{T_A}||_1 -1}{2} \,,
\eeq
where the term $|| \rho^{T_A} ||_1 = \tr [ ((\rho^{T_A})^\dagger  \rho^{T_A})^{1/2}]$ is the trace norm of the partial transpose of $\rho$. The negativity is therefore equal to the absolute value of the sum of all the negative eigenvalues of the partial transpose (whose sum can be different to 1) and it is zero if $A$ and its complement are not entangled. 

A second entanglement monotone is the \emph{logarithmic negativity} \cite{vidal2002}, simply given by
\beq
E_{\mathcal{N}}(\rho,A) = \log_2 || \rho^{T_A}||_1 \,.
\eeq 
This quantity proved to be very useful since it is an upper bound to the entanglement entropy for all pure states
\beq
E_{\mathcal{N}}(\rho,A) \ge S(\rho_A) \,,
\eeq
and it can be computed in a simpler way, since it requires the full spectrum of the density matrix instead of the reduced one that is usually more difficult to extract.

\subsection{Beyond bipartite entanglement}
Entanglement is not only a property of bipartite systems, but it reveals itself also in the more complicated setup of multi-partite systems. Nowadays, despite a very large literature on this topic \cite{amico2008}, we still do not have a meaningful quantification and characterization of multi-partite entanglement \cite{hein2006}. 

In between classical correlations and entanglement, lies another measure of correlations, the \emph{quantum discord}. For the bipartite state $\rho_{AB}$, discord is defined as the difference between quantum mutual information and classical correlations
\beq
D(B|A) \equiv I(A,B) - J(B|A)\,,
\eeq
where $J(B|A)$ is the maximization of the classical correlations measure over all possible POVM measurements $\{E_a\} = M_a^\dagger M_a$, explicitly 
\beq
J(B|A) = \max_{\{ E_a \}} J(B| \{ E_a \}) \,,
\eeq
and $J(B|\{ E_a \}) \equiv S(B) - S(B|\{ E_a \})$ is the difference between von Neumann entropy and conditional entropy. Originally, discord was introduced in \cite{henderson2001, ollivier2002} as an attempt to discriminate between the purely classical correlations and the quantum ones. Since its introduction, the nature of discord has been widely investigated and many physical interpretations have been given, especially in terms of computational advantages for states with positive discord but zero entaglement \cite{brodutch2011c}\footnote{Just before posting this Thesis on the arXiv, a new work by Gheorghiu and Sanders~\cite{GB2014} suggests that non-zero discord is a quantifier for noisy measurements rather than the flagpole of the quantum-classical border.}. 

Although we will not use these concepts in this Thesis, it is important to be aware that quantifying correlations in quantum information theory is still a controversial and open topic. A nice introduction to the argument of multi-partite entanglement and additional references can be found in \cite{eisert2006}, while we suggest to look at \cite{modi2012} for a complete review about quantum discord and related measures.

\section{Stabilizer formalism}
\label{stabIntro}
As we have seen, operations in quantum mechanics can be rather difficult to understand, especially from a physical point of view. For this reason, when dealing with graph states \cite{hein2006} and lattice quantum many-body systems, particular classes of systems that can be described in terms of the geometrical properties of the underlying pattern of the interactions, we will make use of a method known as \emph{stabilizer formalism} \cite{gottesman1997}. This allows to describe quantum states in terms of the action of certain special operators, called \emph{stabilizers}. For a generic state $\ket{S}$, we call stabilizer any operator $\hat{X}$ such that \cite{nielsen2000}
\beq
\label{stacondi}
\hat{K} \ket{S} = \ket{S} \,.
\eeq
Then the stabilizer set $\mathcal{S}=\{ \hat{K}_i \}$ of a quantum state ${\ket{S}}$ composed of $N$ subsystems, is defined as the set of stabilizer operators that have ${\ket{S}}$ as eigenvector with eigenvalue $+1$. The stabilized state ${\ket{S}}$ is uniquely determined if the stabilizer set $\mathcal{S}$ is generated by exactly $N$ independent stabilizer generators. If $\mathcal{S}$ is generated by $n < N$ elements, then it does not stabilize a single state but rather a $2^{N-n}-$dimensional subspace of the global Hilbert space of the system for qubits and $d^{N-n}$ for qudits of dimension $d$. When we discuss the toric code \cite{kitaev2003} in \ref{TTC}, it will be shown that the Hamiltonian ground state is not unique but four-degenerate because the toroidal structure of the system lattice causes two stabilizers to be linearly dependent by the others, see \ref{TTC}.

The stabilizer formalism is a powerful tool to describe topologically ordered systems such as quantum double models \cite{pachos2012}. The ground state of each of these models is in a topological phase and the defining Hamiltonian is constructed as a linear combination of the elements composing the ground state subspace stabilizer set, as explained in Chapter \ref{ChapTO}. Most importantly, the stabilizer formalism allows for a simpler description of the evolution of the state. Given the stabilizer condition from Eq.(\ref{stacondi}), under a unitary transformation of the state, $\ket{S'} = \hat{U} \ket{S}$, $\hat{K}$ transforms as $\hat{K}' = \hat{U} \hat{K} \hat{U}^\dagger$ in order to preserve its stabilizer status $\hat{K}' \ket{S'} = \ket{S'}$. Note that the transformation $\hat{K} \longrightarrow \hat{K}'$ under the action of $\hat{U}$ is \emph{opposite} from the Heisenberg evolution of the observables under the same unitary $\hat{U}$. In fact, when we evolve stabilizers we are not modeling the evolution of observables, but rather evolving the old stabilizers into new stabilizers for the new state. Hence, the unitary evolution applied to the stabilizer must counteract that applied to the state in order to maintain the stabilizer's role as such \cite{demarie2013b}.

In the context of continuous-variable systems, there exists an equivalent way to express the stabilizer relations by using nullifiers \cite{bartlett2002,menicucci2011}. In analogy with Eq.(\ref{stacondi}), an operator $\hat{\eta}$ is called a nullifier for a state $\ket{S}$ when the relation
\beq
\hat{\eta} \ket{S} = 0
\eeq
holds. When the generators of the stabilizer set are elements of a Lie group \cite{tung1985}, then the elements of the Lie algebra that generates the Lie group compose the nullifier set of the state. Note that nullifiers transform under the same transformation rule of the stabilizers.

\section{Relativistic quantum information}
Quantum mechanics deals with physics at small length scales. A fascinating new area in physics is the study of relativistic and gravitational effects on quantum information at scales usually associated to the relativistic regime and, to a lesser degree, the use of quantum information theory in relativistic physics. This is called \emph{relativistic quantum information} \cite{peres2004,terno2005rqi} and deals with the interaction between gravity and the quantum phenomena, analyzing the evolution of quantum systems over very large distances. 


Previously we introduced the concept of qubit. But so far a qubit has only been a mathematical concept, without any physical realization. In practice, a qubit can be encoded by well-chosen degrees of freedom of some physical system. For example \emph{photons}, massless particles of light, are a popular implementation of qubits \cite{scully1997}. In fact, while it is possible to encode a 2-level quantum state into their spin degrees of freedom, they also serve well as information carriers at relativistic lengths, enabling for \emph{gedanken} (and in the near future real) experiments in open space \cite{rideout2012}.

However, these degrees of freedom are not isolated, and they are generally transformed under the effect of gravity during the evolution of the information carriers. Understanding the gravitational effects on photons \cite{brodutch2011} is therefore essential. From a theoretical point of view, it helps to increase our comprehension of the physics of quantum phenomena at relativistic scales, which is particularly important with a theory of quantum gravity in mind, and practically it is a key step toward the implementation of quantum communication protocols between the Earth and satellites \cite{rideout2012,nordholt2002,jennewein2012}.

\subsection{Physics of photons in special relativity}
\label{pop}
In this subsection we aim to give an intuition of the meaning of \emph{polarization rotation} for a photon, looking at the problem in the setting of special relativity. In the following Chapter we will analyze the effects induced by general relativity, motivating how one can construct paths along which the phase introduced is independent of the choice of reference frame.

In a Minkowski space-time, the generic state of a spin-particle is given by some irreducible representation of the Poincar\'e group \cite{weinberg1996-1,tung1985} and can be represented by
\beq
\label{photopacco}
\ket{\phi} = \sum_{\sigma} \int \d \mu( \ak) \phi_{\sigma} (\ak) \ket{\ak , \sigma} \,,
\eeq
where $\ak = (k^0, \bf{k})$ is the momentum four-vector, $\d \mu(\ak) = \d^3 {\bf k} / (2 \pi)^3 (2 k^0)$ is the Lorentz invariant measure and $\sigma$ symbolizes the total spin degrees of freedom. The basis states are complete and labeled by the four-momentum and the spin along a particular direction. Experimentally, a generic single photon state would correspond to a wave packet of the form in Eq.(\ref{photopacco}). In the following, the single photon state of interest $\ket{\varphi}$ is described by a well-defined three-momentum ${\bf k}$, since $\ak = (|{\bf k}|, {\bf k})$, and the helicity eigenvalues $h_{\bf k} = \pm 1 $. Hence, a sharp-momentum state can be written as \cite{peres2004}
\beq
\ket{\varphi} = \sum_{h=\pm 1} \alpha_{h} \ket{{\bf k}, h_{\bf k}}\,, \qquad \text{with} \,\,\,\,\, |\alpha_{+}|^2+|\alpha_{-}|^2 = 1 \,.
\eeq
One can therefore think of the states $\{ \ket{{\bf k}, +1}, \ket{{\bf k}, -1} \}$ as the computational basis of a qubit. Alternatively, it is possible to use a pair of three-vectors to label the same state, the momentum vector ${\bf k}$ and the polarization vector ${\bm \epsilon}^h_{\bf k}$, where ${\bf k} \cdot {\bm \epsilon}^h_{\bf k} = 0 $ and
\beq
\label{bskepsi}
\ket{{\bf k}, h_{\bf k}} \equiv \ket{{\bf k}, {\bm \epsilon}^h_{\bf k}} \, .
\eeq
The descriptions of a quantum state looked by observers associated to two different reference frames connected by a Lorentz transformation $\Lambda$, are related by a \emph{quantum Lorentz transformation}. This is a unitary representation of the Poincar\'e group \cite{weinberg1996-1,tung1985} $\hat{U}(\Lambda)$ that describes the transformation
\beq
\ket{\varphi}' = \hat{U}(\Lambda) \ket{\varphi} \, .
\eeq
States of definite helicity are invariant under Lorentz transformations, while momentum states are generally affected. This motivates the choice of working with sharp-momentum states: Since the basis states in Eq.(\ref{bskepsi}) are direct products of momentum and polarization, the spin states do not entangle with momentum states, although they still acquire a phase. Note that wave packets introduce an additional hurdle. The entangling of the momentum and spin degrees of freedom forbids complete distinguishability of states with different polarizations. As explained in \cite{peres2004}, a Lorentz transformation acts on the generic spin $S$ single-particle state with sharp momentum as
\beq
\ket{\phi} = \ket{\ak, \sigma} \to \hat{U}(\Lambda) \ket{\phi} = \sum_{\sigma} D^S_{\sigma, \sigma'}[ W(\ak , \Lambda) ] \ket{\Lambda \ak, \sigma'} \,,
\eeq
where $D^S_{\sigma, \sigma'}$ is the matrix element of the representation of the Wigner little group element $W(\ak , \Lambda)$, related to the spin $S$ representation of the Lorentz group. In this sense a Lorentz transformation acts as a quantum gate on the particle state. The classical information stored in $\Lambda$ controls how $\ket{\ak}$ transforms, and both control how the spin state $\ket{\sigma}$ changes. 

In order to keep things simple, we are not interested in presenting all the mathematical details of the transformation rules for a massless particle and thus, with much hand-waiving, we outline the idea and show the final result straight-away. For a photon with well-defined momentum ${\bf k}$, the effects of any Lorentz transformation are equivalent to the effects of a single rotation around the $z$-axis by an angle $\vartheta$ \cite{peres2003},
\beq
\Lambda \longrightarrow W(\Lambda, {\bf k}) \approx R_z (\vartheta) \,,
\eeq
where the reference frame is defined by the standard vector, which is a unit light-like vector pointing in the $z$-direction, mathematically $\ak_s = (1,0,0,1)$\footnote{ The standard vector is transformed into the particle momentum by a Lorentz transformation, $L(\ak) \ak_s = \ak$, where $L(\ak) = R_z(\alpha)R_y(\beta)B_z(\sf{b})$, $R$ rotations and $B$ boosts \cite{lindner2003}.}. For any $R_z(\vartheta)$ rotation, the matrix elements are given by 
\beq
D_{\sigma, \sigma'} [R_z(\vartheta)] = e^{i \sigma \vartheta} \delta_{\sigma, \sigma'}\, , \qquad \text{with} \,\, \sigma = \pm 1 \,,
\eeq
and consequently, the photon state transforms as
\beq
\label{rqigate}
\hat{U}(\Lambda) \ket{{\bf k}, \sigma} = \sum_{\sigma} e^{i \sigma \vartheta} \delta_{\sigma, \sigma'} \ket{\Lambda {\bf k}, \sigma'} \,.
\eeq
The physical significance of this last equation is that a Lorentz transformation introduces a relative phase for a photon state written in the helicity basis,
\beq
\label{helphase}
\hat{U}(\Lambda, {\bf k}) [\alpha_+ \ket{{\bf k}, +1} + \alpha_- \ket{{\bf k}, -1}] = e^{i \vartheta} \alpha_+ \ket{{\bf k}, +1} + e^{- i \vartheta} \alpha_- \ket{{\bf k}, -1} \,,
\eeq
and the angle $\vartheta$ is exactly that phase. If one chooses a reference frame to measure the polarization of the photon at the source, then the phase introduced after the Lorentz transformation can be considered as a consequence of the rotation of the polarization frame along the trajectory. 

In a flat space-time, if the sender knows the direction of propagation of the photon it is always possible to setup a standard reference frame as the \emph{triad} of 3-vectors 
\beq
\{ \hbb_1(\hbk), \hbb_2(\hbk), \hbk \}
\eeq
where $\hbk = (0,0,1)$ and the two polarization vectors are respectively $\hbb_1(\hbk) = (1,0,0)$ and $\hbb_2(\hbk) = (0,1,0)$. Then, in principle, one could invert the gate described by Eq.(\ref{rqigate}) to align the receiving detector in accordance with the detector of the sender and measure the correct polarization. We will see in the following Chapter that in general relativity there is no similar procedure to follow and it is therefore necessary to find other ways to extract the information encoded within \cite{brodutch2011}.

\section{Loop quantum gravity}
The two theories we have been discussing in this introduction are the pillars of modern physics: The quantum theory, ranging from quantum mechanics \cite{cohen1977} to quantum field theories \cite{zee2003}, and Einstein's theory of relativity \cite{misner1973}. In general, the gauge theories that describe the fundamental interactions of Nature can be quantized in a canonical way \cite{weinberg1996-1}, i.e. starting from a classical theory and somehow promoting the classical variables to quantum field operators. More deeply, gauge theories are based on the concept of \emph{symmetry}. All forces among particles in Nature are identified by a structural group, which determines the gauge invariance of the theory, in the sense that a solution of the theory equations is still a solution under the action of the gauge group \cite{baez1994}. 

Although this approach works intimately well for particle interactions, something more devious happens when attempting to canonically quantize general relativity \cite{rovelli2004, baez1994}, mainly for two reasons \cite{giesel2013}: First, a theory of quantum gravity seems to require a non-perturbative, or background independent, quantization, since the metric of the manifold becomes itself a dynamical variable that interacts with the presence of mass. Second, it is not a trivial task to identify the gauge group of general relativity, which is usually believed to be the group of space-time diffeomorphisms. Under these assumptions, it becomes rather hard to choose the proper physical observables of the theory \cite{rovelli2011, thiemann2007}. 

Loop quantum gravity (LQG) is a non-perturbative canonical quantization of gravity, where Einstein's equations are described in terms of a SU(2) Yang-Mills gauge theory \cite{rovelli2004}. The most remarkable (and maybe not completely unexpected) result from LQG is that the geometry at the quantum level is discrete \cite{rovelli1995}. In the theory, one defines area and volume operators that have discrete spectra and minimal values. The orthonormal basis states related to the diagonalization of these observables are graphs called spin networks. The vertices of a spin network are labelled by representations of the SU(2) group, and the graph is set on a metric-independent manifold. The dynamics of the theory is specified by a sequence of (allowed) moves that transform the graph from an initial to a final quantum geometrical state. The superposition of all the possible histories determine the quantum state of a space-time \cite{thiemann2007}.

With these ideas in mind, the authors in \cite{ashtekar2003} introduced polymer quantization. Polymer quantization is a toy model proposed to study how semiclassical states can arise from the full theory of quantum gravity, and relies on assumptions similar to the construction of LQG. In practice, it is a representation of the Weyl algebra unitarily inequivalent to the Schr\"odinger's one, which succeeded in describing the kinematics and dynamics of a one-dimensional quantum system on a discretized version of the real line. In Chapter \ref{polymerico} we will study quantum entropy in the context of polymer quantization, investigating whether the entropic predictions of different quantizations are in accordance. 

\section{Discussion}
The short introduction to Loop Quantum Gravity concludes this introductory Chapter. While no new physics has been introduced in these pages, we covered most of the basic definitions that will be employed in the following Chapters and give a taste of the various flavors of the Thesis' topics.

\begin{savequote}[8.5cm] 
\sffamily 
``But Gravity always wins.''  
\qauthor{Radiohead - Fake Plastic Trees} 
\end{savequote}

\chapter{Photons -- Phases and Experiments}
\graphicspath{{Photons}}
\label{GRchapter}

In \ref{pop} we have analyzed the meaning of polarization rotation in Minkowski space-time. In this Chapter we look at photons traveling in a general gravitational field \cite{misner1973,chandrasekar1998}. In particular, gravity causes the polarization of photons to rotate and a meaningful evaluation of the rotation can be achieved only by an appropriate definition of reference frames \cite{brodutch2011b}. The problem of comparison of reference frames is not new, and it has been widely analyzed in the context of quantum information \cite{bartlett2007}. The encoding of quantum information into the degrees of freedom of a physical system always requires a choice of frame where the encoding acquires informational meaning \cite{bruss2007}. Then, exchange of quantum information runs parallel to exchange of information about reference frames between sender and receiver. Partial knowledge of reference frames can lead to loss of communication capacity, and to mistakes in identifying the information content of a physical system \cite{brodutch2011}.

Precise understanding and estimation of the change in photon polarization can be used for tests of relativity \cite{feiler2009}, by sending signals between earth and satellites in orbit. Furthermore, these effects must be accurately evaluated when dealing with the implementation of quantum protocols in space \cite{ursin2009,ursin2007,bonato2009}, such as quantum key distribution \cite{gisin2002}. In principle, in a curved space-time the sender needs to fix a reference frame at each point of the trajectory, which is practically impossible, and exchange this information with the receiver in order to measure the signal appropriately. Therefore, all these implementations hide an expensive price to pay: Finding a realistic way to share a reference frame between the source of the signal and the receiver. 

An answer to this problem was given in \cite{brodutch2011b}, where the authors introduced a natural gauge convention that fixes a set of rules to define reference frames without the need of communication between the parties. Here we present an alternative solution to the problem by determining gauge-invariant trajectories of the photons along which the polarization rotation can be precisely calculated irrespective of the choice of reference frame \cite{brodutch2011}.  

We begin this Chapter explaining concisely the mathematical treatment of photons in general relativity, extending the meaning of polarization rotation to curved backgrounds.  We then show how one can look at this problem from a geometrical point of view and demonstrate gauge-invariant aspects of gravity-induced polarization rotation along closed trajectories both in three-dimensional static projections of the four-dimensional space-time and in the entire four-dimensional manifold. To conclude the analysis we review a number of experimental proposals aimed at testing general relativity at new scales, which are prominently based on some of the concepts derived in this Chapter. 

In the discussion we will make use of abundant terminology and elements taken from differential geometry. While we use \cite{baez1994} and \cite{frankel1999} as primary sources, Appendix \ref{appDiffGeom} provides a short introduction to the relevant concepts. In the following we adopt the $(-,+,+,+)$ signature convention for the metric (\ref{appMetric}).

\section{General relativitistic effects on photon states}
Our kinematic description of photons in a gravitational field relies on the short wave, or geometric optics limit, assumption \cite{born1999}. We assume that the wavelength of the particle is much smaller than the minimum value between the typical curvature radius and some distance taken large enough to ensure that the values of the amplitude, polarization and wavelength vary significantly in its range\footnote{Analogously, the wave period must be much shorter than the time scales involved in the process.}. This allows to adopt the first post-eikonal approximation \cite{misner1973}: Such approximation is an expansion of the source-free Maxwell equations in empty space, based on the geometric optics limit and on the implicit assumption that the electromagnetic field is weak enough not to experience self gravitational interaction \cite{fayos1982}. Then, the more complicated laws of wave propagation in the space-time reduce to the amplitude of the wave being transported along the photons world lines \cite{thorne2014}. At the first-order expansion, photons are approximated to point particles that follow null trajectories with a tangent 4-momentum,
\beq
\ak \to \ak \cdot \ak = \ak^2 = 0\,,
\eeq
and carry a transversal, $\ak \cdot \af =0$, space-like polarization vector,
\beq
\af \to \af \cdot \af = \af^2 = 1\, .
\eeq
The momentum and polarization vectors are parallel transported along the trajectory,
\beq
\label{pte}
\nabla_{\ak} \ak = 0\,,\qquad \nabla_{\ak} \af = 0\,,
\eeq
where a generic $\nabla_{a} b$ corresponds to the covariant derivative of $b$ in the direction of $a$ as explained in \ref{covdef}. An interesting future line of research consists of analyzing the gravitational effects on the light polarization if one where to proceed using the full classical Maxwell equations instead (for an example of such a description in a uniform gravitational field, see \cite{acedo2012}).

A space-time manifold can always be covered by patches that locally look like a Minkowski space-time \cite{baez1994}. Hence, at any point of the space-time, one can define an orthonormal \emph{tetrad}
\beq
\{ \ae_0 , \ae_1, \ae_2, \ae_3  \} \,,
\eeq
where the observer is at rest. Then the local components of the momentum and polarization vectors $\ak^\mu,\, \af^\mu$ are defined in this tetrad. 

In the following discussion we consider stationary gravitational fields. A gravitational field is \emph{stationary}, or \emph{constant}, if it is possible to choose a reference frame where all the components of the metric tensor are independent of the time coordinate $x^0$ \cite{landau1971}. In other words: On a stationary space-time, after a choice of reference frame (or, more poetically, after a choice of \emph{time}) one can select three-dimensional space-like surfaces $\Sigma_3$ all equipped with the same metric. Practically, it is possible from a tetrad, using the $1+3$ Landau-Lifshitz formalism, to construct a \emph{triad} foliating the space-time into space-like surfaces. This three-dimensional projection of the space-time is defined by the map
\beq
\label{pfmap}
\pi_3 : \ma \to \Sigma_3 \, ,
\eeq
where $\ma$ is the four-dimensional space-time manifold and $\Sigma_3$ is a three-dimensional spatial space. Consequently, any four-vector is transformed into the three-vector corresponding to our choice of foliation by a push-forward map $\pi_{{3},*} \ak = {\bf k}$ (see \cite{baez1994,nouri-zonoz1999} for details). This is equivalent to dropping the time-like coordinate of the four-vector. In the following we refer to this as the projection of a vector on the static surface $\Sigma_3$.

Hence, the local (static) description of a photon state can be provided by the standard reference frame
\beq
\{ \hbb_1(\hbk), \hbb_2(\hbk), \hbk \} \, .
\eeq
as explained in \ref{pop}. The polarization vectors $( \hbb_1(\hbk), \hbb_2(\hbk))$ specify a linear polarization basis and are dependent upon the momentum vector $\hbk$. One can rewrite the parallel transport equations in this local representation of the space-time mapping the four-dimensional covariant derivative to the correspondent three-dimensional expression. Then
\beq
\label{3Dequa}
\frac{\text{D} \hbk}{\d \lambda} = \bm{\Omega} \times \hbk \,, \qquad \frac{\text{D} \hbf}{\d \lambda} = \bm{\Omega} \times \hbf \,,
\eeq
where the derivation of these formulas is provided in \cite{fayos1982,nouri-zonoz1999,sereno2004}. The parameter $\lambda$ is called \emph{affine parameter} and determines the trajectory while the symbol $\text{D}$ identifies the three-dimensional covariant derivative. 

The polarization and momentum vector from Eq.(\ref{3Dequa}) undergo an evolution known as \emph{gravitational Faraday rotation} \cite{plebanski1960, sereno2005, ruggiero2007}, which causes the momentum-polarization triad to rotate with angular velocity $\bm{\Omega}$ given by
\beq
\label{Omegatutto}
\bm{\Omega} = 2 \bm{\omega} -(\bm{\omega} \cdot \hbk) \hbk - {\bf E}_g \times \hbk \,.
\eeq
This is analog to the electromagnetic Faraday effect, which explains how a polarized electromagnetic wave that travels through plasma rotates under the action of a magnetic field \cite{rybicki1979}. In contrast with the classical Faraday effect, in this case the rotation is purely a geometric effect where ${\bf E}_g$ is the gravitoelectric field term, and $\bm{\omega} = - {\bf B}_g$ plays the role of the gravitomagnetic field term. The names follow from the quasi-Maxwell form of the Einstein equation for a stationary space-time, derived from the $1+3$ Landau-Lifshitz formalism \cite{nouri-zonoz1999}. Both terms are related to the elements of the projected three-dimensional metric shown in \ref{appMetric} as
\beq
{\bf E}_g = - \frac{\nabla g_{0,0}}{2 g_{0,0} }\, ,\qquad {\bf B}_g = \nabla \times {\bf g} \,,
\eeq
where {\bf g} is a three-dimensional vector with components $g_m = - g_{0,m}/g_{0,0}$.

\section{Geometric phase}
In this section we look at the problem of polarization rotation from a geometric perspective. Starting from the parallel transport equations, we initially derive an alternative equation for the polarization rotation for an arbitrary choice of the polarization basis. We then project it to a static space-time and show that the phase accrued by the photon state in the sense of Eq.(\ref{helphase}) depends on a Machian term and on a reference frame term. The main result of this section is that the phase, along a closed trajectory on $\Sigma_3$, is invariant under the choice of a different reference frame. Moreover, we extend the discussion to general four-dimensional space-times and demonstrate that for closed paths constructed \emph{ad hoc} the same argument still holds.

\subsubsection{An equation for the polarization rotation}
We begin by defining at each point of the trajectory a local orthonormal tetrad (or \emph{vierbein}) $\{ \ae_0, \ae_1, \ae_2, \ae_3 \}$ such that the momentum vector is locally given by 
\beq
\ak = k \ae_0 + k \ae_3 \longrightarrow \ak^\mu = (k,0,0,k) \,.
\eeq
The temporal gauge $\af \cdot \ak = 0$ simplifies things quite a lot, and allows us to set 
\beq
\ab_1 = \ae_1 \,, \ab_2 = \ae_2 \, ,
\eeq
where the local polarization basis $\{ \ab_1, \ab_2 \}$ is chosen according to some procedure. Hence, the general form of the real linear polarization four-vector at some point of the trajectory is 
\beq
\label{generalf}
\af = \cos \chi \ab_1(\ak) + \sin \chi \ab_2(\ak)\, ,
\eeq
and the phase $\chi$ explictly appears in the formulation. We set the initial phase $\chi = 0$ and thus
\beq
\af = \ab_1
\eeq
at the starting point of a trajectory. Using the parallel transport equations from Eq.(\ref{pte}), we can derive a differential equation for the polarization rotation. Since along the trajectory we generically have that
\begin{align}
\sin \chi = \af \cdot \ab_2\,, \qquad \cos \chi = \af \cdot \ab_1 \,,
\end{align}
one can calculate the following covariant derivative 
\begin{align}
\nabla_\ak (\sin \chi) = \nabla_\ak (\af \cdot \ab_2) \,.
\end{align}
Using differential geometry rules this transforms as
\begin{align}
\nabla_\ak (\sin \chi) &= \text{d} \sin \chi (\ak) = \sum_i \ak^i \frac{\partial \sin \chi }{\partial x^i} \nonumber \\
&= \sum_i \frac{\partial x^i}{\partial \lambda} \frac{\partial \sin \chi }{\partial x^i} = \frac{\text{d} \sin \chi}{ \text{d} \lambda}\\
&= \cos \chi \frac{\text{d} \chi}{\text{d} \lambda} = (\af \cdot \ab_1) \frac{\text{d} \chi}{\text{d} \lambda} \nonumber \,,
\end{align}
and
\beq
\nabla_\ak (\af \cdot \ab_2) = (\nabla_\ak \af )\cdot \ab_2 + \af \cdot \nabla_\ak \ab_2 = \af \cdot \nabla_\ak \ab_2\, .
\eeq
The final equation is therefore given by
\beq
\label{4deq}
\frac{\text{d} \chi}{\text{d} \lambda} = \frac{1}{\af \cdot \ab_1} \af \cdot \nabla_\ak \ab_2 \,.
\eeq
\subsubsection{A static equation for the polarization rotation}
One can project the polarization equation derived above to the space-like surface $\Sigma_3$ by constructing the relevant vectors using the push-forward map extracted from Eq.(\ref{pfmap}):
\beq
\pi_{{3},*} \af \to \hbf\, ,\qquad \pi_{{3},*} \ak \to \hbk\,, \qquad \pi_{{3},*} \ab_{1,2} \to \hbb_{1,2}\,.
\eeq
Consequently, in three spatial dimensions we have the following equation:
\begin{align}
\label{seqpr}
\frac{\text{d} \chi}{\text{d} \lambda} &= \frac{1}{\hbf \cdot \hbb_1} \frac{\text{D}(\hbf \cdot \hbb_2)}{\text{d} \lambda}\nonumber\\
&= \frac{1}{\hbf \cdot \hbb_1} \left( \frac{\text{D} \hbf}{\text{d} \lambda} \cdot \hbb_2 + \hbf \cdot \frac{\text{D} \hbb_2}{\text{d} \lambda} \right) \\
&= \bm{\omega}\cdot \hbk + \frac{1}{\hbf \cdot \hbb_1} \hbf \cdot \frac{\text{D} \hbb_2}{\text{d} \lambda}\nonumber \,.
\end{align}
The two terms that compose the phase evolution equation are the Machian effect term $\bm{\omega}\cdot \hbk$, and the reference-frame contribution, respectively. The first quantity corresponds to the original Machian effect that was postulated in \cite{godfrey1970} and specifies how the \textquoteleft gravitational dragging' caused by a rotating mass affects the polarization. The second term is due to the variation of the polarization basis along the trajectory. 

The geometric meaning hidden behind this equation is rather interesting. First introduce a basis of 1-forms $(\eta^1, \eta^2, \eta^3)$, dual to the orthonormal polarization basis (or triad) $\hbb=(\hbb_1, \hbb_2, \hbk)$ at every point of the trajectory, such that $\hbb_i \eta^j = \delta_i^j$. One can write a matrix of connection 1-forms $\b$ using the 1-form basis and with the help of the Ricci rotation coefficients $\beta_{j,l}^{i}$ as \cite{misner1973, frankel1999}
\beq
\b^{i}_{l} = \beta_{j,l}^i \eta^{j} \,.
\eeq
For sake of clarity, remember that the polarization vector can be rewritten as
\beq
\hbf = f^1 \hbb_1 + f^2 \hbb_2 = \cos \chi \hbb_1 + \sin \chi \hbb_2 \longrightarrow f^i = \left( \cos \chi, \sin \chi, 0 \right) \,.
\eeq
We can then calculate the covariant derivative of $\hbf$; we will explicitly show how to do it component-wise. The general formula for the covariant derivative of a vector field ${\bf v}$ in $n$-dimensions along a direction ${\bf k}$ is 
\begin{equation}
(\nabla_{\bf k} {\bf v})^i = \left ( \frac{\partial v^i}{\partial x^j} + \beta^i_{j,k} v^k \right) k^j \,.
\end{equation}
For $i = 1$ and keeping in mind that in this case
\beq
k^j = \frac{\d x^j}{\d \lambda}\,,
\eeq
one easily finds that
\begin{align}
\left(\frac{\text{D} f}{\d \lambda}\right)^1 &= \left( \frac{\partial \cos \chi}{\partial x^j} + \beta^1_{j,k} f^k \right) \frac{\d x^j}{\d \lambda} \nonumber\\
&=\left( -\sin \chi {\frac{\d \chi}{\d x^j}} + \beta^1_{j,k} f^k \right) \frac{\d x^j}{\d \lambda}\\
&= -f^2 \frac{\d \chi}{\d \lambda} + \beta^1_{j,1} f^1 \frac{\d x^j}{\d \lambda} + \beta^1_{j,2} f^2 \frac{\d x^j}{\d \lambda}\nonumber \\
&= - f^2 \left( \frac{\d \chi}{\d \lambda} - \beta^1_{j,2} k^j \right) +  \beta^1_{j,1} f^1 k^j \nonumber \,.
\end{align}
Applying the same argument for the second component, we obtain
\beq
\left(\frac{\text{D} f}{\d \lambda}\right)^2 = f^1 \left( \frac{\d \chi}{\d \lambda} + \beta^2_{j,1} k^j \right) + \beta^2_{j,2} f^2 k^j \,.
\eeq
On the other hand, the third and last component is simply equal to
\begin{align}
\left(\frac{\text{D} f}{\d \lambda}\right)^3 &=  \left( \frac{\partial f^3}{\partial x^j} + \beta^3_{j,k} f^k \right) \frac{\d x^j}{\d \lambda} \nonumber\\
&= \beta^3_{j,1} f^1 k^j + \beta^3_{j,2} f^2 k^j\, .
\end{align}
Taking into account that for any orthonormal frame such as ours, the matrix of the connections is antisymmetric \cite{frankel1999}
\beq
\label{antisc}
\beta^i_j = \beta_{i,j} = - \beta_{j,i}\,,
\eeq
we can use the antisymmetry of the Ricci coefficients, i.e. $\beta^2_{j,1} = - \beta^1_{j,2}$, to rewrite the covariant derivative of $\hbf$ as
\begin{align}
\label{codef}
\frac{\text{D} \hbf}{\d \lambda} = \left( - f^2 \hbb^1 + f^1 \hbb^2 \right) \left( \frac{\d \chi}{\d \lambda} - \beta^1_{j,2} k^j \right) + \beta_{j,1}^1 f^1 k^j \hbb_1 \nonumber \\
+ \beta_{j,2}^2 f^2 k^j \hbb_2 + \left(\beta^3_{j,1} f^1 k^j + \beta^3_{j,2} f^2 k^j \right) \hbk \,.
\end{align}
What is the physical meaning of the $\beta^1_{j,1}$, $\beta^2_{j,2}$ coefficients? From Eq.(\ref{antisc}) it follows straightforwardly that $\beta_{1,1} = \beta_{2,2} =0$ and therefore also
\beq
\beta^1_{j,1} = \beta^2_{j,2} = 0\, .
\eeq
Finally, since $k^i = (0,0,k)$, Eq.(\ref{codef}) becomes 
\begin{align}
\label{dfdl}
\frac{\text{D} \hbf}{\d \lambda} = \left( - f^2 \hbb^1 + f^1 \hbb^2 \right) \left( \frac{\d \chi}{\d \lambda} - \beta^1_{3,2} k \right) + \left(\beta^3_{3,1} f^1 k + \beta^3_{3,2} f^2 k \right) \hbk \,.
\end{align}
Now we compare this last equation with the three-dimensional stationary propagation equation (\ref{3Dequa}) for the polarization vector \cite{fayos1982}
\beq
\label{Ompe}
\frac{\text{D} \hbf}{\d \lambda} = \bm{\Omega} \times \hbf \,.
\eeq
From this equation one finds that
\beq
\left( \frac{\text{D} \hbf}{\d \lambda} \right)^i = \left( - \Omega^3 f^2, \Omega^3 f^1 , \Omega^1 f^2 - \Omega^2 f^1 \right) \,,
\eeq
so the immediate goal now is simply to extract $\Omega^3$ from Eq.(\ref{Omegatutto}), which is rather easy
\beq
\Omega^3 = \bm{\Omega} \cdot \hbk = 2 \bm{\omega} \cdot \hbk - \bm{\omega} \cdot \hbk = \bm{\omega} \cdot \hbk \,. 
\eeq
Using this last expression in Eq.(\ref{dfdl}) we have that
\beq
\Omega^3 f^2 = f^2 \left( \frac{\d \chi}{\d \lambda} - \beta^1_{3,2} k \right)\, ,
\eeq
and an alternative equation for the polarization rotation follows straightforwardly \cite{brodutch2011}:
\beq
\frac{\d \chi}{\d \lambda} = \bm{\omega} \cdot \hbk + \beta^1_{3,2} k \,.
\label{chieq}
\eeq
This equation shows manifestly the dependence of $\Delta \chi$, intended as the phase variation along the trajectory, on the geometrical properties of the space manifold. For completeness, these considerations also lead to the equalities:
\beq
\Omega^1 = \beta_{3,2}^3 k\,, \qquad \Omega^2 = - \beta^3_{3,1} k \,.
\eeq

\subsection{Gauge invariance of the phase $\chi$}
\label{gaugesec}
Having now a convenient way to describe the polarization rotation in a time-independent three-dimensional surface, we can show that any choice of polarization basis along a closed (spatial) trajectory does not affect the value of $\Delta \chi$. On a three-dimensional oriented Riemannian manifold, given a trajectory with a tangent vector $\hbk$, one can always assign at each point a two-dimensional tangent vector space with local basis $\{ \hat{\bf e}_1, \hat{\bf e}_2 \}$ \cite{frankel1999}. Since each two-dimensional real vector $\hat{\bf e}$ belongs to the space $\RR^2$, which is homeomorphic to $\mathbb{C}^1$, it is clear that the transition function between two charts is given by the complex number
\beq
e^{i \psi(\lambda)} \,,
\eeq
that determines a rotation of the basis vectors. In terms of differential geometry this is equivalent to saying that we can define along the trajectory a \emph{complex line bundle} normal to the direction of movement. This is a special case of vector bundle (\ref{bundlesapp}), the connection of which is a single 1-form, in this instance equal to
\beq
\bar{\beta} = \beta^1_{3,2} k \d \lambda \,,
\eeq
defined similarly to the usual treatment of geometric phase \cite{frankel1999, berry1987}. The structure group for this line bundle is thus SO(2). Physically, this last statement means that the freedom of choosing the polarization frame $\left( \hbb_1, \hbb_2 \right)$ at every point of the trajectory is represented by a SO(2) rotation ${\bf R}_{\hbk}(\psi(\lambda)) = \text{exp}(i \psi(\lambda))$ \cite{frankel1999}. 
\begin{figure}[tbp]  
\centering
\setlength{\unitlength}{1cm}
\includegraphics[width=12\unitlength]{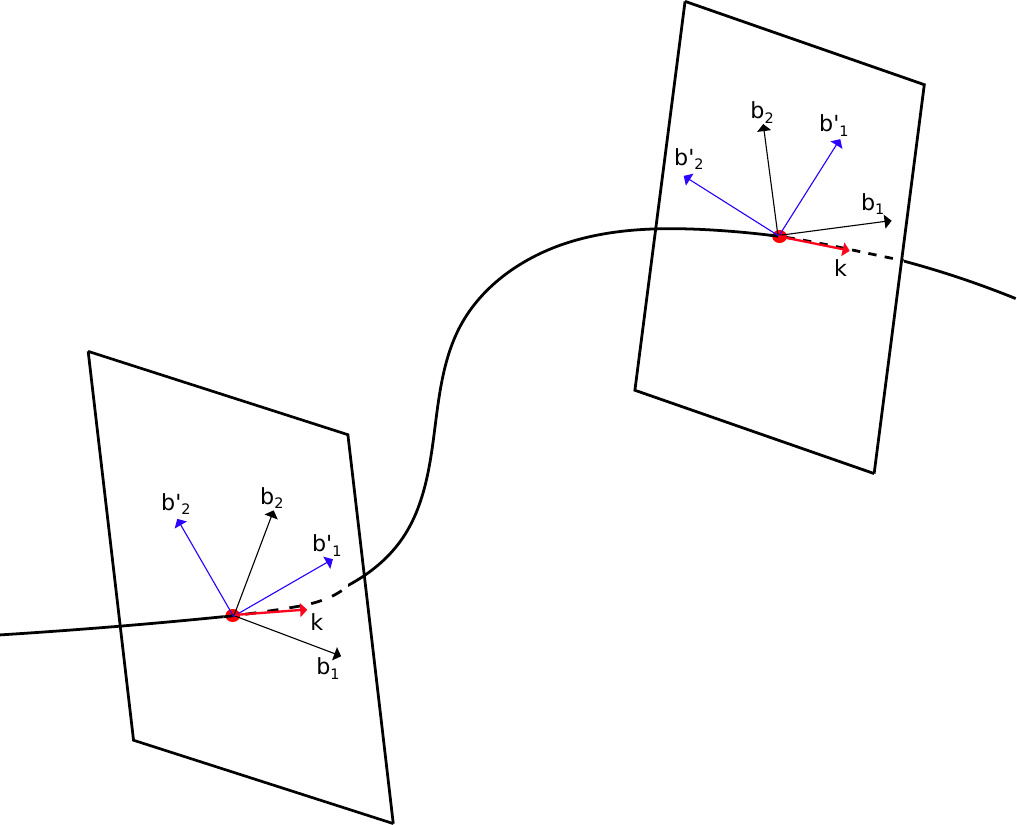}
\caption{Rotation of the basis vectors of the tangent bundle at two different points of the trajectory.}
\label{rotgauge}
\end{figure}

Under the change of basis ${\bf R} \to \hat{\bf e}' =  \hat{\bf e} {\bf R}$, the connection transforms as \cite{frankel1999, baez1994}
\beq
\bar{\beta} \to {\bf R}^{-1} \bar{\beta} {\bf R} + {\bf R}^{-1} \d {\bf R} \,,
\eeq
and for its only component this corresponds to
\beq
\beta^1_{3,2} k \to e^{- i \psi(\lambda)} \beta^1_{3,2} k\, e^{i \psi(\lambda)} + e^{- i \psi(\lambda)} e^{ i \psi(\lambda)} \frac{\d \psi( \lambda)}{\d \lambda} \,. 
\eeq
Therefore, the gauge transformation modifies  Eq.(\ref{chieq}) into 
\beq
\label{gaugechi}
\frac{\d \chi}{\d \lambda} \to \bm{\omega} \cdot \hbk + \beta_{3,2}^1 k + \frac{\d \psi(\lambda)}{\d \lambda} \,.
\eeq
Since we are now dealing with a purely spatial surface, it does no harm to consider a closed trajectory \cite{landau1971}. For a closed trajectory the resulting phase is gauge-invariant, in the sense that $\Delta \chi$ does not depend upon the choice of the frame at any point on the trajectory, as long as one employes the same reference frame to measure polarization at the beginning and at the end of the closed curve. To prove this assertion, it is enough to notice that the last term of Eq.(\ref{gaugechi}) is a total differential and drops out upon the integration on a closed contour $\gamma$, i.e. $\oint_\gamma \d \psi(\lambda) = 0$ and
\beq
\label{connchi}
\Delta \chi = \oint_{\gamma} \bm{\omega} \cdot \hbk \d \lambda + \oint_{\gamma} \bar{\beta} \,.
\eeq
Thus, the polarization phase is a \emph{topological invariant} with respect to the reference frame, arising, as often happens in physics, by integrating a local geometric quantity.

We can formalize this with the help of the bundle 2-form curvature. In general, given a connection $\beta$ the matrix of curvatures is defined as $\theta := \d \beta + \beta \wedge \beta$ (see \ref{curva}), and component-wise this can be rewritten as
\beq
\theta^{i}_{\,\,\, j} = \d \beta^i_{\,\,\, j} + \beta^i_{\,\,\, k} \wedge \beta^k_{\,\,\, j} \,.
\eeq
In the particular case of local orthonormal frames for a two-dimensional Riemannian manifold, since only the off-diagonal terms survive in the matrix of local connections, i.e. $\beta \wedge \beta = 0$ \cite{frankel1999}, the curvature matrix is exactly given by $\theta = \d \beta$ and for the SO(2) bundle it reduces to the single 2-form
\beq
\bar{\theta} = \d \bar{\beta} \,.
\eeq
By using Stokes' theorem, which states that the surface integral of the differential of a $n$-form is equal to the integral of the same $n$-form along the boundary of the surface \cite{baez1994}, the reference-frame term in Eq.(\ref{connchi}) can be rewritten as a surface integral of the bundle curvature as
\beq
\oint_\gamma \bar{\beta} = \int \int_{\text{Sur}_\gamma} \bar{\theta} \,,
\eeq
from which follows:
\beq
\label{chicurv}
\Delta \chi = \oint_{\gamma} \bm{\omega} \cdot \hbk \d \lambda + \int \int_{\text{Sur}_\gamma} \bar{\theta} \,.
\eeq
If we are working in any stationary space where the gravitational body is not rotating and $\bm{\omega} = 0$, the phase will only be a function of the geometrical properties of the manifold embedded inside the closed trajectory of the photon. 

For a Schwarzschild space-time there exists an explicit choice of local reference frames, the \emph{Newton Gauge} \cite{brodutch2011b}, where the standard polarization basis vectors are chosen to prevent the necessity of references to parallel transport or communication between the observers. The Newton gauge is designed to give a zero phase along any trajectory in the Schwarzschild space-time. As a result of the independence of the choice of reference frame, or gauge invariance, of Eq.(\ref{chicurv}), a photon traveling on a closed path in a stationary projection of the Schwarzschild space-time does not accrue a gravitationally-induced phase, regardless of the gauge convention.

\subsubsection{General four-dimensional space-time}
This argument can be taken a step further and generalized to a four-dimensional space-time. From Eq.(\ref{generalf}) we already know that
\beq
\af^\mu = (0, \cos \chi, \sin \chi, 0) \,,
\eeq
and we can calculate the covariant derivative $\nabla_{\ak} \af$ easily, starting with $\mu = 0 \longrightarrow f^0 = 0$,
\beq
\left[ \beta^0_{\nu, \rho} f^\rho \right] k^\nu = \left(\beta^0_{0,1} f^1 + \beta_{0,2}^0 f^2 \right) k^0 + (\beta_{3,1}^0 f^1 + \beta^0_{3,2} f^2)k^3 \,.
\eeq
The calculations for the other indices are maybe boring, but can be comfortably done without too much effort:
\begin{align}
\left( \left[ \frac{\partial f^\mu}{\partial x^\nu} + \beta^\mu_{\nu, \rho} f^\rho \right] k^\nu\right)_{\mu =1} &= -f^2 \frac{\d \chi}{\d \lambda} + (\beta^1_{0,0} f^0 + \beta^1_{0,2} f^2)k^0 + (\beta^1_{3,0} f^0 + \beta_{3,2}^1 f^2) k^3\nonumber \\
&=  -f^2 \frac{\d \chi}{\d \lambda} + (\beta^1_{0,2} f^2)k^0 + (\beta_{3,2}^1 f^2) k^3 \,,\\
\left( \left[ \frac{\partial f^\mu}{\partial x^\nu} + \beta^\mu_{\nu, \rho} f^\rho \right] k^\nu\right)_{\mu =2} &= f^1 \frac{\d \chi}{\d \lambda} + (\beta^2_{0,1} f^1) k^0 + (\beta^2_{3,1} f^1)k^3\, ,\\
\left( \left[ \frac{\partial f^\mu}{\partial x^\nu} + \beta^\mu_{\nu, \rho} f^\rho \right] k^\nu\right)_{\mu =3} &= (\beta^3_{0,1} f^1 + \beta^3_{0,2} f^2) k^0 + (\beta^3_{3,1} f^1 + \beta^3_{3,2} f^2)k^3 \,.
\end{align}
Thus the covariant derivative of $\af$ is given by
\begin{align}
\nabla_{\ak} \af &= \left( - f^2 \bm{\ae}_1 + f^1 \bm{\ae}_2 \right) \left( \frac{\d \chi}{\d \lambda} - (\beta^1_{0,2} + \beta^1_{3,2})k \right) \nonumber \\
&+ (k(\beta_{0,c}^0 + \beta_{3,c}^0)f^c)\bm{\ae}_0 + (k(\beta^3_{0,c} + \beta^3_{3,c})f^c)\bm{\ae}_3 \,,\qquad c=1,2 \,,
\end{align}
where we have used again the antisymmetry (\ref{antisc}) for the elements of the matrix of connection 1-forms. From the parallel transport condition $\nabla_{\ak} \af = 0$, we can see that
\beq
(\beta^0_{0,c} + \beta^0_{3,c})f^c = (\beta^3_{0,c} + \beta^3_{3,c}) f^c = 0\,,
\eeq
and 
\beq
\label{generalchi}
\d \chi = (\beta^1_{0,2} + \beta^1_{3,2}) k \d \lambda := \bar{\Omega} \,,
\eeq
where now $\bar{\Omega}$ plays the role of a bundle connection. 

As long as time-travel remains part of the science-fiction realm, we cannot have a closed trajectory in space-time. Nevertheless, to obtain a gauge-invariant result one can consider two future-directed trajectories that begin and end in the same two space-time points. The careful reader will easily notice that this layout is similar to the two arms of a Mach-Zender interferometer \cite{born1999,scully1997}. To avoid phase variations induced by other factors, the observers align the initial and final propagation directions of the beams and agree on common rules to define the standard polarization basis. In analogy with the previous case, the bundle curvature is the differential of the connection
\beq
\bar{\Theta} = \d \bar{\Omega} \,.
\eeq
Integrating Eq.(\ref{generalchi}) for two open curves $\gamma_1$ and $\gamma_2$ such that $\gamma = \gamma_1 \times (-\gamma_2)$ is closed, and using again the Stokes' theorem, one finds that the phase is equal to
\beq
\Delta \chi = \int_{\gamma_1} \bar{\Omega} + \int_{- \gamma_2} \bar{\Omega} =  \int_{\gamma_1} \bar{\Omega} - \int_{\gamma_2} \bar{\Omega} = \oint_{\gamma} \bar{\Omega} = \int \int_{\text{Sur}_\gamma} \bar{\Theta} \,.
\eeq
Note that in this case it is not necessary to consider a stationary space-time (which is necessary instead for our definition of the $\Sigma_3$ projection) and the result above holds for any general space-time \cite{brodutch2011}.

\section{Discussion and experimental applications}
\label{GRexpe}
In this Chapter we showed that the reference frame term in Eq.(\ref{seqpr}) is responsible for a gauge-independent geometric phase, or polarization rotation, accrued by the photon state on a closed trajectory.

Proper understanding of polarization behavior and polarization rotation measurements is a necessary step when quantum technology is meant to be used for precisions tests of relativity, for quantum information protocols in the presence of gravity and, more at the fundamental level, to design feasible tests for a theory of quantum gravity. A wide variety of potential tests of fundamental physics that can be conducted with satellites in Earth orbit and quantum communication protocols are reviewed in \cite{rideout2012}. The main reason to have experiments performed with the help of satellites is because they allow to reach length-scales and velocities that are practically unreachable on the ground. 

Tasks directly related to the contents of this Chapter include detection of the special and general relativistic effects on the polarization of single photons exchanged between Earth surface and an artificial satellite moving in low earth orbit (LEO), an elliptical orbit about the Earth with altitude up to $2000$ Km. In a number of realistic scenarios, the frame-dragging effects are dominant over the Machian effects, see \cite{brodutch2011b} for the analysis. The expected optical effects are very small. Two examples presented in \cite{rideout2012} illustrate the order of magnitude. Considering the gravitational field produced by the Earth and taking a photon emitted from the LAGEOS satellite orbit ($12,270$ Km) \cite{ciufolini2004} and detected at infinity, one obtains a rotation of $\Delta \chi \approx 39$ arc msec in a single run. In the scattering scenario of a photon emitted along the axis of rotation of the Earth, the resulting phase is minuscule, $\Delta \chi \approx 3 \times 10^{-7}$ arc msec. These results are not gauge-invariant in the sense illustrated previously, in fact to model a closed path one needs at least three nodes. The tiny measurable effects set these experiments beyond reach with the current technology.

We conclude this section describing a proposal for a satellite-based quantum optics experiment conceivable with today's technology, and offering a short overview about more exotic experiments that go beyond the detection of purely relativistic effects.

\subsubsection{The optical COW experiment}
Realistic results in a near-term timeframe are expected to be provided by a satellite-based proposal of the famous Colella, Overhauser, and Werner (COW) experiment. 
The COW experiment \cite{colella1975} is a well-established experiment used to verify the classical equivalence principle in the quantum limit through observation of the phase shift in a neutron beam interferometer induced by a gravitational potential. While the standard experiment is conducted using neutron beams and a rotating interferometer, an equivalent optical setup can be designed making use of optical fibers and a satellite in low earth orbit. This is particularly appealing since a satisfying correspondence between theory and experiment \cite{kaiser2006} has not been achieved yet: Hence, a test conducted at larger length-scale could shade some light on this discrepancy.

To perform the test, at the Earth's surface an optical beam of wavelength $\lambda$ is split by a semitransparent mirror: While one path is delayed on the ground by an optical fibre of length $l \approx 6$ Km, the second path is transmitted to the satellite and then coherently recombined at the end to construct an interferometer. The difference in Newtonian gravitational potential between the two paths causes a different phase shift in the two fibers. Using this optical setup one can analyze gravitational redshift effects in the context of quantum optics. In a weak gravitational field, the phase difference for a wavelength $\lambda = 800$ nm is approximately equal to \cite{rideout2012}:
\beq
\Delta \phi = \frac{2 \pi l}{\lambda} \frac{g h}{c^2} \sim 2 \, \text{rad} \,,
\eeq
where $g$ is the free fall acceleration on Earth's surface, $h \sim 400$ Km is the satellite altitude, and $c$ is the speed of light. It is interesting to point out that the large speed of the satellite may introduce rotational effects since the interferometer path covers an area, caused by the orbital movement of the satellite. Moreover, because of the large distances involved the Earth's gravitational field is not constant and the apparatus would detect a general-relativistic phase-shift, with quantum interference caused by the curvature of the space-time.

\subsubsection{Beyond tests of relativistic effects}
The first and most direct extension to the tests of relativistic effects on discrete quantum systems is probing the physics of quantum field theory in accelerating frames, such as testing the Unruh effect \cite{unruh1976}. Other possible experiments achievable making use of satellites include more exotic topics. For instance, one can consider tests of entanglement at relativistic scales, such as long distance Bell tests \cite{bell1964, clauser1969} and Bell tests with detectors in relative motion. A more practical application is the implementation of quantum cryptography protocols in space \cite{ursin2007, ursin2009,bonato2009}, which could allow to implement secure quantum communication on a global scale. Finally, a fascinating direction to undertake is the analysis of possible manifestations of quantum gravity effects at the Planck scale and how to detect these effects \cite{rideout2012}.

\begin{savequote}[8cm] 
\sffamily
``God does arithmetic.''
\qauthor{Carl Friedrich Gauss}
\end{savequote}

\chapter{Gaussian States}
\graphicspath{{G_states}} 
\label{ChapGS}

The basic primitive of quantum computation is the qubit \cite{nielsen2000}. Qubits and their $d$-dimensional generalizations, the qudits, are the abstraction of discrete physical systems, and their quantum states are defined in finite dimensional Hilbert spaces. 
However, not every physical system can be described by such states, and especially in the field of quantum optics \cite{puri2001, scully1997, born1999}, a generalization to infinitely-dimensional Hilbert spaces is required. With this in mind, here we introduce and analyze continuous-variable (or CV) systems \cite{braunstein2003}, which are often important for quantum computation and information protocols on either practical or theoretical grounds \cite{braunstein2005}. In particular, we will focus on Gaussian states \cite{demarie2012, olivares2012, weedbrook2012}, a subset of the CV states that allow for a very simple and precise mathematical characterization. Furthermore, Gaussian states are widely used today for experimental purposes in quantum optics and quantum information \cite{weedbrook2012, gu2009, menicucci2006, demarie2013b}. They also serve as main ingredients for the CV generalizations of the most famous quantum protocols, for instance Gaussian quantum teleportation \cite{bowen2003,braunstein2003, vanloock2000} and Gaussian quantum key distribution \cite{grosshans2003,jouguet2011}. Last but not least, there exist several proposals to realize quantum computation by means of Gaussian states: At first, in \cite{lloyd1999} it was analyzed how to perform Gaussian quantum computation in the circuit-model setup. This was followed by a second approach, based on encoding qubits into quantum harmonic oscillators in order to simplify error correction and increase the structural fault tolerance \cite{gottesman2001}. A third different scheme is the CV cluster-state quantum computation \cite{menicucci2006}, where the information is processed via local measurements on the modes of a Gaussian lattice of entangled oscillators. 

In this Chapter we start presenting the usual phase-space notation for Gaussian states, showing the equivalence between the density matrix and the Wigner functions description. Then we introduce the basic concepts of covariance matrix, symplectic transformations, Williamson's form and symplectic eigenvalues. After that, we demonstrate how to quantify bipartite entanglement for pure Gaussian states making use of the symplectic spectrum and give an explicit example for two coupled oscillators. To conclude, we introduce the graphical calculus for Gaussian states \cite{menicucci2011}, a powerful method to represent pure Gaussian states in terms of their graph properties. 

This short review serves as a necessary introduction to the following sections of the Thesis. In Chapter \ref{CVTO}, we use Gaussian states to construct the CV analog to the Kitaev surface code \cite{kitaev2003}, a 2-dimensional lattice of highly-entangled harmonic oscillators that exhibits topological order \cite{pachos2012, hamma2005}, through simple Gaussian operations that are experimentally feasible today in the laboratory. But Gaussian states prove to be useful also in the context of polymer quantization \cite{ashtekar2003}, Chapter \ref{polymerico}. Their simple mathematical formalism is the perfect playground to demonstrate the convergence properties of entropy for two coupled quantum harmonic oscillators quantized in unitarily inequivalent representations of the Weyl-Heisenberg algebra \cite{demarie2013}. This Chapter, being a short review of a well established topic, follows the structure of \cite{demarie2012}.

\section{Bosonic systems and Gaussian states representations}
\label{bsgsr}

In general, a continuous-variable state is described in terms of observables with continuous spectra associated with an infinitely dimensional Hilbert space \cite{braunstein2005}. The most typical example is a set of $N$ quantum harmonic oscillators, or bosonic modes. Physically, this is equivalent to the noninteracting quantized electromagnetic field in a cavity, where each $i$-th bosonic mode, or harmonic, is associated to a Hilbert space $\mathcal{H}_i$, and the global space is $\mathcal{H} = \mathcal{H}_i^{\otimes N} = \otimes_{i=1}^N \mathcal{H}_i$. In natural units where $\hbar = 1$, and adopting the convention $\omega_i =1$, we assign to each mode a pair of field operators $(\hat{a}_i, \hat{a}^\dagger_i)$ that fulfill the bosonic commutation relation $[ \hat{a}_i, \hat{a}_j^\dagger ] = \delta_{i,j}$ and define the single-mode Hamiltonian $\hat{H}_i = \hbar \left( \hat{a}_i^\dagger \hat{a}_i + \frac{1}{2} \right)$ \cite{weedbrook2012}.

The set of commutation relations of all the $N$ bosonic operator pairs can be rewritten by collecting the bosonic operators into the vector $\hat{\mathbf{b}} := (\hat{a}_1, ..., \hat{a}_N, \hat{a}^\dagger_1,...,\hat{a}^\dagger_N )$ of length $2N$, and defining the \emph{symplectic form} $\mathbf{\Omega}$ \cite{arvind1995}, whose entries are given by the commutators of the elements of $\hat{\mathbf{b}}:$
\ba
\mathbf{\Omega}_{i,j} = [\hat{\mathbf{b}}_i, \hat{\mathbf{b}}_j] \longrightarrow \mathbf{\Omega} = \left (
\begin{tabular}{c c}
$0$ & $\mathbf{I}_{N}$\\
$- \mathbf{I}_{N}$ & $0$\\
\end{tabular} \right) \, ,
\end{align}
with $i,j = 1,..., N$ and ${\bf I}_N$ the $N \times N$ identity matrix. Note that $\mathbf{\Omega}$ is a skew-symmetric matrix with $\mathbf{\Omega}^T = \mathbf{\Omega}^{-1} = \mathbf{\Omega}$. The Hilbert space $\mathcal{H}_i$ of the $i$-th oscillator is spanned by an infinitely-dimensional Fock basis $\{ \ket{n}_i \}_{n=0}^{\infty}$, whose elements are the eigenstates of the number operator $\hat{n}_i = \hat{a}_i^\dagger \hat{a}_i$, explicitly $\hat{n}_i \ket{n}_i = n \ket{n}_i$. The action of the bosonic operators on the basis states is described by the relations $\hat{a}_i \ket{n}_i = \sqrt{n} \ket{n-1}_i$, and $\hat{a}_i^\dagger \ket{n}_i = \sqrt{n+1} \ket{n+1}_i$. Each single-mode \emph{annihilation} operator $\hat{a}_i$ nullifies the corresponding ground state $\ket{0}_i$, i.e. $\hat{a}_i \ket{0}_i = 0$ \cite{cohen1977}.

We can use an alternative set of field operators to describe a $N$ modes CV state. These are the $N$ pairs of quadrature operators $\{ \hat{q}_i, \hat{p}_i \}$, defined in terms of linear combinations of the annihilation and creation operators:
\begin{equation}
\label{aaad}
\hat{q}_i = \frac{1}{\sqrt{2}} (\hat{a}_i^\dagger + \hat{a}_i)\, , \,\,\,\,\,\, \hat{p} = \frac{i}{\sqrt{2}} (\hat{a}_i^\dagger - \hat{a}_i)\, ,
\end{equation}
with canonical commutation relations
\beq
\label{CCR}
[ \hat{q}_i, \hat{p}_j] = i \delta_{i,j} \,.
\eeq 
In analogy with what done before, one introduces a $2N$ vector $\hat{\mathbf{r}} = (\hat{q}_1,...,\hat{q}_N,\hat{p}_i, ..., \hat{p}_N)$ such that the symplectic form $\mathbf{\Omega}$ can be equally defined as
\ba
\mathbf{\Omega}_{i,j} = - i [\hat{\mathbf{r}}_i, \hat{\mathbf{r}}_j] \longrightarrow \mathbf{\Omega} = \left (
\begin{tabular}{c c}
$0$ & $\mathbf{I}_{N}$\\
$-\mathbf{I}_{N}$ & $0$\\
\end{tabular} \right) \, .
\end{align}
The operators $\hat{q}$ and $\hat{p}$ correspond to observable physical quantities with continuous spectra. Their eigenstates are the non-normalizable basis states $\ket{q}$ and $\ket{p}$, where $\hat{q} \ket{q} = q \ket{q}$ and $\hat{p} \ket{p} = p \ket{p}$, related by a Fourier transform as 
\beq
\ket{q} = \frac{1} {\sqrt{2 \pi}} \int \text{d}p\, e^{- i q p} \ket{p}\, ,
\eeq
and vice-versa.

To give a precise mathematical definition to these statements, one needs to take into account the phase-space description of a CV system. 
Introduce the Weyl operator or phase-space displacement operator \cite{gazeau2009}:
\beq
\label{weylop}
\hat{W}_{\eta} = \text{exp}\{ - i \bm{\eta}^T \mathbf{\Omega} \hat{\mathbf{r}} \} \equiv \bigotimes_{i=1}^N \hat{D}_i(\alpha_i) = \bigotimes_{i=1}^N e^{\alpha_i \hat{a}_i^\dagger - \alpha_i^* \hat{a}_i} \,.
\eeq
The vector $\bm{\eta} = (a_1, ..., a_n, b_1, ... b_n)^T \in \RR^{2N}$ defines the displacement, which can equally be represented in complex form as $\alpha_i = \frac{1}{\sqrt{2}} (a_i + i b_i) \in \mathbb{C}$. The action of the $i$-th displacement operator on the $i$-th mode ground state results in the coherent state $\ket{\alpha}_i = \hat{D}_i (\alpha_i) \ket{0}$, where
\beq
\ket{\alpha}_i = e^{-\frac{1}{2} |\alpha|^2} \sum_{n=1}^\infty \frac{\alpha^n}{\sqrt{n!}} \ket{n}_i \,.
\eeq
The important point here is that any state $\rho$ of $N$ modes can be characterized by its \emph{characteristic function} \cite{puri2001, scully1997}, defined as the expectation value of the Weyl operator
\beq
\chi_\rho (\eta) = \tr [ \rho \hat{W}_\eta ] \,.
\eeq
In fact, $\chi_\rho$ is equivalent to the \emph{Wigner distribution} \cite{puri2001, scully1997}, which is a phase-space representation of the density matrix $\rho$ defined in terms of the $2N$ phase-space quadrature variables $\{q_i,p_i\}$ and eigenvectors $\ket{q}$
\beq
W(q,p) = \frac{1}{\pi^N} \int \d^N q' \bra{q - q'} \rho \ket{q + q'} e^{i q' p} \,.
\eeq
The Wigner function is usually expressed by the Fourier transform of the characteristic function,
\beq
\label{wigchi}
W(q,p) = \frac{1}{(2 \pi)^{2 N}} \int \d^{2 N} \bm{\eta} e^{i \bm{\eta}^T \bm{\Omega} \mathbf{r}} \chi_\rho(\eta) \,,
\eeq
with $\mathbf{r} = (q_1, ..., q_n, p_1, ..., p_n)^T$. The density operator of the quantum state can be written in terms of its characteristic function by means of a Fourier-Weyl relation 
\beq
\rho = \frac{1}{(2 \pi)^N} \int \d^{2 N} \bm{\eta} \chi_\rho(-\bm{\eta}) \hat{W}_\eta \,,
\eeq
where $\hat{W}_\eta$ is the Weyl operator from Eq.(\ref{weylop}), and therefore the state is uniquely determined by $\chi_\rho$ \cite{demarie2012}. This leads to the following important definition: A $N$-mode state $\rho$ is \emph{Gaussian} whenever its characteristic function $\chi_\rho$ is a Gaussian shaped in phase-space, which means that it can be written as \cite{weedbrook2012}
\beq
\label{chigauss}
\chi_\rho(\eta) = \chi_\rho(0) e^{-\frac{1}{4} \bm{\eta}^T \bm{\Omega} \bm{\Gamma} \bm{\Omega} \bm{\eta} - i \bar{\mathbf{r}}^T \bm{\Omega} \bm{\eta}} \,.
\eeq
As a consequence of Eq.(\ref{wigchi}) this is equivalent to saying that a state is Gaussian when its Wigner distribution is a Gaussian function. 

In general (see \cite{ schumaker1986,cohen1977} for a proof), the ground state of any system of N harmonic oscillators is a Gaussian state. In fact, the ground state of any system described by a Hamiltonian linear or quadratic in the canonical operators
\beq
\hat{H} = \frac{1}{2} \sum_i^N \sum_j^N \hat{\mathbf{r}}_i \mathbf{H}_{i,j} \hat{\mathbf{r}}_j \,,
\eeq
defined by a real and positive-definite crossing matrix $\mathbf{H}$, is Gaussian.

The first two statistical moments of a Gaussian state are captured by the vector $ \bar{\mathbf{r}}_i = \tr [\rho \hat{\mathbf{r}}_i] = \6 \hat{\mathbf{r}}_i \9$ of expectation values of the quadrature operators and the $2N \times 2N$ real symmetric matrix $\bm{\Gamma}$, which carries the information about the variances
\beq
\Ga_{i,j} = \text{Re} \tr [\rho(\hat{\mathbf{r}}_i - \6 \hat{\mathbf{r}}_i \9)(\hat{\mathbf{r}}_j - \6 \hat{\mathbf{r}}_j \9)]\, .
\label{gengamma}
\eeq
The matrix $\Ga$ is called \emph{covariance matrix} (or sometimes \emph{noise matrix}) and it plays a central role in the description of the entanglement properties of a Gaussian state \cite{simon1987, simon1988, simon1994}. We can see from the form of the characteristic function $\chi_{\rho}$ in Eq.(\ref{chigauss}) that a Gaussian state $\rho$ is entirely described by the set of its first two statistical moments \cite{holevo1975}, meaning that
\beq
\rho = \rho(\bar{\mathbf{r}}, \Ga) \,.
\eeq
Consequently, all higher-order statistical moments of a Gaussian state can be expressed from $\bar{\mathbf{r}}$ and $\Ga$. Local unitary transformations do not change entanglement \cite{eisert2003} and since displacements $\hat{D}_i$ are single-mode local translations in phase-space, they leave the entanglement properties of the state unaffected. Therefore the elements of $\bar{\mathbf{r}}$ do not contribute to the entanglement and they can all be made zero. Accordingly, the covariance matrix can be rewritten as
\beq
\Ga_{i,j} = \text{Re} \tr [\rho\, \hat{\mathbf{r}}_i \hat{\mathbf{r}}_j] \,.
\eeq
However, a real symmetric matrix $\Ga$ cannot be arbitrary. The canonical commutation relations require the positive definiteness of \cite{simon1987, simon1994}
\beq
\Ga + i \frac{1}{2} \bm{\Omega} \ge 0\,,
\eeq
which really is just another way to rewrite the Heisenberg uncertainty relations.

\subsection{Symplectic transformations}
\label{sympa}
Gaussian states are of fundamental importance in quantum information applications because of the existence of a class of mathematical operators corresponding to common laboratory procedures that preserve Gaussian properties \cite{bartlett2002, olivares2012}. Thus, we define a \emph{Gaussian unitary operation} any unitary transformation that maps a Gaussian state onto a Gaussian state.

There exists a symplectic representation of the Gaussian unitary group \cite{simon1987}. To each Gaussian transformation $\hat{U}$ we can associate a unique \emph{symplectic transformation} $\mathbf{Y} \in \text{Sp}(2 N, \RR)$
\beq
\hat{U}(\Sy) \longleftrightarrow \mathbf{Y}, 
\eeq
where the Lie group $\text{Sp}(2 N, \RR)$ is called the \emph{real symplectic group}\footnote{For more details about this group and its properties, see \cite{arvind1995}.}. The group element $\mathbf{Y}$ describes a linear transformation of the quadrature operators expressed by
\beq
\label{rtransf}
\rho' = \hat{U}(\Sy) \rho\, \hat{U}^\dagger(\Sy) \longrightarrow \hat{\mathbf{r}}' = \mathbf{Y} \hat{\mathbf{r}} = \hat{U}(\Sy)^{-1} \hat{\mathbf{r}} \hat{U}(\Sy) \,.
\eeq
In order to leave the kinematics invariant, these linear homogenous transformations must preserve the canonical commutation relations from Eq.(\ref{CCR}). Hence, in analogy with the classical Hamilton's equations of motion \cite{olivares2012}, the action of any $\mathbf{Y} \in \text{Sp}(2N, \RR)$ on the matrix $\bm{\Omega}$ is given by
\beq
\label{Omtra}
i \bm{\Omega} = [\hat{\mathbf{r}}'_j, \hat{\mathbf{r}}'_k ] = \mathbf{Y} [\hat{\mathbf{r}}_j, \hat{\mathbf{r}}_k ] \mathbf{Y}^T \longrightarrow \bm{\Omega} = \mathbf{Y} \bm{\Omega} \mathbf{Y}^T \, .
\eeq
Furthermore, if $\mathbf{Y}$ is a symplectic transformation then it also satisfies 
\begin{align}
&\mathbf{Y}^T = \bm{\Omega} \mathbf{Y}^{-1} \bm{\Omega}^{-1}\\
&\mathbf{Y}^{-1} = \bm{\Omega} \mathbf{Y}^T \bm{\Omega}^{-1}\, ,
\end{align}
and $\mathbf{Y}^T, \mathbf{Y}^{-1} \in \text{Sp}(2 N, \RR)$. Under the action of a symplectic transformation, the evolution of the covariance matrix is governed by the transformation \cite{olivares2012, weedbrook2012, simon1988}
\beq
\label{gammaSy}
\Ga' = \text{cov} (\mathbf{Y} \hat{\mathbf{r}}) = \mathbf{Y} \text{cov}(\hat{\mathbf{r}}) \mathbf{Y}^T = \mathbf{Y} \Ga \mathbf{Y}^T \,.
\eeq
\subsubsection{Symplectic eigenvalues}
\label{sympaeigolo}
The eigenvalues of a matrix are invariant under a similarity transformation such as $\Ga' = \mathbf{\tilde{Y}}^{-1} \Ga \mathbf{\tilde{Y}}$ \cite{bhatia1997}. However, Eq.~\ref{gammaSy} tells us that the covariance matrix does not transform by a similarity transformation under the action of the symplectic group. Hence, we want to find an alternative form of $\Ga$ such that its eigenvalues are invariant under a symplectic transformation and can be used to uniquely characterize the state. 

To do so, we use Williamson's theorem \cite{williamson1936}. It states that any real symmetric positive-definite $2N \times 2N$ matrix, for instance the covariance matrix $\Ga$ of a $N$-mode Gaussian state $\rho$, can always be made diagonal by means of a suitable symplectic transformation $\mathbf{Y}_w \in \text{Sp}(2N, \RR)$,
\beq
\mathbf{Y}_w \Ga \mathbf{Y}_w^T = \Ga_w \,,
\eeq
where now 
\beq
\Ga_w = \text{diag}(\sigma_1, \sigma_2, ..., \sigma_n, \sigma_1, \sigma_2, ..., \sigma_n)\,,
\eeq
and all the $\sigma_i$ are real. After the transformation $\mathbf{Y}_w$, one finds for the transformed quadrature operators $\hat{\mathbf{r}}'$ that $\text{Re} \6 \hat{\mathbf{r}}'_i \hat{\mathbf{r}}'_j \9 = \delta_{i,j} \sigma_i$. The matrix $\Ga_w$ is called the \emph{Williamson's normal form} of the matrix $\Ga$. It is important to realize that in general the $\{ \sigma_i \}$ are not the eigenvalues of $\Ga$ or of any $\Ga_{\mathbf{Y}} = \mathbf{Y}^T \Ga \mathbf{Y}$ determined by a transformation $\mathbf{Y}$ different from $\mathbf{Y}_w$.

Define a new matrix $\mathbf{M}$ such that $\Ga = - \mathbf{M} \mathbf{\Omega}$ and thus
\beq
\Ga \mathbf{\Omega}^{-1} = - \mathbf{M} \mathbf{\Omega} \mathbf{\Omega}^{-1} \Longrightarrow \mathbf{M} = \Ga \mathbf{\Omega} \,.
\eeq
From (\ref{Omtra}) it is easy to see that $\mathbf{Y}^T \mathbf{\Omega} = \Om \mathbf{Y}^{-1}$ and therefore the action of a symplectic transformation $\mathbf{Y}$ on the matrix $\mathbf{M}$ results in a similarity transformation of $\mathbf{M}$ that preserves its eigenvalues,
\beq
\mathbf{M}' = \Ga' \Om = \mathbf{Y} \Ga \mathbf{Y}^T \Om = \mathbf{Y} \Ga \Om \mathbf{Y}^{-1} = \mathbf{Y} \mathbf{M} \mathbf{Y}^{-1} \,.
\eeq
Hence, every matrix $\bf{M}$ that is determined by varying $\mathbf{Y}$ over the group $\text{Sp}(2 N,\RR)$ shares the same spectrum \cite{simon1987}. In particular, if we take the matrix $\Ga$ to its Williamson form and transform $\bf{M}$ accordingly, the eigenvalues of the matrix 
\beq
\mathbf{M}' = \mathbf{Y}_w \Ga \mathbf{Y}^T_w \Om = \Ga_w \Om
\eeq
will be equal to $\{ \pm i \sigma_j \}$ \cite{vidal2002}. The $N$ absolute values $\{ \sigma_j \}$ of the elements of the spectrum correspond to the $N$ distinct eigenvalues of $\Ga_w$. These are the \emph{symplectic eigenvalues} of the (covariance) matrix $\Ga$ while the set $\{\sigma_j\}$ is called the \emph{symplectic spectrum}. This spectrum characterizes the Gaussian state. By construction it is invariant under any Gaussian transformation. 

This description has immediate practical consequences. Once the covariance matrix $\Ga$ of some Gaussian state is given or calculated, the set of symplectic eigenvalues can be directly obtained from the spectrum of the matrix $\mathbf{M} = \Ga \Om$. In the following section we will show how the symplectic eigenvalues contain the total information about the entanglement properties of the state. This is the reason why this algebraic description of Gaussian states is extremely efficient. It allows to quantify entanglement simply from the symplectic eigenvalues of the matrix of second moments of the state, which are in general much easier to calculate than the eigenvalues of the full density matrix.

\section{Entanglement entropy for Gaussian states}
\label{gauDent}
We have seen previously that for pure states, the entanglement entropy (\ref{entent}) is the preferred way to quantify the entanglement between a subsystem $A$ of some system $AB$, and its complement $B$. Unlike the general case, the von Neumann entropy of Gaussian states has a simple expression in terms of a finite number of the symplectic eigenvalues $\{ \sigma_i \}$ of the covariance matrix. Specifically \cite{demarie2012,eisert2010},
\beq
\label{vnS}
S(\rho) = \sum_{j = 1}^{N_{\text{sub}}} \Big [ \Big( \sigma_j + \frac{1}{2} \Big) \log_2\Big( \sigma_j + \frac{1}{2} \Big) - \Big( \sigma_j - \frac{1}{2} \Big) \log_2\Big( \sigma_j - \frac{1}{2} \Big) \Big] \,,
\eeq
where the index $j$ runs over the modes of one of the two subsystems under examination. Here we present an accessible and precise step-by-step derivation of this formula, showing the link between covariance matrix of a Gaussian state, symplectic eigenvalues and entropy. 

\subsection{Derivation of the formula}
Start with a generic $N$-modes Gaussian state $\rho = \sum_j p_j \ket{\phi_j} \bra{\phi_j}$ with covariance matrix $\Ga$ and $N$ pairs of mass and frequency parameters $\{ m_i, \omega_i \}$. Although rather unphysical, the choice of using equal frequencies for the oscillators is motivated by simplicity, as it is often done in the literature, see for example \cite{audenaert2002}. We still use natural units where $\hbar=1$. Since $\Ga$ is a real symmetric matrix, there always exists a unitary symplectic transformation $\hat{U}(\Sy)$ represented by a matrix $\Sy$ that acts on the state as $\rho^\prime = \hat{U}(\Sy) \rho \hat{U}^\dagger(\Sy)$ and takes $\Ga$ to the normal form $\Ga^\prime = \Sy \Ga \Sy^T$, which in general is not the Williamson form $\Ga_w$ \cite{bhatia1997, simon1994}. As a consequence of the diagonalisation of the covariance matrix, the transformed state $\rho^\prime$ is now described by the direct product of $N$ density matrices
\beq
\label{thermalrho}
\rho^\prime = \rho^\prime_1 \otimes \rho^\prime_2 ... \otimes \rho^\prime_N \, ,
\eeq 
which correspond to uncoupled and non-local thermal quantum oscillators \cite{botero2003}. 

Following the transformation, each oscillator has new mass and frequency $\{m_i^\prime, \omega_i^\prime\}$ and transformed creation and annihilation operators
\ba
\nonumber
\hat{a}^\prime_i &= \sqrt{\frac{m_i^\prime \omega_i^\prime}{2}} (\hq_i^\prime + \frac{i}{m_i^\prime \omega_i^\prime} \hp_i^\prime) \, , \\
\hat{a}_i^{\dagger \prime} &= \sqrt{\frac{m_i^\prime \omega_i^\prime}{2}} (\hq_i^\prime - \frac{i}{m_i^\prime \omega_i^\prime} \hp_i^\prime) \, .
\label{taaad}
\end{align}
From Eq.(\ref{rtransf}) it is easy to see that under a symplectic transformation $\mathbf{Y}$, the bosonic operators transform according to the quadrature operators transformation, an operation more conventionally called a Bogolyubov transformation:
\beq
\hat{\mathbf{b}}' = \mathbf{Y} \hat{\mathbf{b}} \, .
\eeq
Finally, the Hamiltonian of a single thermal oscillator is equal to $\hH_i' = \omega_i' (\hat{a}^{\dagger \prime} \ha_i' + \frac{1}{2})$\footnote{In the following, since we will be always referring to the transformed modes unless stated explicitly, we drop the \emph{primes} to lighten the notation.}. A quantum harmonic oscillator in thermal equilibrium with a bath at temperature $T$ is statistically described by a canonical ensemble \cite{cohen1977}. More explicitly, we can rewrite its density matrix in the Fock basis $\{ \ket{n} \}$ as
\beq
\nonumber
\rho = \sum_n p_n \ket{n} \bra{n} \, ,
\eeq
where the $p_n= Z^{-1} e^{-\frac{E_n}{\kappa_B T}}$ are the probabilities associated to each state $\ket{n} \bra{n}$, with $E_n$ energy of the $n$-th state of the Hamiltonian $\hH \ket{n} = E_n \ket{n}$ and $Z = \tr\Big(e^{-\hH/ \kappa_B T}\Big)$ partition function that normalizes the probabilities \cite{kubo1990}. Considering the decomposition from Eq.(\ref{thermalrho}), a single-mode partition function is equivalent to
\begin{align}
\nonumber
Z_i = \tr\Big( e^{- \hH_i / \kappa_B T} \Big) &= \\
\label{Z}
\sum_{n_i = 0}^\infty &\bra{n_i} e^{-(\hat{a}_i^\dagger \ha_i + \frac{1}{2}) \omega_i /\kappa_B T_i} \ket{n} = \sum_{n_i = 0}^\infty e^{-(n_i + \frac{1}{2}) \beta_i}\, ,
\end{align}
where we used $\hc_i \ha_i \ket{n} = n \ket{n}$ and $\beta_i \equiv \omega_i / \kappa_B T_i$ is a cumulative parameter that depends on the transformed frequency $\omega_i$. Using properties of the geometric series, the last part of Eq.(\ref{Z}) can be rewritten as
\begin{align}
\nonumber
Z_i = e^{-\beta_i / 2} \sum_{n_i = 0}^\infty &e^{-n_i \beta_i} = \\
&e^{-\beta_i/2} [1 + e^{-\beta_i} + e^{-2 \beta_i} + ...] \longrightarrow Z_i = \frac{e^{-\beta_i/2}}{1-e^{-\beta_i}} \,,
\end{align}
and then the density matrix of each uncoupled thermal mode is equal to
\begin{align}
\nonumber
\rho_i = \sum_n Z_i^{-1} e^{-E_n/\kappa_B T} \ket{n} &\bra{n} = \\
&Z_i^{-1} e^{-\hH_i / \kappa_B T_i} = \Big(1 - e^{-\beta_i} \Big) e^{- \hc_i \ha_i \beta_i} \, .
\end{align}
One can further simplify this expression rewriting the density matrix in terms of $\bar{n}_i = \6 n_i \9$, mean occupation number of the transformed modes:
\begin{align}
\nonumber
\bar{n}_i = \6 \hc_i \ha_i \9 = \tr(\rho_i \hc_i \ha_i) = &\sum_{n_i = 0}^\infty \bra{n_i} (1-e^{-\beta_i}) e^{-\hc_i \ha_i \beta_i} \hc_i \ha_i \ket{n_i} = \\
&(1-e^{-\beta_i}) \sum_{n_i = 0}^\infty n_i e^{-n_i \beta_i} = \frac{1}{e^{\beta_i}-1} \longrightarrow e^{\beta_i} = \frac{1+\bar{n}_i}{\bar{n}_i} \, .
\label{betan}
\end{align}
As a result we obtain the following alternative way to write the thermal density matrix,
\beq
\rho_i = \frac{1}{1+ \bar{n}_i} \left( \frac{\bar{n}_i}{1 + \bar{n}_i } \right)^{\hc_i \ha_i} \, .
\eeq
Explicit calculation of the von Neumann entropy $S(\rho_i) = - \tr(\rho_i \log{\rho_i})$ gives:
\begin{align}
\nonumber
-\tr(&\rho_i \log{\rho_i}) = - \sum_{n_i = 0}^{\infty} \bra{n} \frac{1}{1+ \bar{n}_i} \Big( \frac{n_i}{1 + n_i}\Big)^{\hc_i \ha_i} 
\log{\left[ \frac{1}{1+\bar{n}_i} \Big( \frac{\bar{n}_i}{1+\bar{n}_i} \Big)^{\hc_i \ha_i}  \right]} \ket{n_i} = \\
\nonumber
&-\frac{1}{1+\bar{n}_i} \sum_{n_i = 0}^\infty \bra{n_i} \Big( \frac{\bar{n}_i}{1+\bar{n}_i} \Big)^{n_i} \Big[ -\log(1+\bar{n}_i) + n_i \log{\Big( \frac{\bar{n}_i}{1+\bar{n}_i} \Big)} \Big] \ket{n_i} = \\
\nonumber
&-\frac{1}{1+ \bar{n}_i} \Big[ -\log(1+\bar{n}_i) \sum_{n_i = 0}^{\infty} \Big( \frac{\bar{n}_i}{1+\bar{n}_i} \Big)^{n_i} +\log{\Big( \frac{\bar{n}_i}{1+\bar{n}_i} \Big)} \sum_{n_i = 0}^{\infty} n_i \Big(\frac{\bar{n}_i}{1+\bar{n}_i} \Big)^{n_i} \Big] = \\
&(1+\bar{n}_i) \log{(1+\bar{n}_i)} - \bar{n}_i \log \bar{n}_i \, ,
\end{align}
where we used the following summation properties \cite{spiegel1990}:
\begin{align}
\sum_{n_i = 0}^{\infty} \Big(\frac{\bar{n}_i}{1+\bar{n}_i} \Big)^{n_i} = 1+\bar{n}_i \,\,\,\,\,\,\, \text{and} \,\,\,\,\,\,\, \sum_{n_i = 0}^{\infty} n_i \Big( \frac{\bar{n}_i}{1+\bar{n}_i} \Big)^{n_i} = \bar{n}_i (1+ \bar{n}_i) \, .
\end{align}
This is precisely the von Neumann entropy of a single oscillator thermal state expressed in terms of the mean occupation number:
\beq
\label{nS}
S(\rho_i) = (1+ \bar{n}_i) \log_2(1+\bar{n}_i) - \bar{n}_i \log_2{\bar{n}_i} \, ,
\eeq
which is a well-known result in statistical physics \cite{kubo1990}.

It is now very simple to show the connection of the entropy with the symplectic eigenvalue $\sigma_i$ of the state. After the symplectic transformation $\mathbf{Y}$ that takes the (total) covariance matrix $\Ga$ to its normal form, each reduced covariance matrix associated to a single-mode non-local thermal state $\rho_i$ looks like
\begin{align}
\label{quadgamma}
\Ga_i = \left (
\begin{tabular}{c c}
$\6 \hq_i^2 \9$ & $0$\\
$0$ & $\6 \hp_i^2 \9$\\
\end{tabular} \right) \, .
\end{align}
Using the creation and annihilation operators (\ref{taaad}), one easily finds that the expectation values of the transformed quadratures are 
\begin{align}
\nonumber
\6 \hq_i^2 \9 = \tr(\rho_i \hq_i^2) &= Z^{-1} \sum_{n_i = 0}^{\infty} \bra{n} (2 m_i \omega_i)^{-1} (\hc_i + \ha_i)(\hc_i + \ha_i) e^{-\hc_i \ha_i \beta_i} \hc_i \ha_i \ket{n_i} = \\
\nonumber
&(2m_i \omega_i)^{-1} + (m_i \omega_i)^{-1} \frac{1}{e^{\beta_i}-1} = (2 m_i \omega_i)^{-1} \coth{\frac{\beta_i}{2}}\\
\nonumber
&\text{and}&\\
&\6 \hp_i^2 \9 = \frac{m_i \omega_i}{2} + m_i \omega_i \frac{1}{e^{\beta_i}-1} = \frac{m_i \omega_i}{2} \coth{\frac{\beta_i}{2}}\, .
\end{align}
From a simple algebraic argument, the eigenvalues pair $\{ \pm \sigma_i \}$ of the matrix multiplication $i \Ga_i \Om_i$ are:
\beq
\text{eigenvalues}\{i \Ga_i \Om\} = \pm \sqrt{\6 \hq_i^2 \9 \6 \hp_i^2 \9} = \pm \frac{e^{\beta_i}+1}{2 (e^{\beta_i}-1)} = \pm \Big( \bar{n}_i +\frac{1}{2} \Big) \,.
\eeq
The following equivalence shows the explicit connection between the symplectic eigenvalue of a generic thermal Gaussian state and its mean occupation number
\beq
\label{sigman}
\sigma_i = \bar{n}_i + \frac{1}{2} \longrightarrow \bar{n}_i = \sigma_i - \frac{1}{2} \, .
\eeq
We know from Eq.(\ref{Saddit}) that entropy is an additive quantity, and the total entropy of a state $\rho$ equal to the direct tensor product of $N$ states, is simply the sum of the entropies of each state, i.e. for $\rho = \otimes_i \rho_i$ we have $S(\rho) = \sum_i S(\rho_i)$.
Using Eq.(\ref{nS}), the von Neumann entropy for the Gaussian state $\rho$ can therefore be written in terms of the $N$ symplectic eigenvalues $\{ \sigma \}$ of the diagonalised $\Ga$ as
\beq
S(\rho) = \sum_{i=1}^N \Big [ \Big( \sigma_i + \frac{1}{2} \Big) \log_2\Big( \sigma_i + \frac{1}{2} \Big) - \Big( \sigma_i - \frac{1}{2} \Big) \log_2\Big( \sigma_i - \frac{1}{2} \Big) \Big] \,,
\eeq
which corresponds to Eq.(\ref{vnS}). Recall that we are now considering the transformed modes. Since the quantum entropy of a Gaussian state is solely a function of the symplectic eigenvalues, which are invariant under any symplectic transformation $\mathbf{Y}$, entropy is invariant under any $\mathbf{Y}$ itself. 

It is interesting to note from Eq.(\ref{betan}) and Eq.(\ref{sigman}) that the thermal parameter $\beta_i$ of each non-local oscillator is related to the correspondent symplectic eigenvalue as
\beq
\label{betasig}
\beta_i = \ln{\Big ( \frac{1+ \bar{n}_i}{\bar{n}_i} \Big)} = \ln{\Big( \frac{\sigma_i + 1/2}{\sigma_i - 1/2} \Big)} \, .
\eeq

\subsection{An example: Two coupled harmonic oscillators}
\label{ExGS}
As a simple example consider a system composed of two quantum harmonic oscillators with equal mass $m$ and frequency $\omega$ coupled in position, described by the following Hamiltonian:
\beq
\hH = \frac{1}{2m} (\hp_1^2 + \hp_2^2) + \frac{m \omega^2}{2} (\hq_1^2 + \hq_2^2) + \lambda (\hq_1 - \hq_2)^2 \, , 
\eeq
with $\lambda$ positive coupling parameter such that $\lambda < m \omega^2 / 2$. We will use this example again in Chapter \ref{polymerico}, so pay attention now! Since the Hamiltonian is bilinear, the ground state of this system is still Gaussian and we want to exploit this property to calculate the bipartite ground state entanglement between the two oscillators. The global symplectic transformation $\mathbf{Y}$ described by the matrix
\begin{align}
\label{matY}
\mathbf{Y} = \frac{1}{\sqrt{2}} \left (
\begin{tabular}{c c c c}
$1$ & $-1$ & $0$ & $0$\\
$1$ & $1$ & $0$ & $0$\\
$0$ & $0$ & $1$ & $-1$\\
$0$ & $0$ & $1$ & $1$\\
\end{tabular} \right) \, ,
\end{align}
transforms the system into two uncoupled oscillator with new frequencies equal to
\beq
\omega_1' = \omega\,, \,\,\,\,\,\,\,\,\, \omega_2' = \omega \sqrt{1+ \frac{4 \lambda}{m \omega^2}} \equiv \omega \alpha\, .
\eeq
After diagonalizing the Hamiltonian, the corresponding normal modes covariance matrix analog to (\ref{quadgamma}) is
\begin{align}
\label{diagamma}
\Ga' = \frac{1}{2} \left (
\begin{tabular}{c c c c}
$\frac{1}{m \omega}$ & $0$ & $0$ & $0$\\
$0$ & $\frac{1}{m \omega \alpha}$ & $0$ & $0$\\
$0$ & $0$ & $m \omega$ & $0$\\
$0$ & $0$ & $0$ & $m \omega \alpha$\\
\end{tabular} \right) \, .
\end{align}
Inverting Eq.(\ref{gammaSy}) one easily obtains the covariance matrix $\Ga$ of the system
\begin{align}
\Ga = {\bf Y}^{-1} \Ga' ({\bf Y}^T)^{-1} = \frac{1}{4} \left (
\begin{tabular}{c c c c}
$\frac{1+\alpha}{m \omega \alpha}$ & $\frac{1- \alpha}{m \omega \alpha}$ & $0$ & $0$\\
$\frac{1-\alpha}{m \omega \alpha}$ & $\frac{1+\alpha}{m \omega \alpha}$ & $0$ & $0$\\
$0$ & $0$ & $m \omega (\alpha + 1)$ & $m \omega (\alpha - 1)$\\
$0$ & $0$ & $m \omega (\alpha - 1)$ & $m \omega (\alpha +1)$\\
\end{tabular} \right) \, .
\end{align}
In this case, it is straightforward to trace out the complementary degrees of freedom and derive the reduced covariance matrix for the first (or, equivalently, the second) oscillator:
\begin{align}
\Ga_1 = \Ga_2 = \frac{1}{4} \left(
\begin{tabular}{c c}
$\frac{1 + \alpha}{m \omega \alpha}$ & $0$\\
$0$ & $m \omega (\alpha +1)$\\
\end{tabular} \right) \, ,
\end{align}
The last step is to calculate the single symplectic eigenvalue that belongs to the reduced covariance $\Ga_1(\Ga_2)$ matrix from the spectrum of $\Om_{2 \times 2} \Ga_{1,2}$, which is given by
\beq
\sigma_1 = \sigma_2 = \frac{1 + \alpha}{4 \sqrt{\alpha}} \, .
\eeq
Now $\sigma_1$ can be used to quantify the bipartite entanglement between the coupled oscillators using the formula from Eq.(\ref{vnS}).

\subsection{Physical intuition}

We conclude this part mentioning few thermodynamical considerations that highlight a particularly interesting connection between entanglement and thermal properties of the subsystems. It should be understood that through an appropriate choice of a global symplectic transformation $\mathbf{Y}$, a Gaussian state $\rho$ (or $\Ga$) of $N$ coupled oscillators can be decomposed into the tensor product of $N$ single thermal oscillators. If $\rho$ is pure, the new non-local oscillators $\rho_i'$ are also pure, following the decomposition properties of a pure state. Hence, the decoupled oscillators are all in the ground state and a virtual temperature $T_i = 0$ can be assigned to each of them. From the relation (\ref{sigman}) it follows that all the symplectic eigenvalues are equal to $\sigma_i = 1/2$ and therefore the formula in Eq.(\ref{vnS}) confirms the well-known property that the total von Neumann entropy of a pure state is zero. 

Divide the $N$ modes into two sets $A = {A_1, ..., A_a}$ and $B={B_1, ..., B_b}$ such that $a+b = N$ and $A$ belongs to Alice, while $B$ belongs to Bob. In order to calculate the entanglement entropy $S(\rho_A) = S(\rho_B)$ across the bipartition, one needs to know the set of symplectic eigenvalues that belong to one of the two partitions. If Alice wants to study her part of the system, she first notices that, in general, after the division the reduced density matrix $\rho_A = \tr_B \rho$ corresponds to a mixed state. Therefore, locally decomposing $\rho_A$ by means of a local Gaussian unitary $\hat{U}(\mathbf{Y}_A)$ associated to the symplectic transformation $\mathbf{Y}_A$ gives a new state $\rho'_A$ 
\beq
\hat{U}(\mathbf{Y}_A) \rho \hat{U}^\dagger  (\mathbf{Y}_A)  \longrightarrow \rho_A' = \rho_{1,T_1}' ... \otimes \rho_{r, T_r}' \otimes \rho'_{r+1, G_{r+1}} ... \otimes \rho'_{a,G_a} \, ,
\eeq
which contains both thermal (at virtual temperature $T_i \ne0$) and ground state oscillators.

Alice's symplectic spectrum $\{ \sigma_{1, ..., a} \}$ is obtained from the spectrum of $\Ga_a \Om_a$ where $\Ga_a = \tr_B \Ga$ is the reduced covariance matrix of the set $A$. Suppose Alice extracts $s$ symplectic eigenvalues satisfying
\beq
\sigma_{1, ..., s} \ge \frac{1}{2} \, ,
\eeq 
and $a-s$ eigenvalues $\sigma_{s+1, ..., a} = \frac{1}{2}$. This means that Alice's set can be decomposed as
\beq
\label{eqconstra}
\rho_A' = \rho_{1,T_1}' ... \otimes \rho_{s, T_s}' \otimes \rho'_{s+1, G_{s+1}} ... \otimes \rho'_{a,G_a} \, .
\eeq
Physically, this corresponds to having $s$ virtual thermal oscillators with thermal parameter $\beta_i$ given by Eq.(\ref{betasig}) and $a-s$ ground states. \emph{Only the thermal oscillators contribute to the bipartite entanglement entropy}. It is important to be precise at this point and to avoid taking these conclusions too far. The density operator for the virtual modes has the same form as a Gibbs ensemble, however these oscillators are completely isolated and not in thermal equilibrium with any reservoir.

The construction in Eq.(\ref{eqconstra}) was made rigorous in \cite{botero2003,botero2004}, where the authors proved that, after identifying the two sets $A$ and $B$, it is always possible to write the Gaussian pure state $\rho = \ket{\phi}_{A,B} \bra{\phi}$ as
\beq
\ket{\phi}_{A,B} = \ket{\tilde{\phi}_1}_{\tilde{A}_1 \tilde{B}_1} ... \otimes \ket{\tilde{\phi}_s}_{\tilde{A}_s \tilde{B}_s} \otimes \ket{0}_{\tilde{A}_{s+1, ..., a}} \otimes \ket{0}_{\tilde{B}_{s+1, ..., b}} \, ,
\eeq
where $\tilde{A} = \{ \tilde{A}_1, ..., \tilde{A}_a \}$ and $\tilde{B} = \{ \tilde{B}_1, ..., \tilde{B}_b \}$ are transformed modes resulting from the application of some local symplectic transformations on the set $A$ and $B$, and $s$ is equal to the number of symplectic eigenvalues associated to Alice's or Bob's reduced covariance matrix. This means that the state $\ket{\phi}_{A,B}$ can always be rewritten as the direct product of $s$ two-mode squeezed states, where each mode belongs to a different partition of the system, and $N-2s$ oscillator ground states $\ket{0}$ (respectively, $a-s$ in partition $A$ and $b-s$ in partition $B$). 

Each two-mode squeezed state $\ket{\tilde{\phi}_i}_{\tilde{A}_j \tilde{B}_j}$ is described by a sum over the Fock basis of the partitions
\beq
\ket{\tilde{\phi}_j}_{\tilde{A}_j \tilde{B}_j} = \frac{1}{\sqrt{z_j}} \sum_n e^{-\beta_j n/2} \ket{n}_{\tilde{A}_j} \ket{n}_{\tilde{B}_j} \, ,
\eeq
with $\beta_j$ squeezing parameter correspondent to the thermal parameter of the $j$-th thermal oscillator of the local normal-modes decomposition of $\rho_A$ (or $\rho_B$).

\section{Graphical calculus for pure Gaussian states}
\label{gracagas}
We conclude this Chapter with a short introduction to the graphical calculus for pure Gaussian states \cite{menicucci2011}, an elegant graphical representation that allows to describe any Gaussian pure state (and the most common Gaussian transformations) by its graph (and graph transformations). This formalism will provide an important simplification in the subsequent discussion of topologically ordered CV systems \cite{demarie2013b}, Chapter \ref{CVTO}, when describing the connection between the CV cluster-state \cite{menicucci2006, gu2009} and the CV surface code state \cite{demarie2013b}, since both are Gaussian states. Specifically, the measurement patterns that connect the two types of states have definite graph transformation rules, so the connection between the two graphs can be derived using the graphical calculus directly. Before starting, we introduce the concept of graph. 

\subsection{Elements of graph theory}
\label{graphsdef}
A \emph{graph} \cite{diestel2010} is a collection of vertices and connections between vertices, called edges. It is defined by the pair $G = (\mathcal{V},\mathcal{E})$, where $\mathcal{V} = \{1,..., N\}$ is the set of vertices, uniquely labelled by an integer and normally pictured as a set of dots, and $\mathcal{E} = \{(i,j)...\}$ is the set of edges, represented by lines joining pairs of vertices. The structure of the relations among the vertices of a graph $G$ is described by its adjacency matrix ${\bf A}(G)$. This matrix has elements ${\bf A}(G)_{i,j} = 1$, if the edge $(i,j) \in \mathcal{E}$ and ${\bf A}(G)_{i,j} = 0$ otherwise. For a vertex $v$, its neighborhood $\mathcal{N}_v$ is given by all the vertices $v'$ such that $(v,v') \in \mathcal{E}$. Physically, one usually identifies the vertices with quantum systems and the edges with the interaction between the systems. 

A graph is directed if each edge is given a particular orientation, and weighted if all the edges are associated with a number, in which case the corresponding entries in the adjacency matrix are replaced by the values of each weight. Loops are edges that start and end at the same vertex. Undirected graphs without loops and at most one edge between any pair of vertices are called \emph{simple graphs}. The adjacency matrix of a simple graph is thus symmetric with zero diagonal elements. In the description of the graphical calculus we will in general make use of undirected graphs \emph{with} self-loops and complex entries, such that the adjacency matrix is still symmetric but with non-zero elements on the diagonal. Certain properties of a graph can be understood in terms of vertex coloring, a map $c : \mathcal{V} \to \mathcal{C}$, where $\mathcal{C}$ is the set of colors, and $c(v) \neq c(v')$ for all adjacent vertices $v, v'$.

\subsection{The graphical calculus}
From the precedent discussion, a $N$-vacua Gaussian state is described by the covariance matrix
\beq
\Ga_0 = \frac{1}{2} \mathbf{I}_{2N} \,.
\eeq
After a generic Gaussian unitary transformation, represented by the matrix $\mathbf{Y},$ is applied to $\Gamma_0$, the resulting Gaussian state simply has a covariance matrix given by
\beq
\Gamma_{\mathbf{Y}} = \frac{1}{2} \mathbf{Y} \mathbf{Y}^T \, .
\eeq
Exploiting the decomposition properties of symplectic matrices as shown in \cite{simon1988}, we can rewrite the symplectic matrix $\Sy$ using the following unique decomposition
\begin{align}
\Sy = \left(
\begin{tabular}{c c}
$\mathbf{I}$ & $\mathbf{0}$\\
$\mV$ & $\mathbf{I}$\\
\end{tabular} \right)
\left(
\begin{tabular}{c c}
$\mU^{-1/2}$ & $\mathbf{0}$\\
$\mathbf{0}$ & $\mU^{1/2}$\\
\end{tabular} \right)
\left(
\begin{tabular}{c c}
$\mathbf{K}$ & $-\mathbf{J}$\\
$\mathbf{J}$ & $\mathbf{K}$\\
\end{tabular} \right) \, ,
\end{align}
where both $\mU$ and $\mV$ are $N \times N$ symmetric matrices and $\mU = \mU^T > 0$. The last matrix in the decomposition above does not contribute to the $\Sy \Sy^T$ product since it is orthogonal and ${\bf ABC} ({\bf ABC})^T = {\bf ABC} {\bf C}^T {\bf B}^T {\bf A}^T$, therefore one can choose $\mathbf{K} = \mathbf{I}$ and $\mathbf{J} = 0$. After these considerations, we define $\Sy$ as
\begin{align}
\Sy := \left(
\begin{tabular}{c c}
$\mU^{-1/2}$ & $\mathbf{0}$\\
$\mV \mU^{-1/2}$ & $\mU^{1/2}$\\
\end{tabular} \right) \, ,
\end{align}
and the product $\mathbf{Y} \mathbf{Y}^T$ is uniquely specified by $\mathbf{Y}_{\mathbf{U}\mathbf{V}} \mathbf{Y}^T_{\mathbf{U}\mathbf{V}}$, with the covariance matrix $\Ga_{\Sy}$ equal to
\begin{align}
\Ga_{\Sy} = \frac{1}{2} \left(
\begin{tabular}{c c}
$\mU^{-1}$ & $\mU^{-1} \mV$\\
$\mV \mU^{-1}$ & $\mU+ \mV \mU^{-1} \mV$\\
\end{tabular} \right) \, .
\label{gammaZ}
\end{align}
Therefore, it is clear that the complex (adjacency) matrix
\beq
\label{Zmatrix}
\mZ := \mV + i \mU \, ,
\eeq
offers an alternative description for a pure Gaussian state, since it is a linear combination of the matrices that completely determine its covariance matrix. 

The matrix $\mZ$ has a simple transformation rule under a symplectic transformation $\Sy$. If $\Sy$ is decomposed into block form
\begin{align}
\Sy =\left( \begin{tabular}{c c}
$\mathbf{A}$ & $\mathbf{B}$\\
$\mathbf{C}$ & $\mathbf{D}$\\
\end{tabular} \right) \, ,
\end{align}
then the $\mZ'$ matrix associated to the transformed state is given by \cite{menicucci2011}
\beq
\mZ' = (\mathbf{C} + \mathbf{D}\mZ)(\mathbf{A} + \mathbf{B}\mathbf{Z})^{-1}\, .
\eeq
The matrix $\mZ$ represents a graph, in the sense that it is equivalent to the adjacency matrix of an undirected graph with complex-valued edge weights. 

It is easy to establish the connection between the quadrature correlations of the Gaussian state and its graph properties. From Eq.(\ref{gammaZ}) one notices that the left upper block of the covariance matrix (where the $\hq$-correlations are contained), is equal to
\beq
\label{wf1}
\frac{1}{2} \6 {\bf \hq} {\bf \hq}^T \9 = \mU^{-1} \longrightarrow \mU = \frac{1}{2}  \6 {\bf \hq} {\bf \hq}^T \9^{-1}\, .
\eeq
Analogously, from the right upper block the matrix $\mV$ is related to the quadrature correlations as
\beq
\label{wf2}
\mV = \mU \6 \{ {\bf \hq}, {\bf \hp}^T \} \9 = \frac{1}{2}   \6 {\bf \hq} {\bf \hq}^T \9^{-1} \6 \{ {\bf \hq}, {\bf \hp}^T \} \9\,,
\eeq
and therefore, the $\mZ$ matrix can be quickly rewritten as
\beq
\label{wf3}
\mZ = \mV + i \mU = \6 {\bf \hq}{\bf \hq}^T \9^{-1} \6 {\bf \hq} {\bf \hp}^T \9 \,,
\eeq
which shows how the $\mZ$ matrix is related to the correlations of the Gaussian state and vice-versa \cite{menicucci2011}. Because of this connection the matrix $\mZ$ shows up directly in the position-space wavefunction  $\Psi_{\mZ}(\mathbf{q})$ for a $N$-mode Gaussian state $|\Psi_{\mZ}\9$. 

Recall that the most general position-space wavefunction of a single mode Gaussian state with zero means (i.e. $\6 \hq \9 = \6 \hp \9 = 0)$ is equal, up to a phase, to \cite{schumaker1986}
\beq
\psi(q) = \pi^{-1/4} \left(2 \6 \hq^2 \9 \right)^{-1/4} \text{exp}\left(-\frac{1}{2} \tilde{\gamma} q^2 \right)\,,
\eeq
where $\tilde{\gamma} = \frac{1}{2}\6 \hq^2 \9^{-1} - i \6 \hq^2 \9^{-1}(\6 \hq \hp \9)$. Using the relations in Eq.(\ref{wf1}-\ref{wf3}) one obtains the $N$-mode extension of the wave function as a function of $\mZ$
\beq
\Psi_{\mZ}(\mathbf{q})= \pi^{-N/4} (\text{det} \mU)^{1/4} \text{exp} \Big( \frac{i}{2} \mathbf{q}^T \mZ \mathbf{q} \Big)\, ,
\eeq
where $\mathbf{q} = (q1, ..., q_N)^T$ is a column vector of position-space variables.

\subsubsection{Transformation rules and measurements}
\label{trarumea}
The usefulness of the graphical calculus lies in the simple transformation rules of $\mZ$ for the most common laboratory procedures corresponding to Gaussian unitary transformation, as listed in \cite{menicucci2011}, which indeed permit the study of Gaussian states evolution simply in terms of appropriate graph transformations on $\mZ$. To give an example of these concepts, we note that for some generic $N$-mode gaussian state 
\begin{align}
\mZ =\left( \begin{tabular}{c c}
$t$ & $\mathbf{f}^T$\\
$\mathbf{f}$ & $\mathbf{W}$\\
\end{tabular} \right) \, ,
\end{align}
with $t$ a scalar, $\mathbf{f}$ a column vector of length $N-1$ and $\mathbf{W}$ a $N-1 \times N-1$ matrix, any single-mode local Gaussian unitary represented by the matrix
\begin{align}
\Sy =\left( \begin{tabular}{c c}
$a$ & $b$\\
$c$ & $d$\\
\end{tabular} \right) \, ,
\end{align}
acting on the first mode of $\mZ$, returns the transformed matrix:
\begin{align}
\label{trZ}
\mZ' =\left( \begin{tabular}{c c}
$\frac{c + d t}{a + b t}$ & $\frac{\mathbf{f}^T}{a + bt}$\\
$\frac{\mathbf{f}}{a + bt}$ & $\mathbf{W} - \frac{b \mathbf{f} \mathbf{f}^T }{a + b t}$\\
\end{tabular} \right) \, .
\end{align}
The simplest single-mode transformation is the quadrature squeezing, performed by the unitary operator
\beq
\label{sqmatrix}
\hat{S}(s) = e^{-\frac{i}{2} (\log s) (\hq \hp + \hp \hq)} \,, 
\eeq
where $\log s$ is known as the \emph{squeezing parameter}. In the Heisenberg picture, 
\beq
\hS(s)^\dagger \hq \hS(s) = s \hq\,, \qquad \hS(s)^\dagger \hp \hS(s) = \frac{\hp}{s}\,,
\eeq
such that the variance of $\hp$ (or $\hq$), after squeezing, changes by a factor of $s^{-2}$ (of $s^2$) times from its original value. The symplectic matrix associated to the single-mode $\hp$-squeezing ($s > 1$) is
\begin{align}
\Sy_{\text{Sq}}(s) =\left( \begin{tabular}{c c}
$e^{\log s}$ & $0$\\
$0$ & $e^{-\log s}$\\
\end{tabular} \right) \, .
\end{align}
Substituting the matrix elements into Eq.(\ref{trZ}) gives the $\mZ$ matrix of the state after the squeezing.
Another Gaussian operation is the phase shift $\hat{U}(\theta) = e^{- i \theta/2} \text{exp}(-i \theta \hat{a}^\dagger \ha)$, represented by
\begin{align}
\Sy_{\text{PS}}(\theta) =\left( \begin{tabular}{c c}
$\cos \theta$ & $\sin \theta$\\
$- \sin \theta$ & $\cos \theta$\\
\end{tabular} \right) \, .
\end{align}
which is equivalent to a single-mode Fourier transform for $\theta = -\pi/2$. From $\Sy_{\text{PS}}$ and the form of (\ref{trZ}) it is clear that the phase-shift transformation can create additional edges on the graph. 

\begin{figure}[tbp]  
\centering
\setlength{\unitlength}{1cm}
\includegraphics[width=16.5\unitlength]{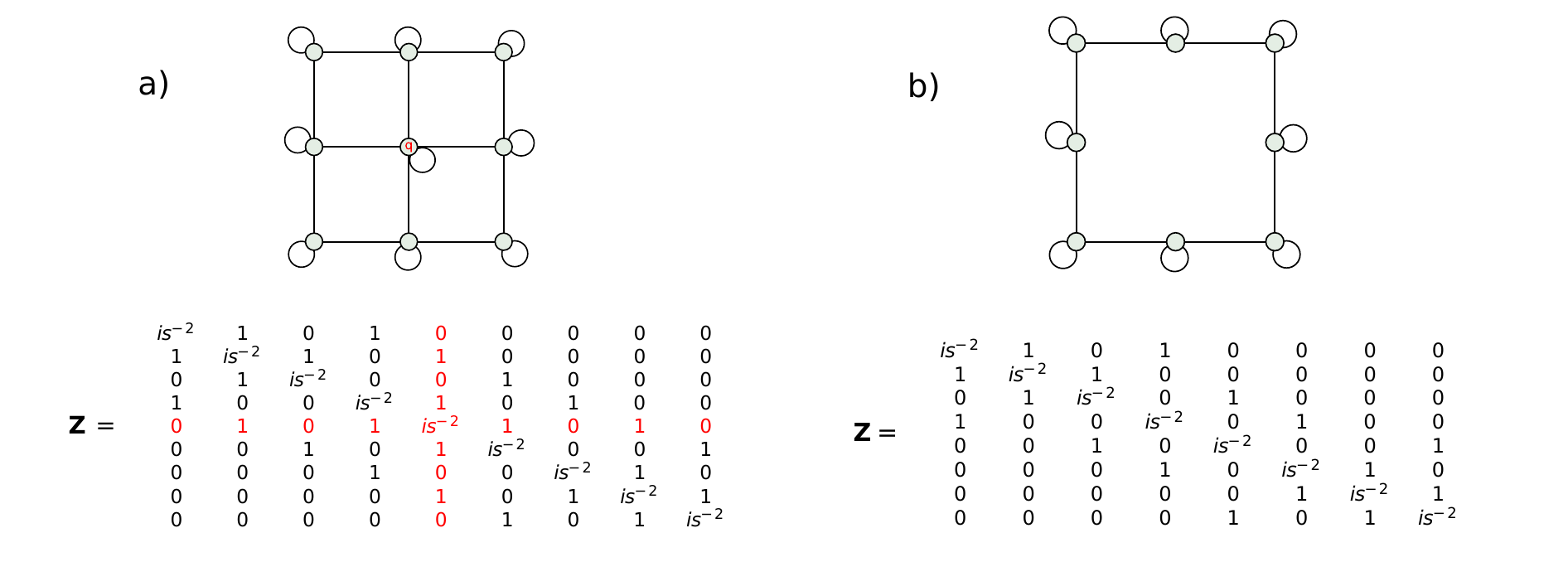}
\caption{Example of a $\hq$ measurement: a) A $3\times 3$ lattice of $\hp$-squeezed modes and its corresponding $\mZ$ matrix. A $\hq$ measurements is performed on the 5th mode and the resulting graph plus the associated $\mZ$ matrix are shown in b).}
\label{qmesfig}
\end{figure}

Following the same logic, one constructs two-mode local Gaussian transformations. The most useful for our purposes is the controlled-Z gate $\cz[g] = \text{exp}(i g \hq_j \hq_k)$, which correlates the two modes $\{ j,k \}$ and adds a weighted edge between the corresponding vertices in the graphical representation. While the symplectic matrix for the $\text{C}_{\hat{Z} (1,2)}$ is
\begin{align}
\Sy_{\text{C}} =\left( \begin{tabular}{c c c c}
$1$ & $0$ & $0$ & $0$\\
$0$ & $1$ & $0$ & $0$\\
$0$ & $g$ & $1$ & $0$\\
$g$ & $0$ & $0$ & $1$\\
\end{tabular} \right) \, .
\end{align}
its action on some state $\mZ$ is expressed by
\begin{align}
\Sy_{\text{C}} =\left( \begin{tabular}{c c}
$\begin{tabular}{c c}
$t_{1,1}$ & $t_{1,2} + g$\\
$t_{2,1} + g$ & $t_{2,2}$\\
\end{tabular}$ & $\mathbf{F}^T$\\
$\mathbf{F}$ & $\mathbf{W}$\\
\end{tabular} \right) \, .
\end{align}

Measurements have a straightforward translation in the $\mZ$ transformation rules language. A $\hq$ measurement on the $k$th mode is equivalent to deleting the $k$th row and column of the $\mZ$ matrix, see Fig.(\ref{qmesfig}) as an example, while a $\hp$ measurement is equivalent to applying a $\pi / 2$ phase shift, a Fourier transform, on the $k$th mode and then measuring the $\hq$ quadrature. At the level of the $\mZ$ matrix, any $\hp$ measurement deletes the measured mode while creating new connections among its nearest neighbors \cite{menicucci2011}. 

\section{Discussion}
In this Chapter we introduced the reader to the relevant concepts of Gaussian states theory, analyzing the phase space representation and demonstrating the dependance of the pure states bipartite entanglement on the symplectic eigenvalues of the covariance matrix. Furthermore, we illustrated the Graphical calculus for pure Gaussian states, which will be used in Chapter \ref{CVTO}, and showed some examples of simple Gaussian transformations.

\begin{savequote}[8.5cm] 
\sffamily
``Kitaev wins \$3M Physics Prize.''
\qauthor{Spiros - Quantum Frontiers blog}
\end{savequote}

\chapter{Topological Order}
\graphicspath{{T_Order}} 
\label{ChapTO}

The Holy Grail of quantum information theory is building a quantum computer \cite{stajic2013}. Nowadays there exist many proposals based on different techniques for achieving this goal. From quantum computing with ion traps \cite{haeffner2008}, superconducting quantum computation \cite{clarke2008}, and all the way up to adiabatic quantum computation \cite{farhi2001} we have a number of paths to start the quest, but so far no one of them has managed to prove to be the fastest and more reliable way to take us to the final destination. The main problem related to quantum computation is the intrinsically ephemeral nature of quantum information. The environment dramatically affects quantum systems, introducing errors that destroy the fragile stability needed to run a quantum algorithm. This effect is known as \emph{decoherence} \cite{zurek2003,schlosshauer2007}. 

The necessity to find a reliable solution against errors in the computation suggested a different approach to quantum computing, based on a class of systems known as \emph{topologically ordered} \cite{nayak2008}. Topologically ordered systems form an unusual class of systems, whose ordered phase cannot be described by Landau's theory of spontaneous symmetry breaking. Typically, transitions between different phases of matter correspond to the loss (or gain) of symmetry by the state of the system. However, the discovery of the \emph{fractional quantum Hall effect} \cite{tsui1982}, where states that belong to different phases have equal symmetries, opened to doors to the concept of topologically ordered systems. Indeed, the physical properties of these systems are not characterized by local order parameters, but rather by appropriate non-local quantities that reflect their global topological properties \cite{pachos2012}.

Thanks to these topological properties, such systems store information non-locally in logically stable structures, allowing for a more efficient protection against external perturbations \cite{nayak2008}. Moreover, in two-dimensions they can serve as platforms where special quasi-particles called \emph{anyons} manifest \cite{brennen2008}. The anomalous statistical properties of anyonic excitations allow for the implementation of quantum gates by braiding-like operations, and more in general can be used to perform quantum computation \cite{stern2013}. 

In this Chapter we examine the main ideas, techniques and subtleties required to comprehend topological order, as a means to prepare the ground for continuous-variable topological order in Chapter \ref{CVTO}. Our starting point is the toric code \cite{kitaev2003}, a two-dimensional multi-spin system defined on a discrete lattice, which is revered as the Godfather of all topological systems. Because of its simple description, the toric code has been widely used as a playground to understand the correlations configuration of topological phases, and how these relate to the entropy scaling for sub-partitions of the system \cite{hamma2005, hamma2005a}. We conclude the Chapter by explaining how to detect topological order harnessing the entanglement properties of the system \cite{kitaev2006, levin2006,iblisdir2010,castelnovo2013}.

\section{Quantum lattice systems}
The toric code belongs to the extended family of quantum double models \cite{kitaev2003}, a particular class of quantum many-body systems structurally based on the properties of finite groups and physically defined on a lattice \cite{eisert2006}. This means that, given an underlying graph, the quantum degrees of freedom are depicted by the vertices of the graph while the edges connecting them determine the interactions. Consider a graph $G = G(\mathcal{V},\mathcal{E})$, defined as in \ref{graphsdef}, with $\mathcal{V}$ set of vertices and $\mathcal{E}$ set of edges. One can introduce a distance $\text{dist}(i,j)$ measure between vertices, which for us is just the shortest path connecting the vertices on the graph. For nearest neighbors $\text{dist} =1$.

In general, quantum systems sitting at the vertices can either be spin degrees of freedom of finite dimension or bosonic (fermionic) degrees of freedom, associated to an infinitely dimensional Hilbert space. 

As for the dynamics, in the following we will always be working with a special class of Hamiltonians called \emph{local} Hamiltonians
\beq
\hH = \sum_i \hat{h}_i \,,
\eeq
where each term $\hat{h}_i$ is supported only on the site $i$ and on a finite number of neighboring sites, typically the nearest-neighbors. Given the Hamiltonian $\hH$ we define the \emph{ground state manifold} as the Hilbert space whose basis states are the ground states, or states of lowest energy, of $\hH$. In general the ground state can be degenerate. 

The energy gap of the Hamiltonian is defined as the energy difference between any ground state and the first excited state of the system
\beq
\Delta E = \text{inf}_{\ket{\psi} \in \ham \setminus \mathcal{G}} \bra{\psi} \hH \ket{\psi} - E_0 \,,
\eeq
where $E_0$ is the energy of the ground states. The energy gap is an extremely important quantity in the study of topological quantum systems: It directly relates to many of their properties, first of all the resilience against external perturbations. It is also related to a more subtle concept, the scaling behavior of the ground state correlations along the subsystems of the lattice. For local operators $\hat{O}(J)$ defined on a finite number of sites $J$, the correlation between two operators is given by
\beq
\6 \hat{O}(A) \hat{O}(B) \9 - \6 \hat{O}(A) \9 \6 \hat{O}(B) \9 \,,
\eeq
where $\6 \hat{O}(J) \9 $ is the ground state expectation value of $\hat{O}(J)$. If $\Delta E > 0$ in the thermodynamic limit of $N \to \infty$ number of vertices of the lattice, then 
the correlations decay exponentially with the operators distance \cite{hastings2006}
\beq
\label{corlen}
| \6 \hat{O}(A) \hat{O}(B) \9 - \6 \hat{O}(A) \9 \6 \hat{O}(B) \9 | \le C e^{- \frac{\text{dist}(A,B)}{\xi}} \,,
\eeq
where $\xi$ is the correlation length determining the distance on the graph on which the correlations go to zero, and dist$(A,B)=\text{dist}(i_A, j_B)$ with $i_A$, $j_B$ vertices where the local operators $A,B$ act upon. However, there is more than that. Consider a subsystem $A \subset \mathcal{V}$, where $A$ now stands for the region of the subsystem on the lattice, of a gapped free bosonic \cite{cramer2006, cramer2006b} or fermionic system \cite{eisert2010} (free means that its Hamiltonian is quadratic in the bosonic or fermionic operators). Then, if $\rho$ represents the (pure) ground state of the system, the von Neumann entropy of the reduced density matrix $\rho_A$ scales as the size of the boundary of $A$ \cite{eisert2010},
\beq
S(\rho_A) = \mathcal{O} (| \partial A |) \,.
\eeq
This is an essential concept in the description of the entropic properties of topological models. Only when an \emph{area law} for the entanglement entropy is satisfied, it is then possible to construct the particular methods utilized to detect topological order by means of linear combinations of entropies, as we discuss below in Section \ref{Woto}.

In the following subsection we describe an example of quantum lattice system, the cluster-state. While this is not related to quantum double models, its generalization to continuous-variable will be used in Chapter \ref{CVTO} as a platform for the implementation of the continuous-variable surface code. 

\subsection{The cluster-state}
\label{qubitCS}
The cluster-state model is a quantum many-body system on a lattice that was introduced as the physical platform to perform measurement-based universal quantum computation \cite{raussendorf2001, raussendorf2002}. A \emph{cluster-state} is a $N$-qubit system associated to a (generally) two-dimensional square graph $G=\{ \mathcal{V}, \mathcal{E} \}$, whose vertices correspond to qubits and whose edges depict the interactions among them \cite{nielsen2005}. The cluster-state is the result of a specific preparation procedure: One starts with a collection of $N$ qubits equally prepared in the superposition state $\ket{+} = \frac{1}{\sqrt{2}} (\ket{0} + \ket{1})$, such that the initial global state is $\ket{\text{G}} = \ket{+}^{\otimes N}$. A controlled-Phase (c-Phase) gate is then applied between pairs of neighboring qubits in accord with a chosen graph, with one qubit per vertex of the graph. Recall that the qubit c-Phase gate between two sites $\6 i,j \9$ is given by the unitary transformation \cite{hein2006}
\beq
\hat{C}_{\text{P}_{(i,j)}} := e^{- i \pi \ket{1}_i \bra{1}\otimes \ket{1}_j \bra{1}} \equiv \ket{0}_i \bra{0} \otimes I_j + \ket{1}_i \bra{1} \otimes \hat{\sigma}^Z_j \,,
\eeq
whose action on two spin states is
\beq
 \hat{C}_{\text{P}_{(i,j)}} \ket{+}_i \ket{+}_j = \frac{1}{\sqrt{2}} (\ket{0}_i \ket{+}_j + \ket{1}_i \ket{-}_j) \, ,
\eeq
with $\ket{-} = \frac{1}{\sqrt{2}} ( \ket{0} - \ket{1})$. The cluster-state has a easier description in terms of stabilizers (\ref{stabIntro}). After the initialization of the spin states, but before the application of the c-Phase gates, the stabilizer group of the global state is simply given by the set $\{ \hat{\sigma}^X_i\}$, since $\hat{\sigma}_j^X \ket{+}_j = \ket{+}_j$. 

On a two-dimensional square lattice, after creating the interactions through c-Phase gates, the stabilizer set is transformed into a new set equal to
\beq
\label{qcsstab}
\{ \hat{K}_j = \hat{\sigma}_j^X \prod_{k \in \mathcal{N}(j)} \hat{\sigma}_k^Z = \hat{\sigma}_j^X \hat{\sigma}_N^Z \hat{\sigma}_S^Z \hat{\sigma}_E^Z \hat{\sigma}_W^Z \} \,,
\eeq
with $\mathcal{N}(j)$ identifying the nearest neighbors of qubit $j$, and with N, S, E, and W indicating the qubit to the North, South, East and West of qubit $j$, respectively. 

Finally, the cluster-state is the ground state of the Hamiltonian constructed by imposing an energy penalty for violating any of the stabilizers conditions:
\beq
\hat{H}_{\text{CS}} = - \sum_j \hat{K}_j \,.
\eeq
The description of experimental realizations of small cluster-states can be found for example in \cite{ walther2005, chen2007, gao2010}. For the sake of clarity, universal quantum computation on a cluster-state is performed by means of a mechanism known as one-way quantum computation. Here, instead of performing logical gates on the individual qubits of the state, as in the circuit-based model of quantum computing, one performs a sequence of single-qubit projective measurements in order to establish the equivalent of a quantum circuit on the state \cite{raussendorf2003}.

\section{Quantum double models}
\label{qdm}
Topological phases of matter reveal themselves in systems usually associated to gapped Hamiltonians with a degenerate ground state subspace, whose degeneracy is determined by the geometric, or topological, properties of the underlying manifold \cite{pachos2012}. 

\emph{Quantum double models} $D(\mathcal{G})$, or \emph{surface codes}, are a broad class of well-studied lattice realizations of topologically ordered systems based on the theory of error-correcting codes and introduced by Kitaev in \cite{kitaev2003}. In particular, they are spin systems on a 2-dimensional lattice, constructed starting from a finite group $\mathcal{G} = \{ g_j \}$ whose elements label an orthonormal basis $\ket{g} : g \in \mathcal{G}$ that define an Hilbert space $\mathcal{H}$ with dimension $|\mathcal{G}|$. Taken the graph of the lattice, with vertex set $\mathcal{V} = \{v_i\}$, edge set $\mathcal{E} = \{ e_i \}$ and face set $\mathcal{F} = \{ f_i \}$, we assign the Hilbert space $\mathcal{H}$ to each edge, in order to have a qudit-particle living on every link with $|\mathcal{G}|$ internal levels. The total Hilbert space of the model is thus $\mathcal{H}_{D} = \mathcal{H}^{\otimes |\mathcal{E}|}$. An orientation is assigned to every edge $e_{j,k} = [v_j, v_k]$, intended as an arrow pointing from one vertex to the other, and a global one is assigned to all the faces (clockwise or counterclockwise) \cite{brennen2008}.

It is possible to define a local Hamiltonian for a quantum double model, defining two set of operators. The first set $\{ \hat{A} (v) \}$ describes the interaction of the particles meeting at a vertex $v$, while the second set $\{ \hat{B}(f) \}$ describes the interaction among particles sitting on the boundary of a face $f$. First recall that a particle state is equivalent to $\ket{j} \equiv \ket{g_j}$. For instance, for the toric code $\mathcal{G} = \mathcal{Z}_2 = \{ 0,1 \}$ and thus the particle state is labeled by the elements $\{ \ket{0}, \ket{1} \}$. Then, introduce the linear operators $L_+^g, L_-^g, T_+^h, T_-^h$, whose action on a generic particle state is \cite{pachos2012}
\beq
L_+^g \ket{z} = \ket{gz}\,, \quad L_-^g \ket{z} = \ket{z g^{-1}}\,, \quad T_+^h \ket{z} = \delta_{h,z} \ket{z}\,, \quad T_-^h \ket{z} = \delta_{h^{-1},z} \ket{z} \,.
\eeq

The Hamiltonian is constructed so that the ground state of the model is invariant under the following local gauge transformations
\beq
\varXi_g(v) = \prod_{e_j \in [v, \star]} L_+^g(e_j) \prod_{e'_i \in [\star,v]} L_-^g(e'_i) \,,
\eeq
and it is equal to \cite{kitaev2003}
\beq
\label{genHam}
\hH = - \sum_v \hat{A}(v) - \sum_f \hat{B}(f) \,.
\eeq
The vertex operators $\{ \hat{A}(v) \}$ project out states that do not satisfy the local gauge invariance condition at $v$, and are given by
\beq
\hat{A}(v) = \frac{1}{| \mathcal{G}|} \sum_{g \in \mathcal{G}} \varXi_g (v) \,.
\eeq
The physical meaning of the face operators is inspired by the interpretation of the quantum double models in terms of lattice gauge theories, where each $\hat{B}(f)$ becomes a magnetic field on the face $f$ \cite{kitaev2003}. Then the operators $\{ \hat{B}(f) \}$ project out states with non-vanishing magnetic flux at the face $f$, 
\beq
\hat{B}(f) = \sum_{\prod_{e_h \in \partial f} h_k = e} \left(\prod_{e_k \in \partial f} T^{h_k}_{o_f(e_k)} (e_k)\right) \,,
\eeq
where we are summing over series of group elements such that, for a face boundary of size $n$
\beq
h_1 h_2 ... h_{n-1} h_n = e\,,
\eeq
with $e$ identity element of $ \mathcal{G}$, and the function $o_f(e_k) = \pm$ whether the face orientation is the same as the orientation of the edge $e_k$ or not. For a square lattice with counterclockwise face orientation, vertical links oriented upwards, horizontal links oriented rightwards and edge labeling as in Fig. \ref{tocode}, we have the following explicit expressions:
\beq
\hat{A}(v) = \frac{1}{| \mathcal{G}|} \sum_{g \in  \mathcal{G}} L_+^g (e_1) L_+^g (e_2) L_-^g (e_3) L_-^g (e_4) \,,
\eeq
and
\beq
\hat{B}(f) = \sum_{h_1 h_2 h_3 h_4 = e} T_-^{h_1} (e_1) T_+^{h_2} (e_2) T_+^{h_3} (e_3) T_-^{h_4} (e_4) \,.
\eeq
It is easy to check that the vertex and face operators are projectors $(\hat{A}(v)^2 = \hat{B}(f)^2 = 1)$ and that they all commute with each other and with the Hamiltonian by construction. Moreover, the ground state $\ket{gs}$ of the Hamiltonian satisfies
\beq
\label{stabcond}
\hat{A}(v) \ket{gs} = \hat{B}(f) \ket{gs} = \ket{gs} \,,
\eeq
and it is thus stabilized by the set of vertex and face operators, which completely determines the ground state manifold.

\subsubsection{Anyone anyons?}
After this basic explanation of the theory of quantum doubles, let us spend few words about anyons and the anyonic properties of such models. Quantum double models are so called because the set of operators $\{ \hat{A}(v) \}$ and $\{ \hat{B}(f) \}$ generate an algebra $\mathfrak{D}$\footnote{There is more! The algebra $\mathfrak{D}$ is a finite-dimensional $\text{C}^*$-algebra, see \ref{csa} for the precise meaning of $\text{C}^*$-algebra and irreducible representation.} that is the quantum double of the original group $ \mathcal{G}$, in the sense of \cite{drinfeld1987}. The ground state of a quantum double model corresponds to the absence of anyons on the lattice, or simply to the \emph{anyonic vacuum}. Excited states are those states that do not satisfy the stabilizer conditions in (\ref{stabcond}) and indicate the presence of quasi-particles on the lattice. The particle's types that are supported by a particular quantum double model correspond one-to-one to irreducible representations of the algebra $\mathfrak{D}$ \cite{kitaev2003}. In this sense the properties of the anyons are dictated by the structure of the group $ \mathcal{G}$ \cite{pachos2012}. 

An elementary excitation occurs when one of the stabilizer conditions is violated. Anyons always exist in pairs and they are created at the endpoints of a string that can be located on two vertices or two faces. Astonishingly enough, anyons do not follow the bosonic or fermionic statistic of particles exchange \cite{stern2008}. In three spatial dimensions and one time dimension, when two point-like particles are exchanged, their wave function will acquire either a plus (bosonic) or a minus (fermionic) sign. In two spatial dimensions, things are different and the wave function can acquire any phase under the braiding of two particles \cite{nayak2008}. Hence anyons obey a fractional statistics and in this sense they elude the usual characterization of particles \cite{wilczek1982}. Anyons are \emph{abelian} if, after having braided one quasi-particle around another (as long as they are different types), the wave-function of the entire system acquires a $-1$ phase that does not depend on the specific form of the paths followed by the particles. 
Conversely, \emph{non-abelian} anyons have a more exotic behavior and realize unitary transformation on the state when wrapped one around the other \cite{stern2013}. These, together with their fusion properties, are at the basis of the so-called topological, or anyonic, quantum computation \cite{pachos2012, brennen2008}. Hence, anyone anyons\footnote{This is A/Prof. Brennen's favorite line about anyons.}?

\subsection{The toric code state}
\label{TTC}
The toric code \cite{kitaev2003} is the simplest topologically ordered system. It serves as a fantastic toy-model for the study of topological properties of most general models and it is usually used to show how quantum information can be stored in a resilient way in a spin-system lattice.
\begin{figure}[tbp]  
\centering
\setlength{\unitlength}{1cm}
\includegraphics[width=6\unitlength]{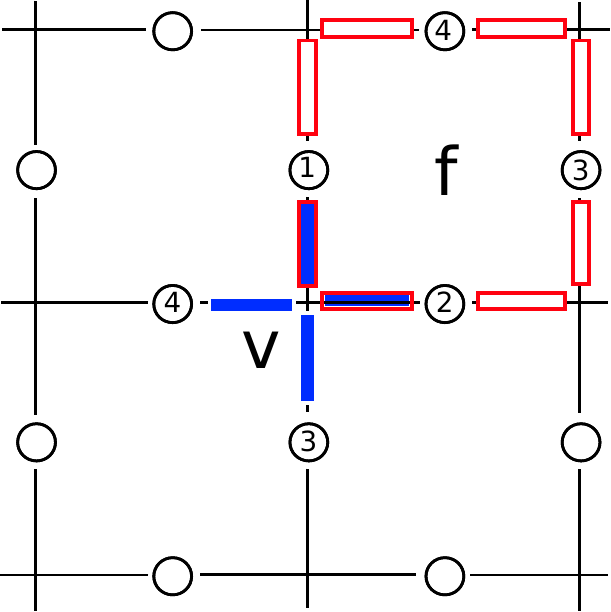}
\caption{Section of the toric code lattice. The vertex and face sites are depicted in blue and red respectively, while sites 1 and 2 overlap. The sites labeling here is not important since the edges are not oriented.}
\label{tocode}
\end{figure}
	
We mentioned earlier that the toric code represents the basic realization of a quantum double model, since it is the quantum double $\mathfrak{D}( \mathcal{Z}_2)$ of the group $ \mathcal{Z}_2 = \{ 0, 1\}$. It is an ensemble of $2 N^2$ qubits, $1/2$-spin particles, sitting on the edges of a $N \times N$ square lattice with toroidal boundary conditions, which does not require any explicit orientation because of the simple group structure. The vertex operators are 
\beq
\label{tcA}
\hA(v) = \prod_{i \in  \mathcal{N}(v)} \hsi_i^X = \hsi_1^X \hsi_2^X \hsi_3^X \hsi_4^X \,,
\eeq
while the face operators become
\beq
\label{tcB}
\hB(f) = \prod_{i \in \partial f} \hsi^Z_i = \hsi_1^Z \hsi_2^Z \hsi_3^Z \hsi_4^Z \,,
\eeq
with $\hsi^{X,Z}$ Pauli-$X$ and Pauli-$Z$ operators respectively. From Fig.(\ref{tocode}) it is clear that each vertex and face can only have 2 or no sites in common and, from the Pauli operators commutation, it correctly follows that all the $\hA$ and $\hB$ commute. Recall that these operators are projectors, and consequently their eigenvalues are $\pm 1$. In analogy with (\ref{genHam}), the Hamiltonian of the system is defined by a linear combination of the stabilizers
\beq
\label{Htc}
\hH = - \sum_v \hA(v) - \sum_f \hB(f) \,,
\eeq
and it can be diagonalized easily. This Hamiltonian commutes with all the stabilizers and defines the qubit toric code. Again, the ground state subspace of (\ref{Htc}) is the space of all states stabilized by the set $\{ \hA(v), \hB(f) \}$, rigorously \cite{kitaev2003}
\beq
\mathcal{GS} =  \{ \ket{g} \in \mathcal{H}_2^{\otimes 2N^2} : \hA(v) \ket{g} = \ket{g}, \,\, \hB(f) \ket{g} = \ket{g}\,\,\, \forall\, v,f \} \,.
\eeq
If there were as many independent stabilizers as there are qubits on the lattice, then the ground state of the toric code would be unique. However, as a consequence of the constraints on the stabilizers
\beq
\prod_v \hA(v) = 1 \,, \,\,\,\,\,\,\, \prod_f \hB(f) = 1 \, ,
\eeq
there are only $2 N^2 -2$ independent stabilizers. This means that the dimension of $\mathcal{GS}$ is $2^2 = 4$ and the ground state is four-degenerate \cite{calderbank1997}. If the code was defined on a Riemannian surface with \emph{genus} $\mathfrak{g}$, then the ground state subspace would be $2^{2 \mathfrak{g}}$-degenerate \cite{hamma2005}. Thus, this should convince the reader that certain properties of the system explicitly depend on the geometry of the underlying manifold. 

Local spin operations create excitation pairs on the lattice, and the group structure of $\mathcal{Z}_2$ defines the families of anyons supported by the code. When in the ground state, the action of a single $\sigma^x$ on a spin creates a pair of $m$-type anyons located inside the adjacent faces, while a $\sigma^z$ creates a pair of $e$-type anyons on the two adjacent vertices. There is a third kind of particle pairs $\epsilon$ which is the composition of the first two excitations, explicitly $\sigma^z \sigma^x = i \sigma^y \to e \times m = \epsilon$ \cite{pachos2012}. 

Closed contractible loops of $\sigma^z$ and $\sigma^x$ operators on the lattice leave the ground state invariant. Physically this corresponds to moving one particle of the pair to the site of the other, annihilating them. However, on the torus it is possible to construct 4 types of non-contractible loops along the two spatial dimensions \cite{hamma2005}. These loops act as logical gates and, while still invariant with respect to the ground state manifold, they transform one ground state basis vector into another. In this sense the toric code can serve as a quantum memory: The basis vectors of the 4-fold degenerate ground state manifold encode two logical qubits in the model of the code. 
 
\section{Witnesses of topological order}
\label{Woto}

Physically, topological order results from non-local long-range correlations that extend across all the spins (or modes, as we will see in the continuos-variable case in the following Chapter) of the system in a topological phase. The resulting lack of a local order parameter makes identifying topological order a difficult task, and a different approach is required to identify it. In this section we deal with this problem: First, we briefly analyze the entropic properties of a topological system and then explain the concept of \emph{topological entanglement entropy} (TEE), an intrinsically non-local parameter that has been appositely defined to characterize topological phases in a variety of systems, including the toric code.

\subsection{Topological effects}
The first hint regarding the topological effects on the bipartite entropy of a region of the toric code (and consequently on the nature of the TEE) was given in \cite{hamma2005, hamma2005a}, where the authors argued that the entanglement entropy $S(\rho) = - \tr (\rho \log_2 \rho)$ could help to detect topological order. By calculating the entanglement entropy of an arbitrary bounded connected region of the lattice for the toric code ground state, they found that its value solely depends on the order of the group generated by the vertex operators that act only on the boundary of the region, which is equivalent to the boundary size, minus a constant ($+1$):
\beq
S(\rho_A) =  |\partial A| -1 \,.
\label{sra-1}
\eeq
Moreover, it was understood that the correction to the area law behavior of the entropy is due to the constraint on the number of independent vertex operators $\hat{A}_v$. Another interesting conclusion is that, although no two spins are correlated, there is no bipartition of the system that gives a zero entropy. Also in this sense the ground states of topological systems are  multi-body, highly entangled states. Note also that the result in \ref{sra-1} is independent of the choice of ground state in case of degeneracy. 

As a remark for the lovers of simplicity and beauty in Nature, in \cite{livine2009} it was found an identical result for a theory based on a different group in a different setup. Specifically, the entanglement entropy of a closed boundary of space for pure three-dimensional Riemannian gravity, formulated in terms of a topological BF theory for the gauge group SU(2), scales in the very same same way as the toric code one:
\beq
S(\rho_A) \propto (V_b -1) \,,
\eeq
where $V_b$ is the number of boundary loops across the cut.

\subsection{Topological entanglement entropy}
\label{subsTEE}
These ideas were later broadened, and unambiguous relationships among the apparently unrelated concepts of entanglement entropy, total quantum dimension of the model, and topological entanglement entropy were established. A topological system can support different types of anyons. One can assign to each anyonic species $\tau_i$ a quantum dimension $d_i$, i.e. the dimension of the Hilbert space associated to the anyon. Alternatively, the quantum dimension is the rate of growth of the Hilbert space when we increase the number of particles $\tau_i$ by one. The sum over all the quasiparticle species is the \emph{total quantum dimension} of the model \cite{brennen2008}:
\beq
\mathcal{D} = \sqrt{\sum_i d_i^2} \, .
\eeq
Both Kitaev and Preskill (KP) \cite{kitaev2006} and Levin and Wen (LW) \cite{levin2006} intuited that, for a topologically ordered phase, the entanglement entropy of a region $A$ scales as
\beq
S_A = \alpha |\partial A| - \gamma + \epsilon \, ,
\eeq
where $\alpha \in \mathbb{R}$, $|\partial A|$ is the size of the boundary of region $A$, $\epsilon$ is a contribution that goes to zero in the limit of $|\partial A| \to \infty $, and the quantity
\beq
\label{gammaD}
\gamma = \log{\mathcal{D}}
\eeq
is exactly the topological entanglement entropy (TEE).

KP, by an appropriate selection of regions of the system, focus on constructing a linear combination of entanglement entropies such that all the terms proportional to the length of the boundaries of the regions cancel out, leaving only the topological term. Specifically, this corresponds to:
\beq
\label{KPTEE}
\gamma \equiv - (S_A + S_B + S_C -S_{AB} - S_{BC} - S_{AC} + S_{ABC}) \, .
\eeq
Once the regions are chosen to be reasonably bigger than the correlation length $\xi$ of the system, the parameter introduced in Eq.~(\ref{corlen}) that determines the exponential decay of the correlations of local operators $\hat{O}$ with respect to the ground state of the system
\beq
\6 \hat{O}(a) \hat{O}(b) \9 - \6 \hat{O}(a) \9 \6 \hat{O}(b) \9 \sim \text{Exp} \left( - \frac{| a - b |}{\xi} \right) \,,
\eeq
it is possible to demonstrate that deformations of the boundaries of the regions, and smooth deformations of the system Hamiltonian, do not change $\gamma$. Physically, this means that $\gamma$ is both invariant, hence only determined by the underlying topological properties, and universal, i.e. local modifications of the Hamiltonian do not affect its value.

\begin{figure}[tbp]  
\centering
\setlength{\unitlength}{1cm}
\includegraphics[width=14\unitlength]{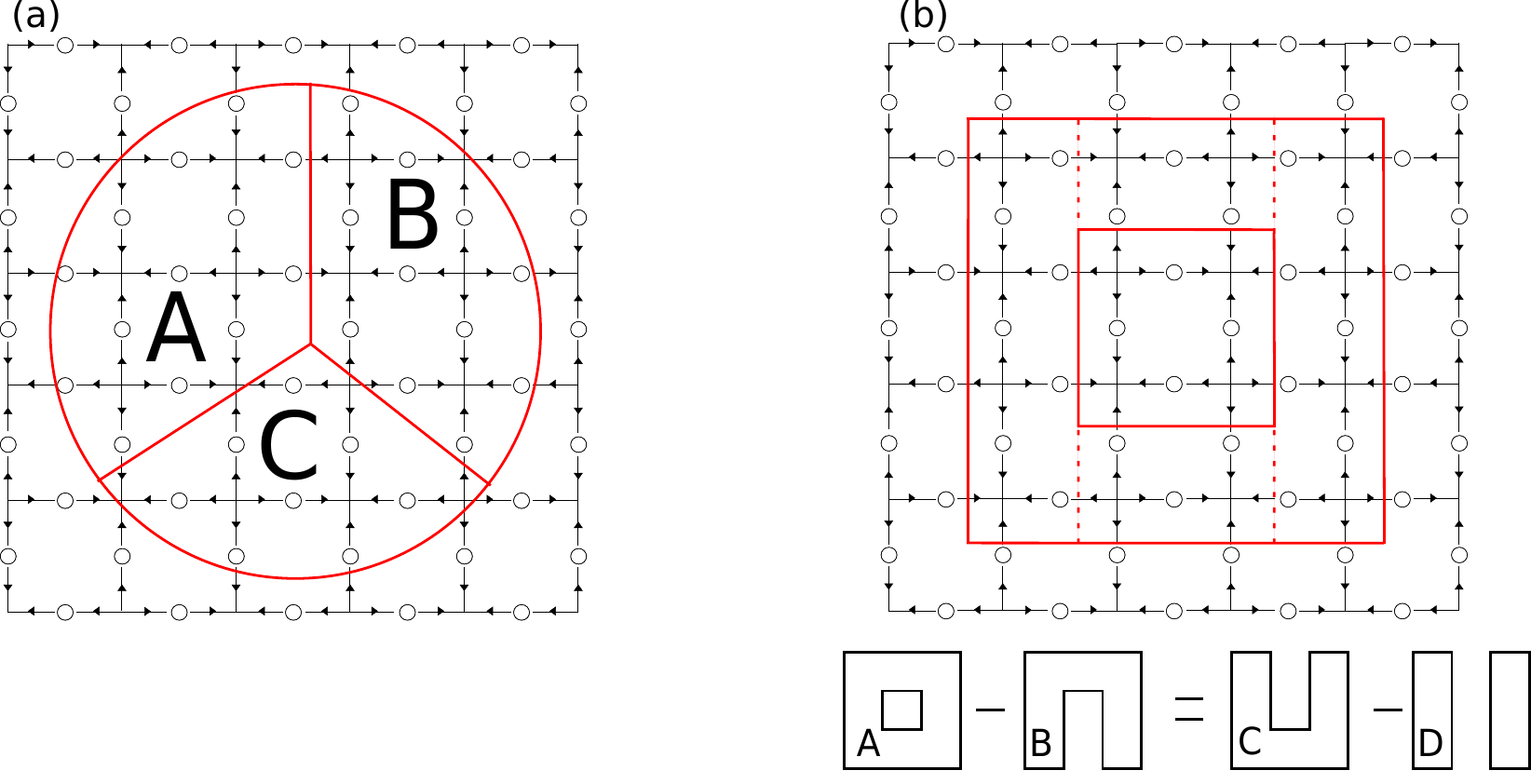}
\caption{Sections used for calculations of topological entanglement entropy by the methods of (a) Kitaev-Preskill \cite{kitaev2006} and (b) Levin-Wen \cite{levin2006}. The areas of the regions in (b) satisfy the equality depicted.}
\label{areasTEE}
\end{figure}

LW moved along a different line of thought, considering partitions of the system with different topologies rather than more general regions as KP, see Fig.~(\ref{areasTEE}). The partitions are constructed such that their pairwise differences are equivalent, and chosen to be large enough, so that short-range correlations decay inside the boundaries. In analogy to Eq.~(\ref{KPTEE}), an explicit expression for the TEE can be derived 
\beq
\label{LWTEE}
\gamma \equiv -\frac{1}{2} [ (S_A - S_B) - (S_C - S_D) ] \, .
\eeq
If the system is not topologically ordered, and if the entanglement entropies only depend on terms proportional to the boundaries, then the differences in both Eq.~(\ref{KPTEE}) and Eq.~(\ref{LWTEE}) are exactly zero (up to exponentially decaying corrections). But while in Eq.~(\ref{KPTEE}) this corresponds to extracting $\gamma$ directly by canceling the area-dependent terms, the nonzero contribution in the LW construction is less trivial. Analyzing the honeycomb lattice with spin-$1/2$ on the edges \cite{kitaev2003}, LW prove that if topological long-range correlations are present, non-local closed string operators with non-vanishing expectation value exist \cite{levin2006} and operators that wind around the region $A$ contribute to a lower value of the entropy $S_A$. This therefore corresponds to a nonzero value of Eq.~(\ref{LWTEE}). They indeed call this contribution TEE. 

The TEE is a witness for topological order in the following sense: the value of $\gamma$ characterizes the global anyonic and entanglement properties of the topological state. When $\gamma = 0$, it follows from Eq.~(\ref{gammaD}) that $\mathcal{D}=1$, which physically means that no anyons are supported by the system, and no long-range topological correlations contribute to the entanglement. The idea behind both the KP and LW derivations of the TEE formulas is to suppress the non-topological correlations and extract only the topological information.

\subsection{Topological order at finite temperature}
\label{tordofit}
For mixed states or states at finite temperature, the von Neumann entropy does not quantify properly the amount of correlations shared by two partitions of a system, and cannot be used as a parameter to detect topological order. Instead, the appropriate signature of topological order is given by a quantity called topological mutual information (TMI), introduced in \cite{iblisdir2010}:
\beq
\gamma_{\text{MI}} = -\frac{1}{2} (I_A + I_B + I_C -I_{AB}-I_{BC}-I_{AC}+I_{ABC}) \, ,
\eeq
This is calculated using the same regions of the KP construction for the TEE, Fig.(\ref{areasTEE}). Now the correlations are measured using (one half of) the quantum mutual information $I_A$, see \ref{secEEMI}, between a region A of a system and its complement $A_c$, i.e. $I_A = S_A + S_{A_c} - S_{A \cup A_c}$, which replaces the von Neumann entropy $S_A$ for a pure state (up to a factor of 2).

In \cite{wootton2012} a modified version of the TMI was proposed to avoid treacherous ambiguities in the recognition of topological order. In fact, certain mixed states (in particular mixture of degenerate ground states of gapped local Hamiltonians) can exhibit long-range non topological correlations and therefore affect the value of the TMI. Together with a new formulation for the TMI, the author also introduced lower and upper bounds that precisely determine the TMI whenever they coincide. Consider now the LW regions as defined in Fig.(\ref{areasTEE}) and label $E = A \setminus B$, set difference of $A$ and $B$, $F_1 = A \setminus C$, and $F_2 = D$ (from which follows that $F = F_1 \cap F_2 = C$). Then the value of the TMI lies between these two bounds:
\beq
\text{max}(I_{E,F} - I_{E,F_1}-I_{E,F_2},0 ) \le \gamma \le I_{E,F} -  \text{max} (I_{E,F_1}, I_{E,F_2}) \, .
\eeq

\subsubsection{Topological logarithmic negativity}
Recently \cite{lee2013, castelnovo2013} it has been proved that, for the toric code, also the negativity and the logarithmic negativity scale with the boundary of the region considered minus a topological constant related to the topological entropy. We will exploit this additional property of topological phases to give stronger evidence that our continuous-variable model of harmonic oscillators on a lattice really exhibits topological order.

\section{Discussion}
In this Chapter we gave a short overview about the theory of topological quantum systems on 2-dimensional lattices, introducing the stabilizers structure of local Hamiltonians and in particular the notion of topological entanglement entropy. In the next Chapter we will extend these concepts to continuos-variable states: Using Gaussian states theory as a toolbox we demonstrate, through the witnesses of topological order introduced here, that also the continuous-variable version of the toric code can exhibit topological order. 

\begin{savequote}[10cm] 
\sffamily 
``To bankrupt a fool, give him information.'' 
\qauthor{Nassim Nicholas Taleb - The Bed of Procrustes} 
\end{savequote}

\chapter{Continuous-Variable Topological Phases}
\graphicspath{{CV_t_Order}} 
\label{CVTO}

We have seen in the previous Chapter that the unusual properties of topologically ordered systems, in particular the resilience against local perturbations, make them attractive candidates for stable quantum memories or processors \cite{pachos2012}. However, probing topological order experimentally is an onerous task. Some possibilities include measuring non-local string operators \cite{jiang2008} or the statistics of anyonic excitations above the ground state, as has been demonstrated experimentally with small photonic networks \cite{lu2009}. However, due to finite correlation lengths of local operators \cite{dusuel2008}, these methods suffer from low visibility if the system is not prepared in a pure phase with vanishing two-point correlations.

An alternative is to study properties of the state itself that are robust to small changes in the correlation length: For instance one could consider the topological entanglement entropy (TEE) \cite{kitaev2006,levin2006} introduced previously in \ref{subsTEE}, which is an intrinsically non-local quantity that characterizes topological phases in a variety of systems, including bosonic spin liquids \cite{isakov2011} such as the qubit toric code \cite{hamma2005}, and fermionic Laughlin states \cite{haque2007}. While useful for numerics, actually measuring TEE in a physical system is a daunting task, since extracting the von Neumann entropy requires knowledge of the complete spectrum of the reduced state. 

In this Chapter we present a possible solution to this problem, studying for the first time topological order in a continuous-variable Gaussian state \cite{weedbrook2012} analog of the discrete-variable surface code states \cite{kitaev2003} introduced in \ref{qdm}. As it happens, rather remarkably, we prove that the TEE of the CV surface code can be easily computed simply from measurements of the quadrature operators (introduced in \ref{aaad}). We also show that unlike its qubit counterpart, the CV surface code state has a parent Hamiltonian that is gapless in the thermodynamic limit. Nevertheless, the quadrature correlations on the lattice still decay exponentially with the distance.

After a short introduction to the CV formalism, we describe how to prepare the CV surface code efficiently using an intermediate mapping to first an ideal (infinitely squeezed) \cite{menicucci2006, gu2009} and then a physical (finitely squeezed) CV cluster-state \cite{menicucci2007, menicucci2008, flammia2009}. The physical CV cluster-state is the ground state of a quadratic Hamiltonian constructed, in analogy with the quantum double models, using the elements of the nullifier sets: We study its spectrum and show that in the thermodynamic limit it is gapless. The central part of the Chapter deals with the problem of detecting topological order: After showing how to derive the covariance matrix $\Ga$ of the CV surface code using the graphical calculus for pure Gaussian states (\cite{menicucci2011} and \ref{gracagas}), we prove the exponential decay of the correlations studying the special form of $\Ga$. We then calculate the topological entanglement entropy and the topological logarithmic negativity \cite{lee2013, castelnovo2013}, following the methods introduced in Chapter \ref{ChapTO}. These quantities are non-zero for any value of the initial squeezing. We continue by analyzing the stability of the CV topological order against two forms of noise: thermalization and noisy input states. To conclude, we propose experimental realizations for this model that are accessible with today's technology.

\section{Continuous-variable generalities}
We have already introduced the Gaussian states formalism in Chapter \ref{ChapGS} and presented the qubit-based cluster-state \cite{raussendorf2001, raussendorf2002} in \ref{qubitCS}. In order to extend the cluster-states and surface codes formalism to continuous-variable states, we need to introduce the language in which continuous-variable logic is spoken. As a first thing we must generalize the Pauli group of single qubit gates (\ref{pauliqubit}) to the Weyl-Heisenberg group of phase-space displacements\footnote{Additional information about this group is presented in \ref{polyquantiz}} \cite{bartlett2002}. For a single qubit, this generalization is most easily accomplished by thinking of the qubit $\hat{\sigma}^X$ and $\hat{\sigma}^Z$ gates as implementing one-unit cyclic shifts in \textquoteleft position\textquoteright
\beq
\hat{\sigma}^X \ket{0} = \ket{1} \, ,
\eeq
and \textquoteleft momentum\textquoteright
\beq
\hat{\sigma}^Z \ket{+} = \ket{-} \, .
\eeq
The CV analogs of the qubit gates are the translation (position-shift) operator $\hat{X}(t)$ and boost (momentum-shift) operator $\hat{Z}(u)$, with $t,u \in \mathbb{R}$. While for qubits the shift is by an element of the cyclic group $\mathcal{Z}_2 = \{ 0,1 \}$, i.e. the gates transform a basis element into the other, in the CV case, one may implement a shift in position or momentum by any real valued amount. In fact, the Weyl-Heisenberg group for continuous-variables is a continuous Lie group, whose generators are the elements of the Lie algebra spanned by the identity operator $\hat{I}$ and the canonical self-adjoint quadrature operators $\hq$, $\hp$, which satisfy the canonical commutation relation $[\hq, \hp] = i$ \cite{tung1985}.

Specifically, the displacement operators are equal to
\beq
\label{contOp}
\hat{\sigma}^X \longrightarrow \hX(t) = e^{- i t \hp}\, , \qquad \text{and} \qquad \hat{\sigma}^Z \longrightarrow \hZ(u) = e^{i u \hq}\, ,
\eeq
with group commutator
\beq
\hX(-t) \hZ(-u) \hX(t) \hZ(u) = e^{- i t u} \, .
\eeq
The action of the displacement operators on the eigenstates (or continuous computational basis) of $\hq$ and $\hp$ is given by \cite{lloyd1999}
\beq
\hX(t) \ket{q}_q = \ket{q + s}_q \qquad \text{and} \qquad \hZ(u) \ket{p}_p = \ket{p + u}_p \, ,
\eeq
where the subscript means that $\hq \ket{y}_{q} = y \ket{y}_{q}$ and $\hp \ket{y}_{p} = y \ket{y}_p$. Then, the Pauli group $\mathcal{P}_N$ for CV quantum computation on $N$ oscillators is simply given by the set $\{ \hX(t)_i, \hZ(u)_i  \}$ with generating algebra $\{ \hq_i, \hp_i , \hat{I}_i \}$ for $i = 1...N$.

The transformations we will use in the following discussion belong to the \emph{Clifford group for continuous-variable} $\mathcal{C}(\mathcal{P}_N)$ \cite{bartlett2002}. This is the group of transformations that preserve $\mathcal{P}_N$ under conjugation, i.e. that given any $\hat{U} \in \mathcal{C}(\mathcal{P}_N)$, then 
$\hat{U} \hat{P} \hat{U}^\dagger \in \mathcal{P}_N$ for every $\hat{P} \in \mathcal{P}_N$. In this way we enforce the requirement that all the operations we perform on the state transform stabilizer operators into stabilizer operators, see \ref{stabIntro}.

The CV Clifford group is the semi-direct product of the Pauli group $\mathcal{P}_N$ and $\text{Sp}(2N, \mathbb{R})$, the real symplectic group met in \ref{sympa}. From the definition, it is clear that CV Clifford operations are Gaussian operations. Moreover, because of the algebraic structure of $C(\mathcal{P}_N)$  one can prove that it is generated only by four operations \cite{bartlett2002,zhang2008}, namely:
\begin{itemize}
\item The translation operator $\hX(t)$;
\item The Fourier transform operator $\hat{F} = e^{\left( i \frac{\pi}{4} (\hq^2 + \hp^2 ) \right)}$, which is the CV analog of the Hadamard operator for qubits and act on the continuous computational basis as $\hat{F} \ket{q}_q = \ket{q}_p$ and $\hat{F} \ket{q}_p = \ket{-q}_q$;
\item The phase gate $\hat{P}(\eta) = e^{\left( \frac{i}{2} \eta \hq^2 \right)}$;
\item The generalization to CV of the controlled-PHASE gate from \ref{qubitCS}, now called controlled-$Z$ gate and represented by the operator $\hat{\text{C}}_{\hat{Z}_{(1,2)}} = e^{i \hq_1 \hq_2}$;
\end{itemize}
Although these operations do not form a universal set of gates for CV quantum computation \cite{bartlett2002, lloyd1999}, together with quadrature measurements they are sufficient to prepare the CV surface codes and detect topological order \cite{demarie2013b}. Note that the single-mode squeezing transformation $\hat{S}(s)$, from \ref{trarumea}, belongs to the Clifford group. 

All the transformations just shown have a representation in the graphical calculus for Gaussian pure states, as explained in \ref{gracagas}. For the full list of graph transformation rules see \cite{menicucci2011}.

\section{Ideal continuous-variable codes}
We start the description of the continuous-variable models introducing the so-called \emph{ideal codes} \cite{gu2009, zhang2008}. These codes are called ideal, or infinitely squeezed, because they represent unphysical states of infinite energy and therefore are not normalizable. Although these codes are mere mathematical approximation without an equivalent physical state, they play the role of theoretical CV analog of the qubit cluster-state and Kitaev's surface codes, being also the limiting case of the physical states that will be introduced in the next section.

\subsection{The ideal continuous-variable cluster-states}
The continuous-variable surface code can be easily described and efficiently prepared using an intermediate mapping to the continuous-variable cluster-state via a simple pattern of quadrature measurements: Hence, this will be our starting point. Intuitively, the CV cluster-state is the CV analog of its qubit-based cousin \cite{raussendorf2001}. There are many ways to construct physical CV cluster-states \cite{menicucci2006, gu2009, menicucci2007, menicucci2008, flammia2009, aolita2011, menicucci2011a}, all of which give slightly different states in the finitely squeezed case \cite{menicucci2011}. Each has important differences that manifest when using them for measurement-based quantum computation. In the ideal, infinitely squeezed case, however, these differences become largely irrelevant, and since infinitely squeezed states are unphysical anyway, we are free to choose for their analysis the method that is simplest. For this reason we choose the \emph{canonical method} \cite{menicucci2006}: Despite its inefficiency when used in practice \cite{vanLoock2007}, this is the most straightforward generalization of the qubit cluster-state preparation procedure described in \ref{qubitCS}. 

Given a square lattice defined on a graph $G = (\mathcal{V}, \mathcal{E})$ (\ref{graphsdef}), we first substitute the $N$ qubits on the vertices with $N$ qumodes initialized in the infinitely-squeezed zero-momentum eigenstate $\ket{0}_p$, which is equivalent to the $\ket{+}$ state for qubits. The global state is therefore $\ket{0}_p^{\otimes N}$, and this is stabilized by the single mode operators $\hat{X}_j(t)$ in the sense that
\beq
\hX_j(t) \ket{0}_p^{\otimes N} = \ket{0}_p^{\otimes N}\, , \quad \forall t \in \mathbb{R}\,,\,\forall j \in (1,..., N) \,.\eeq
Alternatively, the same state is nullified, as expected, by the set $\{ \hp_j \}$ of generators of the stabilizer group since,
\beq
\hX_j(t) \ket{0}_{p_j} = e^{- i s \hp_j} \ket{0}_{p_j} = \ket{0}_{p_j} \longleftrightarrow \hp_j \ket{0}_{p_j} = 0\, .
\eeq
As in the qubit case, the ideal CV cluster-state $\cs$ is the result of the pairwise application of controlled-Z gates $\cz = e^{i \hq_j \hq_k}$ upon all the nearest-neighbor modes $\6 j,k \9$ of the initial state $\ket{0}_p^{\otimes N}$ as depicted by the graph, explicitly 
\beq
\prod_{\6 j,k \9}^{N} \cz \ket{0}_p^{\otimes N} = \cs \, .
\eeq
How does the application of these gates affect the stabilizer set? Under the $\cz$ evolution, the quadrature operators transform as (see \ref{iCVcs} in Appendix \ref{appNullifiers} and \cite{linjordet2013})
\begin{align}
&\cz \hq_j \cz^\dagger = \hq_j \, ,\\
&\cz \hp_j \cz^\dagger = \hp_j - \hq_k \, ,
\end{align}
and the initial state stabilizers $\{ \hX_j(t)\}$ are changed into the cluster-state stabilizers
\beq
\hK_{\text{CS}} = \hX_j(t) \prod_{k \in \mathcal{N}(j)} \hZ_k(t) \, ,
\eeq
which have the same form of the qubit cluster-state stabilizers in Eq.~(\ref{qcsstab}). We can also express these stabilizers by \cite{menicucci2011}
\beq
\hX_j(t) \prod_{k \in \mathcal{N}(j)} \hZ_k(t) = e^{-i t (\hp_j - \sum_{k \in \mathcal{N}(j)} \hq_k)} \, ,
\eeq
equivalent to define the elements of the CV cluster-state nullifier set $\{ \hat{\eta}_j \}$ as
\beq
\hat{\eta}_j = \hp_j - \sum_{k \in \mathcal{N}(j)} \hq_k \,.
\eeq
All the $\hat{\eta}_j$ (and linear combinations of them) commute, and are the elements of the algebra that generates the stabilizer group of $\cs$. In complete analogy with the qubit case, we can construct a Hamiltonian $\hH^{\text{ideal}}_{\text{CS}}$ whose ground state is the CV cluster-state $\cs$ by imposing an energy penalty for violating any of the nullifier conditions:
\beq
\hH^{\text{ideal}}_{\text{CS}} = \sum_{j = 1}^N \hat{\eta}^2_j \, .
\eeq
Since all nullifiers commute and have a continuous spectrum of eigenvalues $(\mathbb{R})$, this Hamiltonian also has a continuous spectrum $[ 0, \infty  )$, and is therefore gapless \cite{gu2009}, see also (\ref{specsHam}) in Appendix \ref{appNullifiers}. 

\subsubsection{Non-unit edge weights}

It is worth pointing out that even at the ideal level there is some additional freedom in the CV cluster-state construction procedure. In particular, ideal CV cluster-state graphs can have any nonzero real-valued weight $g \in \mathbb{R}$ associated with each edge. This modifies the strength of the $\cz$ gate represented by that edge: $\cz [g]:= e^{i g \hq_j \hq_k}$.  These weights were first introduced in \cite{menicucci2007} as a way to enable new methods of construction. They have shown themselves to be very important for the computational properties of these states \cite{alexander2013} and when considering efficient construction of cluster-states with very large graphs \cite{flammia2009,menicucci2011a,menicucci2011}. For the purposes of all future derivations in this Thesis, we set $g = 1$. However, when describing possible experimental implementations of the CV codes at the end of this Chapter, we will show how the results obtained with this assumption apply indistinctly to CV cluster-states constructed with non-unit--but still uniform weight $g$.

\subsection{The ideal continuous-variable surface code}
\label{ticvscs}
In this section we present the description of the continuous-variable surface code. We start introducing the stabilizer description for a general lattice implementation, and then look at the special case of a planar square lattice with defined orientations.

\begin{figure}[tbp]  
\centering
\setlength{\unitlength}{1cm}
\includegraphics[width=15\unitlength]{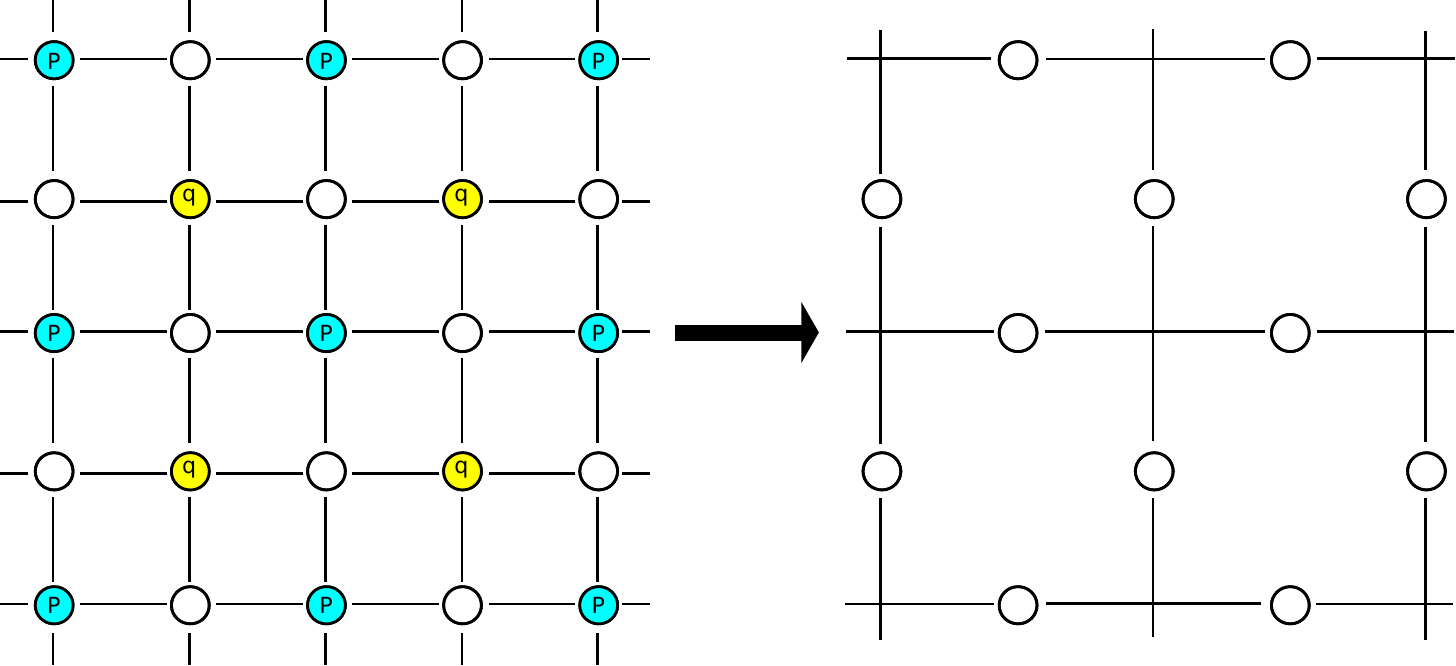}
\caption{Measurement scheme to project the CV cluster-state (left) on a square lattice into the CV surface code (right) described by a square graph $\Lambda$. A $\hq$ measurement removes the measured (yellow) node and all the links departing from it, while a $\hp$ measurement eliminates the corresponding (blu) node but creates new connections among the nearest neighboring nodes.}
\label{scheme}
\end{figure}

The ideal CV cluster-state on a square lattice can be transformed into the corresponding ideal CV surface code by a simple scheme of quadrature measurements \cite{zhang2008}, which was inspired by the dynamical mapping of qubit cluster-states to surface codes \cite{han2007}. In short, start with the CV cluster-state and label vertices by row and column. Then measure in $\hp$ those modes on rows and columns that are both odd and in $\hq$ those that are both even, as in Fig.(\ref{scheme}) for the case of a square lattice.
The measurements change the form of the stabilizers \cite{menicucci2011}, and therefore after the measurements we are left with a state described by a new set of nullifiers that corresponds exactly to the CV analog of Kitaev's surface codes, as we show in Appendix \ref{appNullifiers}.

To continue with the analysis is convenient to borrow notation from the qudit version of surface codes \cite{bullock2007}, which describes the nature of the coupling involved in the nullifiers in terms of a surface code graph $\Lambda = \{ \mathcal{V}, \mathcal{E}, \mathcal{F} \}$. We assume the graph is oriented and that the faces inherit this orientation. Each quantum mode reside on an edge $e_j \in \mathcal{E}$, with the orientation of any edge determined by $e = [v,v']$ for the base of the edge starting at vertex $v$ and the head at vertex $v'$. 

In this very general case, the stabilizers for the CV surface code are given by
\begin{align}
&\hat{A}_v(t) = \prod_{e | v\in\partial e}\hat{Z}_e [o(e,v) t]\, ,
\nonumber\\
&\hat{B}_f(u) = \prod_{e \in \partial f} \hat{X}_e [- o(e,f) u]\, ,
\label{cvscideal}
\end{align}
where the symbols in these expressions denote:
\begin{align}
	 o(e,v) &=
	 \begin{cases}
		+1 & \text{if $e\in [v,\cdot]$}, \\
		-1 & \text{if $e\in [\cdot,v]$},
	\end{cases}
	\nonumber \\
	 o(e,f) &=
	 \begin{cases}
		+1 & \text{if $e$ is oriented the same as $f$}, \\
		-1 & \text{otherwise},
	\end{cases}
\end{align}
and the dot ($\cdot$) stands for any vertex. By construction, the stabilizers commute:
\beq
[\hat{A}_v,\hat{A}_{v'}]=[\hat{B}_f,\hat{B}_{f'}]=[\hat{A}_v,\hat{B}_f]=0 \,.
\eeq
The CV surface code subspace is the $+1$ co-eigenspace of the stabilizers $\hat{A}_v$ and $\hat{B}_f$, in complete analogy with Kitaev's surface codes \cite{pachos2012}.

\begin{figure}[tbp]  
\centering
\setlength{\unitlength}{1cm}
\includegraphics[width=8\unitlength]{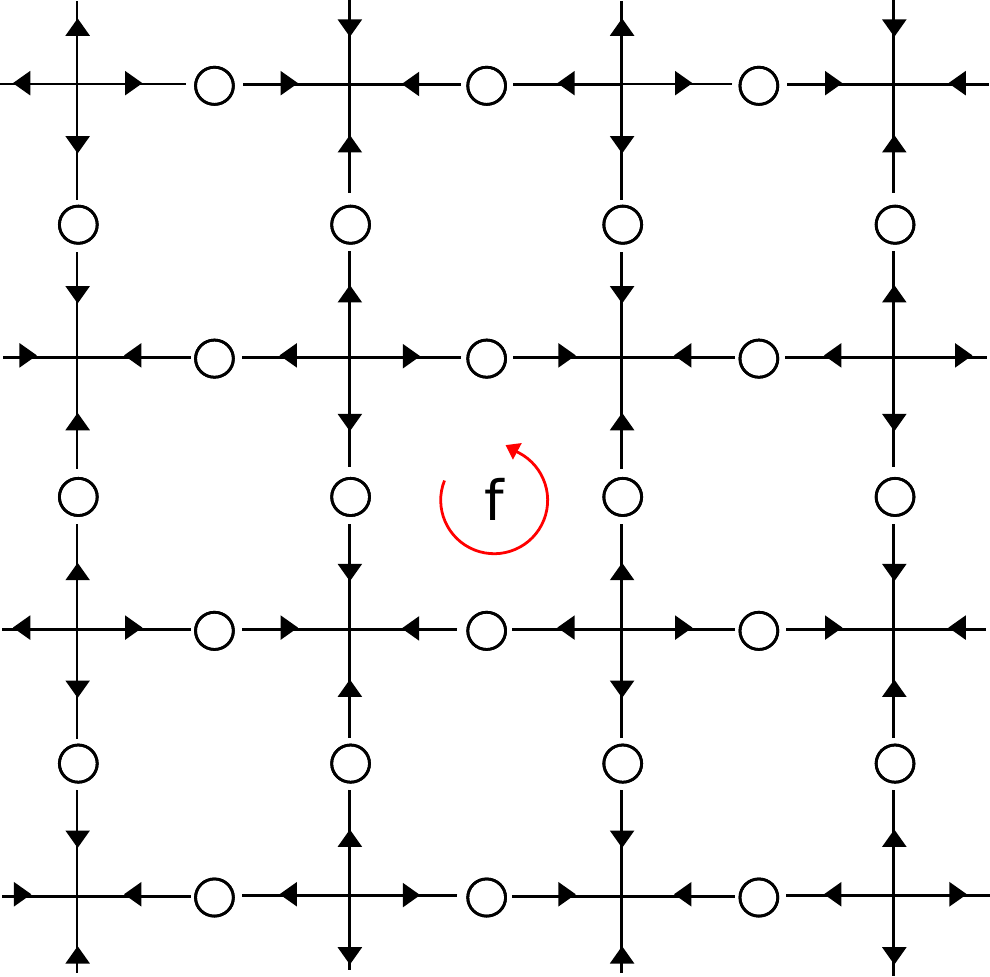}
\caption{Choice of the orientation for the toroidal CV surface code. At each vertex, all the edges are either pointing inward or outward. Consequently, all the faces have the same counterclockwise orientation.}
\label{orienta}
\end{figure}

To simplify the discussion, and to compare our results directly with the toric code state, we specialize to the case of the code graph $\Lambda$ being a toroidal square lattice. In the following discussion, we will always stick with this choice, fixing the edge orientations such that at any vertex $v$, all incident edges point toward $v$, or all point away from $v$ and the faces inherit equal counterclockwise orientation; see Fig.(\ref{orienta}). Under these assumptions, the symbol $o(e,v)$ is a constant $\pm1$ for any vertex $v$ and it only amounts to a sign flip on $t$ in Eq.(\ref{cvscideal}), which has no effect on a stabilizer's role as such, and we can ignore it. 
In this case, the CV stabilizers function as the CV analogs of the surface code stabilizers for qubits from Eq.~(\ref{tcA}) and Eq.~(\ref{tcB}) (up to a $X \leftrightarrow Z$ swap)
\beq
\hat{A}_v(t) \longleftrightarrow \hat{A}^q_v = \prod_{j \in v} \hat{\sigma}_j^X\,, \qquad \hat{B}_f(u) \longleftrightarrow \hat{B}^q_f = \prod_{j \in \partial f} \hat{\sigma}_j^Z \,,
\eeq
and explicitly become
\beq
\hat{A}_v(t) \longrightarrow e^{i \hat{a}_v t}\, , \qquad \, \hat{B}_f(u) \longrightarrow e^{i \hat{b}_f u}\, ,
\eeq
where the stabilizers generators, i.e the nullifiers, are:
\begin{align}
\nonumber
&\hat{a}_v = \sum_{e | v \in\partial e} \hq_e \, ,\\
&\hat{b}_f = \sum_{e \in \partial f} o(e,f) \hp_e \,.
\end{align}
If, on the boundary, one of the edges is missing, then that mode is not included in the nullifiers. It should be clear that the choice of the orientation is purely conventional and does not change the state. Hence, if we were to flip the orientation of every edge, $\hat{b}_f$ simply becomes $-\hat{b}_f$, which is still a nullifier. 

The ideal CV surface code is the non-normalizable ground state of the quadratic Hamiltonian \cite{zhang2008}
\beq
\hH_{\text{SC}}^{\text{ideal}} = \sum_{v\in \mathcal{V}} \hc_v \ha_v + \sum_{f \in \mathcal{F}} \hat{b}^\dagger_f \hat{b}_f \,. 
\label{HidealSC}
\eeq
In analogy with the ideal cluster-state Hamiltonian, the spurious \emph{mode operators} $\hat{a}_v$ and $\hat{b}_f$ are actually Hermitian quadrature operators that all commute, and because of
\begin{align}
[\hat{a}_v, \hH_{\text{SC}}^{\text{ideal}} ] = 0 \,\,\, \forall v \nonumber \,,\\
[\hat{b}_f,\hH_{\text{SC}}^{\text{ideal}}] = 0\,\,\, \forall f \,,
\end{align}
an extensive number of gapless modes exist. Therefore this Hamiltonian has a fully continuous spectrum $[ 0, \infty )$ and is gapless for any number of systems, even on a square lattice with boundary. Anyonic braiding in this model was studied in \cite{zhang2008,morimae2013,milne2012}.

\section{Physical continuous-variable codes}
\label{pCVcodes}
Ideal states are mathematical representations of states that cannot be produced in the laboratory. While the ideal CV surface code is the perfect analog of the qubit toric code, with experimental realization in mind we are required to speak about states that have actual physical significance and can be implemented experimentally. In this section, we consider physical realizations of the ideal states, or more specifically, we analyze the case of finite squeezing. 

\subsection{Physical continuous-variable cluster-state}
Similarly to the previous section, we start considering the CV cluster-state first \cite{menicucci2006}. We follow the canonical method again: starting from $N$ vacuum modes $\ket{0}^{\otimes N}$. The elements of the initial nullifier set are simply the dimensionless single-mode annihilation operators $\hat{a}_j = \frac{1}{\sqrt{2}} (\hq_j + i \hp_j)$, since
\beq
\ha_j \ket{0}^{\otimes N} = 0\, , \quad \forall \, j = 1,...,N  \,.
\eeq
These states are then all squeezed by $\hS(s)$, squeezing operator with squeezing parameter $\log s$ (\ref{sqmatrix}), in contrast with the ideal case, where one starts directly with $N$ zero-eigenstates of the quadratures $\hp_j$, i.e. $\ket{0}_p^{\otimes N}$. The squeezing is followed by the pairwise $\text{C}_{\hat{Z}}$ coupling, yielding the transformed nullifiers (see Appendix \ref{appNullifiers} for the details of this transformation):
\beq
\ha_j \longrightarrow \frac{s}{\sqrt{2}}\left [ \frac{\hq_j}{s^2} + i \left( \hp_j - \sum_{k \in \mathcal{N}(j)} \hq_k \right) \right ] = \hat{\eta}^s_j \,.
\eeq 
These operators satisfy the canonical commutation relations for normal mode operators,
\beq
[\het_j^s, \het_k^s] = 0\, , \qquad [\het^s_j, \het_k^{s\,\dagger}] = \delta_{j,k} \, ,
\eeq
and therefore the CV cluster-state Hamiltonian can be written as \cite{aolita2011}
\beq
\label{hamCS}
\hH_{\text{CS}}(s) = \sum^N_{j=1} \frac{2}{s^2} \Big( (\het_j^s)^\dagger  \het_j^s + \frac{1}{2} \Big) \,.
\eeq
For finite $s$, the system as a gap of $2s^{-2}$. The prefactor provides for finite energy even in the limit of infinite squeezing, where $\lim_{s \to \infty} \hH_{\text{CS}}(s) = \hH_{\text{CS}}^{\text{ideal}}$. 

\subsection{Physical continuous-variable surface code}
\label{pCVsc}
\begin{figure}[tbp]  
\centering
\setlength{\unitlength}{1cm}
\includegraphics[width=8.7\unitlength]{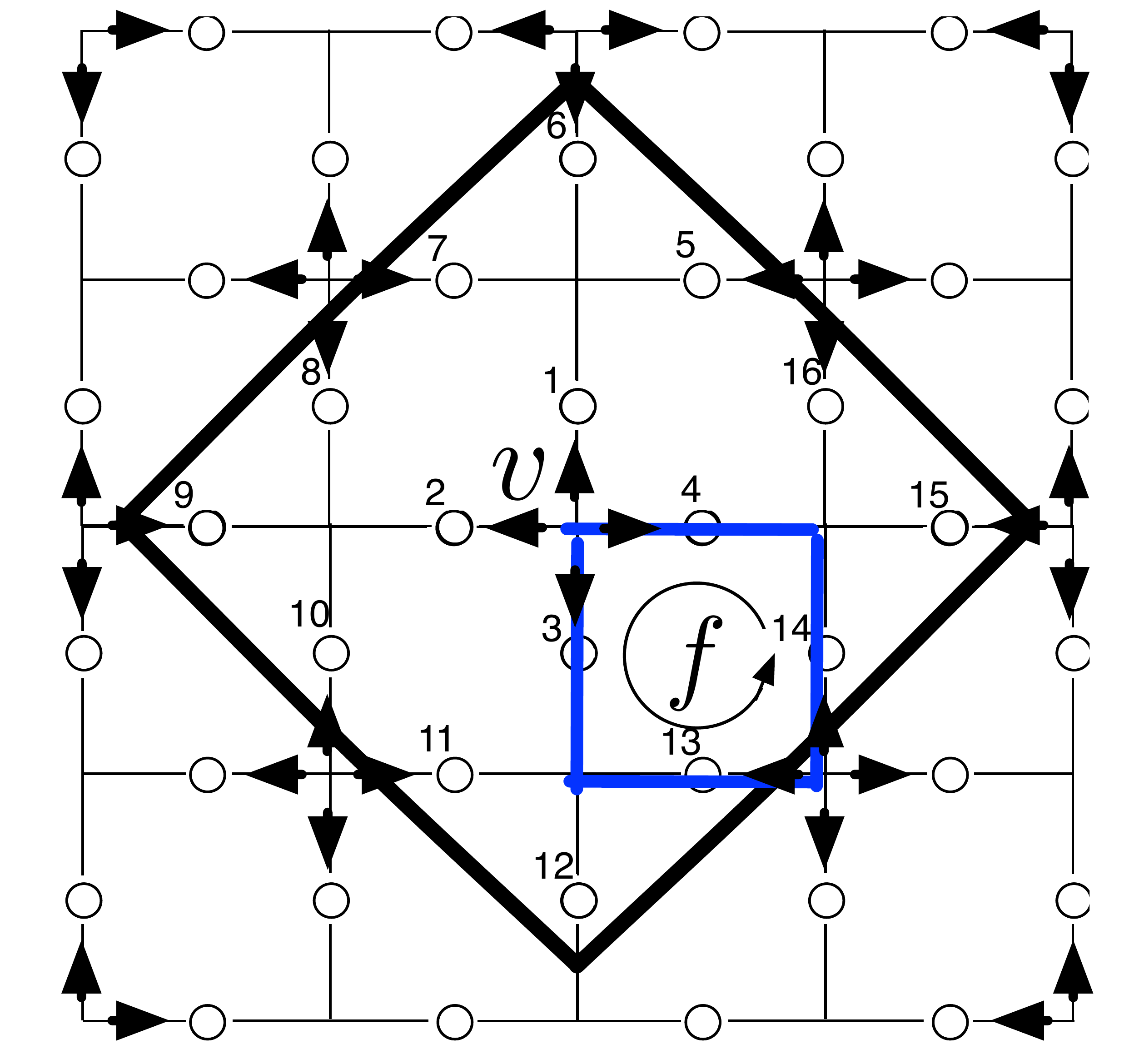}
\caption{Structure of the resultant nullifiers for the finitely squeezed surface code on a square lattice with toroidal boundary conditions. For the vertex $v$ indicated, $\hat{a}_v = \frac{s'}{\sqrt{8}}\Big[ \sum_{j=1}^4 (\hq_j + \frac{i}{s'^2} \hp_j) + \frac{s^2}{s'^2} \sum_{k=5}^{16} \hq_k \Big]$ with $s' = \sqrt{5 s^2 + s^{-2}}$, and for the face $f$, $\hat{b}_f = \frac{s}{\sqrt{8}}[(\hp_3 - \hp_{13} + \hp_{14} - \hp_{4} - \frac{i}{s^2} (\hq_3 - \hq_{13} + \hq_{14} - \hq_{4}))]$.}
\label{nullifier}
\end{figure}

Using the same measurement pattern shown in Fig.(\ref{scheme}), the finitely squeezed CV cluster-state can  be mapped to the finitely squeezed CV surface code. While the necessary steps to derive the nullifiers are once more outlined in Appendix \ref{appNullifiers}, for the case of a square lattice with toroidal boundary conditions, the precise nullifiers follow from Eq.~(\ref{avbf}):
\begin{align}
\label{toronulli}
\hat{a}^s_v &=\frac{s'}{\sqrt{8}}\Biggl[\sum_{e|v\in \partial e} \left(\hat{q}_{e}+\frac{i}{{s'}^2} \hat{p}_{e} \right) +\frac{s^2}{s'^2} \sum_{\substack{v' | [v',v]\in \mathcal{R} \\ e |  v^\prime \in \partial e \wedge v\not\in \partial e}}\hat{q}_e \Biggr], \nonumber \\
\hat{b}^s_f &=\frac{s}{\sqrt{8}}\sum_{e\in \partial f} o(e,f) \left(\hat{p}_{e}-\frac{i}{{s}^2}\hat{q}_{e} \right),
\end{align}
where $s'=\sqrt{5s^2+s^{-2}}$. Note that now also the next-nearest neighbors contribute to the structure of the nullifiers, in contrast to the ideal case, as illustrated in Fig.~(\ref{nullifier}). 
Without loss of generality we can consider a square $n \times m$ lattice: Then the nullifiers commutation relations from Eq.~(\ref{comms}) can be rewritten as
\begin{align}
\label{cmrAB}
	[\hat{a}_v,\hat{a}^{\dagger}_{v'}] &= w\left(d(v,v')\right), \nonumber \\
	[\hat{b}_f,\hat{b}^{\dagger}_{f'}] &=x(d(f,f')), \nonumber \\
	[\hat{a}_v,\hat{a}_{v'}] &= [\hat{b}_f,\hat{b}_{f'}] = [\hat{a}_v,\hat{b}_{f}] = [\hat{a}_v,\hat{b}^{\dagger}_{f}] = 0,
\end{align}
where $d(v,v')$ and $d(f,f')$ are the Euclidean distance between vertices and faces, respectively on the unit-edge-length lattice and dual lattice. The functions $w$ and $x$ are given by:
\begin{align}
\label{cmrA}
	w(0) &=1, \quad w(1)=\frac{(1+ 8s^4)}{4(1+5s^4)}, \quad w(\sqrt{2})= \frac{s^4}{2(1+5s^4)}, \nonumber \\
	w(2) &= \frac{s^4}{4(1+5s^4)}, \qquad w(d>2)=0,
\end{align}
and
\begin{align}
\label{cmrB}
	 x(0)=1, \quad x(1)=\frac{1}{4}, \quad x(d>1)=0. 
\end{align}
Using these nullifiers we can construct a Hamiltonian for the physical CV surface code:
\beq
\label{eq:HSC}
	\hH_{\text{SC}}(s) = \sum_v \frac{8}{s'^2} \hat{a}^{s\,\dagger}_v \hat{a}^s_v+\sum_f \frac{8}{s^2} \hat{b}^{s\,\dagger}_f \hat{b}^s_f.
\eeq
The squeezing dependence of the prefactors is done to ensure the Hamiltonian has finite energy for $s \to \infty$. 

Having described the very basics of the CV surface code, we will proceed exploring the defining properties of this model. We start with an analytical calculation of the energy gap of the Hamiltonian from Eq.~(\ref{eq:HSC}), showing that the system is gapless in the thermodynamic limit. Nonetheless, we will demonstrate analytically that the ground state correlations decay exponentially on the lattice despite the zero gap, and utilize different quantities to numerically prove topological order for the surface code state. Remarkably, all these quantities can be easily computed from quadrature measurements.


\subsubsection{Energy gap}
\label{eGap}
Unlike the discrete-variable case \cite{hamma2005, hamma2005a}, the Hamiltonian $\hH_{\text{SC}}(s)$ is gapless in the thermodynamic limit of infinite modes. This happens because the nullifiers of the most general physical CV surface code in Eq.(\ref{avbf}) do not define normal modes, since the non-trivial commutation relations of the neighboring nullifiers allow for low-energy mode excitations.  

On the $n \times m$ torus, the specifics of the graph are $|\mathcal{E}|=2nm$, $|\mathcal{F}|=nm$, and $|\mathcal{V}|=nm$ \cite{diestel2010}. We first focus on the case where $n \times m$ is odd, so that there are $| \mathcal{E} |$ independent nullifiers that therefore span the space of all the physical-mode annihilation operators. We want to diagonalize the Hamiltonian, and to do so we introduce the normal-mode operators
\beq
\label{normalmodeops}
\hat{c}_j = \sum_{r=0}^{n-1} \sum_{s=0}^{m-1} \alpha_{r,s}^{(j)} \ha^s_{v_{r,s}} \,, \qquad \, \hat{d}_j = \sum_{i=0}^{n-1} \sum_{l=0}^{m-1} \beta_{i,l}^{(j)} \hat{b}^s_{v_{i,l}} \, ,
\eeq
where the vertices at the lattice sites have coordinates $\{ v_{r,s} \}$, and the faces at the dual lattice sites have coordinates $\{ f_{i,l} \}$. In this base the Hamiltonian is
\beq
\hat{H}_{\text{SC}}(s) = \sum_j \frac{8 \omega_j}{s'^2} \hat{c}_j^\dagger \hat{c}_j + \sum_j \frac{8 \delta_j}{s^2} \hat{d}_j^\dagger \hat{d}_j \,.
\eeq
To find the normal-mode frequencies, we need to solve the equations
\begin{align}
\left[ \hat{c}_j, \hH_{\text{SC}}(s) \right] \equiv \left[\hat{c}_j,\sum_v \hat{a}^{\dagger}_v \hat{a}_v \right] = \omega_j \hat{c}_j,
\end{align}
where the equality is ensured by $[\hat{c}_j, \sum_f \hat{b}_f^{s\,\dagger} \hat{b}_f^s]= 0$, and
\begin{align}
\left[ \hat{d}_j, \hH_{\text{SC}}(s) \right] \equiv \left[\hat{d}_j,\sum_f  \hat{b}^{\dagger}_f\hat{b}_f \right] = \delta_j \hat{d}_j\,,
\end{align}
with $[\hat{d_j}, \sum_v \ha^{s\,\dagger}_v \ha_v^s] = 0$. If we introduce the generic vertex label state $\{\ket{r,s}\}$ and the face label state $\{\ket{i,l}\}$,
the two linear equations above can be vectorized and rewritten in these two basis respectively as
\begin{equation}
{\bf M}_v \ket{\alpha^{(j)}}= \omega_j \ket{\alpha^{(j)}},\quad {\bf M}_f \ket{\beta^{(j)}}= \delta_j \ket{\beta^{(j)}} \, ,
\label{eqsnormals}
\end{equation}
where $\ket{\alpha^{(j)}}$ and $\ket{\beta^{(j)}}$ are the vectorized form of the operators~$\hat c_j$ and $\hat d_j$ that result from Eq.~(\ref{normalmodeops}). In the following, we define operators for a periodic square lattice that make the problem solvable. First, define the shift operator $\hat X_x=\sum_{k=0}^{x-1} {\ket{k\oplus_r 1}\bra{k}}$, whose action on the generic vertex label state $\ket{r,s}$ is:
\begin{align}
&\hat{X}_n \ket{r,s} = \ket{r + 1,s} \,,\,\, \hat{X}_n^\dagger \ket{r,s} = \ket{r-1,s}\,, \nonumber\\
&\hat{X}_m \ket{r,s} = \ket{r,s + 1}\,,\,\, \hat{X}_m^\dagger \ket{r,s} = \ket{r,s-1} \,.
\end{align}
The action of the shift operator on the face basis is completely analogous. This allows to rewrite the matrices ${\bf M}_v$ and ${\bf M}_f$, whose elements are non-zero according to the form of the commutation relations in Eq.~(\ref{cmrAB}), in the following elegant way:
\begin{align}
	{\bf M}_v &=\hat{I}_{nm}+w(1)\bigl[\hat{I}_{n}\otimes (\hat X_m+\hat X_m^{\dagger})+(\hat X_n+\hat X_n^{\dagger})\otimes \hat{I}_m\bigr] \nonumber \\
	&\quad+w(\sqrt{2})\bigl[\hat X_n\otimes \hat X_m+\hat X_n^{\dagger}\otimes \hat X_m^{\dagger}+\hat X_n\otimes \hat X_m^{\dagger}+\hat X_n^{\dagger}\otimes \hat X_m\bigr] \nonumber \\
	&\quad+w(2)\bigl[\hat{I}_{n}\otimes (\hat X^2_m+\hat X_m^{2\dagger})+(\hat X^2_n+\hat X_n^{2\dagger})\otimes \hat{I}_m\bigr], \nonumber \\
	{\bf M}_f &=\hat{I}_{nm}+x(1)\bigl[\hat{I}_{n}\otimes (\hat X_m+\hat X_m^{\dagger})+(\hat X_n+\hat X_n^{\dagger})\otimes \hat{I}_m\bigr].
\end{align}
The linear equations in Eq.~\eqref{eqsnormals} can be solved in the Fourier basis via $\hat F_n\otimes \hat F_m$, where
\beq
\hat F_x=\frac{1}{\sqrt{x}}\sum_{j,k=0}^{r-1}e^{i j k \frac{2 \pi}{r}}\ket{j}\bra{k}\,,
\eeq
and the nullifiers in the Fourier basis are decomposed as
\beq
 \hat{a}^s_{v_{r,s}} =\frac{1}{\sqrt{n} \sqrt{m}} \sum_{k_x,k_y} e^{i (r k_r  \frac{2 \pi}{n} + s k_s  \frac{2 \pi}{m})} \hat{\tilde{a}}^s_{v_{k_r,k_s}} \,.
\eeq
In this way, the solutions written in the basis $\{\hat F_n \ket{r}\otimes \hat F_m\ket{s}\}$ and $\{\hat F_n \ket{i}\otimes \hat F_m\ket{l}\}$ are
\begin{align}
	&\{\omega_{j}\} =\left\{1+2w(1)\left[\cos\left(\frac{2\pi j_x}{n}\right)+\cos\left(\frac{2\pi j_y}{m}\right)\right] \right. \nonumber \\
	&\qquad +
2w(\sqrt{2})\left[\cos\left(\frac{2\pi j_x}{n}+\frac{2\pi j_y}{m}\right)+\cos\left(\frac{2\pi j_x}{n}-\frac{2\pi j_y}{m}\right)\right] \nonumber \\
	&\qquad +
\left. 2w(2)\left[\cos\left(\frac{4\pi j_x}{n}\right)+\cos\left(\frac{4\pi j_y}{m}\right)\right]\right\}_{j_x=0, j_y=0}^{n-1, m-1}, \nonumber \\
	&\{\delta_{j}\} = \left\{1+2x(1)\left[\cos\left(\frac{2\pi j_x}{n}\right)+\cos\left(\frac{2\pi j_y}{m}\right)\right]\right\}_{j_x=0,j_y=0}^{n-1,m-1},
\end{align}
treating the normal mode index $j=(j_x,j_y)\in\mathbb{Z}_n\times \mathbb{Z}_m$ as a collective index.  Then, the (squeezing-dependent) gap energy is the lowest-frequency mode energy:
\begin{align}
	\Delta E(s)={\rm min}_{j_x,j_y} \left\{\frac {8s^2\omega_j}{1+5s^4}, \frac {8\delta_j}{s^2} \right\}.
\end{align}
For large system sizes, i.e. $n,m \gg 1$, and choosing freely that $n \le m$, the gap is equal to
\beq
\Delta E (s) \approx \frac{4 \pi^2}{s^2 n^2}\, \, ,
\eeq
and in the thermodynamic limit $\lim n \to \infty$ the gap goes to zero. 

If $n$ and $m$ are even, then not all the face nullifiers are independent. To see this simply bicolor all the lattice faces and assign a plus sign to face operators of one color and a minus sign to faces of the other, then add them to get zero. Thus, the Hamiltonian $\hH_{\text{SC}}(s)$ is underconstrained, and there exists an exact gapless zero mode. For a square lattice with planar boundaries (and not toroidal as we discussed so far), there are boundary effects, but these make only a small modification to the gap, which still scales like the inverse of the system size.

To conclude, we have proved that in distinction to the cluster-state Hamiltonian $\hH_{\text{CS}}$ (\ref{hamCS}), the surface code Hamiltonian $\hH_{\text{SC}}$ (\ref{eq:HSC}) is gapless in the thermodynamic limit, although for infinite squeezing both models are gapless. This may not sound promising, since the qubit toric code state is gapped \cite{kitaev2003}, and usually topologically ordered states are associated with gapped parent Hamiltonians that provide dynamical stability. However, there are other instances of topologically ordered gapless models in two dimensions, such as quantum loop gases: They have been investigated in different contexts, see for example \cite{troyer2008,zhang2006}. In the following section we prove that, although our CV surface code model happens to be gapless, it still obeys an area law and exhibits topological order. 

\section{Detecting topological order in the continuous-variable surface code}
\label{dtopocvcs}
Our aim in this Chapter is to prove that the CV cluster-state, in analogy with the qubit-based cluster-state \cite{hamma2005, hamma2005a}, is topologically ordered. To accomplish this objective we exploit the Gaussian properties of the physical CV cluster-state: In fact, the ground state of the Hamiltonian in Eq.~(\ref{eq:HSC}) is Gaussian, since the procedure to construct it makes use only of Gaussian transformations and quadrature measurements.

The easiest approach to test topological order is by computing the topological entanglement entropy (TEE) for the state and check whether it is bigger than zero or not. Computation of the TEE requires calculating the entanglement entropy for different regions of the system, which is generally very complicated. However, as we have seen in Chapter \ref{ChapGS}, for Gaussian states this reduces to having complete knowledge of the state covariance matrix, from which calculating the entanglement entropy of any subsystem is as easy as stealing candies from a mathematician. 

In this section we first show how to derive the covariance matrix of the CV surface code state making use of the graphical calculus for pure Gaussian states. Then we study the behavior of the correlations, proving analytically that they decay exponentially with the graph distance. Under this assumption we are allowed to use the two formulas for the TEE. Finally we present our numerical results and confirm that the state is topologically ordered with a TEE that asymptotically grows linearly with the squeezing parameter $\log (s)$. To give additional evidence to this result, we also test the state against two other witnesses of topological order, the topological mutual information and the topological logarithmic negativity.

\subsection{Graphical representation}
We start the discussion by showing how to represent the physical CV cluster-state using the graphical calculus \cite{menicucci2011}. This will provide us with a straightforward method to calculate the covariance matrix of the state.

Both qubit cluster-states and the qubit surface code state can be represented by graphs. However, their definitions in terms of graphs are incompatible. In particular, qubit cluster-states (or qubit ``graph states", as they are often confusingly called in the literature, although cluster-states are just a specific form of graph states) are represented by graphs that act as a well-defined \emph{recipe} for creating the state, in the sense that nodes represent qubits prepared in $\ket{+}$ state, and the edges denote controlled-PHASE gates between them. While it is not necessary to make cluster-states this way, any cluster-state can be so constructed. 

\begin{figure}[tbp]  
\centering
\setlength{\unitlength}{1cm}
\includegraphics[width=8.5\unitlength]{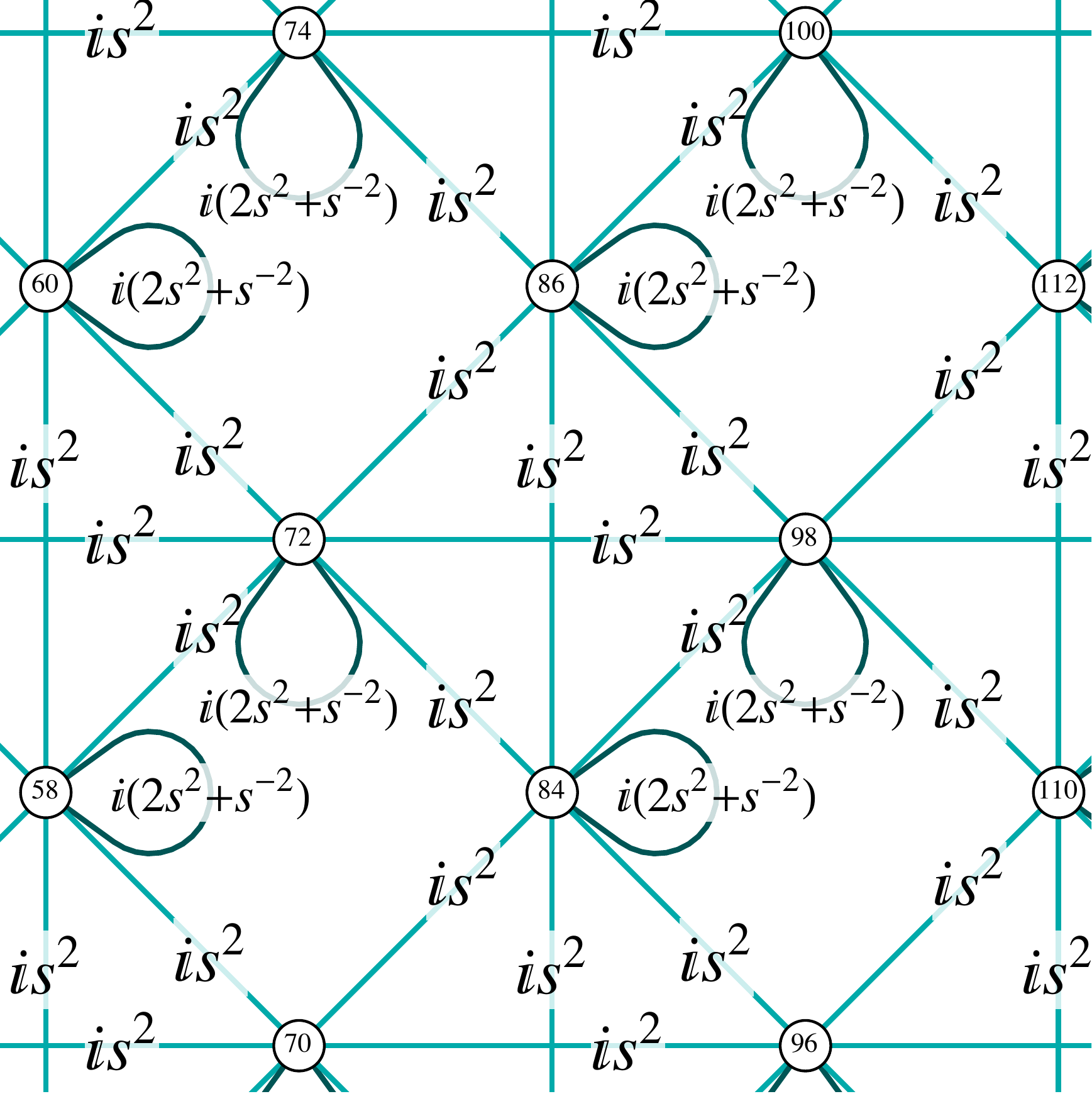}
\caption{Gaussian pure-state graph $\mZ_{\text{SC}}$ for a section of the canonical CV surface code state on a square lattice. Similar to before, $\log s$ is the squeezing parameter and cyan links or self-loops indicate positive imaginary weights.}
\label{graphSC}
\end{figure}

Nonetheless, if this recipe is applied to the graphs for qubit surface code states, one fails because for the surface code the qubits live on the edges, while faces and vertices define stabilizers in terms of all $\hat{\sigma}^z$'s or all $\hat{\sigma}^x$'s. Hence there is no analog for this in the graph recipe used to define qubit cluster-states, and although one can \emph{define} by fiat a connection between the two types of graphs, it is at best a patch-up job. 

In contrast with the discrete-variable case, we have seen in \ref{gracagas} that the graphical calculus for Gaussian pure states offers a \emph{unified} graphical representation of any Gaussian pure state in terms of a recipe for its creation. Again, one does not have to follow this recipe to construct a Gaussian state, but any Gaussian state can be prepared following the instructions defined uniquely by its graph. As such, since both the CV cluster-states and the CV surface code are Gaussian, only one type of graph is needed. Furthermore, the measurement pattern that connects these two types of states has definite graph transformation rules (\ref{trarumea}), so the connection between the two graphs can be directly \emph{derived} using the graphical calculus instead of \emph{invented ad hoc patches} between two inequivalent kinds of graphs, as in the qubit case\footnote{One can use the Clifford group representation for the qubit case in an analogous sense.}. 

\begin{figure}[tbp]  
\centering
\setlength{\unitlength}{1cm}
\includegraphics[width=8.5\unitlength]{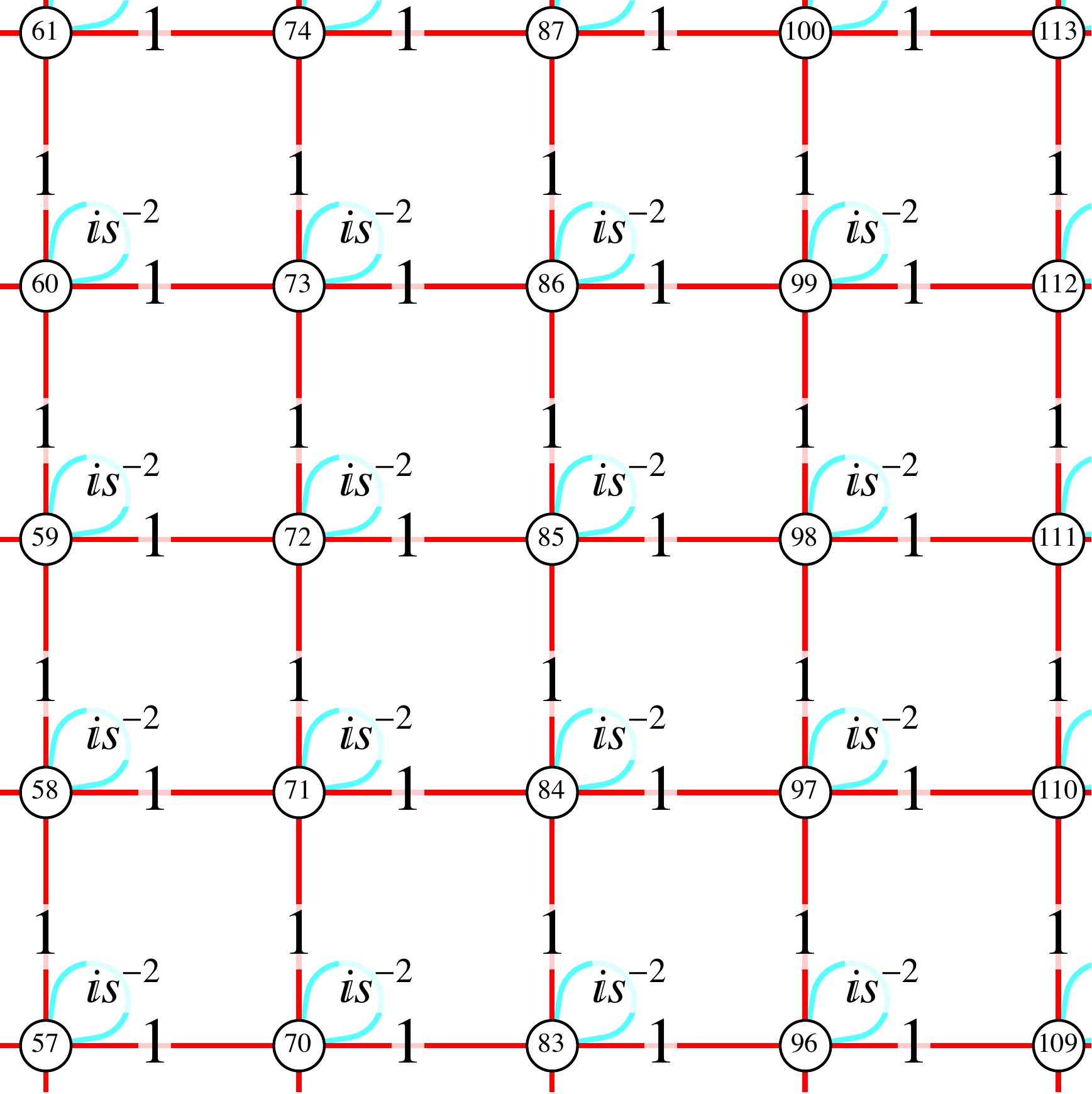}
\caption{Gaussian pure state graph $\mZ_{\tcs}$ for a section of the canonical CV cluster-state on a square lattice with equivalent edge weights $g=1$. The self-loops on the modes represent the squeezing and the color of the lines indicate the phase, red = positive real, cyan = positive imaginary. The squeezing parameter is $\log s$.}
\label{graphCS}
\end{figure}

The $N$-mode physical CV cluster-state can be directly and uniquely represented by the undirected, complex-weighted graph whose adjacency matrix is $\mZ_{\tcs}$, making solely use of the additional adjacency matrix $\mAd$ that describes the square-lattice pattern of interactions among the modes \cite{menicucci2011, demarie2013b}. Explicitly:
\beq
\mZ_{\tcs} (s) := \mAd + is^{-2} \mI_{N} \, ,
\eeq
with squeezing parameter $\log s$ and $\mI_N$ the $N \times N$ unit matrix. This is illustrated in Fig.(\ref{graphCS}). 
To derive the graph of the finitely squeezed CV cluster-state with open boundary conditions, we start with $\mZ_{\tcs}$ and perform the measurement scheme as shown in Fig.(\ref{scheme}), with the advantage that measurements have a straightforward translation in the $\mZ$ transformation rules language as explained in \ref{trarumea}. We repeat them, noting that a $\hq$ measurement on the $k$-th mode is equivalent to deleting the $k$-th row and column of the $\mZ$ matrix, while a $\hp$ measurement is equivalent to applying a $\pi/2$ phase shift on the $k$-th mode and then measuring (on the same mode) the $\hq$ quadrature. 

Then the $\mZ_{\tcs}$ matrix transforms into the CV finitely squeezed surface code state $\mZ_\text{SC}$, shown in Fig.(\ref{graphSC}) \cite{demarie2013b}. In this case, it is a purely imaginary matrix whose entries are given by
\beq
\mZ_{\text{SC}} (s) = i \mU_{\tsc}(s) \, ,
\eeq
where 
\beq
\label{usc}
\mU_{\tsc} (s) = s^2 \mathbf{A}_{\tsc} + (s^{-2} + 2s^2 ) \mI_N \,,
\eeq
and $\mathbf{A}_{\tsc}$ is the unweighted adjacency matrix of the surface code (without self-loops, see Fig.~\ref{graphSC}). Note that the $\mV_{\tsc}$ component of the $\mZ_{\tsc}$ matrix, see Eq.~(\ref{Zmatrix}), is zero for the surface code. Using the connection between $\mZ$ matrix and covariance matrix from Eq.(\ref{gammaZ}), we finally derive the covariance matrix of the physical CV surface code state:
\begin{align}
\label{cov}
\Ga_{\text{SC}}(s) = \frac{1}{2} \left (
\begin{tabular}{c c}
$\mathbf{U}^{-1}_{\text{SC}}(s)$ & $0$\\
$0$ & $\mathbf{U}_{\text{SC}}(s)$\\
\end{tabular} \right) \, .
\end{align}
Complete knowledge of the covariance matrix makes calculations of entanglement entropy for different regions of the system straightforward.

\subsection{Quadrature correlations on the lattice}
\label{quacorre}
\begin{figure}[tbp]  
\centering
\setlength{\unitlength}{1cm}
\includegraphics[width=12.5\unitlength]{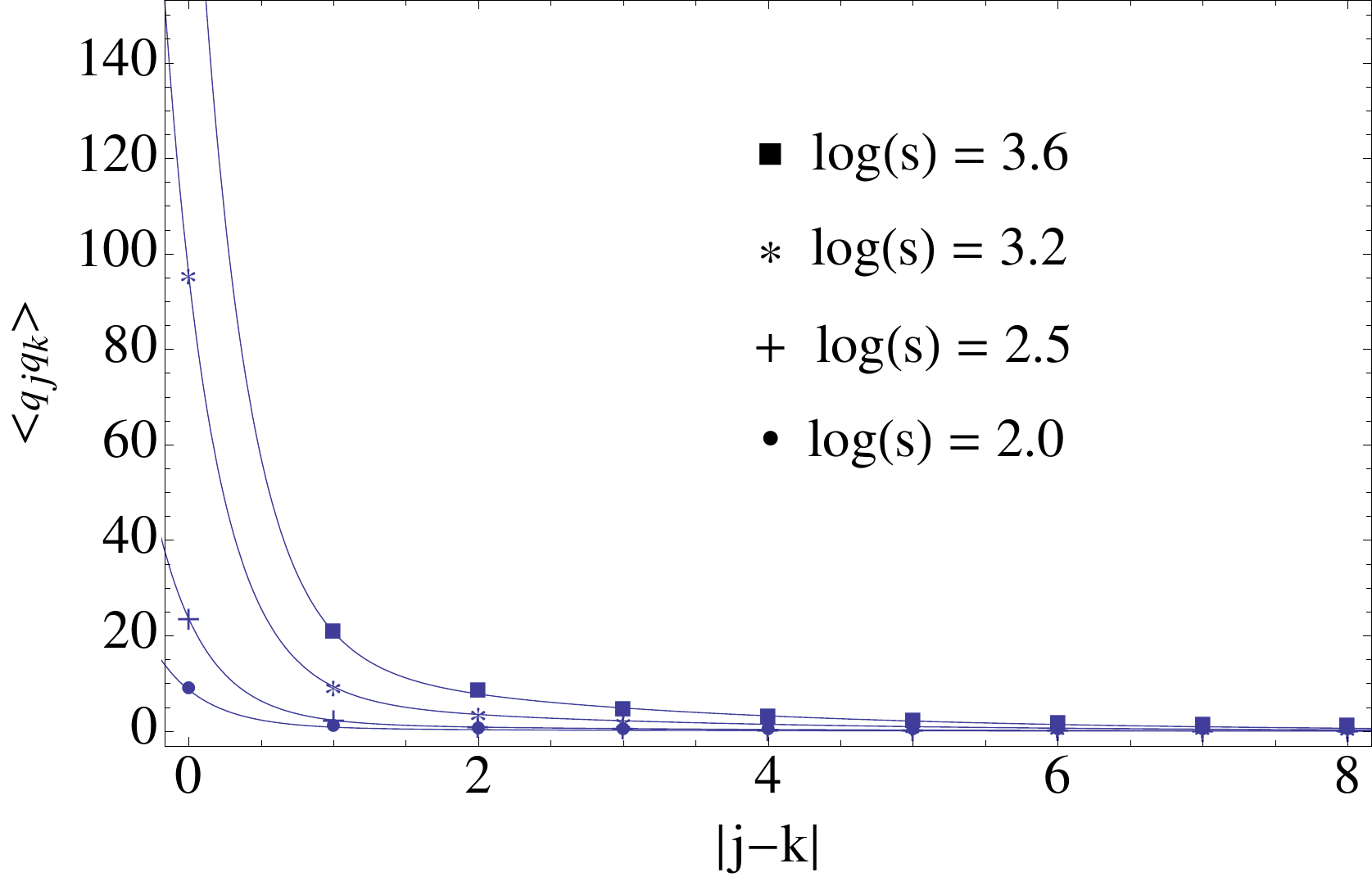}
\caption{$\hq$-correlations for the qumodes along the main axes of the CV surface code for different values $log(s)$ of the initial squeezing. They decay as $a e^{-\frac{|j-k|}{\xi_a}} + b e^{- \frac{|j-k|}{\xi_b}}$ and for $\log(s)>2$ the correlation length scales converge quickly to $\xi_a=0.33,\xi_b=2.42$.  In contrast, $\hp$-correlations immediately drop to zero beyond one unit separation.}
\label{corre}
\end{figure}

The covariance matrix $\Ga_{\tsc}(s)$ allows us to study the $\hq$ and $\hp$-correlations for the modes of the surface code \cite{demarie2012}. In fact, if we have a look at the generic formulation for the covariance matrix, Eq.(\ref{gengamma}):
\begin{align}
\label{gencovmatrix}
\Ga = \left (
\begin{tabular}{c c}
$\6 \hat{\mathbf{q}}^2 \9$ & $\6 \hat{\mathbf{q}} \hat{\mathbf{p}} \9$\\
$\6  \hat{\mathbf{p}}  \hat{\mathbf{q}} \9$ & $\6 \hat{\mathbf{p}}^2 \9$\\
\end{tabular} \right) \, ,
\end{align}
and compare it with the expressions in Eq.~(\ref{usc}) and Eq.~(\ref{cov}), one can immediately notice that the $\hp$-correlations are non-zero exclusively in the nearest neighborhood of each mode, due to the particular form of the matrix $\mU_{\tsc}$. Furthermore, the form of the covariance matrix tells us that the $\hat{q}$-$\hat{p} $-correlations are zero everywhere. On the other hand, because of the presence of the $\mathbf{U}^{-1}_{\text{SC}}(s)$ term, the $\hq$-correlations require a more elaborate analysis. The exponential decay of the quadrature correlations is a necessary assumption to use the topological entropy formulas, as explained in Chapter \ref{ChapTO}. Numerical values of the $\hq$-correlations along the main axes of the system are shown for a few values of the initial squeezing in Fig.~(\ref{corre}). In the following we provide bounds on the correlations for arbitrary squeezing.

In \cite{demarie2013b} we show that the spectral range of $\mathbf{U}_{\text{SC}}(s)$, denoted $\sigma(\mathbf{U}_{\text{SC}}(s))$, satisfies $\sigma(\mathbf{U}_{\text{SC}}(s)) \subset [ a, b]$ where $a=s^{-2}$ and $b=s^2(8 + s^{-4})$. On a generic $n \times m$ lattice, the $nm \times nm$ matrix $\mathbf{U}_{\text{SC}}(s)$ is block tridiagonal with $n$ identical $m \times n$ matrices ${\bf A}$ on the diagonal and identical ${\bf B}$ on the immediate upper and lower blocks. Now the matrix coordinates $(i,j)$ correspond to Euclidean coordinates $((i_x, i_y),(j_x, j_y))$ on the lattice, where $i = m i_x + i_y$ for $i_x \in \{ 0, ..., n-1 \}$ and $i_y \in \{ 0, ..., m-1 \}$, etc. It is convenient to define a \emph{graph distance} $d(i,j) = \text{max} \{ |i_x - j_x|, |i_y-j_x| \}$ between coordinates $(i_x, i_y)$ and $(j_x, j_y)$. Since, away from the edges, the bulk of the $\mZ$ graph is the union of a square graph with a graph having two diagonal edges passing through every other face, the graph distance is the number of edges on the shortest path between $(i_x, i_y)$ and $(j_x, j_y)$, and it satisfies 
\beq
ed(i,j)/\sqrt{2}\leq d(i,j)\leq ed(i,j)\,,
\eeq
where $ed(i,j)=\sqrt{(i_x-j_x)^2+(i_y-j_y)^2}$ is the Euclidean distance. The matrix ${\bf A}$ is itself tridiagonal with elements $\alpha=2s^2+s^{-2}$ on the main diagonal and $\beta=s^2$ on the immediate upper and lower diagonal. The matrix ${\bf B}$ is also tridiagonal with diagonal elements $\delta=s^2$ and immediate upper and lower diagonal elements either equal to zero or $\delta$. A theorem of Demko, Moss and Smith \cite{demko1984, molinari2012} shows that banded matrices of a certain class have inverses with matrix elements that decay exponentially with the distance from the diagonal.  Specifically they show for matrices ${\bf M}$ of size $N\times N$ and spectral range $\sigma({\bf M})\subset [a,b]$ with $a>0$ [Ref. \cite{demko1984} Proposition 5.1]:
\begin{equation}
\sup\{|{\bf M}^{-1}_{i,j}|:(i,j)\in D_n({\bf M})\}\leq C_0 q^{n+1} \,,
\label{decaybound}
\end{equation}
where the \emph{decay sets} are
\[
D_n({\bf M})=(\{1,\ldots N\}\times \{1,\ldots N\})\setminus S_p({\bf M}) \,,
\]
and the \emph{support sets} are
\[
S_p({\bf M})=\bigcup_{k=0}^p\{(i,j):{\bf M}^k_{i,j}\neq 0\}.
\]
Here $C_0=\frac{(1+\sqrt{b/a})^2}{2b}$ and $q=\frac{\sqrt{b/a}-1}{\sqrt{b/a}+1}<1$.

The matrices relevant to our problem are in this class.  The matrix power $\mathbf{U}^k_{\text{SC}}(s)$ is a banded block symmetric matrix with blocks of size $m\times m$ and block band width $2k+1$.   Furthermore, each such block is banded with band width $2k+1$.     Thus the support set $S_p(\mathbf{U}_{\text{SC}}(s))$ is the set of those matrix coordinates $(i,j)$ such that the graph distance $d(i,j)$, is no more that $2p+1$.   Similarly, the decay set is all matrix coordinates outside the support set.  The statement in Eq.~(\ref{decaybound}) means that for nodes separated in graph distance $d(i,j)>2p+1$, with associated matrix coordinates $(i,j)$, the inverse matrix element $\mathbf{U}^{-1}_{\text{SC}}(s)_{i,j}$ falls off exponentially with graph distance as $q^{(d(i,j)+1)/2}$. 

This implies that the position correlations between nodes $i$ and $j$ separated in graph distance by $d(i,j)$ satisfy 
\begin{equation}
\langle \hat{q}_i \hat{q}_j\rangle\leq Ce^{-(d(i,j)+1)/\xi} \,,
\end{equation}
where the constant is given by
\[
C=\frac{(1+\sqrt{8s^{4}+1})^2}{4(8s^2+s^{-2})} \,,
\]
and the correlation length is
\begin{equation}
\xi=\frac{2}{\ln \Big[\frac{\sqrt{8s^4+1}+1}{\sqrt{8s^4+1}-1}\Big]}\,,
\end{equation}
as we wished to prove.

\subsection{Proof of topological order}
\label{prtoder}
After this long and slightly tedious introduction, we are ready to give a proof that the CV surface code is topologically ordered. A key feature of topological phases is the absence of local parameters that allow to detect them. The exponential decay of the correlations ensures area law behavior of the entropy, which is the necessary condition for using the topological entanglement entropy (TEE) formulas in Eq.~(\ref{KPTEE}) and Eq.~(\ref{LWTEE}), while the complete knowledge of the covariance matrix allows calculations of the entanglement entropy between different regions of the system. For instance, consider a generic set of modes $A$, which corresponds to a region of the lattice. To calculate the entanglement entropy $S(A)$, it is enough to derive the proper reduced covariance matrix $\Ga_{\tsc_A}$ deleting from $\Ga_{\tsc}$ the rows and columns corresponding to the modes in $A^c$, the complementary set of $A$. Then one extracts the symplectic eigenvalues and applies the formula in Eq.~(\ref{vnS}), as explained in Section \ref{gauDent}. Also recall that the TEE formulas we use are \cite{kitaev2006}
\beq
\label{f1}
S_{\text{topo}}^{\text{KP}}  \equiv - ( S_A + S_B + S_C - S_{AB}- S_{BC}- S_{AC} + S_{ABC}) = \gamma \, ,
\eeq
and \cite{levin2006}
\beq
\label{f2}
S_{\text{topo}}^{\text{LW}}  \equiv - \frac{1}{2} [ (S_A - S_B) - (S_C - S_D) ] = \gamma \, ,
\eeq
with regions shown in Figs.~\ref{areasTEE}a and~\ref{areasTEE}b, respectively.
If the system is not topologically ordered, these combinations (i.e. $\gamma$) are exactly zero. In our simulations, we use the $36 \times 36$ square bulk of a larger CV surface code state. 
Selecting from the covariance matrix $\Ga_{\tsc}$ the reductions corresponding to regions of the lattice chosen as prescribed by the formulas in Eq.(\ref{f1}) and Eq.(\ref{f2}), we evaluate the TEE as a function of the initial squeezing. 

\begin{figure}[tbp]  
\centering
\setlength{\unitlength}{1cm}
\includegraphics[width=13.5\unitlength]{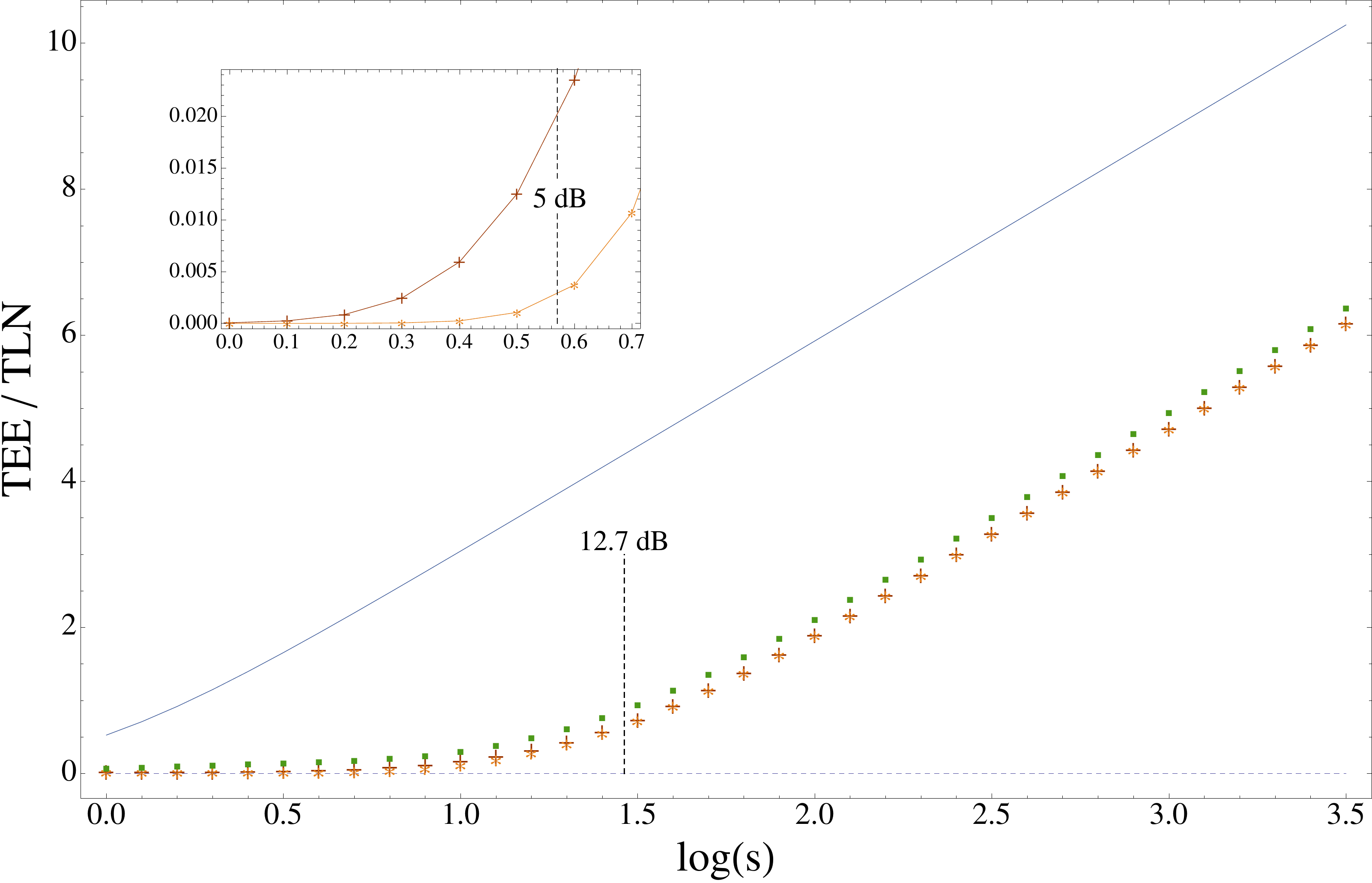}
\caption{Topological entanglement entropy (TEE) $\gamma$ as a function of the squeezing parameter $\log s$ for the CV surface code on a square lattice with 1296 modes. The continuous line represents the values of an analytically obtained upper bound (see main text and Appendix \ref{boundsA}); the asterisks (\ora$\ast$\blk) indicate the TEE calculated using Eq.~\eqref{f1}; the crosses (\red$+$\blk) label the TEE using Eq.~\eqref{f2}; the squares (\grn$\blacksquare$\blk) depict the TLN. The main graph shows the maximum single-mode squeezing of $12.7$ dB achieved to date \cite{mehmet2011, eberle2010}. Finally the dashed line corresponds to TEE for the CV cluster-state, which is always zero. \emph{Inset}: TEE for levels of multimode squeezing with 5 dB marked as achievable with current optical technology \cite{yokoyama2013}. The TLN is too large to be visible on the scale of the inset. Strictly speaking, $\gamma >0$ even for $\log{s}=0$, which happens because the controlled-$Z$ gates contain additional squeezing (see text).}
\label{Stopo}
\end{figure}

In Fig.(\ref{Stopo}) we show our results \cite{demarie2013b}. While the caption of the figure explains in detail what the symbols indicate, we wish to comment on these findings. First of all, note that even for $s \to 1$, which corresponds to starting (in the CV cluster-state preparation) with vacuum states instead of momentum-squeezed states, the TEE is very small but non-zero. This is \emph{not} a numerical artifact but a consequence of applying the controlled-$Z$ gates, which introduce additional squeezing \cite{vanLoock2007, yoshikawa2008} to input vacuum states. It is worth noting that the values of the TEE calculated using the KP formula (\ref{f1}) and the LW formula (\ref{f2}) are extremely close for relatively large squeezing, and the difference accounts for numerics approximation, as expected. However, on the inset the difference is greater that the absolute values of $S_{\text{topo}}^{\text{KP}}$ and $S_{\text{topo}}^{\text{LW}}$: This is due to the different formulas used, which result in different values of $\gamma$ for small squeezing. As a sanity check, we also calculated the TEE for a large ($36 \times 36$ modes\footnote{The CV cluster-state TEE must be zero for any system size}) CV cluster-state, which proves to be zero for any value of the squeezing. This is in complete analogy with the qubit-based cluster-state, which does not exhibit topological order.

Further, we plot the Topological Log-Negativity (TLN) for the KP regions, because of recent results that show this quantity to be a good witness of topological order for stabilizer states of Abelian anyon models \cite{castelnovo2013, lee2013}. For a global state with covariance matrix $\Ga = \Ga_q \oplus \Ga_p$, as in our case (\ref{cov}), the log-negativity of the reduced state with support on a subsystem $A$ is \cite{vidal2002,audenaert2002}\footnote{Note that in \cite{audenaert2002} the authors define the logarithmic negativity in the same way we do in Eq.~(\ref{lognegeq}), but forget an important $1/2$ overall factor in the definition.}:
\beq
\label{lognegeq}
\mathcal{N}(\rho_A) = -\frac{1}{2} \sum_{i = 1}^N \log_2 [\text{min}(1, \lambda_i (\Ga_q {\bm \mu}_A \Ga_p {\bm \mu}_A))] \,,
\eeq
where ${\bm \mu}_A = {\bf P}_{A^{\perp}} \oplus (- {\bf P}_A)$, with ${\bf P}_X$ being the projector onto the modes in region $X$ \cite{audenaert2002} and $\lambda_i$ being the $i$-th eigenvalue of the matrix argument. Then the TLN is constructed replacing the von Neumann entropy with the logarithmic negativity corresponding to the same region,
\beq
\gamma_{\text{LN}} \equiv - ( \mathcal{N}_A + \mathcal{N}_B + \mathcal{N}_C - \mathcal{N}_{AB}- \mathcal{N}_{BC}- \mathcal{N}_{AC} + \mathcal{N}_{ABC}) \, .
\eeq
Remarkably, we find that the TLN is a tight upper bound on the TEE with the same asymptotic slope. 

To obtain this slope, we derived an upper bound  (details can be found in Appendix \ref{boundsA}) to the TEE by considering the smallest meaningful portion of the surface code. Specifically, the upper bound is the entanglement entropy of a one-mode subsystem of a three-mode correlated state. The squeezing-dependent symplectic eigenvalue of the reduced covariance matrix corresponding to the chosen mode is given by $\sigma_1 = \frac{1}{2}(1+3s^4 + 2s^8)^{1/2}(1+3s^4)^{-1/2}$, and the exact value of the upper bound for the TEE is represented by the straight line in Fig.~(\ref{Stopo}), with an asymptotic slope of $\lim_{s \to \infty} \frac{d \gamma(s)}{d (\log s)} = 2/\text{ln}(2) \approx 2.8854$.

\subsection{Noise model}
\label{noisynoise}
To conclude the analysis, we check the resilience of the CV topological phase against noise in the state preparation \cite{demarie2013b}. Errors in the construction of the finitely-squeezed CV cluster-state can be modeled by considering a thermal state with respect to the CV cluster-state Hamiltonian
\beq
\label{thermalCS}
\rho_{\tcs} (\beta) = \frac{e^{- \beta \hH_{\tcs}}}{\tr\Big[ e^{-\beta \hH_{\tcs}} \Big]} \,,
\eeq
as the pre-measurement initial state. Such a state could be generated by several different physical mechanisms. One scenario is in a network of non-interacting photons: One could start with $N$ separable thermal modes as input states instead of the momentum-squeezed vacuum states considered in the theory explained before. Each one of these thermal states is described by 
\beq
\rho( \beta) = \frac{\prod_j e^{- \beta \frac{2}{s^2} \hc_j \ha}}{\tr \Big[ \prod_j e^{- \beta \frac{2}{s^2} \hc_j \ha} \Big]}\, ,
\eeq
with inverse temperature parameter $\beta$. It follows that the thermal cluster-state can be generated by subsequent unitary operations via squeezing, beams splitters, and phase shifters in analogy with the procedure for the pure CV cluster-state. Another scenario is to engineer the Hamiltonian $\hH_\tcs (s)$, which is gapped for finite squeezing, and then wait until the system reaches equilibrium with an environment at temperature $\beta^{-1}$. Eventually, the thermal CV surface code is derived from the thermal cluster-state by quadrature measurements, exactly as explained for the pure case. 

We have seen in \ref{tordofit} that for such mixed states topological order can be detected by making use of the topological mutual information (TMI) \cite{iblisdir2010}. The TMI is constructed replacing in Eq.~(\ref{f1}) the von Neumann entropy $S_X$ with (half of) the mutual information $I_X = S_X + S_{X_c} - S_{X \cup X_c}$ between a region $X$ and its complement $X_c$, precisely:
\beq
\label{TMI}
\gamma_{\text{MI}} \equiv - \frac 1 2 ( I_A + I_B + I_C - I_{AB}- I_{BC}- I_{AC} + I_{ABC})\, .
\eeq
Calculations of the values for the TMI still require knowledge of the covariance matrix of the system (together with its reductions) and its symplectic spectrum. Unfortunately, for mixed states we cannot use the graphical calculus for Gaussian states to derive the covariance matrix, since it only applies to pure states. In spite of that, for this particular system things are easier than one might expect and the covariance matrix of the thermal CV surface code state can be calculated in a rather simple way. First of all, it is straightforward to see that any initial state composed of $N$ equal non-interacting thermal modes is a Gaussian state with covariance matrix given by 
\begin{align}
\Ga_0 = \bigoplus_{i=1}^N \frac{1}{2} \left (
\begin{tabular}{c c}
$\kappa$ & $0$\\
$0$ & $ \kappa$\\
\end{tabular} \right) = \frac{\kappa}{2} \, \mathbf{I}_{2N \times 2N} \, ,
\end{align}
and characterized by the parameter
\beq
\kappa = \coth{\frac{\beta \epsilon_0}{2}} = \coth{\frac{\beta}{s^2}}\, ,
\eeq
with $\epsilon_0 = 2/s^2$ energy gap of the finitely squeezed CV cluster-state. Any Gaussian operation $\mathbf{S}$ performed on $\Ga_0$ is easily computed noticing that 
\beq
\mathbf{S} \Ga_0 \mathbf{S}^T = \frac{\kappa}{2} \mathbf{S} \mathbf{S}^T \, ,
\eeq
and therefore the mixed CV cluster-state covariance matrix corresponding to the density matrix (\ref{thermalCS}) is just given by $\kappa \Ga_{\tcs}$, with $\Ga_{\tcs}$ covariance matrix of the pure CV cluster-state. In order to check how the covariance matrix $\Ga_{\tcs}$ maps to the covariance matrix for the mixed surface code state after the quadrature measurements, we have to translate the effect of measurements into the covariance matrix language. Take a $2N \times 2N$ covariance matrix $\Ga$: If we perform a measurement on the $N$-th mode (this can be easily generalized to any mode), then $\Ga$ is
\begin{align}
\Ga =  \left (
\begin{tabular}{c c}
$\mathbf{A}$ & $\mathbf{C}$\\
$\mathbf{C}^T$ & $\mathbf{B}$\\
\end{tabular} \right) \, ,
\end{align}
where $\mathbf{A}$ is the reduced $(2N -1) \times (2N-1)$ covariance matrix of the first $N-1$ modes, $\mathbf{B}$ is the reduced matrix for the measured mode $N$, and $\mathbf{C}$ is the $(2N \times 2)$ matrix that keeps track of the intra-modes correlations. A $\hq$ measurement on the $N$ mode results in a new covariance matrix given by \cite{spedalieri2013}
\beq
\Ga \longrightarrow \Ga_{\hq} = \mathbf{A} - \mathbf{C}(\mathbf{\Pi} \mathbf{B} \mathbf{\Pi})^{-1}\mathbf{C}^T \,,
\eeq
where 
\begin{align}
\mathbf{\Pi} =  \left (
\begin{tabular}{c c}
$1$ & $0$\\
$0$ & $0$\\
\end{tabular} \right) \, ,
\end{align}
and $(\mathbf{\Pi} \mathbf{B} \mathbf{\Pi})^{-1}$ is the pseudo-inverse (or Moore-Penrose inverse \cite{horn1990}) of the singular matrix $\mathbf{\Pi} \mathbf{B} \mathbf{\Pi}$. For a $\hp$ measurement one has instead \cite{spedalieri2013}
\beq
\Ga \longrightarrow \Ga_{\hq} = \mathbf{A} - \mathbf{C}((\mathbf{I}-\mathbf{\Pi}) \mathbf{B} (\mathbf{I}-\mathbf{\Pi}))^{-1}\mathbf{C}^T \,.
\eeq
Since the covariance matrix of the mixed CV cluster-state is $\Ga  = \kappa \Ga_{\tcs}$, it follows immediately that the special form of this matrix provides with an easy solution to the problem of representing measurements in the mixed case. For either quadrature measurement, using the formulas above shows that $\kappa \Ga_{\tcs}$ under the same evolution of the pure states case, maps to the surface code state covariance matrix equivalent to the zero-temperature covariance matrix times $\kappa$, explicitly:
\beq
\kappa \Ga_{\tcs} \longrightarrow \text{measurements} \longrightarrow \kappa \Ga_{\tsc} \,.
\eeq
It is really important (before someone turns white thinking that I am claiming topological order is resilient against temperature) to note that \emph{this is not} a thermal state for $\hH_{\tsc}(s)$. More naively, here we are modeling noise in the preparation of the CV surface code considering input thermal states instead of pure states. As a result, we have a final state that is not in thermal equilibrium, created via gaussian operations performed on an initial thermal state.

\begin{figure}[tbp]  
\centering
\setlength{\unitlength}{1cm}
\includegraphics[width=7\unitlength]{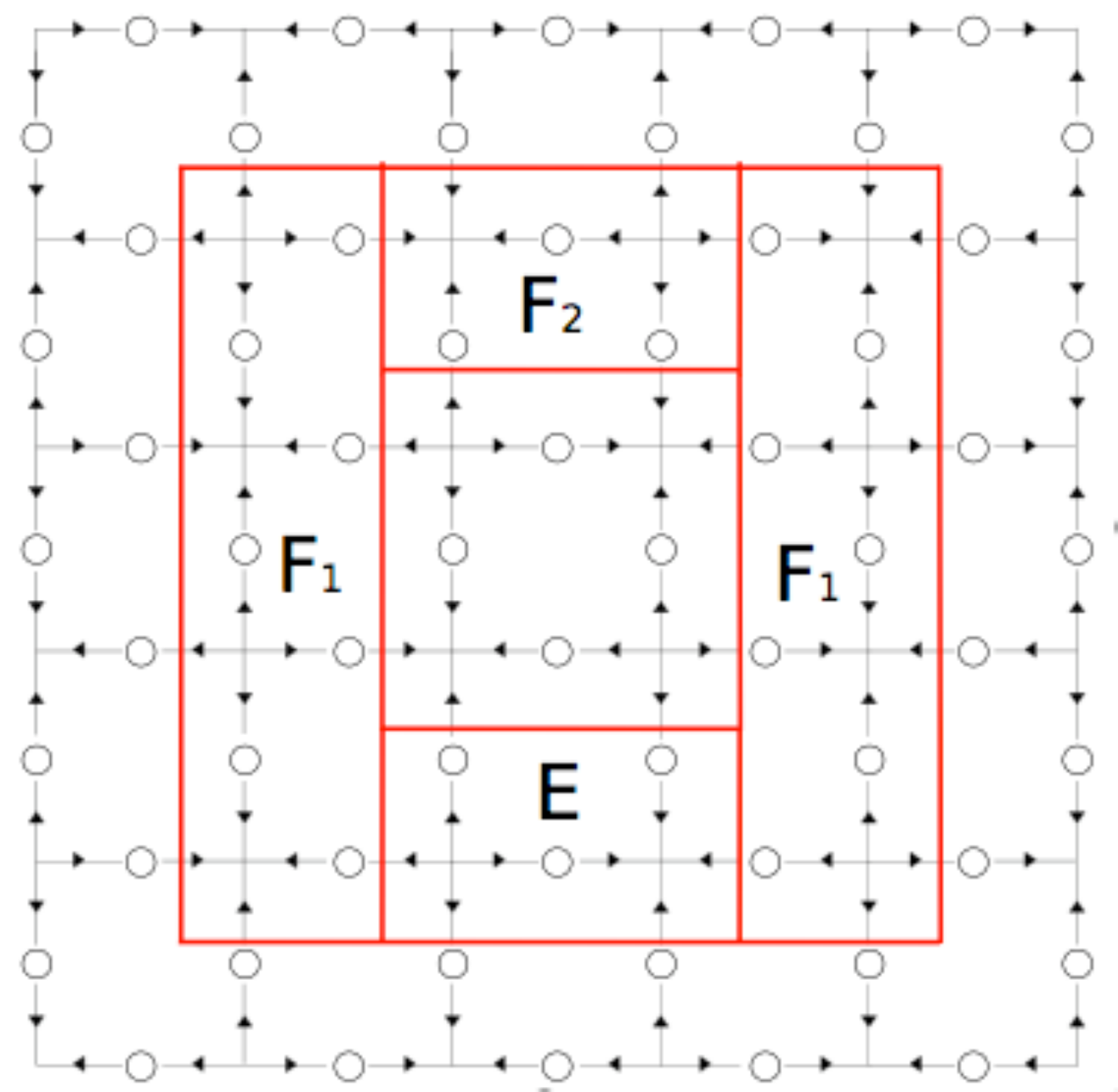}
\caption{Regions used for the calculations of the TMI lower and upper bounds.}
\label{tmibounds}
\end{figure}

\begin{figure}[tbp]  
\centering
\setlength{\unitlength}{1cm}
\includegraphics[width=13.5\unitlength]{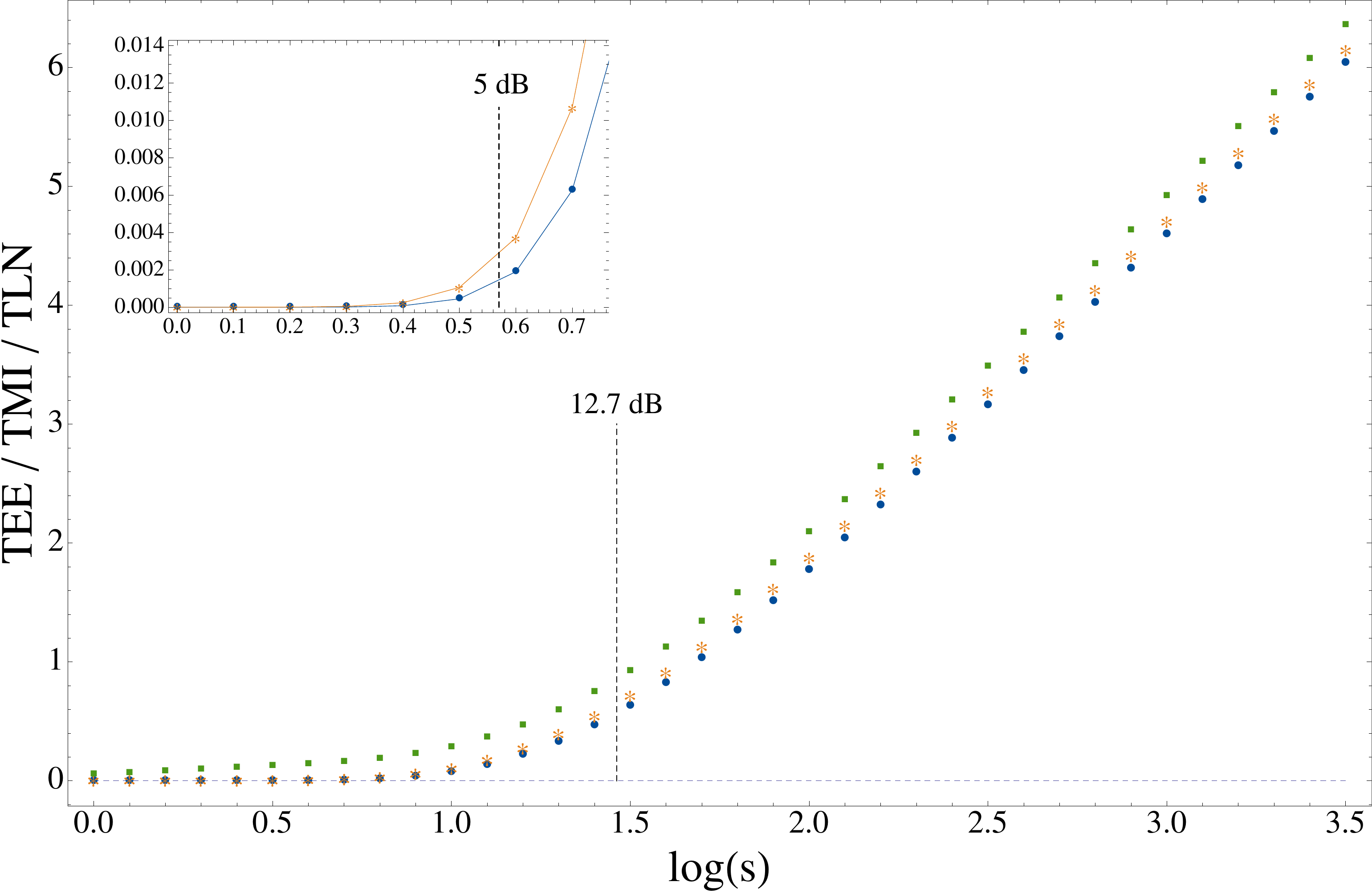}
\caption{Topological entanglement entropy (TEE) $\gamma$, topological logarithmic negativity (TLN) and  topological mutual information (TMI) as a function of the squeezing parameter $\log s$ for the CV surface code on a square lattice with 1296 modes. The asterisks (\ora$\ast$\blk) indicate the TEE calculated using Eq.~\eqref{f1}; the squares (\grn$\blacksquare$\blk) depict the TLN; the circles (\blu$\bullet$\blk) indicate the lower bound for the topological mutual information~(TMI) $\gamma^{l}_{\rm MI}$ for noisy state preparation modeled as a mapping from a thermal state of the cluster-state Hamiltonian in the high-temperature limit (note that this is \emph{not} the TMI for a thermal state of the surface code Hamiltonian.) The main graph shows the maximum single-mode squeezing of $12.7$ dB achieved to date \cite{mehmet2011, eberle2010}. Finally the dashed line corresponds to TEE for the CV cluster-state, which is always zero. \emph{Inset}: TEE and TMI for levels of multimode squeezing with 5 dB marked as achievable with current optical technology \cite{yokoyama2013}. The TLN is too large to be visible on the scale of the inset.}
\label{StopoTMI}
\end{figure}

As explained in \ref{tordofit}, the TMI is limited by lower and upper bounds \cite{wootton2012}. Considering divisions into regions as in Fig.~(\ref{tmibounds}), we find that the value of the TMI is bounded by:
\beq
\text{max} (I_{E,F} - I_{E,F_1} - I_{E,F_2},0) \le \gamma_{\text{MI}} \le I_{E,F} - \text{max} (I_{E,F_1}, I_{E,F_2}) \,,
\eeq
where $F = F_1 \cap F_2$. The TMI for this class of mixed states is lower bounded by the value computed for the worst-case scenario of $\kappa \to \infty$. We then need to calculate the corresponding lower bound for $\gamma_{\text{MI}}^l = \lim_{\kappa \to \infty} \gamma_{\text{MI}}$. In fact, this value for the bound is the lowest possible bound when creating the CV surface code state from a thermal CV cluster-state. It therefore illustrates the maximum extent to which the TMI can sink below the TEE for any given value of the squeezing parameter for this particular construction of the noise model. The value of the lower bound for the 1296 modes mixed CV surface code is depicted by the circles $(\bullet)$ in Fig.~(\ref{StopoTMI}). 

Surprisingly, the difference between the TEE and the lowest possible value of the TMI is very small. Intrigued by this behavior, we have studied analytically the $\kappa \to \infty$ case and found that all the $\kappa$-dependent contributions in Eq.~(\ref{TMI}) cancel out. As shown in Appendix \ref{boundsA}, the TMI for $\kappa \to \infty$ is given by
\beq
\gamma_{\text{MI}}^l = -\frac{1}{2} \sum_X \zeta(X)\sum_i' \log_2(e \sigma_i^X) \,,
\eeq 
where $X$ runs over all the regions as in Fig.~(\ref{areasTEE})a and their complements, $\zeta(X) = \pm 1$ in accordance with the TMI formula in Eq.~(\ref{TMI}), and the prime on the sum indicates that we include only the zero-temperature symplectic eigenvalues $\sigma_i^X$ (for each region $X$) for which $\sigma_i^X > 1/2$ holds. This confirms the results of our numerical analysis. 

It is worth adding a comment about our error model. This is a very general noise model, which acts as an appropriate playground to test the robustness of the topological phase against preparation errors. Nevertheless, this is not the most likely source of error for a quantum optical realization of such a scheme. Since it is not possible to perform arbitrarily accurate quadrature measurements, a physical motivated error model would be mixing the relevant mode with a vacuum mode through a beam splitter having reduced transmittivity and \emph{then} perform perfect quadrature measurements. This procedure would mimic imperfect measurements and allow for additional investigations of the order resilience.

\section{Experimental implementations for the continuous-variable surface code}
\label{expimpCV}
We conclude this Chapter presenting potential experimental realizations of the CV surface code. The CV cluster-states considered in this Chapter are canonical CV cluster-states~\cite{menicucci2011}, so named because they are the states that would result if one were to use the canonical method of constructing them~\cite{zhang2006, menicucci2006, gu2009}. This method generates Gaussian states with graphs of the form $ {\bf Z} = {\bf V} + i s^{-2} {\bf I}$, where the entries of ${\bf V}$ are either 0 or~1. While this method is straightforward theoretically, the $\hat{\text{C}}_{\hat{Z}}$ gates are experimentally difficult and inefficient in an optical setting, where the most progress has been made~\cite{vanLoock2007,yoshikawa2008}.

There exist more efficient and scalable optical construction methods although they produce cluster-states with uniform non-unit edge weight \cite{vanLoock2007,menicucci2007,menicucci2008,flammia2009,menicucci2010,menicucci2011a}. By using these optical methods one can produce medium-sized or even very large states \cite{armstrong2012, roslund, pysher2011}, including a recently demonstrated $10,000-$mode cluster-state with linear topology \cite{yokoyama2013}. Very large square lattices with toroidal \cite{menicucci2008}, cylindrical \cite{menicucci2011} and planar topology can also be made, as well as higher-dimensional lattices \cite{wang2013}.

Another procedure to create similar states is by cooling a circuit-QED system to the ground state of the Hamiltonian $\hH_{\text{cs}} (s)$ in Eq.~(\ref{hamCS}) \cite{aolita2011}. It has been shown \cite{gerardo2009} that by using superconducting co-planar waveguides coupled pairwise via dissipative Cooper pair boxes, one can engineer an effective $\hq-\hq$ interaction between the microwave modes in neighbouring waveguides. Similarly, one can also generate the $\hp-\hp$ coupling by changing the location of the box in the waveguides. While the cluster-state Hamiltonian in Eq.~(\ref{hamCS}) also has $\hq-\hp$ couplings, there exist parent Hamiltonians for CV cluster-states (up to phase shifts) that consist only of $\hq-\hq$ and $\hp-\hp$ couplings \cite{menicucci2011}.

All of these methods can be used to efficiently produce square-lattice CV cluster-states with uniform edge weight $g$, with $0 < g < \frac{1}{4}$. In the optical case, these states can be very large (thousands of modes) \cite{flammia2009} and all produce a Gaussian state with a graph of the form ${\bf Z}_g = g {\bf V} + i \epsilon {\bf I}$. It is easy to notice that such a state can be transformed into a canonical CV cluster-state with effective initial squeezing $\tilde{s} = \sqrt{g/\epsilon}$ by squeezing each mode in $\hq$ by a factor of $\sqrt{g}$,
\beq
{\bf Z}_g \longrightarrow g^{-1} {\bf Z}_g = {\bf V} + i (\epsilon/g) {\bf I} \,,
\eeq
despite being constructed by a completely different method. Since entanglement measures are local-unitary invariant, all of our results apply to these states if we take $s = \tilde{s}$. Furthermore, we do not need to actively perform the single-mode squeezing before measuring the TEE/TMI/TLN. We can simply rescale the outcomes of measurements on the original $g$-weighted) state \cite{demarie2013b}.

\section{Discussion}
In this Chapter we presented a model of a topologically ordered continuous-variable surface code state, analog of the toric code, constructed using only correlated quantum harmonic oscillators and Gaussian operations. Besides allowing for a simple experimental preparation, this model manifests the remarkable property that its topological entropy $\gamma$ can be observed simply via quadrature measurements. 

Moreover, in contrast to discrete-variable systems, the topological entanglement entropy $\gamma$ now is a continuous function of a physical system parameter, specifically the initial squeezing; however, this is not surprising since the parent Hamiltonian for the system is gapless. 

It is also shown here that the CV surface code state can be prepared beginning in a gapped CV cluster-state phase, and that the topological phase is robust against preparation errors modeled as thermal input and can be detected by the topological mutual information. This provides a practical path to observe topological order in systems of interacting bosons using technology available today.

\begin{savequote}[10cm] 
\sffamily 
``Von Neumann told me: You should call it entropy, for two reasons. In the first place your uncertainty function has been used in statistical mechanics under that name, so it already has a name. In the second place, and more important, nobody knows what entropy really is, so in a debate you will always have an advantage.''  
\qauthor{Claude Elwood Shannon} 
\end{savequote}

\chapter{Entropy in Polymer Quantization}
\graphicspath{{Polymer}}
\label{polymerico}

\section{Introduction}
Entropy brings on its shoulders the weight of most of the classical and quantum information theory \cite{nielsen2000, bruss2007, peres2004}. Historically, entropy was considered to have an extensive behavior, thus growing or decreasing proportionally to the volume of the system under analysis. As a matter of fact, the entropy of Gibbs states in statistical mechanics scale extensively with the size of the volume of the system. On the other hand, black hole thermodynamics \cite{bekenstein1973, hawking1974} introduced the idea that an upper bound on the entropy of a system (and entanglement between the subsystems) scales with the area, and not with the volume. This idea was later formalized as a \emph{holographic principle} \cite{bousso2002}, suggesting that a $d$ dimensional theory can be described by a $d-1$ dimensional theory that lives on the boundary of the region considered. 

Such an area-law behavior has been demonstrated in a variety of situations \cite{eisert2010}, which include string theory and loop quantum gravity (LQG) \cite{rovelli2004, thiemann2007}, calculations of black hole entropy \cite{bekenstein1973,rovelli1996}, and related models \cite{livine2008,livine2009}. These properties seem to suggest a deeper connection among the entropic properties of massive objects and the behavior of correlations in entangled quantum many-body systems.

Loop quantum gravity and the spin networks/spin foams formalisms introduced the idea of fundamental discreteness of Nature. Investigation of the problem of emergence of the semiclassical states of quantum gravity and the associated mathematical issues were one of the motivations to study a procedure known as \emph{polymer quantization} of a single non-relativistic particle \cite{ashtekar2003} as a useful toy model to test the predictions of LQG. 

{\red  }


In this framework and considering recent advances in coupling gravity and matter \cite{giesel2013} in LQG, where it was necessary to re-consider holography in these systems, one could wonder wether compatible predictions for observables in two unitarily inequivalent quantization schemes  correspond to  ``close'' values of the predicted entropies or not. While the desirable answer is \emph{yes}, its validity is neither obvious nor guaranteed. The polymer quantization provides a convenient model setting to study this problem, which becomes particularly interesting after the results of Chapter \ref{CVTO}, where  area laws are exploited to characterize quantum systems. 

With these questions in mind, in this Chapter it will be demonstrated that if the Schr\"odinger and the polymer schemes give \emph{close} (to be precisely defined below in the sense of physical equivalence) predictions for fundamental observables, their predictions of entropy are also close \cite{demarie2013}. Moreover, the two entropies coincide in the continuum limit of the polymer quantization. We start introducing the algebraic approach to quantum theory, which relies on the concept of $\text{C}^*$-algebras \cite{baez1992}. Then we describe the polymer quantization of the Weyl algebra, pointing out its differences from the Schr\"odinger quantization, before analyzing the behavior of entropy in the different representations. Finally, the formal proof is illustrated by comparing the ground state entanglement entropy of two coupled harmonic oscillators in the two schemes. The task is easily accomplished using once more the properties of Gaussian states. 

\section{Algebraic quantum theory}
In the usual Hilbert space formulation of quantum mechanics, physical observables are elements of an algebra represented as a set of operators acting on a Hilbert space, whose unit vector elements are identified with physical states. A different approach consists in promoting this algebra to the fundamental object of the theory. In this sense the physical theory is inscribed into the properties of the algebra and quantum states are characterized independently from the representation of the algebra of observables on the Hilbert space \cite{baez1992}. This procedure was first introduced in \cite{haag1964} for quantum field theories defined on a Minkowski space-time. The first step consists in assigning to a quantum system a $\text{C}^*$-algebra that implicitly describes the relations among observables and contains all the relevant physical structure (i.e. locality, Lorentz invariance). One then returns to the usual Hilbert space formulation after defining quantum states via the Gelfand-Naimark-Segal (GNS) construction \cite{halvorson2006}, which establishes a 1-to-1 correspondence (bijection) between vectors of a Hilbert space and linear positive functionals, or states, of a $\text{C}^*$-algebra.

If the algebra is the basic structure of the theory, one may ask whether different representations of the same algebra lead to different values of the observables. While there are theorems that relate observables in unitarily inequivalent representations, this question raises more subtle issues. What about quantities like entropy that are not observables? Are they affected by the choice of a different algebra representation? And if they are, to what extent? This is our motivation to study the behavior of entropy in quantization schemes that are unitarily inequivalent to the usual Schr\"odinger quantization.

\subsection{$\text{C}^*$-algebras}
\label{csa}
Here we clarify the terminology used in the previous section. A complex or real \emph{algebra} $\mathfrak{A}$ \cite{haag1964} is a linear space such that, for each $A,B \in \mathfrak{A}$, there exists an element $AB \in \mathfrak{A}$ called their product, which is bilinear, associative and satisfy certain additional properties as listed in \cite{baez1994} (these are not important for the following discussion). Then, an algebra $\mathfrak{A}$ is a $*$-algebra if for all elements $A \in \mathfrak{A}$ there exists $A^* \in \mathfrak{A}$, the adjoint of $A$, such that the map $()^*$ satisfies $(A^*)^* = A$ and $(AB)^* = B^* A^*$. In addition, an algebra is normed if we can associate to each element $A$ a non-negative number $||A||$ called the \emph{norm} of $A$, such that $||A|| > 0$ if $A \ne 0$. When the additional condition $||A^*|| = ||A||$ holds for all $A \in \mathfrak{A}$, the algebra $\mathfrak{A}$ is a $*$-normed algebra. 

A \emph{Banach algebra} is a $*$-normed and \emph{complete} algebra, where complete means that all the Cauchy sequences of elements in the algebra, topologically defined in terms of the distance between elements $|| A - B||$, converge. Finally, the set of \emph{$\text{C}^*$-algebras} is defined by the subset of Banach algebras whose elements satisfy the condition $||A^*A|| = ||A||^2$ and $||AB|| \le ||A|| \cdot ||B||$. 

In order to define the representation of an algebra, one needs to introduce a couple of concepts more: A \emph{homomorphism} is a linear mapping between two algebras that preserves the product. If the homomorphism also preserves the adjoint, then it is a $*$-homomorphism. A bijective $*$-homomorphism is a \emph{$*$-isomorphism} between algebras. 

Finally, the \emph{representation} of a given algebra $\mathfrak{A}$ is a homomorphism
\beq
\mathcal{R} : \mathfrak{A} \to \mathcal{R}(\mathcal{H}) \,,
\eeq
that maps the elements of $\mathfrak{A}$ to the elements of $\mathcal{R}(\mathcal{H})$, the algebra of linear operators on the linear space $\mathcal{H}$ (note, not necessarily a Hilbert space). If the map $\mathcal{R}$ is a isomorphism, it is called \emph{faithful}. If the only subspaces of the space $\mathcal{H}$ invariant under the action of $\mathcal{R}(\mathcal{H})$ are the zero element set $\{ 0 \}$ and the space $\mathcal{H}$ itself, the representation is \emph{irreducible}. Two representations $\mathcal{R}$ and $\mathcal{S}$, defined over the spaces $\mathcal{H}$ and $\mathcal{H}'$, are unitarily equivalent if, for all $A \in \mathfrak{A}$, there exists a unitary transformation $U : \mathcal{H} \to \mathcal{H}'$ such that $U \mathcal{R}(A) = \mathcal{S}(A) U$ \cite{halvorson2006}. 

But what is so special about $\text{C}^*$-algebra? These algebras are interesting because, as it was proved by Gelfand and Naimark in \cite{gelfand1943}, each (abstract) $\text{C}^*$-algebra is isomorphic to the algebra $\mathfrak{B}(\mathcal{H})$ of linear bounded operators on a Hilbert space $\mathcal{H}$. More precisely, the operators are bounded if for all vectors $x$ of unit norm in $\mathcal{H}$ one has $\6 A x, Ax \9^{1/2} \le ||A||$. Therefore, if we start from a $\text{C}^*$-algebra, we can always describe the associated physical theory in the usual terms of vectors, operators and Hilbert space, and the converse is also true. For this reason $\text{C}^*$-algebras are at the ingredients of the algebraic description of quantum theory.

As a final remark, states on a $\text{C}^*$-algebra $\mathfrak{A}$ are defined as positive linear functionals $\varphi$, in the sense that
\beq
\varphi(A) \in \mathbb{C}\,,\,\, \varphi(A^* A) \ge 0\,, \quad \forall A \in \mathfrak{A} \,.
\eeq
States are normalized, i.e. $\varphi(I_{\mathfrak{A}}) = 1$, where $I_{\mathfrak{A}}$ is the identity of the algebra. Pure states are states that cannot be written as linear combinations of two functionals $\varphi \ne \varphi_1 + \varphi_2$, and mixed otherwise. It is straightforward to see that states here belong to the (positive) dual space $\mathfrak{A}^*$ of the algebra (not to be confused with the set of adjoints). The duality between states and operators is the reason why the Heisenberg and the Schr\"odinger representations are equivalent despite their apparent dichotomy\footnote{I am extremely pleased I managed to use this word in the Thesis, eventually.} in the description of the evolution of the system.

\subsection{Inequivalent representations of an algebra and physical equivalence}
\label{IRAA}
In the algebraic description of quantum theory, states are determined by the expectation values of the elements of the algebra, which determine the physical significance of the theory. Given a $\text{C}^*$-algebra, assuming there are different representations, how can one choose the physically relevant ones or, alternatively said, those representations that describe the same physics? This question is easy to answer in the case of unitarily equivalent representations, since they leave the eigenspectra of the observables invariant and therefore describe the same physics. The problem is far less trivial when one considers unitarily inequivalent representations, where the eigenvalues normally differ between representations. Haag and Kastler \cite{haag1964} understood that limited experimental accuracy in measurements of observables relaxes the conditions for physical equivalence, defining new criteria that are expressed below.

Assuming a state is described by the collection of probabilities 
\beq
P_i(\varphi, A_i) = \frac{\varphi(A_i^* A_i)}{\varphi(I_{\mathfrak{A}})}\,, \,\,\, \forall A_i \in \mathfrak{A}\,,
\eeq 
in the laboratory, the experimentalists can only extract values of the observables correspondent to
\begin{align}
|\varphi(I_{\mathfrak{A}}) - 1| < \epsilon_0 \,, \nonumber \\
|\varphi(A_i^* A_i) - P_i| < \epsilon_i \,,
\end{align}
for the set of positive tolerances $(\epsilon_0, ... \epsilon_i, ...)$. Then, for all practical purposes, the state $\varphi$ is not a point of the algebra $\mathfrak{A}^*$ but it is identified by a neighborhood of a point in $\mathfrak{A}^*$, whose size depends on the given tolerances. Therefore, two representations $\mathcal{R}_1(\mathfrak{A})$ and $\mathcal{R}_2(\mathfrak{A})$ are physically equivalent if each state neighborhood of $\mathcal{R}_1(\mathfrak{A}^*)$ contains an element of $\mathcal{R}_2(\mathfrak{A}^*)$ and vice-versa.

This condition of physical equivalence is equivalent to \emph{Fell's theorem} \cite{fell1960}, which guarantees that, although two representations may be unitarily inequivalent, it is impossible to distinguish between them by using finite-precision expectation values of a finite number of observables. Let us state this concept more precisely. For a state $\rho_1$ of one of the representations, a finite set of operators $\{ A_i \}$ that belong to the chosen algebra $\mathfrak{A}$ and whose expectations are calculated on that state, and the set of tolerances $\{\epsilon_i\}$, then there exists a \textquoteleft physically equivalent' state $\rho_2$ on another representation resulting in the expectation values that differ from the first set by less than the prescribed tolerances. Mathematically, this corresponds to
\beq
|\tr[\rho_1 A_i] - \tr[\rho_2 A_i] | < \epsilon_i \,, \,\,\, \forall \, A_i \in \mathfrak{A} \,.
\eeq
Unfortunately, just like Haag and Kastler's statement, Fell's theorem's proof is not constructive, therefore an explicit construction of states and operators is needed in each case.

In the following section we first define the Weyl algebra, and then introduce the polymer quantization, a unitarily inequivalent representation of the Schr\"odinger quantization of the Weyl algebra, which shares many similarities with the structure of loop quantum gravity. The unitarily inequivalence of the representations will be demonstrated using the Stone-von Neumann theorem \cite{rosenberg2004}, which states the condition for two different irreducible representations of the canonical commutation relations to be unitarily inequivalent. In this frame we investigate the behavior of entropy for one-dimensional quantum systems. 

\section{Polymer quantization}
\label{polyquantiz}
The relevant $\text{C}^\star$-algebra for the following discussion is the \emph{Weyl algebra}, uniquely defined by the Weyl relation,
\beq
\hW(q_1, k_1) \hW(q_2, k_2) = e^{-i/2 (q_1 k_2 - q_2 k_1)} \hW(q_1 +q_2, k_1 + k_2) \, ,
\eeq
and the unitary condition
\beq
\hW(q,k)^\star = \hW(-q, -k) \, .
\eeq
The algebra elements $\hat{W}(q,k)$ are known as Weyl or displacement operators, where $q$ and $k$ are the translation and boost parameters, respectively. Generically, the Weyl operator can be expressed as a product of the translation and boost operators, respectively $\hV(q)$ and $\hU(k)$,
\beq
\label{CCRsvn}
\hW(q,k) = e^{i/2 q k}\hat{V}(q) \hat{U}(k) = e^{-i/2 q k} \hat{U}(k) \hat{V}(q) \, .
\eeq
We have already encountered these operators before, see Eq.~(\ref{contOp}), when generalizing the Pauli operators to the continuous-variable description of quantum states. In order to differentiate the contexts, here let me stick with the traditional formalism of polymer quantization, using $\hU$ and $\hV$ to label them.

\subsubsection{Schr\"odinger representation of the Weyl algebra}
In the standard Schr\"odinger representation of the Weyl algebra, the $\hU(q)$ and $\hV(k)$ operators are continuous functions of the real parameters $q$ and $k$ and are thus generated by self-adjoint operators \cite{tung1985}, specifically the canonical position and momentum operators,
\beq
\label{contUV}
\hV(q) = e^{- i q \hp}\, , \,\,\,\,\,\,\,\,\, \hU(k) = e^{i k \hat{x}}\, .
\eeq
Hence, the Weyl operator is defined as
\beq
\hat{W}(q,k) := e^{i (q \hx + k \hp)} \,,
\eeq 
and it acts on wave functions as 
\beq
\hW (q,k) \psi(x) = e^{-\frac{i}{2}k q} e^{i k x} \psi(x + q) \, .
\eeq
There exists different representations of the Weyl algebra but, curiously enough, the Schr\"odinger's is the only representation where the translation and boost operators are continuous functions of their parameters and can both be written as in Eq.~(\ref{contUV}).
\subsubsection{Polymer representation of the Weyl algebra}
The \emph{polymer quantization} is an alternative irreducible representation of the Weyl algebra 
where quantum states are represented by countable sums of plane waves. In this sense the polymer Hilbert space $\mathcal{H}_{\text{poly}}$ is built around (Harald) Bohr compactification of the real line \cite{halvorson2004, gamelin1969}. A slightly different construction is possible \cite{ashtekar2003}, defining first a countable set of points $\{ x_i \}$ on the real line (a \textquoteleft graph' in the following), and noting then that the (position) space $\mathcal{H}_\poly$ is spanned by the basis states $\ket{x_i}$,
\beq
\label{pqbasis}
\psi_{x_i} (x_j) = (x_j \ket{x_i} = \bra{x_j} k_i \9 = \delta_{x_j, x_i} \, ,
\eeq
where $\delta$ is the Kronecker's delta, and the various bracket signs are explained below. Countable superpositions of these basis states,
\beq
\ket{\Psi} = \sum_{x_i \in \gamma} \psi(x_i) \ket{x_i}\, ,
\eeq
where $\gamma$ is a graph and the functions $\psi$ satisfy certain fall-off conditions, form a space of cylindrical functions $\text{Cyl}_\gamma$. The generalized infinite-dimensional space Cyl is a space of functions that are cylindrical with respect to some graph. The Hermitian inner product on Cyl follows from (\ref{pqbasis}),
\beq
\bra{\Psi} \Psi' \9 = \sum_{x \in \gamma \cap \gamma'} \psi^*(x) \psi'(x) \, ,
\eeq
notice that the inner product is performed solely on the intersection set of the two graphs. The Cauchy completion of Cyl is the space $\mathcal{H}_\poly$, and the triplet of spaces
\beq
\text{Cyl} \subset \mathcal{H}_\poly \subset \text{Cyl}^* \, ,
\eeq
where $\text{Cyl}^*$ is an (algebraic) dual of Cyl, shares similarities with the construction of the physical space of LQG: However, this extends well beyond the scope of this Thesis. Nevertheless, it is interesting to point out that, similarly to LQG, polymer quantization is built on a background independent Hilbert space of discrete sum of plane waves. In LQG it is not possible to define directly the curvature of the connection, just like the momentum operator in the polymer scheme. Other questions regarding the topological problems are discussed in \cite{ashtekar2003, fredenhagen2006, corichi2007a, corichi2007b, velhinho2007} 

Dual elements belonging to $\text{Cyl}^*$ are denoted by $( \varUpsilon |$, and their action on the elements of $\mathcal{H}_\poly$ is denoted by
\beq
( \varUpsilon | \Psi \9 \equiv \varUpsilon (\Psi) \in \mathbb{C} \,.
\eeq
The inner product defines a dual element by
\beq
\Phi (\Psi) \equiv \bra{\Phi} \Psi \9 \, .
\eeq
We introduce now the momentum dual states $( p| \in \text{Cyl}^\star$ by
\beq
(p | x \9 \equiv e^{- i p x}\, ,
\eeq
hence
\beq
(p| = \sum_{x \in \mathbb{R}} e^{- i p x} (x| \, , \qquad \text{and} \qquad \psi(p) = (p| \Psi \9 \, .
\eeq
In the \textquoteleft position' representation that was described so far, the operator $\hU(k)$ acts on a basis state as
\beq
\hU(k) \ket{x_i} = e^{i k x_i} \ket{x_i}
\eeq
and, being weakly continuous in $k$, it has a self-adjoint generator $\hx$ that acts by multiplication and has normalized eigenstates,
\beq
\hU (k) = e^{i k \hx}\,, \,\, \longrightarrow \,\, \hx \ket{x_i} = x_i \ket{x_i} \, ,
\eeq
just like in Schr\"odinger quantization. On the other hand, the action of the $\hV(\mu)$ operator is given by
\beq
\hV(q) \ket{x_i} = \ket{x_i - q} \, ,  
\eeq
and it is easy to see that since $\ket{x_i}$ and $\hV(q) \ket{x_i}$ are orthogonal to each other independently from the size of the translation, the translation operator is not weakly continuous and the momentum operator is undefined since there is no self-adjoint operator generating $\hV(q)$. Notice also that
\beq
\hV^\dagger(q) = \hV(-q) \,.
\eeq
Similarly, while it is easy to show that the usual relation
\beq
( p| \hV(q) = e^{i p q} (p| 
\eeq
holds, it does not correspond to any state of Cyl. 

Clearly, a polymer momentum representation is spanned by the states $\ket{p_j}$. In this case there exists a self-adjoint momentum operator with normalizable eigenstates, but the position operator is undefined. 

\subsubsection{The Stone-von Neumann theorem}
The relationship between the polymer and Schr\"odinger representation is determined by the \emph{Stone-von Neumann theorem} \cite{bratelli1992}. This theorem asserts that: Given pairs of different irreducible representations of the Weyl algebra $(\hat{U}, \hat{V})$ on a Hilbert space, satisfying Eq.~(\ref{CCRsvn}) and weakly continuous in the parameters $q$ and $k$, then they are equivalent to the standard Schr\"odinger representation on the Hilbert space $L^2(\RR^n)$ \cite{rosenberg2004}. Since the polymer representations are not weakly continuous, they do not satisfy the conditions of the Stone-von-Neumann theorem and thus are \emph{not unitarily equivalent} to the Schr\"odinger representation \cite{ashtekar2003,halvorson2004}. 




\subsubsection{Polymer kinematics}
While a mathematically rigorous construction of the polymer states and operators involves a multi-scale lattice \cite{fredenhagen2006}, for sake of simplicity we follow an easier approach and will work with a regular lattice $\gamma$, where the neighbouring points have a fixed spacing $\mu$. This means that in our discussion, for a graph $\gamma$, the following
\beq
| x_j - x_{j \pm 1} | = \mu \, , \,\,\,\, \forall \, x_j \in \gamma 
\eeq
always holds. We have seen that a momentum operator cannot be defined in the same way as in Schr\"odinger quantization. Nevertheless, it is possible to introduce an \emph{effective} momentum operator through a finite difference of translation operators
\beq
\hp_\mu \equiv - \frac{i}{2 \mu} \left(\hV (\mu) - \hV^\dagger(\mu) \right) \,.
\eeq
The limit $\mu \to 0$ in Schr\"odinger representation gives the usual momentum operator $\hp$.
Similarly, its square is
\beq
\label{psquare}
\hp_\mu^2 \equiv (\hp_\mu)^2 = \frac{1}{4 \mu^2} \left( 2 - \hV(2 \mu) - \hV(-2\mu) \right) \,.
\eeq 
It is easy to prove that these operators are self-adjoint \cite{fredenhagen2006}. However, we still need to be careful when working with $\hp^2_\mu$, because this operator is not positive on a general space $\text{Cyl}_\gamma$, even if it is defined on a regular lattice. Consider for example the state (we now drop the $i$ lower index for the position states labeling whenever is not needed)
\beq
\ket{\Psi} = \sum_{x \in \gamma} e^{i k x} e^{-\frac{x^2}{2d^2}} \ket{x} \,,
\eeq
then the expectation value $\6 \Psi| \hp_\mu^2 |\Psi \9$ is given by:
\begin{align}
\6 \Psi| \hp_\mu^2 |\Psi \9 = \frac{1}{4 \mu^2} \6 \Psi | 2& -\hV(2 \mu) - \hV(-2\mu) | \Psi \9 \nonumber \\
= \bra{y} \sum_y e^{- i k y} e^{-\frac{y^2}{2d^2}} &\sum_x \left[ 2 e^{i k x} e^{-\frac{x^2}{2d^2}} - e^{i k (x - 2\mu)} e^{-\frac{(x-2\mu)^2}{2 d^2}} - e^{ik(x+2\mu)} e^{-\frac{(x + 2\mu)^2}{2d^2}}  \right] \ket{x} \nonumber \\
= \sum_x e^{-\frac{x^2}{d^2}} \Big[ 2 - \Big( & e^{-i2k\mu} e^{\frac{2 x \mu}{d^2}} + e^{i2k\mu} e^{-\frac{2 x \mu}{d^2}}  \Big) e^{-\frac{2 \mu^2}{d^2}} \Big] \nonumber \\
= \sum_x e^{-\frac{x^2}{d^2}} 2 \Big[ 1 - ( & \cos(2 k \mu) \cosh(2x\mu/d^2) - i \sin(2 k \mu) \sinh(2x\mu/d^2)  ) e^{-\frac{2 \mu^2}{d^2}} \Big] \,,
\end{align}
which is real only for a symmetric graph, i.e. when $x \in \gamma \Leftrightarrow -x \in \gamma$. Hence, we restrict ourselves to such graphs. 

There is another subtlety concerning this operator. From (\ref{psquare}) it follows that $\hp^2_\mu$ skips the neighboring lattice sites when acting on states, hence its eigenfunctions can have support on either even- or odd-numbered sites $n \mu$, $n \in \mathbb{Z}$. As a result, using $\hp_\mu^2$ as the kinetic term for a Hamiltonian $\hH_\mu$ leads to a double degeneracy of the eigenstates \cite{ashtekar2003, husain2007, kunstatter2012} when compared with the Schr\"odinger representation. Without appealing to it, one can also note that the state counting gives twice the semiclassical result $\int \text{d}p \text{d} x / 2\pi$. Depending on one's goals, it is possible either to adjust the state counting or to introduce an \textquoteleft adjusted' kinetic operator as in \cite{ashtekar2003},
\beq
\label{polykin}
\hK_\mu \equiv \widehat{p^2}_\mu \equiv \frac{1}{\mu^2} \left(2 - \hV(\mu) - \hV^\dagger(\mu) \right)\, ,
\eeq
which resolves the double degeneracy.

The commutation relations between the operators are given by:
\begin{align}
&[\hx, \hV(\mu)]= - \mu \hV(\mu) \,, \nonumber\\
&[\hx, \hp_\mu] = \frac{i}{2} \left( \hV(\mu) + \hV^\dagger(\mu) \right) = i \left( 1 - \frac{1}{2} \mu^2 \hK_\mu \right) \,,\\
&[\hp_\mu, \hK_\mu] = 0 \nonumber\,.
\end{align}

It is important to emphasize that certain modifications introduced by the polymer quantization are a consequence of the form of the effective momentum operator. Since it is defined using the translation operator $\hV$, it carries within a direct dependence upon the choice of the scale of the lattice. 

A key step in extracting physical predictions in the polymer quantization is to consider certain states, called shadow states \cite{ashtekar2001}, which realize the physically equivalent state on $\mathcal{H}_\poly$ \cite{ashtekar2003}. Given a state $|\Psi\9 \in \text{Cyl}_\gamma$, a shadow state corresponding to the dual element $( \Psi |$ is constructed such that its action on $\ket{\Psi}$ is equivalent to the scalar product between its shadow and the same state, i.e.
\beq
( \Phi \ket{\Psi} = \bra{\Phi_\gamma^\s} \Psi \9 \, ,
\eeq
physically this means that the shadow states are projections of the elements of $\text{Cyl}^\star$ onto $\text{Cyl}_\gamma$,
\beq
(\Phi| = \sum_{x \in \mathbb{R}} \Phi^*(x) (x| \longrightarrow \ket{\Phi_\gamma^\s} = \sum_{x \in \gamma} \Phi (x) \ket{x} \, .
\eeq
Using these states it is possible to explicitly demonstrate physical indistinguishability of the predictions of the polymer and Schr\"odinger quantizations, as mandated by Fell's theorem, in the sense that the expectation values of the observables of interest, using shadow states, fall within a prescribed range of tolerance from the Schr\"odinger predictions. 

With the definition of the shadow states we concluded this short review of all the necessary concepts for the understanding of the following section, where we address the convergence of entropy in polymer quantization and compare it to known results in \textquoteleft canonical'  quantum mechanics. 

\section{Entropy in different quantization schemes}
\label{entropypolymerico}
Our goal in this Chapter is to find a relationship between the von Neumann entropy of a state $\rho$ (in the Schr\"odinger representation), 
\beq
S(\rho) = -\tr \rho \log \rho \,,
\eeq
and of its polymer analogue state $\rho^\mu$. There are several reasonable ways to relate the states but, unlike the case of observables, there is no theorem that guarantees equivalence of the entropy predictions! Moreover, entropy is not a continuous function, and without additional restrictions there are states of infinite entropy in a neighborhood of any state \cite{wehrl1978}. 

Throughout the following we will use certain properties of entropy as described in Section \ref{entropyprop} (see \cite{wehrl1978, ohya2004} for more details). The concavity of the entropy of a convex combination of one-dimensional projectors $\rho = \sum_i \omega_i \rho_i$, with $\sum_i \omega_i = 1$ and $\forall \omega_i >0$,
\beq
S(\rho) \ge \sum_i \omega_i S(\rho_i) \,,
\eeq
together with its upper bound
\beq
\label{concavity2}
S(\rho) \le - \sum_i \omega_i \log \omega_i \,.
\eeq
and the upper bound given by a set of density matrices $\rho_k$ weakly convergent to the density matrix $\rho$
\beq
S(\rho) \le \lim \text{inf}\,S(\rho_k) \,.
\eeq
Keep in mind that the relation $\tr | \rho_k - \rho| \to 0$ is ensured by the weak convergence.

It can be shown \cite{wehrl1978,eisert2002} that the lower semicontinuity of the relative entropy $S(\sigma|\rho)$ in $\rho$ implies the lower semicontinuity of the free energy $F$ at the temperature $\beta^{-1}$
\beq
\label{fenergy}
F(H, \rho, \beta) := \tr \, \rho H - S(\rho) / \beta\,,
\eeq
where $H$ is the Hamiltonian of the system, and the state $\rho$ is not necessarily thermal.

This point is crucial for our goals: The lower semicontinuity of the free energy has an important consequence. If in addition to the weak convergence of the states (\ref{stwc}), also the energy expectation values converge, i.e.
\beq
\tr \rho_k H \to \tr \rho H\,,
\eeq
then, from (\ref{fenergy}), the entropies converge as well,
\beq
\label{Sconv}
S(\rho) = \lim S(\rho_k)\,.
\eeq
Now that the hypothesis have been set, we can analyze entropy in polymer quantization. The simplest case to study in order to establish the relationship between the state $\rho$ and its polymer analogue is when the Schr\"odinger state
\beq
\rho = \sum_i \omega_i \ket{\psi_i} \bra{\psi_i}\,, \quad \sum_i \omega_i = 1\,, \quad \forall\, \omega_i >0
\eeq
is a mixture of the eigenstates of some operator, and its polymer counterpart $\rho^\mu$ has the corresponding eigenbasis $\ket{\Psi_i}$. Then the decomposition
\beq
\rho^\mu = \sum_i \omega_i \ket{\Psi_i} \bra{\Psi_i}
\eeq
trivially has the same entropy. In this case, it is natural to argue that the expectation values of all the observables of interest in two representations are close, but it should be established in the case-by-case analysis.

Consider now the analogue
\beq
\rho^\mu = \sum_i \omega_i \ket{\Psi(\mu)_i} \bra{\Psi(\mu)_i} \,,
\eeq
where $\ket{\Psi(\mu)_i}$, the normalized \textquoteleft close approximation' of the states $\ket{\psi_i}$ in the sense of Fell's theorem, are mixed with the same weights. We assume that the states $\ket{\Psi(\mu)_i}$ are pure, but not necessarily orthogonal.

In this case, the inequality (\ref{concavity2}) that results from the concavity of entropy reads as
\beq
S(\rho_\s^\mu) \le - \sum_i \omega_i \log \omega_i = S(\rho) \, ,
\eeq
therefore the entropy of the state in the Schr\"odinger quantization bounds the entropy of the polymer analog. This result holds for any two representations where the equivalent state of any pure $\ket{\psi}$ is also pure. 

A closely related result is derived as follows \cite{demarie2013}. In the following we will use the harmonic oscillator eigenstates both in the Schr\"odinger and polymer representations, labeling the states as $\ket{n}$ and $\ket{n^\mu}$, respectively. If the Schr\"odinger state is projected on the regular lattice with spacing $\mu$, resulting in $\ket{n^\mu_\s}$, then
\beq
\ket{n^\mu} = \ket{n^\mu_\s} + \ket{\Delta n^\mu}\,, \qquad \, \bra{\Delta n^\mu} \Delta n^\mu \9 \overrightarrow{_{\mu \rightarrow 0}}\, 0\, . 
\label{apshad}
\eeq
In these bases the two physically equivalent states $\rho$ and $\rho^\mu$ can be written as
\beq
\rho = \sum_{k\, l} \omega_{k l} \ket{k} \bra{l} \Longleftrightarrow \rho^\mu = \sum_{k \, l} \left( \omega_{kl} + \Delta \omega^\mu_{kl} \right) \ket{k^\mu} \bra{l^\mu} \,,
\eeq
for some $\Delta \omega_{k l}^\mu$. This quantity converges to zero as to ensure the agreement for the probabilities for projecting on the states $\alpha \ket{m} + \beta \ket{n}$ and their polymer analogues. 

Consider now a classical Hamiltonian of the form
\beq
H = \frac{1}{2m} p^2 + V(x) \,.
\eeq
We can test the convergence of the energy, establishing the separate convergence of expectations of the kinetic and potential energy terms as follows. The polymer kinetic term is given by Eq.~(\ref{polykin}), and is built from the difference of two Weyl algebra operators. Hence, the direct use of the Fell's theorem guarantees that
\beq
\lim_{\mu \to 0} \tr \rho^\mu \hK_\mu = \tr \rho \hp^2 \, .
\eeq
On the other hand, using Eq.~(\ref{apshad}) for all potentials $V(x)$ such that the value $\6 l |V|m \9$ is finite, the projection onto the lattice and the subsequent summation form the Riemannian sum for the above Schr\"odinger integral expression. Then, for $V(x)$ growing sub-exponentially with $x$ we have that
\beq
\bra{\Delta l^\mu} V \ket{m^\mu_\s} \to 0\, \qquad \text{with} \qquad \mu \to 0\,,
\eeq
hence
\beq
\6 l^\mu | V(\hx) \ket{m^\mu} \to \bra{l^\mu_\s} V \ket{m^\mu_\s} \to \bra{l} V(\hx) \ket{m} \,,
\eeq
where the potential in the first term refers to polymer quantization and in the last term to Schr\"odinger quantization. This ensures the convergence of the potential term.

Since both the matrix elements of the density matrices and the energy expectation values converge, Eq.~(\ref{Sconv}) applies and we can establish the desired convergence
\beq
\label{Sproof}
\lim_{\mu \to 0} S(\rho^\mu) = S(\rho) \,,
\eeq
proving indeed that, for lattice size $\mu \to 0$, the value of the entropy in polymer quantization reduces to the same value in Schr\"odinger quantization \cite{demarie2013}.

\subsection{An example: Convergence of the entropy of two coupled harmonic oscillators}
\label{conventropy}
We illustrate the result from Eq.~(\ref{Sproof}) by considering the entanglement of the ground state of two position-coupled harmonic oscillators, in the simplest case of the quadratic Hamiltonian
\beq
H = \frac{1}{2 m} (p_1^2 + p_2^2) + \frac{m \omega^2}{2} (x_1^2+x_2^2) + \lambda(x_1 - x_2)^2\, ,
\eeq
where the positive constant $\lambda$, which is bounded by $\lambda < \frac{m \omega^2}{2}$, determines the coupling. Just like in the previous Chapters, we arrange the position and momenta (either classical or quantum) of the $N$ particles (in this example $N=2$) into a single $2N$-dimensional vector $\mathbf{r}^T=(x_1,\ldots, x_N,p_1,\ldots, p_N)$ and make use of the Gaussian states formalism.

In the following, the parameter that determines the closeness of the physical predictions in the two different quantization schemes is the ratio of the lattice size $\mu$ to the oscillator scale $d=(m \omega)^{-1/2}$. Since we want to model quantum geometry, we always assume that $\mu \ll d$, i.e.
\beq
\frac{\mu}{d} = \mu \sqrt{m \omega} \ll 1\, .
\eeq 

In the Schr\"odinger representation, it is very easy to calculate the covariance matrix of the ground state of a single harmonic oscillator. Using a covariance matrix definition physically equivalent to the one in Eq.~(\ref{gengamma}), but slightly different in the mathematical form to take care of any modification induced by the polymer quantization scheme,
\beq
\Ga_{i,j} = \frac{1}{2} \6 \{ \hat{\mathbf{r}}_i - \6 \hat{\mathbf{r}}_i \9,\, \hat{\mathbf{r}}_j - \6 \hat{\mathbf{r}}_j \9 \} \9\, ,
\eeq
where $\{, \}$ is the anticommutator, then we are simply required to calculate the second statistical moments of the canonical observables \cite{demarie2012}. This is an easy-to-achieve task, and using some old good first-year calculus it is immediate to check that for the ground state of a free quantum harmonic oscillator the following holds \cite{cohen1977}:
\begin{align}
\mathbf{\Gamma} = \left (
\begin{tabular}{c c}
$\6 \hx^2 \9$ & $\frac{1}{2} (\6 \hx \hp \9 + \6 \hp \hx \9)$\\
$\frac{1}{2}(\6 \hp \hx \9 + \6 \hx \hp \9)$ & $ \6  \hp^2 \9$\\
\end{tabular} \right ) =
 \frac{1}{2} \left (
\begin{tabular}{c c}
$d^2$ & $0$\\
$0$ & $ d^{-2}$\\
\end{tabular} \right ) \, .
\end{align}
Recall that for a Gaussian state $\rho$ described by a covariance matrix $\Ga$, the von Neumann entropy of any subsystem $\rho_A$ (which is still Gaussian) is simply given by the formula of Eq.~(\ref{vnS})
\beq
S(\rho_A) =\sum_{j=1}^n\left[ \Big( \sigma_j +\frac12 \Big) \log_2\Big( \sigma_j +\frac12 \Big) - \Big( \sigma_j -\frac12 \Big)\log_2\Big( \sigma_j -\frac12 \Big)\right]\,,
\label{ent-cor}
\eeq
with $\{ \sigma_j \}_A$ set of the symplectic eigenvalues of the reduced covariance matrix $\Ga_A$.

Calculations in the polymer representation are a bit trickier because one needs to take into account the corrections due to the introduction of the discrete spatial lattice. Let me start recalling that the wave function of the harmonic oscillator ground state in the Schr\"odinger quantization, using the $ \{\ket{x}\} $ representation and units where $\hbar = 1$, is equal to
\beq
\label{schgs}
\phi_0(x) = \bra{x} 0 \9 = \left( \frac{m \omega}{\pi} \right)^{1/4} e^{-\frac{m \omega}{2} x^2} = \left( \frac{1}{d^2 \pi} \right)^{1/4} e^{-\frac{1}{2} \frac{x^2}{d^2}} \,. 
\eeq
Then the corresponding harmonic oscillator ground state in the polymer quantization is described by the state
\beq
( {\bf \Phi}_0 | \in \text{Cyl}^\star \, .
\eeq
To extract the physical information from $( {\bf \Phi}_0 |$ we use its shadow state 
\beq
\ket{{\bf \Phi}_0^\s} \in \text{Cyl}_\gamma \qquad \text{such that} \qquad ( {\bf \Phi}_0 | f_\gamma \9 = \bra{{\bf \Phi}_0^\s}  f_\gamma \9 \, ,
\eeq
for any generic element $\ket{f_\gamma} \in \text{Cyl}_\gamma$. On an infinite regular lattice with spacing $\mu$, each point of the lattice $x_n$ is given by $x_n = n \mu $. We assume for simplicity that the lattice is centered on $x_0 = 0$, but the discussion is equivalent for any center point. From Eq.~(\ref{schgs}), the explicit form of the state $\ket{{\bf \Phi}_0^\s}$ is
\beq
\ket{{\bf \Phi}_0^\s} = c \sum_{n \in \mathbb{Z}} e^{-\frac{n^2 \mu^2}{2 d^2}} \ket{n \mu} \, ,
\eeq
where now $c$ is an arbitrary normalization constant and $\{\ket{n \mu}\}$ is a orthonormal basis on $\text{Cyl}_\gamma$. To be pedantic, we should label the shadow state as $\ket{{\bf \Phi}_{0,\mu}^\s}$, but we drop the under-script $\mu$ when it is clear what we are referring to. The norm of the shadow ground state is
\begin{align}
\label{norms}
\6{\bf \Phi}_0^\s \ket{{\bf \Phi}_0^\s} &= |c|^2 \sum_{n = -\infty}^\infty \sum_{m = -\infty}^\infty e^{-\frac{n^2 \mu^2}{2 d^2}} e^{-\frac{m^2 \mu^2}{2 d^2}} \bra{n \mu} m \mu \9 \nonumber \\
&= |c|^2 \sum_{n = -\infty}^\infty  e^{-\frac{n^2 \mu^2}{d^2}} \,,
\end{align}
since $\bra{n \mu} m \mu \9 = \delta_{n,m}$. To simplify this expression, we can make use of the \emph{Poisson summation formula} \cite{spiegel1990,ashtekar2003}. It states that for sufficiently well-behaved real functions $f(n)$, with $n \in \mathbb{Z}$, the following equality holds:
\beq
\label{poissonf}
\sum_{n = - \infty}^\infty f(n) = \sum_{n = - \infty}^\infty \tilde{f}(n)\, ,
\eeq
where $\tilde{f}(n)$ is the Fourier transform $\mathcal{F}(f(n))$ of the function $f(n)$, defined by
\beq
\forall \, n \in \mathbb{R}\, , \qquad \tilde{f}(n) = \int_{- \infty}^\infty f(y)e^{- 2 \pi i n x} dx\, .
\eeq
After rewriting Eq.~(\ref{norms}) using Eq.~(\ref{poissonf}), we have (with $f(n) =e^{-\frac{n^2 \mu^2}{d^2}} $)
\beq
\label{step1}
|c|^2 \sum_{n = -\infty}^\infty  e^{-\frac{n^2 \mu^2}{d^2}} = |c|^2 \sum_{n = -\infty}^{\infty} \int_{-\infty}^{\infty} e^{-\frac{y^2 \mu^2}{d^2}} e^{- 2 \pi i y n} dy\, .
\eeq
The easiest way to solve this integral is by completing the square in the exponent, i.e.
\beq
\left( -\frac{y^2 \mu^2}{d^2} - 2\pi i y n + \frac{\pi^2 n^2 d^2}{\mu^2} \right)- \frac{\pi^2 n^2 d^2}{\mu^2} = -\left(\frac{y \mu}{d} + \frac{\pi i n d}{\mu} \right)^2 - \frac{\pi^2 n^2 d^2}{\mu^2}\, ,
\eeq
such that the right-side of Eq.~(\ref{step1}) becomes
\beq
|c|^2 \sum_n \int_{\mathbb{R}} e^{-\left(\frac{y \mu}{d} + \frac{\pi i n d}{\mu} \right)^2} e^{- \frac{\pi^2 n^2 d^2}{\mu^2}} dy\, .
\eeq
Using the variable substitutions 
\beq
s = \frac{y \mu}{d} \qquad \text{and} \qquad z = s + \frac{\pi i n d}{\mu}\, ,
\eeq
we can immediately derive the following expression
\beq
|c|^2 \sum_n  e^{-\frac{\pi^2 n^2 d^2}{\mu^2}} \int_{\mathbb{R}} \frac{d}{\mu} e^{-z^2} \text{d}z = |c|^2 \frac{\sqrt{\pi} d}{\mu} \sum_n e^{-\frac{\pi^2 n^2 d^2}{\mu^2}} \, ,
\eeq
where we exploited the well-known equality $\int_{\mathbb{R}} e^{- z^2} \text{d}z = \sqrt{\pi}$. The assumption $\mu / d \ll 1$ allows us to keep only the $n=0,\, n= \pm 1$ terms of the series, therefore the norm of the shadow ground state is approximately equal to
\beq
\6{\bf \Phi}_0^\s \ket{{\bf \Phi}_0^\s} \approx |c|^2 \frac{\sqrt{\pi}d}{\mu} \left( 1 + 2e^{-\frac{\pi^2 d^2}{\mu^2}} \right)\, .
\eeq 
Calculations in the polymer representation also give zero expectations for the ground state of the harmonic oscillator, $\6\hx\9=0$, $\6\hp_\mu\9=0$, but the variances now carry modifications dependent on the lattice size $\mu$ \cite{ashtekar2003}:
\beq
\6\hx^2\9\approx \frac{d^2}{2}\left(1-\frac{4\pi^2d^2}{\mu^2}e^{-\pi^2 d^2\!/\mu^2}\right), \qquad \6\hp_\mu^2\9\approx\frac{1}{2d^2}\left(1-\frac{\mu^2}{2d^2}\right).
\eeq
Furthermore, the correlation terms vanish exactly 
\beq
\6 \hx \hp_\mu \9 + \6 \hp_\mu \hx \9 = 0 \, .
\eeq
For the complete details of these calculations, the reader can refer to Appendix \ref{expPolymer}. Hence, keeping only the leading terms in $\mu/d$  the correlation matrix for one oscillator in polymer quantization becomes:
\beq
\mathbf{\Gamma}_{0,\mu}=\frac{1}{2} \left(
\begin{tabular}{cc}
$d^2$ & $0$ \\
$0$ & $\frac{1}{d^2}\left(1-\frac{\mu^2}{2d^2}\right)$
\end{tabular}\right) \, .
\eeq
One of the consequences of the polymer quantization is that now the product of the uncertainties is less than its standard normalized value of $\frac{1}{4}$, in fact:
\begin{align}
(\Delta x)^2 \approx \frac{d^2}{2} \qquad \text{and} \qquad (\Delta p_\mu)^2 \approx \frac{1}{2d^2} \left( 1 - \frac{\mu^2}{2 d^2} \right) \, ,\\ 
\text{such that} \qquad (\Delta x)^2 (\Delta p_\mu)^2 = \frac{1}{4} \left[ 1 - \frac{\mu^2}{2 d^2} + O(\frac{\mu^4}{d^4}) \right] \,.
\end{align}
This means that the state is no longer Gaussian: The correlation matrix violates the defining inequality ${\bf \Gamma}_{0,\mu} + i {\bf J} \ge 0$, and the entropy calculated using Eq.~(\ref{ent-cor}) becomes complex for a pure state, explicitly $S \thicksim i \mu^2/d^2$.

Going back to the coupled oscillators, to quantify the amount of entanglement among them we use again the von Neumann entropy of either of the reduced density matrices. It was shown in the example (\ref{ExGS}) at the end of Chapter \ref{ChapGS} how to calculate the von Neumann entropy of two coupled oscillators transforming the system to the normal modes by means of a symplectic transformation. This gives two uncoupled oscillators with new frequencies $\omega_1'$ and $\omega'_2$:
\beq
\omega_1' = \omega\,, \,\,\,\,\,\,\,\,\, \omega_2' = \omega \sqrt{1+ \frac{4 \lambda}{m \omega^2}} \equiv \omega \alpha > \omega \, .
\eeq
The (Schr\"oedinger) correlation matrix in the normal coordinates is 
\beq
\Ga' = \text{diag}(d^2,d^2 \alpha^{-1},d^{-2},d^{-2}\alpha)\,,
\eeq
and the only symplectic eigenvalue of the reduced correlation matrix is 
\beq
\sigma_1 = \frac{1+\alpha}{4 \sqrt{\alpha}} \,.
\eeq
The polymer quantization is treated similarly. As it is done in the usual quantization process, one needs to perform the symplectic transformation before quantizing (\'a la polymer), since these two procedures do not commute. Using $\alpha >1$ and similarly to Eq.~(\ref{diagamma}), the covariance matrix of the two uncoupled oscillators in polymer quantization looks like:
\begin{align}
\Ga'_\mu = \frac{1}{2} \left (
\begin{tabular}{c c c c}
$d^2$ & $0$ & $0$ & $0$\\
$0$ & $\frac{d^2}{ \alpha}$ & $0$ & $0$\\
$0$ & $0$ & $\frac{1}{d^2}\left(1-\frac{\mu^2}{2d^2} \right)$ & $0$\\
$0$ & $0$ & $0$ & $\frac{\alpha}{d^2}\left(1-\frac{\mu^2}{2d^2} \right)$\\
\end{tabular} \right) \, .
\end{align}
The original covariance matrix can be reconstructed using the (inverse of the) appropriate symplectic transformation $\mathbf{Y}$ described in Eq.~(\ref{matY}). Then we find
\begin{align}
&\Ga_\mu = {\bf Y}^{-1} \Ga'_\mu ({\bf Y}^T)^{-1} \nonumber \\ 
&=\frac{1}{4} \left (
\begin{tabular}{c c c c}
$\frac{d^2(1+\alpha)}{\alpha}$ & $\frac{d^2 (1- \alpha)}{\alpha}$ & $0$ & $0$\\
$\frac{d^2(1-\alpha)}{\alpha}$ & $\frac{d^2(1+\alpha)}{\alpha}$ & $0$ & $0$\\
$0$ & $0$ & $\frac{d^2(1+\alpha)-(1+\alpha^2)\mu^2}{d^4}$ & $\frac{(-1+\alpha)(-2d^2+(1+\alpha)\mu^2)}{d^4}$\\
$0$ & $0$ & $\frac{(-1+\alpha)(-2d^2+(1+\alpha)\mu^2)}{d^4}$ & $\frac{d^2(1+\alpha)-(1+\alpha^2)\mu^2}{d^4}$\\
\end{tabular} \right) \, ,
\end{align}
and the reduced covariance matrix for either the first or second oscillator is simply
\begin{align}
\Ga_{1,\mu} = \Ga_{2,\mu} = \frac{1}{4} \left(
\begin{tabular}{c c}
$\frac{d^2(1+\alpha)}{\alpha}$ & $0$\\
$0$ & $\frac{d^2(1+\alpha)-(1+\alpha^2)\mu^2}{d^4}$\\
\end{tabular} \right) \, .
\end{align}
Keeping only the leading terms in the powers of $\mu/d$, the symplectic eigenvalue determining the bipartite entanglement in the polymer quantization is
\begin{align}
\sigma_1^{\text{poly}} &= \frac{1+\alpha}{4\sqrt{\alpha}} - \frac{(1+\alpha^2)}{16 \sqrt{\alpha}} \frac{\mu^2}{d^2} \nonumber \\
&= \sigma_1 - \frac{(1+\alpha^2)}{16 \sqrt{\alpha}} \frac{\mu^2}{d^2}\, ,
\end{align}
and the von Neumann entropy, calculated using $\sigma_1^{\text{poly}}$ into Eq.~(\ref{ent-cor}) is equal to
\beq
S^{\text{poly}} = S - \frac{1+\alpha^2}{32 \sqrt{\alpha}}\left( \log_2 \frac{1 - 4\sqrt{\alpha} + \alpha}{1 + 4\sqrt{\alpha} + \alpha} \right) \frac{\mu^2}{d^2} \,,
\eeq
with $S$ von Neumann entropy in the Schr\"odinger quantization, in agreement with the results about entropy derived in previous sections of this Chapter \cite{demarie2012}. 

\section{Discussion}
The main result in this Chapter is the derivation of a general bound that relates entropies of physically equivalent states in unitarily inequivalent representations of the Weyl-Heisenberg algebra. In particular, we proved that the entropies of two states described in the Schr\"odinger and the polymer quantizations agree. 

The lattice effects modify the expectation values of physical observables, the commutation relations and the entropy values such that shadow states of semi-classical states do not satisfy the Gaussian property exactly, but only up to the terms of the order of $\mu^2/d^2$. The results derived here give also physical justification to the entropy corrections derived by Chacon-Acosta et al. \cite{chacon2011} for the statistical thermodynamic properties of one-dimensional polymer quantum systems. 

Since the convergence of entropy of the shadow states to the Schr\"odinger representation value is not necessarily uniform, the following scenario is plausible. Remember that both the discrepancy in the expectation value of the momentum variances and symplectic eigenvalues are of the order of $\mu^2/ d_j^2$ for each (uncoupled) oscillator. We showed that this is also the order of magnitude of the corresponding change in entropy contribution if the formula (\ref{ent-cor}) is used. Therefore, even if the expectations of the observables agree, for a fixed value of $\mu$ and a large number $N$ of oscillators the two entropies will differ by $~ n \mu^2/ d^2$, which may be a significant amount. We leave this as an open question. 

\begin{savequote}[10cm] 
\sffamily
``My mind rebels at stagnation. Give me problems, give me work, give me the most abstruse cryptogram or the most intricate analysis, and I am in my own proper atmosphere. I can dispense then with artificial stimulants. But I abhor the dull routine of existence. I crave for mental exaltation. That is why I have chosen my own particular profession.'' 
\qauthor{Sherlock Holmes - The Sign of Four}
\end{savequote}

\chapter{Conclusions}

\label{concluded}

In this Thesis we investigated the interplay of geometry and quantum information in various physical situations. In the first part we looked at the effects of gravity on photon polarization, identifying paths that induce a gauge-invariant phase of the photon state. In the second part we studied topological order in the context of continuous-variable systems. While the first two parts of this work are driven by the effort of simplifying the experimental realizations of certain quantum information tasks, the third part is mainly theoretical and deals with the problem of entropy in different quantization schemes. In this conclusive Chapter we review the results obtained in each part of the Thesis and wind up this work by tickling the reader's attention with a list of open questions.

\section{Relativistic quantum information}
Quantum information tasks in space rely on the sapient use of reference frames to make the processing of information meaningful. 

In Chapter \ref{GRchapter} we dealt with this problem, describing the effects of gravity on photon polarization. We investigated the possibility to design particular paths of propagation along which the polarization phase accrued is gauge-invariant, i.e. independent of the choice of reference frame in the course of the trajectory. We first focused on the spatial three-dimensional projection of a general four-dimensional space-time, which is an excellent approximation of the near-Earth environment. After deriving an equation for the phase polarization in terms of the geometrical properties of the underlying spatial manifold, in section \ref{gaugesec} we showed that along a closed path that phase is invariant under a change in the polarization reference frame. 

Interestingly enough, the same argument can be extended to general four-dimensional space-times. 
Constructing a closed path by means of mirrors carefully posed, we proved that the total variation in the polarization of two photons moving along the two paths of a closed trajectory that resembles a Mach-Zender interferometer, depends again solely on the geometrical bundle curvature enclosed by the curve. 

Gauge-independence of the phase bypasses the problem of having to communicate reference frames between observers at the two ends of a trajectory, simplifying for instance the setup of an experiment aimed at testing the relativistic effects of gravity on quantum systems. We discussed this and other related experimental proposals at the end of Chapter \ref{GRchapter} in section \ref{GRexpe}. 

\section{Continuous-variable topological order and Gaussian states}
Topologically ordered systems are the platforms for implementing stable quantum memories or universal quantum computers. However, topological phases are usually associated with discrete-variable systems, which are extremely challenging to prepare and control in the laboratory. Here we extended this concept to continuous-variables, proving that it is indeed possible to detect topological order on a lattice of interacting quantum harmonic oscillators, which is based on two-body interactions and offers an easier experimental implementation. 

In Chapter \ref{CVTO} we describe the properties of the physical continuous-variable surface code state, analog of Kitaev's surface codes, identifying the appropriate form of the stabilizer set in \ref{pCVsc} and using them to construct the defining Hamiltonian, whose ground state determines the code state. 

Although the parent Hamiltonian of our surface code model is gapless (\ref{eGap}), which might affect the topological stability, we proved that the entropy on the lattice follows an area law. Besides, we provided with an analytical proof of the exponential decay of the correlations in the system. In section \ref{quacorre} we showed that the momentum-momentum correlations are finite, and the momentum-position correlations are zero. Then, the position-position correlations are completely determined by the inverse of a matrix ${\bf U_{\text{SC}}}$ that is proportional to the adjacency matrix with self loops for the graph describing the surface code state. The matrix ${\bf U_{\text{SC}}}$ is banded block-diagonal, positive and has bounded spectrum, which means that the position-position correlations decay exponentially. We also derived an upper bound on the correlations. 

A prominent feature of our model, made evident in the numerics presented in Fig.~(\ref{Stopo}), section \ref{prtoder}, is that the topological order can be verified via quadrature measurements on the state. Moreover, errors in state preparation modeled as thermal inputs do not affect the size of the topological order as adequately quantified by the topological mutual information, see Fig.~(\ref{StopoTMI}). If an experiment were to follow our protocol, even for highly mixed states, the output surface code would have topological mutual information only slightly lower than the pure state case. This means the order in the state is well protected against local errors during preparation, as it should be. 

Continuous-variable surface codes have not received proper attention in the literature, and when they did, it was done under the unphysical case of infinite squeezing. The results we derived here open up new possibilities for topologically protected operations with continuous degrees of freedom. Given the tremendous advance in technologies to construct large-scale continuous-variable cluster-states, as described in \ref{expimpCV}, we expect that this will motivate further progress in this direction. 

\section{Entropy in polymer quantization}
Loop quantum gravity is a background-independent theory that attempts to find a solution to the problem of the quantization of gravity. The basic feature of the theory is quantum geometry, and the fundamental spatial discreteness is one of its predictions. 

Polymer quantization provides a simple mathematical setup in a purely quantum mechanical context to investigate different aspects of loop quantum gravity, such as the emergence of semiclassical states of quantum gravity. Similarly to loop quantum gravity, in the polymer quantization of a usual non-relativistic system there is a well-specified position operator with exact eigenstates, but no momentum operator -- only an effective analog that acts as a translation operator on the underlying lattice.
In this sense polymer quantization is different from the canonical Schr\"odinger quantization of the Weyl-Heisenberg algebra.

Motivated by the absence of a conceptual analysis of entropy across different quantizations, in Chapter \ref{polymerico} we studied the relation between the value of the quantum entropy in polymer and Schr\"odinger quantization schemes. While the two theories are physically equivalent in terms of physical observables, whose closeness of the values is ensured by Fell's theorem, we were interested in investigating eventual discrepancy between the values of the entropy. In \ref{entropypolymerico} we proved that for mixture of pure states, the entropy of the state in the Schr\"odinger quantization is a general bound for the polymer analog. 

Moreover, exploiting the continuity properties of the entropy, we showed the convergence of the value of the polymer entropy to the canonical Schr\"odinger one in the limit of lattice size going to zero. 

To conclude this analysis, in \ref{conventropy} we examined the case of two coupled harmonic oscillators quantized in the polymer sense. We derived the covariance matrix of the ground state and showed the form of the corrections induced by the discreteness of the space. These corrections are propagated to the value of the entanglement entropy of any of the two oscillators, reducing it by a series of terms proportional to the lattice size: This confirmed our theoretical predictions. 

\section{Open Problems:}
\begin{itemize}
\item An important extension of our work on photonic qubit probes of space-time curvature is the analysis of gravitational effects on massive spin $1/2$ particles. This is a particular cumbersome task, since in this case we have to take care of the coupling between spin and momentum explicitly, without having the advantage of classical analysis. While these effects have been studied under Lorentz transformations, an extension to account for general relativistic effects is wanting. 
\item We still do not have a complete mathematical characterization of the continuous-variable surface code state in terms of the quantum double of a continuous group, for example the $U(1)$ group. It would be fascinating to establish a connection between CV topological phases and gauge theories, in analogy to generalized Kitaev's surface codes and discrete gauge theories. 
\item An interesting open question is whether it is possible to perform local error correction following errors in the preparation of the state, analogous to what is done for toric codes, for example by local measurements of the hermitian quadratic form of the nullifiers.
\item One could study whether it is possible to utilize the toroidal version of this model to store quantum information, in analogy with the qudit-version of the toric code, and additionally investigating how the string-net operators introduced in \cite{demarie2013b} would affect such information. 
\item Further investigations involve the existence of a meaningful correspondence between the entanglement spectrum on a cut of the CV surface code and a thermal state of a boundary Hamiltonian, as shown for the discrete case in \cite{li2008}.
\item A future expansion of the work related to the study of entropy in polymer quantization involves the precise estimation of the discrepancy in entropies for a fixed lattice scale and systems composed of a large number of particles. In fact, while the polymer correction for a single oscillator is too small to be observed in an experiment, it might be that for a large number of particles the discrepancy grows and becomes significantly large. 
\end{itemize}

\appendix

 \begin{savequote}[8cm] 
\sffamily 
``If you wish to make an apple pie from scratch, you must first invent the universe.''  
\qauthor{Carl Sagan - Cosmos} 
\end{savequote}

\chapter{Elements of Differential Geometry}
\label{appDiffGeom}

In this Appendix we review the basic concepts of differential geometry that are used in the main text of this Thesis, in particular throughout Chapter \ref{GRchapter}. As a comment for the students or the interested readers, this short set of notes is mainly based on the book by Baez and Muniain~\cite{baez1994}, in our opinion one of the clearest books on the subject. 

\section{Manifolds}
Physics does not happen on white sheets of paper, it has to happen somewhere \emph{physical}. The conceptual abstraction where the modern field theories of physics, and in particular General relativity, come to life is a space-time. The mathematical description of our space-time is given by a pair ($\ma$,$g$), with $\ma$ a smooth 4-dimensional manifold and $g$ a metric. Physically, each point of $\ma$ is an \emph{event} and the metric is used to calculate the distance (in a manifold sense) between events. Physical processes are intended as continuous lines connecting events over a manifold. It is always possible to assign a set of coordinates to the events, and indeed necessary when dealing with applied physics. But different observers may assign different reference frames to the same physical events. How can they communicate their findings if they are speaking seemingly different languages? This is where differential geometry comes to help. 

Differential geometry contains the set of rules necessary to expand Euclidean geometry to any general space-time (or manifold). The underlying physical idea is that the laws of physics can be rewritten in a form independent of any choice of coordinates or reference frame or, more specifically, the laws of physics are covariant. 

Therefore, the mathematical stage where physics takes place is a manifold. Without worrying too much about the rather technical mathematical properties, let us define a $n$-dimensional smooth manifold as a (topological) space that can be covered by open sets $O_\alpha$ that locally look like the Euclidean space $\mathbb{R}^n$. Furthermore, there must exist continuous functions $\varphi_\alpha$ such that $\varphi_\alpha : O_\alpha \to \mathbb{R}^n$ and each transition function $\varphi_{\alpha,\beta}=\varphi_\alpha \circ \varphi_\beta^{-1}:\mathbb{R}^n \to \mathbb{R}^n$, defined over the overlap of two such sets $O_{\alpha, \beta} = O_\alpha \cap O_\beta$, is smooth, or infinitely differentiable. Each pair $(O_\alpha,\, \varphi_\alpha)$ is called a \emph{chart}. The collection of these charts is an \emph{atlas} and Einstein's principle of \emph{general covariance} states that all observers must observe the same physics independently of their choice of atlas. 

\section{Calculus on manifolds}
Since we are not dealing with physics defined on the usual $\mathbb{R}^n$ spaces anymore, we need to extend the calculus elements to manifolds. For the sake of this Thesis, there is no need to define each and every concept of differential geometry, but we will clarify the meaning of the entities used in Chapter \ref{GRchapter} to describe the gauge-invariance of the polarization in closed paths. These entities are \emph{vector fields}, \emph{differential forms}, \emph{sections of fiber-bundles} and the functions of these sections, familiarly known as \emph{connections}.

\subsection{Vector fields}
To define vector fields we need to introduce the concept of real scalar function on a smooth manifold. This is a function that takes a point in $\ma$ and assigns to it a real number, namely
\beq
f : \ma \to \mathbb{R} \,.
\eeq
Locally on a chart $(O_\alpha, \varphi_\alpha)$, $f$ looks like
\beq
f_\alpha :=  f \circ \varphi_\alpha^{-1}\, \qquad \text{such that} \,\qquad f_\alpha : \mathbb{R}^n \to \mathbb{R}\,.
\eeq
We can only define the mathematical properties of a function for its local description; For $N$ patches covering the manifold, there will be $N$ local description of $f_{\alpha = 1,...,N}$. If each of these functions is (locally) smooth, i.e. $f_\alpha \in C^\infty(O_\alpha),\, \forall \, \alpha$, and one can glue them together using the charts transition function for each overlapping $O_\alpha$, $O_\beta$
\beq
f_\beta = f_\alpha \circ \psi_{\alpha, \beta}\,,
\eeq
then we say that $f$ itself is infinitely-differentiable and $f \in C^\infty(\ma)$.

\emph{Vector fields} are the generalization of the derivative for manifolds. Specifically, a vector field $v$ on $\ma$ is a linear function acting on $f$
\beq
V(\ma) \ni v : C^\infty (\ma) \to C^\infty (\ma)\,, 
\eeq
which also satisfies the Leibniz rule $\forall \, f,g \in C^\infty(\ma)$:
\beq
v(fg) = v(f)g + f v(g) \,.
\eeq
We name the space of all vector fields $V(\ma)$. Things can be easily understood when the manifold is $\mathbb{R}^n$. Then we can use the usual cartesian coordinates $ x^\mu = \{ x^0, x^1, ... x^n \}$ to label events and for a function $f : \mathbb{R}^n \to \mathbb{R}$, a vector field looks like
\beq
v f = v^\mu \partial_\mu f = v^1 \partial_1 f + v^2 \partial_2 f +... v^n \partial_n f\,,
\eeq
where $v^\mu \in C^\infty(\mathbb{R}^n)$ are the \emph{components} of the field and the elements $\partial_\mu = \frac{\partial}{\partial x^\mu}$ form a basis for every vector field on $\mathbb{R}^n$.

\subsubsection{Tangent vector fields}
Take a point $p \in \ma$, for $\ma$ a $n$-dimensional manifold. Then, intuitively, a small neighborhood of the point can be considered locally equivalent to $\mathbb{R}^n$, just like a soccer field on the Earth is practically equivalent to a flat 2-dimensional plane. 

A \emph{curve} on $\ma$ is a continuous smooth map from some interval in $\mathbb{R}$ to $\ma$
\beq
\gamma(t) : t \in [0,a] \subset \mathbb{R} \to \ma \,.
\eeq
Fixing $\gamma(0) = p$, one can see that each vector in $\mathbb{R}^n$, the local representation of the neighborhood of $p$, determines the directions a curve can depart from $p$. Then a tangent vector in $p$ is a map that takes a function and gives a real number
\beq
v_p : C^\infty(\ma) \to \mathbb{R}\, ,
\eeq 
or, given some curve $\gamma$, where $p = \gamma(t) \in \ma$ then
\beq
v_p f = v (f)(\gamma(t))\,.
\eeq
We therefore define \emph{tangent space} $T_p \ma$ at $p$ as the vector space of all the tangent vectors $v_p$. By construction the space $T_p \ma$ is isomorphic to $\mathbb{R}^n$. 

Now consider a chart $O_{x}$ such that the point $p$ has local coordinates $x^\mu$. Let us explain the difference between having the coordinates index $\mu$ on the top $^\mu$ or the bottom $_\mu$ of a symbol. Assume the vector field basis is $\{ \partial_\mu \}$, then every $v_p \in T_p \mathbb{R}^n$ can be written as
\beq
v_p = v^\mu (\partial_\mu)_p \,,
\eeq
where we call $v^\mu$ the vector components in $O_x$ of the tangent vector $v_p$. If we describe the same point from another chart $O_y$, with $p =y^\nu$, then the same vector is written as
\beq
v_p = v'^\nu \partial'_\nu \,, 
\eeq
with $\partial'_\nu = \partial / \partial y^\nu$, and the two basis must be related by some linear transformation $T_\mu^\nu$
\beq
\partial_\mu = T_\mu^\nu \partial'_\nu \,.
\eeq
From this, it immediately follows that there exists a transformation rule for the two sets of components, explicitly given by
\beq
\label{convvec}
v'^\nu = v^\mu \frac{\partial y^\nu}{\partial x^\mu} \,. 
\eeq
Any vector $( x^1, x^2, ... x^n )$ that transforms in such a way is a \emph{contravariant vector}. Hence, the components of a tangent vector are said to be contravariant and they deserve an index up $^\mu \to v^\mu$.

\subsection{Differential forms}
One of the goals of differential geometry is to extend the concepts of differential calculus as gradient, cross product, and such to the coordinate-free manifolds world. The technical entities that allow us to do so are known as \emph{differential forms}, together with their calculus rules. A differential form is some mathematical beast that belongs to the \emph{Algebra} $\Omega(\ma)$, a vector space with more special properties, defined over $C^\infty(\ma)$ and generated by the elements of the sub-algebra $\Omega^1(\ma)$. We are aware this may sound very confusing, so let us try to explain better this concepts in a concise but precise way. Please keep in mind that our intention here is not generalising differential calculus, but rather defining what differential forms are, in order to clarify our findings in the main body of this Thesis. 

\subsubsection{Cotangent vector fields: The 1-forms.}
Take $\mathbb{R}^3$ as our reference manifold with global chart coordinates $\{ x^\mu \}$: then $f,g$.. are functions $\in \RR^3$ and $v,w..$ generic elements of $V(\RR^3)$, the vector space in $\RR^3$ with basis $\{ \partial_\mu \}$. We then introduce a new object $\d f$, call it 1-form and define it as a map
\beq
\d f : V(\RR^3) \to C^\infty(\RR^3)\, .
\eeq
So a 1-form is something that takes a vector field $v$ and produces a function instead, i.e. $\d f(v) = v f \in C^\infty(\RR^3)$. It is created by acting with the \emph{exterior derivative} $\d$ on any function $f$, where $\d : C^\infty(\RR^3) \to \Omega^1(\RR^3)$, and $\Omega^1(\RR^3)$ is the algebra of 1-forms on the manifold $\RR^3$. It easily follows that in this special case, 
\beq
\d f = \partial_\mu f \d x^\mu = \partial_x f \d x + \partial_y f \d y + \partial_z f \d z \,, 
\eeq
and $\{ \d x, \d y, \d z \}$ is a basis for $\Omega^1(\RR^3)$. For $\RR^n$ the basis generalizes to $\{ \d x^\mu \}$ and
\beq
\d x^\mu (\partial_\nu) = \delta_\nu^\mu \,.
\eeq
Thus, in the comfort of $\RR^n$, $\d f$ reduces to the well-known total \emph{differential} of a function. Generalizing now to a manifold $\ma$, we say that $\omega$ is a \emph{1-form} (or a covector field) satisfying
\beq
V^*(\ma) \ni \omega : V(\ma) \to C^\infty(\ma) \,,
\eeq
with $V^*(\ma)$ space of all 1-forms on $\ma$. Note that on a coordinate chart, a 1-form can be expressed as a linear combination of the coordinate differentials
\beq
\label{covvec}
\omega = \omega_\mu \d x^\mu \,.
\eeq
Consider again the usual point $p \in \ma$; It is always possible to construct a vector space $T_p^*\ma$ of \emph{cotangent vectors} $\omega_p$ such that 
\beq
T_p^*\ma \ni \omega_p :   T_p \ma \to \RR \,.
\eeq 
The equations above explain the concept of \emph{duality}. We say that on $\RR^n$ the basis $\{ \d x^\mu \}$ is a 1-form basis dual to $\{ \partial_\mu \}$. On the qubit Hilbert space $\mathcal{H}_2$, the kets $\{ \ket{0}, \ket{1} \}$ span the space $T_p(\mathcal{H}_2)$ and the bras $\{ \bra{0}, \bra{1} \}$ span $T^*_p(\mathcal{H}_2)$, where $p$ is the centre of the Bloch's sphere. 

From Eq.(\ref{covvec}) we see that the components of a 1-form carry a lower index. Why is that? By considering again two charts describing $p$ with coordinates $x^\mu$ and $y^\mu$, one can prove that the transformation rule for the $\omega_\mu$ is given by
\beq
\omega'_\nu = \omega_\mu \frac{\partial x^\mu}{\partial y^\nu}\, .
\eeq
For this reason, the components $\omega^\mu$ of $\omega$ form a \emph{covariant vector}, and their transformation rule is the inverse of the one for contravariant vectors Eq.(\ref{convvec}).

\subsubsection{2-forms, 3-forms.. \emph{n}-forms}
For a manifold $\ma$ with 1-form space $\Omega^1(\ma)$ and dimension $n$, it is possible to build 2-forms, 3-forms and so on up to n-forms starting from the 1-forms. Defining the $\wedge$ symbol, which combines two 1-forms and satisfies 
\beq
\label{extal}
\zeta \wedge \omega = - \omega \wedge \zeta\,, 
\eeq
for a local chart $\{ \partial_\mu \}$ and $\{ dx^\mu \}$, a n-form space $\Omega^n(\ma)$ is the \emph{exterior algebra}\footnote{An exterior algebra is an algebra equipped with the anti-commutative property from Eq.(\ref{extal}).} of n-forms 
\beq
\omega = \omega_{\mu_1, ... \mu_n} \d x^{\mu_1} \wedge \d x^{\mu_2} \wedge ... \wedge \d x^{\mu_n} \,,
\eeq
with $\omega_{\mu_1, ... \mu_n} \in C^\infty(\ma)$. It follows that the total space of the \emph{differential forms} is the algebra over $C^\infty(\ma)$
\beq
\Omega (\ma) = \oplus_n \Omega^n (\ma) \,.
\eeq

\section{The metric}
\label{appMetric}
The concept of distance between two points of a manifold, or rather two events in a space-time, is summarized by the concept of a \emph{metric}. For a vector space $V$, a metric $g$ is a map that takes two vectors in $V$ and produces a real number
\beq
g: V \times V \to \RR\, .
\eeq
For a manifold, a metric determines at every point $p \in \ma$ a local metric $g_p$ that defines an inner product for the tangent space $T_p\ma$. The easiest example for a metric is the Minkowski metric\footnote{We do not discuss here the problem of the signature of a metric, but simply adopt the usual convention $(-,+,+,+)$.} $\eta$, globally defined for a Minkowski space-time:
\beq
\eta( v, w) = -v^0 w^0 + v^1 w^1 + v^2 w^2 + v^3 w^3 \,.
\eeq
The quantity above defines an \emph{interval} in such a space-time. More generally, an interval is locally defined in terms of the differential of the coordinates as
\beq
\d s^2 = g_{\mu, \nu} \d x^\mu \d x^\nu \, ,
\eeq 
and the elements $g_{\mu, \nu}$ are the components of the metric $g_p$. For our purposes, given a four-dimensional manifold $\ma$, a generic interval, and therefore its metric $g_p$, can be written as
\beq
\d s^2 = - h (\d x^0 - g_m \d x^m)^2 + \d l^2
\eeq
where $h = -g_{0,0}$ is a scalar, $\bf{g}$ is a  three-dimensional  vector with components $g_m = -g_{0,m}/g_{0,0}$ and $l_{m,n} \to \d l^2 = l_{m,n} \d x^m \d x^n$ determines the metric of the three-dimensional static space $\Sigma_3$ that can be assigned to each point of $\ma$ and described by the Landau-Lifshitz $1+3$ formalism \cite{landau1971}.

For the definition-lovers, we call a manifold \emph{Riemannian} \cite{frankel1999} iff $g(v,w) \ge 0$ at each point of the manifold. In the most general case, the manifold is pseudo-Riemannian, which simply means that it comes equipped with a non-degenerate metric (i.e. $g(v,w) = 0$ for all $w$ only if $v = 0$). The space-times used in relativity (and the Minkowski space-time as a special case) are Lorentzian manifolds, a subset of the pseudo-Riemannian that exhibits a special form of the metric signature \cite{baez1994}. 

\section{Fiber bundles}
\label{bundlesapp}
Modern field theories are expressed in terms of \emph{sections} and \emph{connections} on \emph{fiber bundles}. These concepts are also central in general relativity and for that reason we are going to deal with them now. The main idea is the following.

Consider a manifold $\bma$, a base manifold of dimensions $b$ corresponding to a physical space-time, and a second manifold $\ta$, the total manifold with dimensions $b+t$. Then take a continuous and non-invertible map
\beq
\pi : \ta \longrightarrow \bma \,,
\eeq
which projects the total manifold to the base manifold. We then associate a sub-manifold $\pi^{-1}(p) \subset \ta$ of dimension $t$ to each point $p \in \bma$, composed of all those points that are projected into $p$ by the map $\pi$. We name this sub-manifold the \emph{fibre over $p$},
\beq
F_p := \{ x \in \ta: \pi(x) = p\} \,.
\eeq
For example, we can think of the tangent space $T_p \bma$ as a fiber over $p \in \bma$ and thus the total manifold will be the \emph{tangent bundle} of all the tangent spaces of each point in our space-time. It is important to understand the point here: Locally, for a chart $O_\alpha$ of $\bma$, the sub-manifold $\pi^{-1}(O_\alpha)$ is homeomorphic\footnote{An homeomorphism is the equivalent of isomorphism for manifolds.} to the manifold produced by the direct product of the chart itself and the \emph{standard fibre} $F$ with $\text{dim} F = t$,
\beq
\pi^{-1}(O_\alpha) \sim O_\alpha \times F \,.
\eeq
If this condition is always satisfied, then we say that the bundle $\pi$ is locally trivial. 

We can make things easier, assuming that the standard fiber $F$ is a n-dimensional vector space. Then we call the bundle $\pi : \ta \to \bma$, with $\pi^{-1}(O_\alpha) \sim O_\alpha \times \RR^n$, a n-dimensional \emph{real vector bundle}. 

\subsubsection{Bundle sections}
Given a bundle $\pi$, a \emph{section} is a map $s$ that associates to each point $p$ of the base manifold a point $s(p)$ in the fiber above $p$, i.e.
\beq
s : \bma \to \ta \qquad \text{such that} \qquad \forall p \in \bma,\, s(p) \in F_p \subset \ta \,.
\eeq
Intuitively, a section is the projection of a path from the base manifold to the fibers above the points of the base path. The set of all sections is denoted by $\varGamma(\ta,\bma)$ and locally we can always define a basis $\{ e_1, ... e_b \}$ such that each section on a patch is given by $ s = c^i e_i $ with $c^i \in C^\infty(\bma)$.

In modern physics, sections of certain special bundles represent fields and are used to describe the different interactions in Nature. Such bundles are vector bundles called \emph{G-bundles}, where $G$ is a Lie group \cite{tung1985}. For a $G$-bundle, the standard fiber is a vector space $F$ where the group $G$ has a representation $\mathfrak{g}$. This allows to study the symmetrical properties of the sections of a $G$-bundle, which are determined by \emph{gauge transformations} of the sections. With some mathematical freedom, we define a gauge transformation as a linear map $T$ that acts over the set $\varGamma(\ta,\bma)$ with $\ta$ some $G$-bundle
\beq
T: \varGamma(\ta,\bma) \to \varGamma(\ta,\bma) \,,
\eeq
and $T(p)$ element of the gauge group $G$ for any $p \in O_\alpha\,, \forall \alpha$. In the main text, we apply these concepts showing the gauge invariance of the photon polarization under rotation of the polarization axis in closed curves. 

\subsubsection{Connection and curvature}
\label{conncurvapp}
The problem that arises almost immediately after introducing the sections, is how to describe the variation of a section, since it crosses different vector fields at different points. It should be clear from the previous discussion that a section is nothing else but a vector field defined on the base manifold! Therefore its variation corresponds to calculating the derivative of a vector field at two different points, each described by two different patches and set of coordinates. How can we \emph{connect} these descriptions together? Through the introduction of the \emph{covariant derivative} of a vector field. 

Consider a point $p$ over a manifold $\ma$, whose neighborhood is covered by the chart $O_\alpha$ with local coordinate frame $\{x^\mu \}$. Then, the covariant derivative in the direction $k$ of a vector field $v$, locally defined around $p \in \ma$, whit $k$ tangent vector in $p$, is
\beq
\label{covdef}
(\nabla_k v)^{\mu} = \Big(\frac{\partial v^\mu}{\partial x^\nu} + \beta^\mu_{\nu, \sigma} v^\sigma\Big) k^\nu \,.
\eeq 
The $\beta^\mu_{\nu, \sigma}$ are called \emph{Ricci coefficients} or the \emph{coefficients of the affine connection}. For a local vector field basis $\{ e_i \}$, these coefficients are defined as
\beq
\nabla_{e_\mu} e_\nu = e_{\sigma} \beta^\sigma_{\mu, \nu} \,. 
\eeq
The theory of general relativity is \emph{torsion free}. This means that a specific quantity, the \emph{torsion tensor} $T^\mu_{\nu, \sigma}$, is equal to zero
\beq
T^\mu_{\nu, \sigma} = \beta^\mu_{\nu,\sigma} - \beta^\mu_{\sigma, \nu} = 0\,,
\eeq 
which implies the symmetry of the connection's coefficients:
\beq
\beta^\mu_{\nu,\sigma} = \beta^\mu_{\sigma, \nu} \,.
\eeq
Note that it is always possible, given a chart with a local vector field basis $\{ e_i \}$ and a dual 1-form basis $\{ \sigma^i \}$, to build a local matrix of connection 1-forms as
\beq
\beta^\mu_{\,\,\,\,\nu} := \beta_{\gamma, \nu}^\mu \sigma^\gamma \,.
\eeq

Using the covariant derivative one can define the meaning of a \emph{geodesic} and \emph{parallel transport}. Take a curve $\gamma(\lambda)$ on a manifold $\ma$, where $\lambda$ is an affine parameter that describes the curve, then a vector $v(\gamma(\lambda))$ is parallel transported along $\gamma$ iff
\beq
\nabla_{v(\gamma(\lambda))} v(\gamma(\lambda)) = 0\,.
\eeq
When this condition is satisfied, the curve $\gamma(\lambda)$ is a geodesic. 

A curved space-time is characterized by a quantity that captures the difference from a flat space-time. This is called, unsurprisingly, \emph{intrinsic curvature} and it is described by a local matrix of curvature 2-forms
\beq
\label{curva}
\theta^\mu_{\,\,\,\, \nu} := \d \beta^\mu_{\,\,\,\,\nu} + \beta^\mu_{\,\,\,\,\gamma} \wedge \beta^\gamma_{\,\,\,\,\nu} \,.
\eeq
While we are not interested in explicitly showing the mathematical derivation of the formula above, the analogy between Eq.(\ref{curva}) and the notion of intrinsic curvature is given by the coefficients expansion of $\theta$ in terms of some local 2-forms basis $\{ \sigma^\mu \wedge \sigma^\nu \}$, i.e.
\beq
\theta^\mu_{\,\,\,\, \nu} = \frac{1}{2} R^\mu_{\nu, \gamma, \eta} \sigma^\gamma \wedge \sigma^\eta \,.
\eeq
When the local vector field basis is $\{ \partial_\mu \}$, then the coefficient $R^\mu_{\nu, \gamma, \eta}$ is the Riemann curvature tensor, a quantity that only depends on the metric of the manifold and, in this sense, a fully intrinsic property of the space-time. 

 \begin{savequote}[10cm] 
\sffamily 
``And I will show you something different from either\\
Your shadow at morning striding behind you\\
Or your shadow at evening rising to meet you\\
I will show you fear in a handful of dust.''
\qauthor{Thomas S. Eliot - The Waste Land} 
\end{savequote}

\chapter{Nullifiers transformations}
\label{appNullifiers} 

Here we present details of the derivation of the nullifier sets for the continuous-variable cluster state and surface code, both in the ideal and physical case. 

\section{Ideal continuous-variable cluster state}
\label{iCVcs}
In the ideal case of $N$ free quantum harmonic oscillators the initial global state is
\beq
\ket{\psi} = \lim_{s \to \infty} \prod_j^N \hat{S}_j(s) \ket{0}_{j} = \ket{0}^{\otimes N}_{p} \,,
\eeq
where $\ket{0}$ is the ground state of the quantum harmonic oscillator, or the vacuum state, and $\ket{0}_{p_j}$ is the zero-eigenstate of the operator $\hp_j$. This state is nullified by the collection of operators $\{ \hat{p}_j \}$. To study the evolution of the nullifiers after the application of the controlled-$Z$ gates
\beq
\cs = \prod_{\6 j,k \9}^{N} \cz \ket{0}_p^{\otimes N}\,,
\eeq
we really only need to calculate the transformation:
\beq
\cz \hp_j \cz^\dagger \,.
\eeq
This is easily done by defining
\beq
\cz(g) = e^{i g \hq_j \hq_k}  \longrightarrow \hp_j(g) = \cz(g) \hp_j(0) \cz(g)^\dagger\,.
\eeq
Hence, the derivation over the evolution parameter is:
\begin{align}
\frac{\d \hp_j(g)}{\d g} &= i \hq_j \hq_k \cz(g) \hp_j(0) \cz(g)^\dagger - i \cz(g) \hp_j(0) \hq_j \hq_k \cz(g)^\dagger \nonumber \\
&= i \hq_k \cz(g) [ \hq_j, \hp_j ] \cz(g)^\dagger  \\
&= - \hq_2 \,. \nonumber
\end{align}
Since $\hp_j(0) = \hp_j$ and $\hp_j(g) = \hp_j(0) - g \hq_j$, for $g=1$ one finds that
\beq
\cz \hp_j \cz^\dagger = \hp_j - \hq_j \,.
\eeq
Additional controlled-$Z$ gates applied between the mode $j$ and its neighboring sites repetitively add the corresponding $-\hq_x$ factor since
\beq
\cz \hq_j \cz^\dagger = \hq_j \,,
\eeq
for the obvious commutativity of the operators. Therefore, on a square lattice the new set of nullifiers for the $\cs$ state is equal to $\{ \eta_j \}$ with
\beq
\label{etabeta}
\het_j = \hp_j - \sum_{k \in \mathcal{N}(j)} \hq_k \,.
\eeq

\subsection{Spectrum of the $\hH_{\text{CS}}$ Hamiltonian}
\label{specsHam}
To prove that the Hamiltonian whose ground state is the ideal CV cluster state
\beq
\hH_{\text{CS}} = \sum_{j = 1}^N \hat{\eta}^2_j \, 
\eeq
is gapless, we consider the case where all but one qumode states are initialized to $\ket{0}_{p_j}$. Then the initial global state is equal to
\beq
\ket{\psi_N} = \ket{0}_p^{\otimes (N-1)} \ket{s}_{p_N} \,,
\eeq
where $\ket{s}_{p_N} = \hat{S}(s) \ket{0}$ with $s \ll 1$ and the label $N$ is chosen for simplicity. After the application of the controlled-$Z$ gates, the resulting state
\beq
\cs_N = \prod_{\6 j,k \9} \cz \ket{\psi_N}  \,,
\eeq
fails to satisfy the nullifier condition imposed by the element $\hat{\eta}_N$ and corresponds to an excited state of $\hH_{\text{CS}}$. The parameter $s$ determines the distance of $\cs_N$ from the ground state $\cs$. Then, using the relation
\beq
\hat{\eta}_N = \prod_{\6 j,k \9} \cz \hp_N (\prod_{\6 j,k \9} \cz)^\dagger \,,
\eeq
one can calculate  
\beq
\begin{array}{ll}
\hat{\eta}_N \cs_N &= \hat{\eta}_N \prod_{\6 j,k \9} \cz \ket{\psi_N} \\
&=\prod_{\6 j,k \9} \cz \hp_N (\prod_{\6 j,k \9} \cz)^\dagger \prod_{\6 j,k \9} \cz \ket{\psi_N} \\
&= \prod_{\6 j,k \9} \cz  \hp_N \ket{\psi_N}\\
&= s \prod_{\6 j,k \9} \cz \ket{\psi_N} \\
&= s \cs_N \,.
\end{array}
\eeq
Hence, it immediately follows that $\cs_N$ is indeed an eigenstate of $\hH_{\text{CS}}$ with eigenvalue
\beq
\hH_{\text{CS}} \cs_N = s \cs_N \,.
\eeq
and this confirms that the Hamiltonian $\hH_{\text{CS}}$ is critical in the limit of infinite squeezing.

\section{Physical continuous-variable cluster state}
For the finitely squeezed cluster state the approach is similar. The only difference lies in the initial state, which is now given by a collection of momentum-squeezed states,
\beq
\ket{\psi}_s = \prod_j^N \hat{S}_j(s) \ket{0}_j \,.
\eeq
Since the operators $\hp_j$ do not nullify this state anymore, one has to start from the very initial nullifier set of (dimensionless) annihilation operators $\{ \hat{a}_j \} \to \hat{a}_j \ket{0}_j = 0$. Then, following the same reasoning as before, the nullifier set $\{ \hat{\eta}_j^s \}$ for the cluster state derives from 
\begin{align}
\cz \hat{S}_j(s) \hat{a}_j \hat{S}_j(s)^\dagger \cz^\dagger &=  \cz \hat{S}_j(s) \frac{1}{\sqrt{2}} (\hq_j + i \hp_j) \hat{S}_j(s)^\dagger \cz^\dagger \nonumber \\
&= \frac{1}{\sqrt{2}} \cz \left( \frac{\hq_j}{s} + i s \hp_j \right) \cz^\dagger \\
&= \frac{1}{\sqrt{2}} \left( \frac{\hq_j}{s} + i s (\hp_j - \hq_j) \right) \nonumber \,,
\end{align}
and a generic element of the set is
\beq
\hat{\eta}^s_j = \frac{s}{\sqrt{2}}\left [ \frac{\hq_j}{s^2} + i \left( \hp_j - \sum_{k \in \mathcal{N}(j)} \hq_k \right) \right ] \,.
\eeq

\section{Ideal continuous-variable surface code}
The derivation of the nullifier set for the CV surface code requires more attention and implies knowledge of the nullifiers transformation rules under quadrature measurements, which we are going to review briefly here. 

The explicit form of the surface code nullifiers for a square lattice (while allowing for smooth or open boundaries) are obtained by taking linear combination of neighboring nullifiers of the CV cluster state. 
A quadrature measurement on a qumode removes it from the cluster \cite{gu2009}. Given a set of exact nullifiers for a Gaussian state \cite{menicucci2011}, one can obtain new nullifiers by a three-step process \cite{demarie2013b}: 
\begin{itemize}
\item[1.] Given a quadrature measurement $\hat{x}_j$ to be made on mode $j$, where $\hat{x}_j \in \{\hq, \hp\}$, using linear combinations of the original nullifiers, write a new set of nullifiers (remember, they need to commute) such that the canonically conjugate local quadrature $\hat{y}_j$ (where $[\hat{x}_j, \hat{y}_j] = \pm i)$ appears only in one nullifier in the new set.
\item[2.] In each new nullifier, replace $\hat{x}_j$ with the real-valued measurement outcome.
\item[3.] Eliminate the nullifier that contains $\hat{y}_j$.
\end{itemize}
We always assume that the outcome of the measurement is zero because any other outcome would merely result in the same state up to displacements in phase space. These displacements can always be undone by local unitaries and therefore do not change any entanglement measure we might want to calculate \cite{demarie2012}, as explained in \ref{bsgsr}. 

As an example consider the simplest case of the square $3 \times 3$ ideal CV cluster state, with the vertices labeled from 1 to 9, row by row. The initial nullifier set is $\{ \hat{\eta}_j \}$ for $j = 1,..,9$ from Eq.~(\ref{etabeta}). We follow the measurement scheme \cite{zhang2008} illustrated in Fig~(\ref{scheme}) and start performing the measurement $\hp_1 = \ket{p}_1 \bra{p}$. 

Step one of the process described above makes us note that both $\het_2$ and $\het_4$ contain the canonical conjugate $\hq_1$. Thus, we substitute $\het_2$ with the linear combination $\het_2 - \het_4 = (\hp_2 - \hq_3 - \hp_4 + \hq_7)$ and leave $\het_4$ as it is. We then jump to step two and substitute $\hp_1$ with its (zero) measurement outcome, \emph{de facto} replacing $ \het_1 \to \het_1 - \hp_1$. Finally, we remove $\het_4$ from the set, and obtain the nullifier set for the post-measurement state:
\beq
\{ \het_1 - \hp_1, \het_2 - \het_4, \het_3, \het_5, \het_6, \het_7, \het_8, \het_9  \} \,.
\eeq
This procedure, while tedious, can be efficiently repeated for the other measurements. After measuring $\hp_3$ one is left with the new set:
\beq
\{ \het_1 - \hp_1, \het_2 - \het_4 - \het_6, \het_3 -\hp_3, \het_5, \het_7, \het_8, \het_9 \} \,.
\eeq
The $\hq_5$ measurement simply disconnects the node from the cluster and we are left with:
\beq
\{ \het_1 - \hp_1, \het_2 - \het_4 - \het_6, \het_3 - \hp_3, \het_7, \het_8, \het_9 \}
\eeq
To save some ink, we collapse measurement $\hp_7$ and $\hp_9$ into one line and the final set of nullifiers is:
\beq
\{\het_1 - \hp_1, \het_3 - \hp_3, \het_7 - \hp_7, \het_9 - \hp_9, \het_2 + \het_8 - \het_4 - \het_6 \} \,.
\eeq
The form of the nullifiers in this case, expressed by the signs in the linear combinations, corresponds to the special case of a square lattice with counterclockwise face orientation and all edges pointing toward or away from the vertices, alternatively. For a square lattice with generic orientation, the construction just described tells us that the CV ideal cluster state nullifiers are:
\begin{align}
\hat{a}_v &= \sum_{e | v \in \partial e} \hq_e \nonumber \,,\\
\hat{b}_f &= \sum_{e \in f} o(e,f) \hp_e \,,
\end{align}
where $o(e,f) = \pm 1$ if the edge $e$ is oriented with $(+)$ or against $(-)$ the face $f$. A more general expression that holds for any kind of lattice is given in the main text in \ref{ticvscs} and it is based on the same derivation presented here. 

\section{The physical continuous-variable surface code}
For the physical CV surface code the procedure to follow is the same of the precedent case, but more elaborate. To obtain the face nullifiers $\hat{b}_f^s$, one sums the cluster state nullifiers immediately adjacent to the $\hq$-measured mode (along the cardinal directions) with orientation-dependent signs (e.g. $\het^s_N + \het^s_S - \het^s_E - \het^s_W$). The vertex nullifiers $\hat{a}_v^s$ are more complex and require next-nearest-neighbor nullifiers to be added to the sum of neighboring nullifiers around the $\hp$-measured mode in order to achieve step one in the procedure above. Then, the general form of the CV surface code nullifiers for a generic lattice with possibly incomplete vertices and faces is \cite{demarie2013b, linjordet2013}
\begin{align}
\label{avbf}
	\hat{a}^s_v &= \frac{s_v}{\sqrt{2V(v)(1+(s/s_v)^2)}}\nonumber \\
	&\qquad \times \Biggl[\sum_{e|v\in \partial e}
\left(\hat{q}_{e}+\frac{i}{{s_v}^2} \hat{p}_{e} \right) + \frac{s^2}{s_v^2}\sum_{\substack{v' | [v',v]\in \mathcal{R} \\ e |  v^\prime \in \partial e \wedge v\not\in \partial e}}\hat{q}_e \Biggr] \,, \nonumber \\
	\hat{b}^s_f &=\frac{s}{\sqrt{2|\partial f|}}\sum_{e\in \partial f}o(e,f) \left(\hat{p}_{e}-\frac{i}{{s}^2}\hat{q}_{e}\right),
\end{align}
with $s_v = \sqrt{V(v) s^2 + s^{-2}}$, $V(v)$ valence of vertex $v$ and $|\partial f|$ boundary size of the lattice face. It is important to realize that now there is a dependence upon the position on the lattice of the vertex or face that we consider. The form of the commutation relations for these nullifiers is not as straightforward as in the previous cases. They are determined by the distance between the vertices and the boundary size of the faces as \cite{demarie2013b}
\begin{align}
\label{comms}
	[\hat{a}^s_v,\hat{a}^{s\,\dagger}_{v'}] &= 
	\begin{cases}
		1 & \text{if $d(v,v')=0$,} \\
		\frac{(s_v^2+s_{v'}^2+s^2(V(v)+V(v')))/2s_vs_{v'}}{[V(v)V(v')(1+(s/s_{v})^2)(1+(s/s_{v'})^2)]^{1/2}} & \text{if $d(v,v')=1$,} \\
		\frac{2s^2/s_vs_{v'}}{[V(v)V(v')(1+(s/s_{v})^2)(1+(s/s_{v'})^2)]^{1/2}} & \text{if $d(v,v')=\sqrt{2}$,} \\
		\frac{s^2/s_vs_{v'}}{[V(v)V(v')(1+(s/s_{v})^2)(1+(s/s_{v'})^2)]^{1/2}} & \text{if $d(v,v')=2$,} \\
	0 & \text{if $d(v,v')>2$,}
	\end{cases}
	\nonumber \displaybreak[0] \\
	[\hat{b}^s_f,\hat{b}^{s\,\dagger}_{f'}] &=
	\begin{cases}
		1 & \text{if $f=f'$,} \\
		\frac{1}{\sqrt{|\partial{f}||\partial{f'}|}} & \text{if $[f,f'] \in \mathcal{E}$,} \\
		0 & \text{otherwise,}
	\end{cases}
	\nonumber \displaybreak[0]  \\
	[\hat{b}^s_f,\hat{b}^s_{f'}] &= [\hat{a}^s_v,\hat{a}^s_{v'}] = [\hat{a}^s_v,\hat{b}^s_{f}] = [\hat{a}^s_v,\hat{b}^{s\,\dagger}_{f}] = 0.
\end{align}
Now $d(v,v')$ is the Euclidean distance between the two vertices $v$ and $v'$, where the edge lengths of the graph are unit length. Then the most generic CV physical surface code Hamiltonian is given by
\beq
\label{HscPHYS}
\hH_{\text{SC}} (s) = \sum_{v\in \mathcal{V}} \frac{2 V(v) (1 + s^2/s^2_v)}{s_v^2} \hat{a}^{s\,\dagger}_v \ha^s_v + \sum_{f \in \mathcal{F}} \frac{2 |\partial f|}{s^2} \hat{b}_f^{s\,\dagger} \hat{b}^s_f\, ,
\eeq
and the squeezing dependence of the prefactors for the vertex and face parts ensures the Hamiltonian has finite energy in the infinitely squeezed limit:
\begin{equation}
\label{finenlimit}
\lim_{s\rightarrow \infty} \hH_{\text{SC}}(s) =\sum_{v \in \mathcal{V}} \Biggl(\sum_{e|v\in \partial e} 
\hat{q}_{e}\Biggr)^2+\sum_f \Biggl(\sum_{e\in \partial f} o(e,f)\hat{p}_{e}\Biggr)^2\, .
\end{equation}
Here we used the fact that for infinite squeezing, each vertex nullifier involves a sum of $\hq$'s around that vertex and its four neighboring vertices, and since they all commute, the parent Hamiltonian is simply the squared sum of $\hq$'s around each vertex.

\begin{savequote}[9cm] 
\sffamily 
``Why is it that when one man builds a wall, the next man immediately needs to know what's on the other side?''  
\qauthor{George R.R. Martin - A Game of Thrones} 
\end{savequote}

\chapter{Entropy bounds}
\label{boundsA}

In Chapter \ref{CVTO} we have shown how to calculate the topological entanglement entropy (TEE) $\gamma$ and the topological mutual information (TMI) $\gamma_{\text{MI}}$ for the finitely squeezed continuous-variable (CV) surface code. In this Appendix we first compute an upper bound to the TEE, calculating the subsystem entropy of a simpler network of entangled modes and then, analyzing the structure of the contributions to the TMI in the limit of high temperature noisy input states, we derive a lower bound to the TMI.

\section{Topological entanglement entropy upper bound}

To derive the TEE upper bound, we invoke the calculation of subsystem entropy appropriate to stabilizer states, which have vanishing two-point correlation functions. Consider stabilizer states that are quantum doubles of a finite group $\mathcal{G}$ (such as the toric code, the quantum double of $\mathcal{G}=\mathcal{Z}_2$) as explained in Chapter \ref{ChapTO}. As shown in \cite{hamma2005}, the TEE is calculated by dividing the system into two subsystems $A,\,B$ and identifying the redundant gauge transformations defined on the boundary between the two regions. Then the entanglement entropy of subsystem $A$ is the logarithm of the number of the (all equivalent) Schmidt coefficients of the state. Exploiting the group properties of $\mathcal{G}$ allows one to write the entropy as $S(A) = (|\partial A| - 1) \log_2| \mathcal{G} |$, implying the TEE is $\gamma = \log_2|\mathcal{G}| = \log_2 \mathcal{D}$.

For the CV surface codes (or for CV systems in general), it is complicated\footnote{For the interested researchers: Maybe impossible?} to extract an analogous exact expression for the entropy of a subsystem because the Schmidt coefficients are not all equal as in the discrete case. Furthermore, the TEE is infinite for infinitely squeezed (ideal) CV surface code states, and the definition of quantum dimensions is not so clear for finitely squeezed CV surface code states since we do not yet have a description of this model in terms of the quantum double of a group. Nevertheless we can go ahead and compute the subsystem entropy in the same way that would be done for the discrete case and treat this as a bound for the TEE of the CV surface code state. It is simply an upper bound because we are ignoring longer-range correlations that degrade the topological order, but since the correlations length is bounded for any finite amount of squeezing, this should be a reasonable tight bound. 

\begin{figure}[tbp]  
\centering
\setlength{\unitlength}{1cm}
\includegraphics[width=15.6\unitlength]{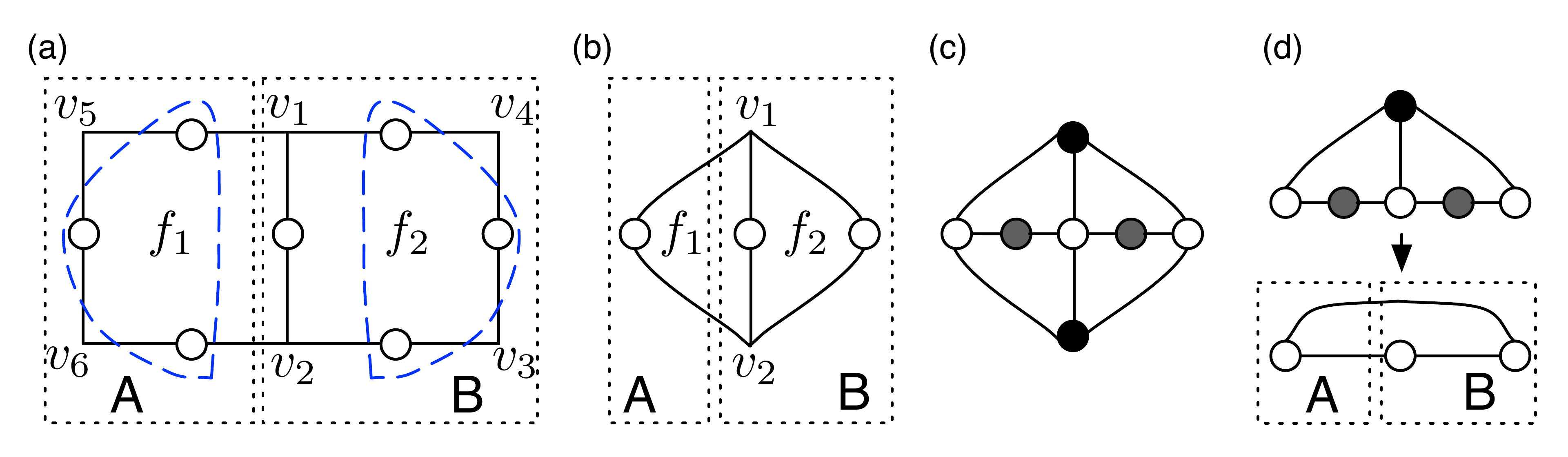}
\caption{Simplified network of modes with which to derive an upper bound on the TEE for the CV surface code:  (a)  This shows a surface code state for a quantum double model with discrete variables.  There are six vertices, two faces, and seven edges where physical modes reside. Not all the vertex stabilizers are independent. Rather, their product is the identity, so the number of independent stabilizers is $5+2$, equal to the number of physical modes.  We can deform the lattice while preserving the topological order by replacing three modes of each face, excluding the mode on the shared boundary, with one mode as shown in (b).  This network has two independent vertex stabilizers, and two independent face stabilizers which equals the number of physical modes.  For the toric code, the network represents a ground state that is a GHZ state.  (c)  A finitely squeezed cluster-state graph with seven modes maps to the three-mode network by measuring the grey modes in the $\hq$ basis and the black modes in the $\hp$ basis.  In fact, the nullifiers on the top and bottom black modes of the graph act equivalently on the white modes meaning one of them is redundant. Therefore, an even simpler CV cluster-state graph suffices as depicted in (d).  Upon measurements, the reduced network has three modes with a correlation matrix that can be computed exactly to yield an upper bound to the CV surface code TEE.}
\label{ubounds}
\end{figure}

A simple configuration to start with is a quantum double model with a discrete group on a lattice with two faces. This can be realized using a graph with just three edges (that correspond to physical modes) and two vertices, as shown in Figs.(\ref{ubounds}a,b). For the toric code, the ground state of such a configuration would be the GHZ \cite{greenberg2007,bouwmeester1999} state $(\ket{000}+\ket{111})/\sqrt{2}$ since both vertices implement the stabilizer $\hat{\sigma}^x_1 \hat{\sigma}^x_2 \hat{\sigma}^x_3$, and one face enforces $\hat{\sigma}^z_1 \hat{\sigma}_2^z$, while the other face enforces $\hat{\sigma}^z_2 \hat{\sigma}^z_3$.

Identifying two qubits on one of the faces with subsystem $B$, the subsystem entropy is $S(A) = S(B) = 2- \gamma = 1$, where 2 comes from the size of the boundary of region $B$ and therefore $\gamma = 1$. There is a simple way to obtain this simplified surface code network of three modes from a finitely squeezed CV cluster state with six modes, after measuring out three of the modes (Fig.(\ref{ubounds}c,d)). The resultant CV network has a correlation matrix that can be computed exactly, following the method explained in Chapter \ref{CVTO}. The symplectic spectrum for the single-mode subsystem $A$, complement of $B$, has two symplectic eigenvalues $\{ \pm \sigma_1 \}$ (\ref{sympaeigolo}) with
\beq
\sigma_1 = \frac{1}{2} \sqrt{\frac{1 + 3s^4 + 2s^8}{1 + 3s^4}} \, ,
\eeq
thus, using the formula in Eq.(\ref{vnS}), the entanglement entropy $S(A)$ is
\beq
S(A) = \Big[ \Big( \sigma_1 +\frac{1}{2} \Big) \log_2 \Big( \sigma_1 +\frac{1}{2} \Big) -  \Big( \sigma_1 - \frac{1}{2} \Big) \log_2 \Big( \sigma_1 - \frac{1}{2} \Big) \Big]
\eeq
and the upper bound for the TEE can be expressed as
\beq
\gamma \le S(A) \,.
\eeq

\section{Surface code topological mutual information from high-temperature cluster-states}
In \ref{noisynoise} we introduced a noise model for the CV surface code state, modeling a CV cluster state at thermal equilibrium with an environment at temperature $\beta^{-1}$ and preparing the CV surface code by quadrature measurements on the mixed CV cluster state.  Here we derive a lower bound for the topological mutual information (TMI) of this (noisy) CV surface code state, analyzing the limit $\kappa \to \infty$ (with $\kappa = \coth{\beta \epsilon_0 / 2}$, $\epsilon_0 = 2/s^2$ gap of the CV physical cluster state), which corresponds to the strongest possible decrease of the TMI from the TEE for the noise model considered. 

Recall that for a reduction $\rho_A$ of a pure state $\rho$, the von Neumann entropy $S(\rho_A)$ determines the entanglement entropy of the subsystem with respect to its complement. We should know by now that for Gaussian states one can use the formula
\beq
\label{vnAPP}
S(\rho_A) = \sum_{i=1}^{n_A^>} \Big[ \Big(\sigma_i + \frac{1}{2}) \log_2 \Big( \sigma_I + \frac{1}{2} \Big) -  \Big(\sigma_i - \frac{1}{2}) \log_2 \Big( \sigma_I - \frac{1}{2} \Big)\Big] \, ,
\eeq
where $\{ \sigma_i \}_A$ is the collection of $n_A^>$ symplectic eigenvalues associated to the reduced covariance matrix $\Gamma_A$ of the subsystem $\rho_A$. About the notation (proudly invented by the author of this Thesis), in the following $n_X^{\ge} = n_X$ indicates the total number of symplectic eigenvalues $\ge \frac{1}{2}$ associated to a region $X$, $n_X^>$ indicates the number of symplectic eigenvalues $> \frac{1}{2}$, and $n_X^=$ denotes those $= \frac{1}{2}$. 

Since the normal-mode energies of the thermalized CV cluster state Hamiltonian we are considering are all equal, this is physically equivalent to squeezing identical thermal single-mode states (still defined by the parameter $\kappa =\coth{\beta/s^2}$) instead of vacuum states in the canonical construction procedure as explained in Chapter \ref{CVTO}. To detect the topological order of the resulting mixed CV surface code state, we use the TMI (\ref{tordofit}):
\beq
\label{tmiAPP}
\gamma_{\text{MI}} = -\frac{1}{2} (I_A + I_B + I_C - I_{AB} - I_{BC} - I_{AC} + I_{ABC}) \, \,
\eeq
with $I_X$ mutual information for a subsystem $X$.

The numerical simulations show that the TMI does not decrease significantly with an increment of the initial value of $\kappa$. This is interesting and rather unintuitive since we expected that higher temperature of the initial states would quickly kill the value of the TMI, hence an analytical expression is required to confirm our findings. 

First of all, recall that the covariance matrix $\Ga$ of the resulting mixed CV surface code state is equivalent to the ``pure'' one ($\Ga_0$) times $\kappa$, mathematically
\beq
\Ga = \kappa \Ga_0 \, .
\eeq
As a consequence, the symplectic eigenvalues of $\Ga$ (or any reduced section of it) are simply given by the pure symplectic eigenvalues multiplied by the overall $\kappa$ factor, $\{ \sigma_i \} = \{ \kappa \sigma_i^0 \}$. This simple transformation of $\Ga$ is a special case that only arises due to the fact that all normal modes of the CV cluster state Hamiltonian are identical, resulting in equal symplectic eigenvalues $\kappa / 2$.

\begin{figure}[tbp]  
\centering
\setlength{\unitlength}{1cm}
\includegraphics[width=6.5\unitlength]{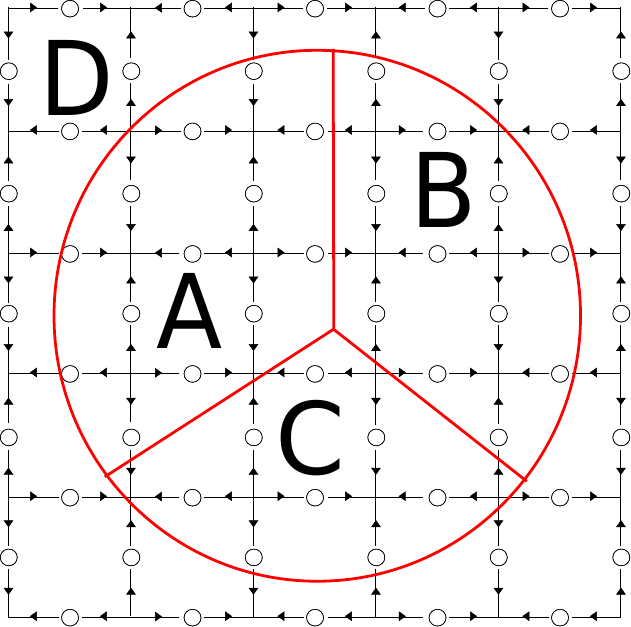}
\caption{Regions used for the TMI calculations. Note that they are the same as those used by Kitaev and Preskill \cite{kitaev2006} for the definition of the TEE, but since the state is mixed now also the external region D needs to be considered.}
\label{tmiRegions}
\end{figure}

For a large value of $\kappa$, we find the following asymptotic expression for each eigenvalue contribution to Eq.(\ref{vnAPP}):
\beq
\label{approxform}
\left(\kappa\sigma + \frac{1}{2}\right) \log_2\left(\kappa\sigma + \frac{1}{2}\right) - \left(\kappa\sigma - \frac{1}{2}\right)\log_2\left(\kappa\sigma - \frac{1}{2}\right) \nonumber \approx \log_2(e\kappa \sigma) \, .
\eeq
We can use this to show the behavior of the TMI as $\kappa \to \infty$. To start, consider the first term of the TMI formula, Eq.(\ref{tmiAPP}):
\beq
I_A = S_A + S_{BCD} - S_{ABCD} \, ,
\eeq
where the regions used are shown in Fig.(\ref{tmiRegions}).
If the total state $ABCD$ has $N$ modes, then, for region $A$, we have $n_A$ modes and for its complement $BCD$ we have $n_{BCD}$ modes, such that $n_A + n_{BCD} = N$. Hence, the von Neumann entropy for $A$ is given by
\beq
S_A =\sum_{i=1}^{n_A} \left[ \left(\kappa \sigma_i^A + \frac{1}{2}\right)\log_2\left(\kappa \sigma_i^A + \frac{1}{2}\right) \right. - \left. \left(\kappa \sigma_i^A - \frac{1}{2}\right)\log_2\left(\kappa \sigma_i^A - \frac{1}{2}\right)   \right] \, .
\eeq
In the limit of very high temperature, corresponding to $\kappa \to \infty$, we can use the approximation (\ref{approxform}) and rewrite the von Neumann entropy $S_A$ as
\beq
\label{MIA}
\lim_{\kappa \to \infty} S_A \approx S_A^l = \sum_{i=1}^{n_A} \log_2(e \kappa \sigma_i^A) = \sum_{i=1}^{n_A} \log_2(e \sigma_i^A) + n_A \log_2 \kappa\, .
\eeq
Divide now the $n_A$ symplectic eigenvalues into the two sets $n_A^>$ and $n_A^=$, such that $n_A^> + n_A^= n_A$. Hence, we can rewrite $S_A^l$ as:
\beq
S_A^l =  \sum_{i=1}^{n_A^>} \log_2(e \sigma_i^A) + \sum_{i=1}^{n_A^=} \log_2 \left(\frac{e}{2} \right)  + n_A \log_2 \kappa \,.
\eeq
The same argument can be repeated for the region $BCD$ and find that
\beq
S_{BCD}^l =  \sum_{i=1}^{n_{BCD}^>} \log_2(e \sigma_i^{BCD}) 
+ \sum_{i=1}^{n_{BCD}^=} \log_2 \left(\frac{e}{2} \right)  + n_{BCD} \log_2 \kappa \,.
\eeq
The $N$ symplectic eigenvalues of $ABCD$ are all equal to $\kappa/2$, consequently:
\beq
S_{ABCD}^l = \sum_{i=1}^N \log_2(\kappa \frac{e}{2}) = \sum_{i=1}^N \log_2\kappa + \sum_{i=1}^N \log_2 \frac{e}{2} = N \log_2\kappa + N \log_2 \left(\frac{e}{2} \right) \, ,
\eeq
and the value for the mutual information $I_A^l$ in the $\kappa \to \infty$ limit is given by
\begin{align}
I_A^l &= S_A^l + S_{BCD}^l - S_{ABCD}^l \nonumber \\
&= \sum_{i=1}^{n_A^>} \log_2(e \sigma_i^A)
+\sum_{i=1}^{n_{BCD}^>} \log_2(e \sigma_i^{BCD}) + (n_A^= + n_{BCD}^= - N) \log_2 \left(\frac{e}{2} \right) \, .
\end{align}
The miracle is that all the $\kappa$-contributions cancel out exactly and, asymptotically, $I_A^l$ is not a function of $\kappa$. Using $n_A^= + n_{BCD}^= N - n_A^> - n_{BCD}^>$ and the area law behavior for the entropy, i.e., for regions sufficiently big, $n_A^> = n_{BCD}^>$ (although this does not mean that the sets $\{ \sigma_A^i \}$ and $\{ \sigma_i^{BCD} \}$ are the same), we can rewrite the mutual information as
\beq
I_A^l = \sum_{i=1}^{n_A^>} \log_2(e \sigma_i^A) + \sum_{i=1}^{n_{BCD}^>} \log_2(e \sigma_i^{BCD}) - 2 n_A^>  \log_2 \left(\frac{e}{2} \right)\,.
\eeq
The same argument applies for each other contribution to the TMI, for example, for region $C$ we have:
\beq
I_C^l = \sum_{i=1}^{n_C^>} \log_2(e \sigma_i^C) + \sum_{i=1}^{n_{ABD}^>} \log_2(e \sigma_i^{ABD}) - 2 n_C^>  \log_2 \left(\frac{e}{2} \right)\,. 
\eeq
When substituting these expressions for the mutual information into the TMI formula in Eq.(\ref{tmiAPP}), all the elements $2n_X^> \log_2(\frac{e}{2})$, which depend on the size of the region boundary, sum to zero for the same reason why all the area-dependent elements cancel out in the KP argument \cite{kitaev2006} for the topological entanglement entropy. 

Consequently, the lower limit for the TMI is simply given by
\beq
\gamma_{MI}^{l} = -\frac{1}{2} \sum_{X} \zeta(X) \sum_{i=1}^{n_X^>} \log_2 (e \sigma_i^X) \, ,
\eeq
where $X$ runs over all the possible combinations of regions, and the function $\zeta(X)$ is defined as
\begin{align}
	\zeta(X) &= 
	\begin{cases}
		+1 & \text{if $X \in \{A,B,C,D,ABC,ABD,ACD,BCD\}$} \\
		-1 & \text{if $X \in \{AB,AC,AD,BC,BD,CD\}$} 
	\end{cases} \,,
	\nonumber \\
\end{align}
for a partitioning of the system as in Fig. \ref{tmiRegions}. This formula confirms our numerics and prove that that the value of the TMI, for the particular case of a CV surface code state, is not affected by $\kappa$ in the limit of thermal input states at high-temperature.

\begin{savequote}[10cm] 
\sffamily 
"It's the job that's never started as takes longest to finish." 
\qauthor{John R.R. Tolkien - The Lord of the Rings} 
\end{savequote}

\chapter{Expectation values in polymer quantization}
\label{expPolymer}

In this Appendix we show the calculations to derive the expectation values and the variances of the position and momentum-analog operators in polymer quantization \cite{ashtekar2003}. As pointed out in Chapter \ref{polymerico}, when we speak about expectation value of an observable for a state $( \Psi | \in \text{Cyl}^\star$, what we really mean is extracting a physical prediction using the shadow state $\ket{\Psi^{\text{shad}}}$ that realizes the physically equivalent state to $( \Psi|$ on $\mathcal{H}_{\text{poly}}$.\\
~\\
~\\
Before starting we list the definitions and results needed to perform the calculations. First of all, recall that the variance of an operator $\hat{A}$ for some generic state $\ket{\phi}$ with norm $|| \phi ||^2 = \bra{\phi} \phi \9$ is given by:
\beq
(\Delta \hat{A})^2 := \frac{\bra{\phi} \hat{A}^2 \ket{\phi}}{|| \phi ||^2} - \left( \frac{\bra{\phi} \hat{A} \ket{\phi}}{|| \phi ||^2} \right)^2 \, .
\eeq
In polymer quantization, physical states are defined as $( {\bf \Phi}_0 | \in \text{Cyl}^\star$ and for practical calculations we need to substitute the usual $\bra{\phi} \hat{A} \ket{\phi}$\footnote{\`A la Schr\"odinger.} with $(\phi | \hat{A} \ket{\phi^{\text{shad}}}$ \cite{demarie2013}. The space dual to $\text{Cyl}^\star$ is the space $\text{Cyl}$ of cylindrical functions with respect to some graph, in our case the real symmetric line with spacing $\mu$. An orthonormal basis on $\text{Cyl}$ is given by $\{ \ket{n \mu} \}$, which is the set of the eigenstates of the position operator $\hx$:
\beq
\hx \ket{n \mu} = n \mu \ket{n \mu}\, .
\eeq
Intuitively, the dual basis $( x |$ is defined by the product
\beq
( x | n \mu \9 = \delta_{x, n\mu} \, ,
\eeq
and the action of the translation operator $\hat{V}(\mu)$ on a generic state $\ket{n \mu}$ is given by
\beq
\hat{V}(\mu) \ket{n \mu} = \ket{n \mu - \mu}\, .
\eeq
Consider now a 1-dimensional harmonic oscillator in polymer quantization on the real symmetric line. The shadow ground state of the harmonic oscillator is given by
\beq
\ket{{\bf \Phi}_0^\s} = c \sum_{n \in \mathbb{Z}} e^{-\frac{n^2 \mu^2}{2 d^2}} \ket{n \mu} \, ,
\eeq
and, as shown in the main text, its norm is equal to
\beq
\6{\bf \Phi}_0^\s \ket{{\bf \Phi}_0^\s}= |c|^2 \sum_{n = -\infty}^\infty  e^{-\frac{n^2 \mu^2}{d^2}} \, .
\eeq
Then, the expectation value of the position operator $\hx$ is calculated as
\begin{align}
( {\bf \Phi}_0 | \hx \ket{{\bf \Phi}_0^\s} &=|c|^2 \sum_{n=-\infty}^{\infty} \sum_{m=-\infty}^{\infty} e^{-\frac{n^2 \mu^2}{2 d^2}} (m \mu) e^{-\frac{m^2 \mu^2}{2 d^2}} (n \mu | m \mu \9 \\
&= |c|^2 \sum_{n = -\infty}^{\infty} (n \mu) e^{-\frac{n^2 \mu^2}{d^2}} = 0\, ,
\end{align}
where we used the definition for the dual basis $( x_n|$ in $\text{Cyl}^\star$, $(x_n| x_m \9 = \delta_{n,m}$ and the obvious antisymmetry in the expression above, i.e. $( n \mu \, e^{- \frac{n^2 \mu^2}{d^2}})_{n} = - (n \mu \, e^{- \frac{n^2 \mu^2}{d^2}})_{-n}$. Interestingly enough, this is equivalent to the $\hx$ expectation value for the oscillator ground state in Schr\"odinger quantization. To calculate the variance we also need the expectation value of the fluctuations $\hx^2$, which can be derived exploiting the useful Poisson summation formula
\beq
\sum_{n = -\infty}^\infty f(n) = \sum_{n = -\infty}^{\infty} \int_{-\infty}^{\infty} f(y) e^{- 2 \pi i y n} \text{d}y \,.
\eeq
Then we have:
\begin{equation}
\begin{array}{ll}
( {\bf \Phi}_0 | \hx^2 \ket{{\bf \Phi}_0^\s} &= |c|^2 \sum_{n=-\infty}^{\infty} (n \mu)^2 e^{-\frac{n^2 \mu^2}{d^2}}\\
&=|c|^2 \sum_n \int_{-\infty}^{\infty} y^2 \mu^2 e^{-\frac{y^2 \mu^2}{d^2} - 2 \pi i y n} \text{d} y \\
&=|c|^2 \sum_n \mu^2 \int_{-\infty}^{\infty} y^2 e^{-\left( \frac{y \mu}{d} + i\frac{\pi n d}{\mu} \right)^2 - \frac{\pi^2 n^2 d^2}{\mu^2}} \text{d} y\\
&= |c|^2\sum_n e^{- \frac{\pi^2 n^2 d^2}{\mu^2}} \mu d \int_{-\infty}^{\infty} e^{-z^2} \left( \frac{d}{\mu}z - i \frac{\pi n d^2}{\mu^2} \right)^2 \text{d} z\\
&= |c|^2\sum_n e^{- \frac{\pi^2 n^2 d^2}{\mu^2}} \mu d \int_{-\infty}^{\infty} e^{-z^2} \left(\frac{d^2}{\mu^2}z^2 -\frac{2 \pi i n z d^3}{\mu^3} - \frac{\pi^2 n^2 d^4}{\mu^4} \right) \text{d}z\\
&= |c|^2\sum_n e^{- \frac{\pi^2 n^2 d^2}{\mu^2}} \mu d \left( \frac{d^2}{\mu^2} \frac{\sqrt{\pi}}{2} - \frac{\pi^2 n^2 d^4}{\mu^4}\sqrt{\pi} \right)\\
&= |c|^2  \frac{d^3}{\mu} \frac{\sqrt{\pi}}{2} \sum_n e^{- \frac{\pi^2 n^2 d^2}{\mu^2}} \left( 1 - \frac{2 \pi^2 n^2 d^2}{\mu^2} \right) \, ,
\end{array}
\end{equation}
where we used the substitution
\beq
z = y \frac{\mu}{d} + i \frac{\pi n d}{\mu} \,,
\eeq
and the Gaussian integral formulas \cite{spiegel1990}:
\beq
\int_{-\infty}^\infty e^{-z^2} = \sqrt{\pi} \, , \qquad \int_{-\infty}^\infty z e^{-z^2} = 0 \, , \qquad \int_{-\infty}^\infty z^2 e^{-z^2} = \frac{\sqrt{\pi}}{2} \, .
\eeq
After having obtained the expectation value for $\hx^2$, we can calculate the fluctuations in $\hx$:
\begin{align}
(\Delta x)^2 = \frac{( {\bf \Phi}_0 | \hat{x}^2 \ket{{\bf \Phi}_0^\s}}{|| {\bf \Phi}_0^\s ||^2} - \left( \frac{( {\bf \Phi}_0 |  \hat{x} \ket{{\bf \Phi}_0^\s}}{|| {\bf \Phi}_0^\s ||^2} \right)^2 \, ,
\end{align}
keeping in mind that $\6 \hx \9 = 0$; for the first terms, i.e. $n= 0, n = \pm 1$, we have:
\begin{align}
\begin{array}{ll}
(\Delta x)^2 &\approx \left[ |c|^2 \frac{d^3 \sqrt{\pi}}{2 \mu} \left( 1 + 2 e^{-\frac{\pi^2 d^2}{\mu^2}} \left( 1 - \frac{2 \pi^2 d^2}{\mu^2} \right) \right) \right] \left[ |c|^2 \sqrt{\pi} \frac{d}{\mu} \left( 1 + 2e^{-\frac{\pi^2 d^2}{\mu^2}} \right) \right]^{-1}\\
&= \frac{d^2}{2} \left[ 1 + 2e^{-\frac{\pi^2 d^2}{\mu^2}} - \frac{4 \pi^2 d^2}{\mu^2} e^{-\frac{\pi^2 d^2}{\mu^2}} \right] \left( 1 + 2 e^{-\frac{\pi^2 d^2}{\mu^2}} \right)^{-1}\\
&\approx \frac{d^2}{2} \left(1 - \frac{4 \pi^2 d^2}{\mu^2} e^{-\frac{\pi^2 d^2}{\mu^2}} \right) \,, 
\end{array}
\end{align}
equivalent to the expectation value in the Schr\"odinger quantization $(\Delta x)^2 = \frac{d^2}{2}$ plus a correction induced by the discretization of the real line.\\
In the main text we defined the effective momentum operator by the finite difference
\beq
\hp_{\mu} \equiv - \frac{i}{2 \mu} \left(  \hat{V}(\mu) - \hat{V}(\mu)^\dagger \right) \, ,
\eeq
and to obtain the expectation value $\6 \hp_\mu \9$ one first calculates
\begin{align}
( {\bf \Phi}_0 | \hat{V}(\mu) \ket{{\bf \Phi}_0^\s}\, ,
\end{align}
which is equal to
\begin{align}
( {\bf \Phi}_0 | \hat{V}(\mu) \ket{{\bf \Phi}_0^\s} = |c|^2 \sum_n e^{-\frac{n^2 \mu^2}{2 d^2}} e^{-\frac{(n-1)^2 \mu^2}{2 d^2}} = |c|^2 e^{-\frac{\mu^2}{2 d^2}} \sum_n e^{-\frac{n^2 \mu^2}{d^2} + \frac{n \mu^2}{d^2}}\, .
\end{align}
Thanks to the always useful Poisson re-summation formula and integral tricks as shown before in this Appendix, we can rewrite the equation above in a more suitable form:
\begin{equation}
\begin{array}{ll}
( {\bf \Phi}_0 | \hat{V}(\mu) \ket{{\bf \Phi}_0^\s} &= |c|^2 e^{-\frac{\mu^2}{2 d^2}} \sum_{n = -\infty}^{\infty} \int_{-\infty}^\infty e^{- 2 \pi i y n}e^{-\frac{y^2 \mu^2}{d^2} + \frac{y \mu^2}{d^2}} \text{d} y \\
&= |c|^2 e^{-\frac{\mu^2}{2 d^2}} \sum_n \int_{-\infty}^\infty e^{-\left[ \frac{y \mu}{d} +\left( \frac{\pi i n d}{\mu} - \frac{\mu}{2 d} \right) \right]^2} e^{\left( \frac{\pi i n d}{\mu} - \frac{\mu}{2 d} \right)^2} \text{d} y\\
&= |c|^2 e^{-\frac{\mu^2}{2 d^2}} \frac{d}{\mu} \sum_n e^{\left( \frac{\pi i n d}{\mu} - \frac{\mu}{2 d} \right)^2}  \int_{-\infty}^\infty e^{-\left[ z +\left( \frac{\pi i n d}{\mu} - \frac{\mu}{2 d} \right) \right]^2} \text{d} z\\
&= |c|^2 e^{-\frac{\mu^2}{4 d^2}} \frac{d \sqrt{\pi}}{\mu} \sum_n e^{-\frac{\pi^2 n^2 d^2}{\mu^2}-i \pi n} \\
&=  |c|^2 e^{-\frac{\mu^2}{4 d^2}} \frac{d \sqrt{\pi}}{\mu} \sum_n e^{-\frac{\pi^2 n^2 d^2}{\mu^2}} \left( \cos{\pi n} - i \sin{\pi n} \right) \,.
\end{array}
\end{equation}
After keeping only the first few terms thanks to the ratio $\frac{d}{\mu} \gg 1$, we have
\beq
( {\bf \Phi}_0 | \hat{V}(\mu) \ket{{\bf \Phi}_0^\s} = |c|^2   e^{-\frac{\mu^2}{4 d^2}} \frac{d \sqrt{\pi}}{\mu} \left( 1 - 2 e^{-\frac{\pi^2 d^2}{\mu^2}} \right) \, .
\eeq
To normalize this value, we divide by the norm $||\mathbf{\Phi}_0^\text{shad}||^2$ of the state:
\begin{align}
\begin{array}{ll}
\label{expectV}
\6 \hat{V}(\mu) \9 &= \frac{|c|^2 \frac{d \sqrt{\pi}}{\mu}e^{-\frac{\mu^2}{4 d^2}}  \left( 1 - 2 e^{-\frac{\pi^2 d^2}{\mu^2}} \right)}{|c|^2 \frac{d \sqrt{\pi}}{\mu} \left( 1 + 2 e^{-\frac{\pi^2 d^2}{\mu^2}} \right)} \\
&\approx e^{-\frac{\mu^2}{4 d^2}} \left( 1 - 4 e^{-\frac{\pi^2 d^2}{\mu^2}} \right) \,.
\end{array}
\end{align}
One can notice at first sight that the expectation value of the operator $\hat{V}(\mu)^\dagger =  \hat{V}(-\mu) $ is equal to 
\beq
\6 \hat{V}(\mu)^\dagger \9 = \6 \hat{V}(\mu) \9\,,
\eeq
since in the last expression, the $\mu$ parameter only appears squared and a sign flip is irrelevant. Therefore, the expectation value of the polymer momentum operator vanishes similarly to what happens in Schr\"odinger quantization:
\beq
\6 \hp_\mu \9 = -\frac{i}{2 \mu} \left( \6 \hat{V}(\mu) \9 - \6 \hat{V}(\mu)^\dagger \9 \right) = 0\, .
\eeq
In the main text we defined in two different ways the kinetic operator. Canonically, it is represented by the square of the momentum operator, which, in polymer quantization, corresponds to
\beq 
\hat{p}_\mu^2 \equiv (\hat{p}_\mu)^2 = \frac{1}{4 \mu^2} \left( 2 - \hat{V}(2 \mu) - \hat{V}(-2\mu) \right) \,.
\eeq
It is easy to calculate the fluctuations of $\hat{p}_\mu^2$ by substituting $\mu$ with $\pm 2 \mu$ in Eq.(\ref{expectV}):
\begin{align}
(\Delta p_\mu)^2 = \frac{( {\bf \Phi}_0 |  \hat{p}_\mu^2 \ket{{\bf \Phi}_0^\s}}{|| {\bf \Phi}_0^\s ||^2} - \left( \frac{( {\bf \Phi}_0 |  \hat{p}_\mu \ket{{\bf \Phi}_0^\s}}{|| {\bf \Phi}_0^\s ||^2} \right)^2 \, ,
\end{align}
where
\begin{align}
\begin{array}{ll}
\6 \hat{p}^2_\mu \9 &= \frac{1}{4 \mu^2} \left( 2 - \6 \hat{V}(2 \mu) \9 - \6 \hat{V}(-2 \mu) \9 \right)\\
&\approx \frac{1}{4 \mu^2} \left( 2 - 2 e^{-\frac{4 \mu^2}{4 d^2}} \right)\\
&\approx \frac{1}{2 \mu^2} \left( 1 - 1 + \frac{\mu^2}{d^2} - \frac{\mu^4}{2 d^4} \right)\\
&= \frac{1}{2 d^2} \left( 1 - \frac{\mu^2}{2 d^2} \right) \,.
\end{array}
\end{align}
Therefore one immediately derives that:
\beq
(\Delta p_\mu)^2 \approx \frac{1}{2 d^2} \left( 1 - \frac{\mu^2}{2 d^2} \right) \,.
\eeq
For simplicity, a renewed version of the kinetic operator was introduced, in order to explain how to avoid some conceptual problems deriving from the use of the $\hat{p}^2_\mu$ operator. This is given by
\beq
\hat{K}_\mu \equiv \widehat{p^2}_\mu \equiv \frac{1}{\mu^2} \left( 2 - \hat{V}(\mu) - \hat{V}(-\mu) \right) \, ,
\eeq
whose fluctuations are calculated in an identical fashion:
\beq
( \Delta K_\mu)^2 \approx \frac{1}{4 d^2} \left( 1 - \frac{\mu^2}{2 d^2} \right) \,.
\eeq
Things become a bit more complicated when trying to calculate the expectation values for the position-momentum correlations $\6 \hx \hp_\mu \9$ and $\6 \hp_\mu \hx \9$. Surprisingly, the polymer approximation does not affect the total value and one obtains that the equality $\6 \hx \hp_\mu \9 + \6 \hp_\mu \hx \9 = 0$ holds exactly, in accordance with Schr\"odinger quantization predictions. Let us show the explicit calculations for this result. For our purposes, it is sufficient to calculate the following quantity:
\beq
( {\bf \Phi}_0 |  \hx \hp_\mu + \hp_\mu \hx \ket{{\bf \Phi}_0^\s} \, ,
\eeq
which reduces to calculating
\beq
\label{correv}
( {\bf \Phi}_0 |  \hx \hat{V}(\mu)^\dagger - \hx \hat{V}(\mu) + \hat{V}(\mu)^\dagger \hx - \hat{V}(\mu) \hx  \ket{{\bf \Phi}_0^\s} \, .
\eeq
Let us consider the first half of the line above first. It is equal to:
\begin{align}
\begin{array}{ll}
( {\bf \Phi}_0 |  &\hx \hat{V}(\mu)^\dagger - \hx \hat{V}(\mu) \ket{{\bf \Phi}_0^\s} = |c|^2 \sum_n n \mu e^{-\frac{n^2 \mu^2}{2d^2}} \left( e^{-\frac{(n+1)^2 \mu^2}{2d^2}} - e^{-\frac{(n-1)^2 \mu^2}{2d^2}} \right)\\
&= |c|^2 \sum_n n \mu e^{-\frac{\mu^2 (1 + 2n + 2n^2)}{2 d^2}} \left(1 - e^{\frac{2 n \mu^2}{d^2}}\right)\\
&= |c|^2 \mu \sum_n \int_{-\infty}^{\infty} y e^{-\frac{\mu^2 (1 + 2y + 2y^2)}{2d^2}} \left( 1 - e^{\frac{2 y \mu^2}{d^2}} \right) e^{- 2 \pi i y n} \text{d} y\\
&= - |c|^2 \mu  \sum_n \left( \frac{d \sqrt{\pi} e^{-\frac{\mu^2}{4d^2} + i n \pi - \frac{d^2 n^2 \pi^2}{\mu^2} } (\mu^2 + 2 i d^2 n \pi)}{2 \mu^3} + \frac{d \sqrt{\pi} e^{-\frac{\mu^2}{4d^2} - i n \pi - \frac{d^2 n^2 \pi^2}{\mu^2} } (\mu^2 - 2 i d^2 n \pi)}{2 \mu^3} \right) \\
&= - |c|^2 \mu  \sum_n \frac{(-1)^n d \sqrt{\pi} e^{-\frac{\mu^4 + 4d^4 n^2 \pi^2}{4 d^2 \mu^2}}}{\mu} \, .
\end{array}
\end{align}
Proceeding in the same way, we can solve the second part of Eq.(\ref{correv}):
\begin{align}
\begin{array}{ll}
( {\bf \Phi}_0 |  & \hat{V}(\mu)^\dagger \hx - \hat{V}(\mu) \hx \ket{{\bf \Phi}_0^\s} = |c|^2 \sum_n \mu e^{-\frac{n^2 \mu^2}{2d^2}} \left( (n+1) e^{-\frac{(n+1)^2 \mu^2}{2d^2}} - (n-1) e^{-\frac{(n-1)^2 \mu^2}{2d^2}} \right)\\
&= |c|^2 \sum_n \mu e^{-\frac{\mu^2 (1 + 2n + 2n^2)}{2 d^2}} \left(1 - e^{\frac{2 n \mu^2}{d^2}} (n-1) + n \right)\\
&= |c|^2 \mu \sum_n \int_{-\infty}^{\infty} e^{-\frac{\mu^2 (1 + 2y + 2y^2)}{2d^2}} \left( 1 - e^{\frac{2 y \mu^2}{d^2}}(y-1) + y \right) e^{- 2 \pi i y n} \text{d} y\\
&= |c|^2 \mu  \sum_n \frac{(-1)^n d \sqrt{\pi} e^{-\frac{\mu^4 + 4d^4 n^2 \pi^2}{4 d^2 \mu^2}}}{\mu} \, ,
\end{array}
\end{align}
and we immediately notice that
\beq
\6 \hx \hp_\mu + \hp_\mu \hx \9 = 0\, .
\eeq
 
\backmatter

\chapter{List of Symbols}



Here we present a list of symbols used throughout the thesis. The tables are labeled according to the chapters they refer to. Whenever a symbol is used throughout the thesis with the same meaning, it is only mentioned once in the chapter where it has been introduced for the first time. 

As a general rule, matrices and vectors are depicted in \emph{bold}, for example ${\bf A}$. On the other hand, four-dimensional vectors are written in sans serif, i.e. $\ak$. Operators have a caret on top, $\hat{H}$ (with the exception of density matrices, i.e. $\rho$, which always correspond to greek letters). Bold vectors with a caret identify unit-length vectors, for instance $\hbk$. Note that some symbols are repeated in different chapters with a different meaning. Refer to the tables below for the precise distinctions.

\begin{table}[ht]
\centering
Chapter \ref{CHintro2} \\
\begin{tabular}{|c |c|}
\hline
Symbol & Definition \\ [0.5ex]
\hline
$\mathcal{H}$ & Hilbert space \\
$\ket{\psi}$ & pure state \\
${\bf \hat{H}}$ & Hamiltonian matrix \\
$ {\bf \hat{A}}$ & operator \\
$ \hat{\sigma}^X , \hat{\sigma}^Y, \hat{\sigma}^Z$ & 1-qubit Pauli operators \\
$\rho, \sigma$ & density matrices\\
$H(X)$ & Shannon entropy of a variable $X$\\
$S_\alpha(X)$ & R\'enyi-$\alpha$ entropy of $X$\\
$S(X||Y)$ & relative entropy between $X$ and $Y$\\
$I(X,Y)$ & mutual information of $X$ and $Y$\\
$S(\rho)$ & quantum entropy of $\rho$\\
$S(\rho||\sigma)$ & quantum relative entropy of $\rho$ and $\sigma$\\
$S_\alpha(\rho)$ & quantum R\'enyi-$\alpha$ entropy of $\rho$\\
$I(\rho,\sigma)$ & quantum mutual information between $\rho$ and $\sigma$\\
$E(\rho_{A,B})$ & entanglement entropy of $\rho_A$ or $\rho_B$\\
$\mathcal{N}(\rho,A)$ & negativity of $\rho_A$\\
$E_{\mathcal{N}}(\rho,A)$ & logarithmic negativity of $\rho_A$\\
$D(B|A)$ & quantum discord between $\rho_a$ and $\rho_B$\\
$\hat{K}$ & stabilizer operator\\
$\hat{\eta}$ & nullifier operator\\
\hline
\end{tabular}
\end{table}

\begin{table}[ht]
\centering
Chapter \ref{GRchapter}\\
\begin{tabular}{|c |c|}
\hline
Symbol & Definition \\ [0.5ex]
\hline
$\nabla_{{\sf x}}$ & Covariant derivative in direction of {\sf x} \\
$\ma $ & 4-dimensional space-time manifold\\
$\Sigma_3$ & 3-dimensional space manifold \\
$\lambda$ & affine parameter of the trajectory \\
$\chi$ & polarization phase \\
\ak & Photon 4-momentum vector\\
\af & Photon 4-polarization vector\\
$\ab_1$ & Photon 4-momentum polarization basis \\
$\ae_\mu$ & $\mu$-basis vector of the local orthonormal tetrad\\
$\hbk$ & Photon 3-momentum vector \\
$\hbf$ & Photon 3-polarization vector \\
$\hbb_i$ & $i$-th photon 3-polarization basis vector \\
$\hat{\bf e}_i$ & $i$-th basis vector of the local orthonormal triad\\
$\bm{\Omega}$ & Angular velocity of rotation of the momentum polarization triad \\
$\bm{\omega}$ & gravitomagnetic field term \\
$\eta_i$ & $i$-th basis 1-form \\
$\beta_{j,l}^{i}$ & Ricci rotation coefficient \\
$\beta$ & matrix of connection 1-forms on $\Sigma_3$ \\
$\theta$ & matrix of curvature 2-forms on $\Sigma_3$ \\
$\Omega$ & matrix of connection 1-forms on $\ma$ \\
$\Theta$ & matrix of curvature 2-forms on $\ma$ \\
\hline
\end{tabular}
\end{table}

\begin{table}[ht]
\centering
Chapter \ref{ChapGS} \\
\begin{tabular}{|c |c|}
\hline
Symbol & Definition \\ [0.5ex]
\hline
$\hat{a}_i, \hat{a}^\dagger_i$ & $i$-th bosonic field operators \\
$\mathbf{\Omega}$ & symplectic form\\
$\hat{q}_i, \hat{p}_i$ & Hermitian quadrature operators\\
$\hat{W}_\eta$ & Weyl operator\\
$\chi_\rho(\eta)$ & characteristic function of $\rho$\\
$W(q,p)$ & Wigner distribution\\
$\bar{\mathbf{r}}$ & vector of first moments\\
$\Ga$ & covariance matrix\\
$\mathbf{Y}$ & symplectic transformation\\
$\Ga_w$ & Williamson's normal form\\
$\sigma_i$ & $i$-th symplectic eigenvalue\\
$m$ & mass of a quantum oscillator\\
$\omega$ & frequency of a quantum oscillator\\
$Z$ & partition function\\
$\beta$ & thermal parameter\\
$\bar{n}$ & mean occupation number\\
$\lambda$ & coupling parameter\\
$G$ & graph \\
$\mathcal{E}, \mathcal{V}, \mathcal{F}$ & edge, vertex and face set of a graph G\\
${\bf A}(G)$ & adjacency matrix of a graph $G$\\
$\mathcal{N}_v$ & neighborhood of a vertex v\\
${\bf Z}$ & matrix representation of a pure Gaussian state\\
${\bf V}, {\bf U}$ & matrices composing ${\bf Z} = {\bf V} + i {\bf U}$\\
$\hat{S}(s)$ & squeezing operator\\
$\log s$ & squeezing parameter\\
\hline
\end{tabular}
\end{table}

\begin{table}[ht]
\centering
Chapter \ref{ChapTO} \\
\begin{tabular}{|c |c|}
\hline
Symbol & Definition \\ [0.5ex]
\hline
$\Delta E$ & energy gap of the Hamiltonian \\
$\xi$ & correlation length \\
$| \partial A |$ & boundary of a region $A$\\
$\hat{C}_{\text{P}_{(i,j)}}$ & qubit c-PHASE gate between two sites $\6 i,j\9$\\
$\mathcal{G}$ & group with order $|\mathcal{G}|$\\
$D(\mathcal{G})$ & quantum double model of a group $\mathcal{G}$\\
$g$ & element of the group $\mathcal{G}$\\
$\hat{A}(v)$ & vertex stabilizer operator\\
$\hat{B}(f)$ & face stabilizer operator \\
$ \mathfrak{D}$ & algebra of a group $ \mathcal{G}$ \\
$\mathcal{GS}$ & ground state manifold of a quantum double model\\
$\mathfrak{g}$ & genus of a surface \\
$\mathcal{D}$ & total quantum dimension of a topological model\\
$\gamma$ & topological entanglement entropy\\
$\gamma_{\text{MI}}$ & topological mutual information \\
\hline
\end{tabular}
\end{table}

\begin{table}[ht]
\centering
Chapter \ref{CVTO} \\
\begin{tabular}{|c |c|}
\hline
Symbol & Definition \\ [0.5ex]
\hline
$\hX(t), \hZ(u)$ & translation and boost operator \\
$\ket{q}_q$ & position eigenstate \\
$\ket{p}_p$ & momentum eigenstate \\
$\mathcal{C}(\mathcal{P}_N)$ & Clifford group for continuous-variable\\
$\hat{\text{C}}_{\hat{Z}_{(i,j)}}$ & controlled-Z gate between sites $i$ and $j$ \\
$\cs$ & continuous-variable cluster-state \\
$\hat{a}_v$ & vertex nullifier \\
$\hat{b}_f$ & face nullifier \\
$\hH^{\text{ideal}}_{\text{CS}}$ & Hamiltonian of the ideal CV cluster-state\\
$\hH^{\text{ideal}}_{\text{CS}}$ & Hamiltonian of the ideal CV surface code state\\
$\hH_{\text{CS}}(s)$ & Hamiltonian of the physical CV cluster-state  \\
$\hH_{\text{SC}}(s)$ & Hamiltonian of the physical CV surface code state \\
$\hat{c}, \hat{d}$ & normal mode operators  \\
$\hat{X}_x$ & lattice shift operator in direction $x$ \\
${\bf Z}_{\text{CS}}(s)$ & ${\bf Z}$ matrix for the CV cluster-state\\
${\bf Z}_{\text{SC}}(s)$ & ${\bf Z}$ matrix for the CV surface code state\\
$\Ga_{\text{CS}}(s)$ & covariance matrix of the CV cluster-state \\
$\Ga_{\text{SC}}(s)$ & covariance matrix of the CV surface code state \\
$\epsilon_0$ & energy gap of the physical CV cluster-state \\
$\kappa$ & thermal parameter of the CV cluster-state \\
\hline
\end{tabular}
\end{table}

\begin{table}[ht]
\centering
Chapter \ref{polymerico} \\
\begin{tabular}{|c |c|}
\hline
Symbol & Definition \\ [0.5ex]
\hline
$\mathfrak{A}$ & real or complex algebra \\
$A,B$ & elements of an algebra\\
$\mathcal{R}(\mathfrak{A})$ & representation $\mathcal{R}$ of the algebra $\mathfrak{A}$ \\
$\varphi$ & states on a algebra \\
$\hW(q,k)$ & Weyl operators \\
$q,k$ & translation and boost parameters \\
$\hV(q), \hU(k)$ & translation and boost operators \\
$\hx$ & position operator\\
$\mathcal{H}_{\text{poly}}$ & Polymer Hilbert space\\
$\ket{x_i}$ & position basis states on $\mathcal{H}_{\text{poly}}$\\
$\gamma$ & 1-dimensional graph on the real line\\
$\mu$ & fixed spacing of the graph $\gamma$ \\
$\text{Cyl}_\gamma$ & space of cylindrical functions on a graph $\gamma$ \\
$\text{Cyl}^*$ & algebraic dual of Cyl \\
$( \varUpsilon |$ & dual elements belonging to $\text{Cyl}^*$\\
$ \hp_\mu$ & effective momentum operator \\
$\hat{K}_\mu$ & adjusted kinetic operator \\
$ \ket{\Phi_\gamma^\s}$ & shadow state of $ ( \Phi |$\\
$F$ & free energy \\
$\rho^\mu$ & polymer density matrix\\
$d$ & scale of the quantum harmonic oscillator \\
\hline
\end{tabular}
\end{table}


\end{document}